\documentclass[preprint,authoryear,11pt]{elsarticle}

\usepackage{placeins}
\usepackage{eurosym}
\usepackage{amsmath}
\usepackage{amsthm}
\usepackage{amssymb}
\usepackage{mathptmx}
\usepackage{bm}
\usepackage{graphicx}
\usepackage{threeparttable}
\usepackage{longtable}

\newcommand{\be}{\begin{equation}}
\newcommand{\ee}{\end{equation}}
\newcommand{\beq}{\begin{eqnarray}}
\newcommand{\eeq}{\end{eqnarray}}

\newcommand{\vx}{\mbox{\bf {x}}}

\newcommand{\vk}{\mbox{\bf {k}}}

\newcommand{\Planck}{{\it Planck}}
\newcommand{\C}{{\cal C}}
\def\Mpl{M_{\rm pl}}

\newcommand\arcdeg{\mbox{$^\circ$}}%
\newcommand\arcsec{\mbox{$^{\prime\prime}$}}%
\newcommand\sun{\odot}%
\newcommand\diameter{\ooalign{\hfil/\hfil\crcr\mathhexbox20D}}%
\newcommand\sq{\mbox{\rlap{$\sqcap$}$\sqcup$}}%
\newcommand\degr{\arcdeg}%

\newcommand\aj{\ref@jnl{AJ}}%
\newcommand\araa{\ref@jnl{ARA\&A}}%
\newcommand\apj{\ref@jnl{ApJ}}%
\newcommand\apjl{\ref@jnl{ApJ}}%
\newcommand\apjs{\ref@jnl{ApJS}}%
\newcommand\ao{\ref@jnl{Appl.~Opt.}}%
\newcommand\apss{\ref@jnl{Ap\&SS}}%
\newcommand\aap{\ref@jnl{A\&A}}%
\newcommand\aapr{\ref@jnl{A\&A~Rev.}}%
\newcommand\aaps{\ref@jnl{A\&AS}}%
\newcommand\azh{\ref@jnl{AZh}}%
\newcommand\baas{\ref@jnl{BAAS}}%
\newcommand\jrasc{\ref@jnl{JRASC}}%
\newcommand\memras{\ref@jnl{MmRAS}}%
\newcommand\mnras{\ref@jnl{MNRAS}}%
\newcommand\pra{\ref@jnl{Phys.~Rev.~A}}%
\newcommand\prb{\ref@jnl{Phys.~Rev.~B}}%
\newcommand\prc{\ref@jnl{Phys.~Rev.~C}}%
\newcommand\prd{\ref@jnl{Phys.~Rev.~D}}%
\newcommand\pre{\ref@jnl{Phys.~Rev.~E}}%
\newcommand \prl{\ref@jnl{Phys.~Rev.~Lett.}}%
\newcommand\pasp{\ref@jnl{PASP}}%
\newcommand\pasj{\ref@jnl{PASJ}}%
\newcommand\qjras{\ref@jnl{QJRAS}}%
\newcommand\skytel{\ref@jnl{S\&T}}%

\newcommand\solphys{\ref@jnl{Sol.~Phys.}}%
\newcommand\sovast{\ref@jnl{Soviet~Ast.}}%
\newcommand\ssr{\ref@jnl{Space~Sci.~Rev.}}%
\newcommand\zap{\ref@jnl{ZAp}}%
\newcommand\nat{\ref@jnl{Nature}}%
\newcommand\iaucirc{\ref@jnl{IAU~Circ.}}%
\newcommand\aplett{\ref@jnl{Astrophys.~Lett.}}%
\newcommand\apspr{\ref@jnl{Astrophys.~Space~Phys.~Res.}}%
\newcommand\bain{\ref@jnl{Bull.~Astron.~Inst.~Netherlands}}%
\newcommand\fcp{\ref@jnl{Fund.~Cosmic~Phys.}}%
\newcommand\gca{\ref@jnl{Geochim.~Cosmochim.~Acta}}%
\newcommand\grl{\ref@jnl{Geophys.~Res.~Lett.}}%
\newcommand\jcp{\ref@jnl{J.~Chem.~Phys.}}%
\newcommand\jgr{\ref@jnl{J.~Geophys.~Res.}}%
\newcommand\jqsrt{\ref@jnl{J.~Quant.~Spec.~Radiat.~Transf.}}%
\newcommand\memsai{\ref@jnl{Mem.~Soc.~Astron.~Italiana}}%
\newcommand\nphysa{\ref@jnl{Nucl.~Phys.~A}}%
\newcommand\physrep{\ref@jnl{Phys.~Rep.}}%
\newcommand\physscr{\ref@jnl{Phys.~Scr}}%
\newcommand\planss{\ref@jnl{Planet.~Space~Sci.}}%
\newcommand\procspie{\ref@jnl{Proc.~SPIE}}%

\newcommand{\lya}{Ly$\alpha$}
\newcommand\ha{H$\alpha$}
\newcommand\hb{H$\beta$}
\newcommand\oiii{\hbox{[O{\scriptsize$\rm III$}]}}
\newcommand\oii{\hbox{[O{\scriptsize$\rm II$}]}}
\newcommand\nii{\hbox{[N{\scriptsize$\rm II$}]}}
\newcommand\sii{\hbox{[S{\scriptsize$\rm II$}]}}
\newcommand\oi{\hbox{[O{\scriptsize$\rm I$}]}}
\newcommand\hi{\hbox{H{\scriptsize$\rm I$}}}

\def\eps@scaling{1.0}%
\newcommand\epsscale[1]{\gdef\eps@scaling{#1}}%
\newcommand\plotone[1]{%
 \centering
 \leavevmode
 \includegraphics[width={\eps@scaling\columnwidth}]{#1}%
}%

\newcommand\plottwo[2]{%
 \centering
 \leavevmode
 \columnwidth=.45\columnwidth
 \includegraphics[width={\eps@scaling\columnwidth}]{#1}%
 \hfil
 \includegraphics[width={\eps@scaling\columnwidth}]{#2}%
}%
\newcommand\plotfiddle[7]{%
 \centering
 \leavevmode
 \vbox\@to#2{\rule{\z@}{#2}}%
 \includegraphics[%
  scale=#4,
  angle=#3,
  origin=c
 ]{#1}%
}%

\def\plotone#1{\centering \leavevmode       
\epsfxsize=\columnwidth \epsfbox{#1}}

\def\refe@jnl#1{{#1}}
\def\aj{\refe@jnl{Astron.~J.}}                  
\def\araa{\refe@jnl{Annu.~Rev.~Astron.~Astrophys.}}
\def\apj{\refe@jnl{Astrophys.~J.}}                 
\def\apjl{\refe@jnl{Astrophys.~J.~Lett.}}          
\def\apjs{\refe@jnl{Astrophys.~J.~S.~S.}}          
\def\aap{\refe@jnl{Astron.~Astrophys.}}            
\def\mnras{\refe@jnl{Mon.~Not.~R.~Astron.~Soc.}}   
\def\prd{\refe@jnl{Phys.~Rev.~D}}        
\def\fcp{\refe@jnl{Fund.~Cos.~Phys.}}  
\def\physrep{\refe@jnl{Phys.~Rep.}}
\def\physlett{\refe@jnl{Phys.~Lett.}}
\def\nat{\refe@jnl{Nature}}                  
\def\jcap{\refe@jnl{JCAP}}                  

\DeclareTextSymbol{\degre}{OT1}{23}

\begin{document}
\begin{frontmatter}

\title{J-PAS: The Javalambre-Physics of the Accelerated Universe Astrophysical Survey}
\tnotetext[$^*$]{This documents describes J-PAS and outlines its main scientific goals}

\author[1,2]{N.~Ben\'{\i}tez}
\author[2,3,4]{R.~Dupke}
\author[5,1]{M.~Moles}
\author[6]{L.~Sodr\'e}
\author[5]{A.~J.~Cenarro}
\author[5]{A.~Mar\'\i n-Franch}
\author[2]{K.~Taylor}
\author[5]{D.~Crist\'obal}
\author[12]{A.~Fern\'andez-Soto}
\author[6]{C.~Mendes de Oliveira}
% ...

\address[1]{Instituto de Astrof\'{\i}sica de Andaluc\'{\i}a-CSIC, Granada, Spain}
\address[2]{National Observatory, Rio de Janeiro, Brazil}
\address[3]{The University of Michigan, Ann Arbor, MI, USA}
\address[4]{The University of Alabama, Tuscaloosa,AL, USA}
\address[5]{Centro de Estudios de F\'\i sica del Cosmos, Teruel,  Spain} 
\address[6]{Instituto de Astronomia, Geof\i sica e Ci\^encias  Atmosf\'ericas, Universidade of Sa\~o Paulo, Brazil} 
% ...
\address[12]{Instituto de Fisica de Cantabria, Spain} 
% ...

\begin{abstract}
   The Javalambre-Physics of the Accelerated Universe Astrophysical
 Survey (J-PAS) is a narrow band, very wide field Cosmological Survey to be 
carried out from the Javalambre Observatory in Spain with a 
purpose-built, dedicated 2.5m telescope and a $4.7\sq\degr$ camera with 1.2Gpix. 
Starting in 2015, J-PAS will observe $8500\sq\degr$ of Northern Sky 
and measure $0.003(1+z)$ precision photometric redshifts for $9\times10^7$ LRG and ELG galaxies 
plus several million QSOs, about $50$ times more than the largest current 
spectroscopic survey, sampling an effective volume of $\sim 14$ Gpc$^3$ up to $z=1.3$. 
J-PAS will be the first radial BAO experiment to reach Stage IV. 

 J-PAS will also detect and measure the mass of $7\times 10^5$ galaxy clusters and groups, setting 
constrains on Dark Energy which rival those obtained from BAO measurements. 
Thanks to the superb characteristics of the Javalambre site 
(seeing $\sim 0.7\arcsec$), J-PAS is expected to obtain a deep, sub-arcsec 
image of the northern sky, which combined with its unique photo-z precision 
will produce one of the most powerful cosmological lensing 
surveys before the arrival of Euclid. In addition, J-PAS unprecedented 
spectral time domain information will enable a self-contained SN survey 
that, without the need for external spectroscopic follow-up, will detect, 
classify and measure $\sigma_z\sim 0.5\%$ redshifts for $\sim 4000$ SNeIa 
and $\sim 900$ core-collapse SNe.   

  The key to the J-PAS potential is its innovative approach: the combination 
of $54$ $145\AA$ filters, placed $100\AA$ apart, and a multi-degree
field of view (FOV) is a 
powerful ``redshift machine'', with the survey speed of a 4000 multiplexing 
low resolution spectrograph, but many times cheaper and much faster to build. 
Moreover, since the J-PAS camera is equivalent to a very large,
$4.7\sq\degr$ ``IFU'', 
it will produce a time-resolved, 3D image of the Northern Sky with a very wide 
range of Astrophysical applications in Galaxy Evolution, the nearby Universe and the 
study of resolved stellar populations. J-PAS will have a lasting legacy value in 
many areas of Astrophysics, serving as a fundamental dataset for future 
Cosmological projects. 

\end{abstract}


\end{frontmatter}

\vfill\eject 

\tableofcontents 

\vfill\eject

\section{Introduction}

  The last decade has seen an accumulation of very large field Astrophysical 
Surveys (area $>5000 \sq\degr$). A key factor in this development has been the undoubted 
success of the Sloan Digital Sky Survey which has spawned significant advances in almost 
all the fields in Astrophysics. The quest for the origin of Dark Energy has also been a powerful 
motivator, fostering many of the current projects 
like Pan-STARRS \citep{2002SPIE.4836..154K}, DES \citep{des}, 
and BOSS \citep{BOSS}, and also being one of the main goals of the very
large extragalactic surveys planned to start around the beginning of
the next  decade  LSST \citep{lsst}, Euclid \citep{euclid} and DESI
\citep{levi2013}. 

 All these surveys are based on two traditional, one century-old, 
astronomical methods: broad-band imaging ($R \sim 6$)\footnote{For
  imaging, we define the wavelength resolution as
  $R_\lambda=\lambda/\Delta_\lambda$, where $\Delta_\lambda$ is the
  filter width. Another alternative definition would be 
$R_z=(1+z)/\delta_z$, the inverse of the redshift error. This is
usually much higher than $R$ for photometric redshifts, for instance, 
for broadband imaging $R_z\sim 25$, for J-PAS-like Narrow Band(NB) imaging $R_z\sim 333$} 
 supplemented by moderate resolution spectroscopy ($R \sim 500$).  

  Optical broad-band imaging with traditional astronomical filter systems is observationally 
efficient but yields very limited redshift information, with, typically, $dz/(1+z) \gtrsim 3\%$, see e.g. 
\citet{2010A&A...523A..31H} and references therein. Spectroscopy for cosmological purposes provides higher resolution, 
$dz/(1+z)\sim 0.0005-0.001$ but to be competitive requires very high
object multiplexing $\gtrsim 1000$, making state-of-the-art spectrographs extremely expensive 
and very complex to develop. In addition, the information provided by low-resolution, cosmologically-oriented spectroscopy is relatively limited for other purposes, since the spectra are usually low $S/N$ and for efficiency reasons only objects of 
direct cosmological interest are systematically targeted. 

 Several projects like COMBO-17 \citep{wolf08}, ALHAMBRA \citep{molino13} and COSMOS \linebreak \citep{ilbert09} 
have carried out medium band imaging over a few square degrees,
hinting at the potential of this approach. These surveys, with $\sim
300\AA$ medium band filters reach precisions of $dz/(1+z)\approx
0.8\%$ and of $0.6\%$ for the highest  quality photo-z. This is
already not far from the $0.35\%$ precision required for radial BAO measurements.  

  \citet{Benitez2009b} showed that medium band ($R \sim 20$) and
  narrow band ($R \sim 60$) filter systems 
are much more effective, in terms of {\it photometric redshift} depth, than what a naive extrapolation from pure {\it photometric} 
depth would imply. However systematic, multiple-narrow band wide field imaging has not been 
attempted so far. One objective reason is that until quite recently, it was not possible to build homogeneous 
filters with a large enough scale. But perhaps the main objection is that NB imaging is quite inefficient for 
individual objects, since it requires repeated observations to cover a
large spectral range; if prompted to consider a NB cosmological survey
many would dismiss the idea out of hand \citep{tversky}. However, as explained below, when 
NB imaging is combined with a large enough FoV, the result is a
redshift machine  more 
powerful than any existing spectrograph.   

  Moreover, with a system of contiguous $\sim 100$\AA-width filters it is possible to 
reach $\approx0.3\%$ redshift precisions for enough LRGs to competitively measure the radial Baryonic Acoustic 
Oscillation (BAO) scale at $z<1$ \citep{B2009}.  J-PAS will observe
with an improved version of that system, with the goal of maximizing the effective 
volume over which we can measure the BAO scale using not only LRGs
($z<1.1$), but also blue galaxies ($z<1.35$) and QSOs ($z<3$), 
while presenting several features which make the data much more
powerful for a wide range of Cosmological and Astrophysical goals.

\subsection{{\it Quasi}-Spectroscopy: Wide field Narrow Band Imaging as a Redshift Machine}

  To understand the power of the J-PAS approach, it is instructive to look at the ``raw'' relative efficiencies of imaging and 
spectroscopy when observing an object's SED within a particular wavelength range. Let's assume that we have a source with a spectral 
energy distribution flux $F_\lambda$, observed against a background $B_\lambda$. The imaging system is defined by an average throughput $\eta_I$, 
a filter width $\Delta\lambda$ and a number of filters $n_f$. The
spectrograph is defined by a throughput $\eta_S$. In both cases, we consider a detector 
with a spatial pixel scale $p_s$ and readout noise $\sigma_{ron}$. The full total exposure time is $t$, divided into $n_r$ individual read outs, and the covered wavelength range is $L=\lambda_{max}-\lambda_{min}$. The central wavelength is thus $\bar\lambda = (\lambda_{max}+\lambda_{min})/2$. 
 
  For the spectrograph, we have that the signal-to-noise ($S/N$) reached for a fixed  time $t$, defined as $q_S=(S/N)_S$ is 
\begin{equation}
q_S=\frac{\bar{F}L\eta_St}{\sqrt{\bar{B}L\eta_St+A_Sn_r\sigma_{ron}^2}}
\end{equation}
where $\bar{F}$ and $\bar{B}$ are, respectively, the average object
flux and background flux in the spectral range $L$ and $A_S$ is the number of pixels covered by the object spectrum on the CCD. For a spectral resolution $R=\lambda/\delta_\lambda$, the pixel scale 
in the wavelength direction will be $p_\lambda\sim  \bar{\lambda}/(2R)$, where we assume at least two pixels per resolution element. 
If we use a slit of spatial scale $D$, the total number of pixels covered by the spectrum will be $A_S=(L/p_\lambda)\times(D/p_s) \approx 2RD/p_s$ 
for the typical wavelength range considered here ($L\approx\bar{\lambda}$). 

 For the imaging case, we are using $n_F$ different filters, and we don't cover the whole spectrum in a single shot, only a section $\Delta_\lambda$. Therefore the exposure time for each wavelength segment 
will be smaller, $t_{exp}=t/n_f$, and $q_I=(S/N)_I$ will be equal to: 

\begin{equation}
q_I=\frac{\bar{F}\Delta_\lambda\eta_It}{\sqrt{\bar{B}\Delta_\lambda\eta_It+A_In_r\sigma_{ron}^2}}
\end{equation}

 Here we assume that $\bar{F}\eta_I \approx \sum(F_i\eta_i)/n_f$ and $\bar{B}\eta_I \approx \sum (B_i\eta_i)/n_f$, where the $i$ values correspond to the individual filters. The total number of pixels covered by the $n_f$ apertures of diameter $D$ is $A_I=n_f\pi(\frac{D}{2p_s})^2$. 

  Let's examine the limit case in which both spectroscopy and imaging are totally background dominated. Then 
\begin{equation}
q_{SB}\approx \frac{\bar{F}}{\sqrt{\bar{B}}}\sqrt{L\eta_St}
\end{equation}
and 
\begin{equation}
q_{IB}\approx \frac{\bar{F}}{\sqrt{\bar{B}}}\sqrt{\Delta_\lambda\eta_It}
\end{equation}
Therefore, the relatively $S/N$ ratio of spectroscopy vs. imaging for the background-dominated observation of a single object is 
\begin{equation}
\frac{q_{SB}}{q_{IB}}\approx \sqrt{\frac{L\eta_S}{\Delta_\lambda\eta_I}} 
\end{equation}

  If we take as fiducial values $L=9100-3600=5500\AA$,
  $\Delta_\lambda=145\AA$, $\eta_I=0.7$, $\eta_S=0.25$, we have
  $\frac{q_{SB}}{q_{IB}}\approx 3.7$\footnote{To make a comparison which 
  focuses on the effects of the filter width, we have not taken into 
  account fiber aperture effects which are not present in imaging, and which 
in practice would reduce the ratio by a factor of $\sim 2$ in favor of 
NB imaging}. Thus, as most astronomers will
  say intuitively, spectroscopy is significantly more efficient than
  NB imaging (as described here) for a single object observation,
  since it requires $13.5$ times longer to reach the same $S/N$.

  The inclusion of readout noise barely changes this result. Let's define, for the imaging case, the ratio $r_I$ 
between the total readout and background noise 
\begin{equation}
r^2_I=\frac{A_In_r\sigma^2_{ron}}{\bar{B}\Delta_\lambda\eta_It}
\end{equation}

Then 
\begin{equation}
q_{I}\approx \frac{\bar{F}}{\sqrt{\bar{B}}}\sqrt{\frac{\Delta_\lambda\eta_It}{1+r^2_I}}
\end{equation}

  For the spectroscopic case, since we are assuming the same number of readouts per wavelength segment $n_r$ and the same readout noise $\sigma_{ron}$, 
we can write 
\begin{equation}
r^2_S=\frac{A_Sn_r\sigma^2_{ron}}{\bar{B}L\eta_St}=\frac{A_s\Delta_\lambda\eta_I}{A_IL\eta_S}r^2_I
\end{equation}

And therefore 

\begin{equation}
q_S=  \frac{\bar{F}}{\sqrt{\bar{B}}}\sqrt{\frac{L\eta_St}{1+\frac{A_S\Delta_\lambda\eta_I}{A_IL\eta_s}r^2_I}}
\end{equation}

  For our fiducial values, the ratio $(L\eta_S)/(\Delta_\lambda\eta_I)\approx 13.5$. The ratio $A_S/A_I=(8Rp_s)/(n_f\pi D)\approx (R_S/100)$, 
where we have assumed $D=2\arcsec$, $p_s=0.4\arcsec$ and $n_f=54$ 

 Therefore we have 
\begin{equation}
\frac{q_S}{q_I}=\sqrt{\frac{L\eta_S}{\Delta_\lambda\eta_I}}\sqrt{\frac{1+r^2_I}{1+0.07(R_S/100)r^2_I}} 
\end{equation}

 In the J-PAS case, assuming $\sigma_{ron}\approx 6e^-$ (as we have done, to be conservative, for all the calculations and mocks presented throughout the paper, although the goal for the camera is $4e^-$), we get $r_I\approx 0.5$. Thus,  the inclusion of readout noise for realistic cases does not change things significantly, since the $r_I$-containing factor on the right goes from $1.10$ to $0.85$ for resolutions $R_s=250-4000$. 

  However, the relevant quantity to decide which approach is better as a redshift machine is not the efficiency for an individual object, but the survey speed $v$, which is defined as the total number of objects which can be observed per unit time, with the same signal-to-noise $q$. 
\begin{equation}
v=\frac{N}{t_q}
\end{equation}

And thus
\begin{equation}
\frac{v_I}{v_S} \propto \frac{N_It_S}{N_St_I} 
\end{equation}
where $t_I$ and $t_I$ are, respectively, the time required for a spectrograph and the imaging system to reach $S/N=q$. 
Disregarding the readout noise factors, we have $t_I/t_S=(L\eta_S)/(\Delta_\lambda\eta_I)$, therefore 
\begin{equation}
\frac{v_I}{v_S} \propto \frac{N_I\Delta_\lambda\eta_I}{N_SL\eta_S}
\end{equation}

  The most efficient spectrographs, like BOSS \citep{BOSS}, have $N_S \sim 1000$. In the case of NB imaging, the effective ``multiplexing'' can be extremely large depending on the camera FOV and the density of objects of interest. J-PAS can estimate highly precise photo-z, valid to measure line-of-sight BAOs  for $N_I=52,000$ galaxies in $4.7\sq\degr$.  

 Thus, for $<=0.3\%$ photo-z, an instrument like the one which will be used by J-PAS 
is about 4 times faster, in terms of survey speed, than a 1000-x spectrograph, and it is comparable with a 4000-x spectrograph. 
\begin{equation}
\frac{v_I}{v_S} \approx 4\left(\frac{n_g}{11000 gals /\sq\degr}\right)\left(\frac{FOV}{4.7 \sq\degr}\right)\left(\frac{1000}{N_s}\right) 
\end{equation}
where $n_g$ is the galaxy density per square degree and $FOV$ is the Field of view of the camera in square degrees.

We have endeavored to compare both imaging and spectroscopic
techniques in a numerically balanced way, 
however, there are several "hidden variables" favoring imaging
techniques which are more 
difficult to quantify but are nevertheless important to be aware of. 
 We list some of these here:

1. Selection effects in spectroscopy:  The necessity of object
pre-selection for multi-object spectrographs 
introduces many unintended biases to a spectroscopic survey.  
These include effects due to morphology, magnitude and surface density
limits all of which have effects on completeness and window function
uncertainties 
which are largely eliminated in an imaging survey.

2. Astrometry:  Astrometric errors which can be induced by systematic
epoch effects and proper motion uncertainties may lead to small
position errors which critically impact aperture coupling efficiencies.
Uniform magnitude selection:  Spectroscopic optimization of detector
real estate requires selection of objects of uniform brightness which
can introduce further selection biases.

3. Sky subtraction:  Object multiplex has always to be traded with
sampling of the sky both for multi-fibre and multi-slit spectroscopy.  
Sky subtraction for imaging, if done carefully, is generally a much
more robust technique which minimally impacts both stochastic and
systematic errors.

4. Acceptance aperture:  Spectroscopic apertures are always limited by
the design of the spectrograph and cannot generally be optimized for
particular atmospheric conditions.  The size of the instrument scales
linearly with aperture; the cost is a much steeper function.  There is
thus always a strong driver for limiting the size of the aperture
which is especially demanding for large telescope spectrographs.  
For point sources this simply means reduced aperture coupling ratios 
which are determined by variable seeing conditions, while for 
marginally resolved sources this leads to incomplete sampling 
of the intrinsic morphology. 

6. Atmospheric dispersion and differential atmospheric refraction:  
These effects can strongly perturb spectroscopic efficiency as a
function of wavelength thus producing errors in the observed source
SEDs which are very difficult to impossible to calibrate out.

7. Fibre effects:  Spectrograph efficiencies are a function of fibre
properties such as throughput, focal ratio degradation and
non-telecentric feeds.  The first two effects can be a function of
fibre placement and telescope orientation (and consequently time) and
are very difficult to quantify.  Non-telecentric feeds can be well
determined but inevitably lead to reductions in efficiency.  Many
large telescope fibre systems also require multi-fibre connectors
which again impact efficiency; it is indeed a fact that very few
instrumentation papers actually quote the overall fibre spectrograph
efficiency as measured on the sky.

  Of course there are negatives for multi-band imaging surveys the 
most prominent of which is the fact that different wavelength samples 
are taken sequentially which introduce variable efficiencies and PSFs.  
However, this information can be recovered from the data itself with 
carefully monitoring of the photometric conditions and by tying the 
photometry to an all-sky photometric calibration survey, supplied in 
the case of J-PAS by the T80 telescope.  Another limitation is in the 
effective spectral resolution given by wave-band limitations.  However
a concerted effort has been made to optimize the band-pass of the J-PAS 
survey through exhaustive S/N modeling; it is incidental but fortunate 
that this optimization has led to band-passes which are obtainable
through a standard interference filter fabrication processes.

 But perhaps the main advantage of NB imaging lies in the relative
 simplicity and low cost (about an order of magnitude lower) of the required instrumentation, 
specially when compared with a spectrograph with several thousand multiplexing.    

\section{J-PAS survey description}

\label{survey}
\subsection{The filter system}

 As it was shown by \citet{B2009}, a contiguous set of filters spaced by $\sim 100\AA$ width is able to produce photometric redshifts with a precision of $0.003(1+z)$ for Luminous Red Galaxies. The original and main scientific goal of PAU-Consolider project, the origin of J-PAS, was measuring 
the radial scale of the Baryonic Acoustic Oscillations using these LRG population. However, the photo-z for Blue Galaxies are 
also precise enough to use them as BAO probes, significantly increasing the effective volume of the survey. QSOs can 
also be detected and their redshifts measured with excellent precision \citep{2012MNRAS.423.3251A}. The J-PAS filter system and 
observing strategy has also been carefully optimized to maximize the returns of the survey in other areas of Astrophysics without compromising 
the main goal of constraining the Dark Energy equation of state. 
 
 These are the main modifications we have introduced with respect to the filter system of \citet{B2009}: 

1. The extension of the NB observations blue-wards to $3785$\AA: this will enable a detailed and unique study of the galaxy properties in the local Universe, and the central wavelengths of the filters (and therefore of the whole system) have been chosen accordingly. It also improves slightly the photo-z precision of the higher redshift galaxy population. 

2. Extension of the NB filter range to $9100$\AA. This increases significantly the effective volume covered by the Survey by extending the redshift range at which we are able to measure photo-z for both LRG and ELGs. Going much further to the red becomes inefficient because of the fast increase with wavelength 
of the sky background. 
 
\begin{figure}[H]
\centering
{\includegraphics[width=0.8\textwidth,keepaspectratio]{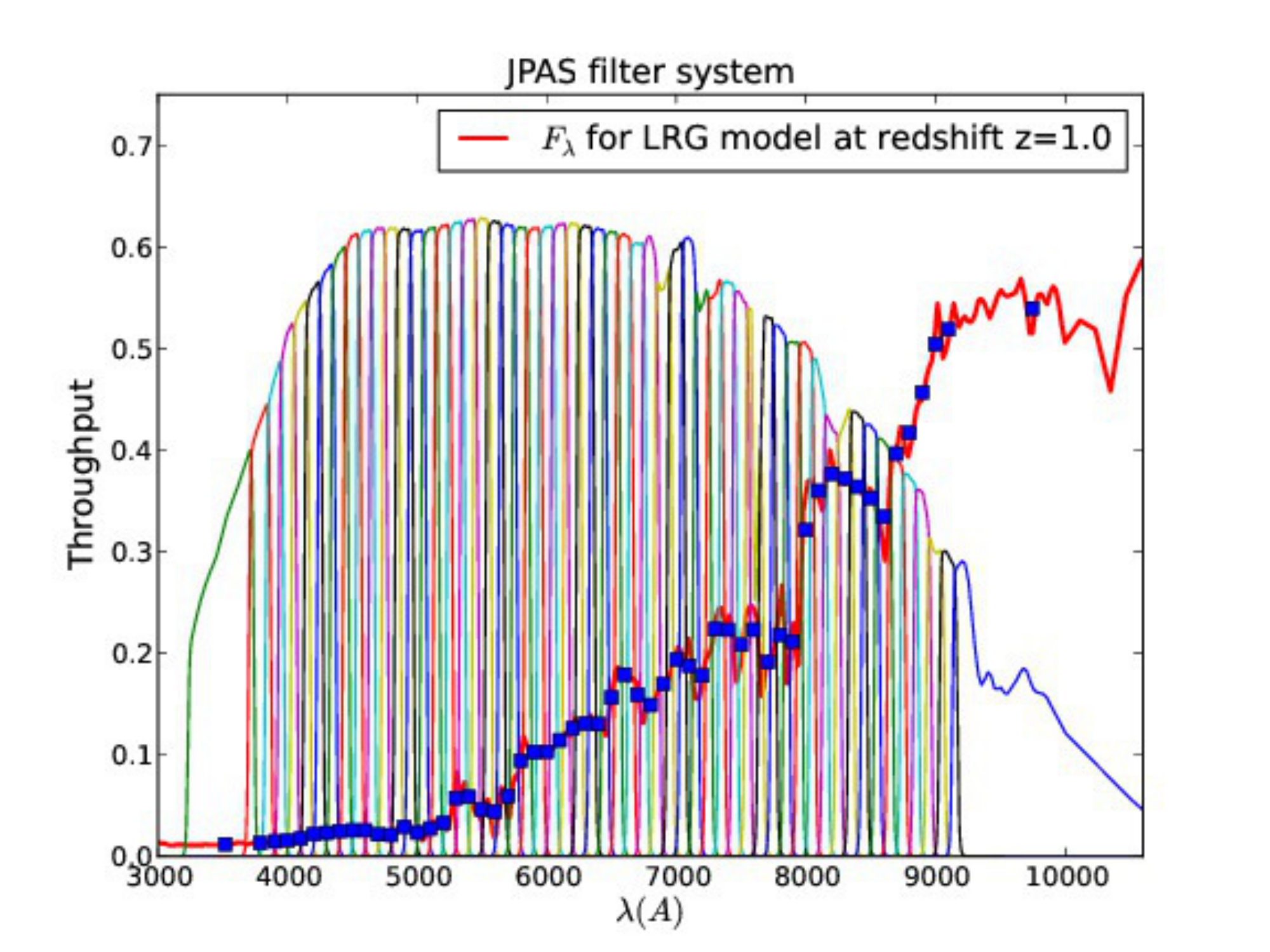}}
\caption{The J-PAS filter system. We have included the redshifted spectrum of an early type
galaxy at z=1.0 from Polleta et al. 2007. The filters are spaced by about 100~\AA\ but have FWHM of $145~\AA$, what produces a significant overlap 
among them. The blue squares represent the flux which would be observed through the filters. Note that many spectral features apart from the 4000~\AA\ break are resolved, that is why the precision in redshift is much larger than that which would be produced by a single break, $\Delta z/(1+z) \sim \Delta\lambda/\lambda \sim 0.02$ }
\label{fig:filters}
\end{figure}

3.To avoid duplications, our total number of filters has to be a multiple of 14, the number of CCDs in our camera. Putting together all the above considerations we arrive to 56 main filters, 54 of them NB, 1 medium-band (which covers the UV edge) and 1 broad band (which covers the interval red-wards of $9100$\AA), with the NB filters spaced by $100$~\AA. Our simulations (Ben\'\i tez et al., in preparation) show that photo-z precision and depth are more sensitive to the filter spacing than to the filter width (provided it is narrow enough). The width of our filters is set to $145$\AA, the minimum width required by the manufacturer, to ensures filter homogeneity across our field of view. The resulting filter system is shown in Figure \ref{fig:filters}. 

  In addition, we include three regular broad band filters in our observations, $u,g$ and $r$. The first filter has a redder cutoff
 that the SDSS $u$ band, the other two are similar to the ones used in that survey. These BB observations will be quite deep compared to the NB 
imaging ($5\sigma$ limiting magnitude in a $3\arcsec$ aperture of $23.1$, $23.7$ and $\approx 24$ respectively). The $r-$band plays a special role, because it will be the main filter used for detection and weak lensing measurements. The BB filters are all contained in a single tray, and will be used only when the image quality is in the top $10\%$ of the observatory range. Given the superb seeing at the Javalambre site, and the exquisite care being 
taken to make sure that neither dome nor camera significantly degrade it, we expect to get a deep $<0.8\arcsec$ imaging of $\sim 8500\sq\degr$ of the Northern Sky which will be extremely useful for lensing analyses.

\subsection{Observing strategy}

\subsubsection{Exposure time and filter distribution}

  Despite the use of drift scan by previous surveys, as SDSS, 
we have decided to use a traditional ``point and shoot''  
mosaicking strategy. We found that drift scan is only marginally 
more efficient (once all the factors, as overlaps, etc. are taken into account) and 
therefore does not warrant the extremely strict ---and therefore 
risky---technical requirements it imposes on both camera and 
telescope design. The simpler imaging strategy we have adopted 
gives us more flexibility in the observing strategy and makes possible to re-use 
the reduction software of other surveys like ALHAMBRA, which was mostly 
developed by members of the J-PAS collaboration \citep{molino13,cristobal09}. 

 We are using a $14$ CCD, $4.7\sq\degr$ camera, and 56+3 filters. It is not
 currently possible to build a NB filter of the required width and
 homogeneity over the whole $4.7\sq\degr$; even covering all the
 CCDs simultaneously with copies of the same individual filter, would require us to 
 purchase 826 filters, surpassing the full cost of the camera. 
We therefore employ a single copy of the main 56
 filter system, spreading them into 4 different trays (T1-4), each
 with 14 different filters. An additional broad band filter tray (T5)
 contains 8 $r-$band filters, 3 $g-$band filters and 3 $u-$ band
 filters. Within the NB trays, T1-4, the filters are
 distributed as contiguously as possible in wavelength, although some
 scrambling among trays is needed to minimize the presence of image ghosts due to 
 reflections. Tables 3.,4. and 5. 
list the filters, indicating which trays they belong to. 
  Since the filters will be distributed parallel to each other, 
 is quite straightforward to cover the whole observing area homogeneously 
 with all of them. It is sufficient to follow a pattern which tiles the 
full J-PAS sky with the CCD having the smallest effective area: that ensures that 
 all the filters in each tray will also tile the sky with no gaps. 

   Our basic exposure is 60 seconds, and we will carry out at least 4 exposures in each filter, following a $2+1+1$ 
pattern whenever is possible, i.e. taking initially 2 almost simultaneous exposures , and then leaving an interval of a month before 
the third  and approximately 20 days between the 3rd and 4th exposures. This is close to the optimum strategy for SNeI detection,  
giving J-PAS the opportunity of carrying out one of the most powerful ground-based SN surveys. 
For the filters in the 4th, and reddest tray, we will repeat the 4x60 observations. The $u,g$ and $r$ filters, 
which will be included in a 5th tray will be exposed by, respectively $225,225$ and $600$ seconds, using $6,6$ and $16$ exposures. 

 The total effective exposure time on any point of the sky will therefore be $1050s$ (the BB tray) + $56 \times 240s$ 
(one pass with all the 56 filters in T1-4) + $14\times 240$ (the 2nd pass with T4) = $4.96h$. Since the J-PAS main camera has $4.7\sq\degr$, 
the survey speed is, therefore, $\sim 1\sq\degr h^-1$, and it will require $\sim 9000$ h on target to complete the full survey.  

 Of course, the total observing time will be  higher when we include predictable overheads: 

\begin{itemize}

\item {\it Readout time} We will take a total of $56\times 4+14\times 4+ 6+6+16=308$ exposures, which for a estimated readout time of $11s$, 
are equivalent to $3388$s or $0.94$h. 

\item {\it Overlap} Our observations will overlap about $5\%$, to ensure an homogeneous photometric calibration across different pointings. 

\item {\it Visibility} Due to the inconvenient (from the point of view of Extragalactic Astronomy) 
presence of most of the Milky Way in the Northern Hemisphere, we will need an additional $10\%$ of survey time, which will 
be used to repeatedly observe some areas in the sky (creating a J-PAS deep field in an area almost coincident with Stripe 82). 

\end{itemize}

  Therefore, combining all these effects, we would actually need a factor of $1.19\times 1.05\times 1.10 = 1.38$ more on-target time,
 or about $12400$h of observations. 

 Taking into account the experience of similar observatories we conservatively expect to be able to observe on-target effectively for at least $1800$ h/yr (this is equivalent to $\sim 48\%$  useful time, similar to the values at e.g. Calar Alto). Therefore, the full completion of J-PAS will require $6.88$ yrs. The expected  initial date for the survey is mid-2015, so the full survey will finish in 2021, around the start of other Stage IV projects like Euclid, LSST or DESI. As we will see below, the observing strategy of J-PAS is designed in such a way that it will obtain highly competitive cosmological constraints 
from its 2nd-3rd year of operation. 

\subsubsection{Limiting magnitudes}

  To generate mock observations we have written a python Exposure Time Calculator which is included in the BPZ 2.0 python distribution (Ben\'\i tez 2014) 
As usual, it generates the expected $S/N$ given a certain observational set-up and object magnitude. It takes as inputs the expected full filter transmission curve (assumed to take into account the CCD, optics and the atmosphere at the desired airmass), the pixel size, mirror area, sky background, total exposure and the number of readouts and the readout noise. Given an object area and magnitude, it calculates all the relevant parameters as the sky and readout noise within the aperture, $S/N$, etc. Included in the calculation is a further degradation of the theoretical results by factor of ~$0.1-0.25$, as measured empirically when working with real instruments (as the one used in the ING ETC http://catserver.ing.iac.es/signal/). Table~\ref{ETC} lists the global parameters used to generate the mock observations. Our NB imaging will be binned by a $2\times2$ factor into effective $(0.456\arcsec)^2$ pixels and 
we assume a readout noise of $6e^{-}$ for our mocks, despite the fact that the purported goal for JPcam is $4e^{-}$. 

\begin{table}
\centering 

\caption{ETC global parameters used for the synthetic observations}
\label{ETC}
\begin{tabular}{cc}
\hline
Mirror area & $3.89$m$^2$ \\
Readout noise & $6e^{-}$ \\
Min. Number of exposures & 4 \\ 
Binned pixel size & $0.456^2\sq\arcsec$ \\
Aperture area & $7.07\sq\arcsec$ \\
\hline
\end{tabular}
\end{table}

  A crucial factor in the ETC is a realistic estimation of the sky background. The sky at the OAJ site at the Pico del Buitre was directly measured in 2009, during the last solar cycle minimum, and it is extremely dark \citep{Moles2010}, clearly comparable (at similar altitudes) with other superb sites as 
Mauna Kea or La Silla. However since the former measurements only sample a small part of the solar cycle, we estimate the sky brightness for our mock observations using the Mauna Kea 2500m results of \citet{Krisciunas}, which cover a full solar cycle and are comparable to the OAJ measurements at solar minimum. We take the average of the 1992-1996 years, which approximately correspond to the same part of the solar cycle as the future 2015-2021 J-PAS observations; this is quite conservative because there is evidence that the current solar cycle will be much milder that the previous one (http://solarscience.msfc.nasa.gov/SunspotCycle.shtml). Since Krisciunas et al. (1997) only provide  measurements for the B and V bands, we calculate the expected colors in other bands by using the average dark sky color of all the observatories listed in Patat et al. (2003). In addition we brightened the sky by 0.12 mags to take into account that we assume that the average airmass will be 1.2. 

\begin{figure}[h]
\centering 
{\includegraphics[width=0.8\textwidth,keepaspectratio]{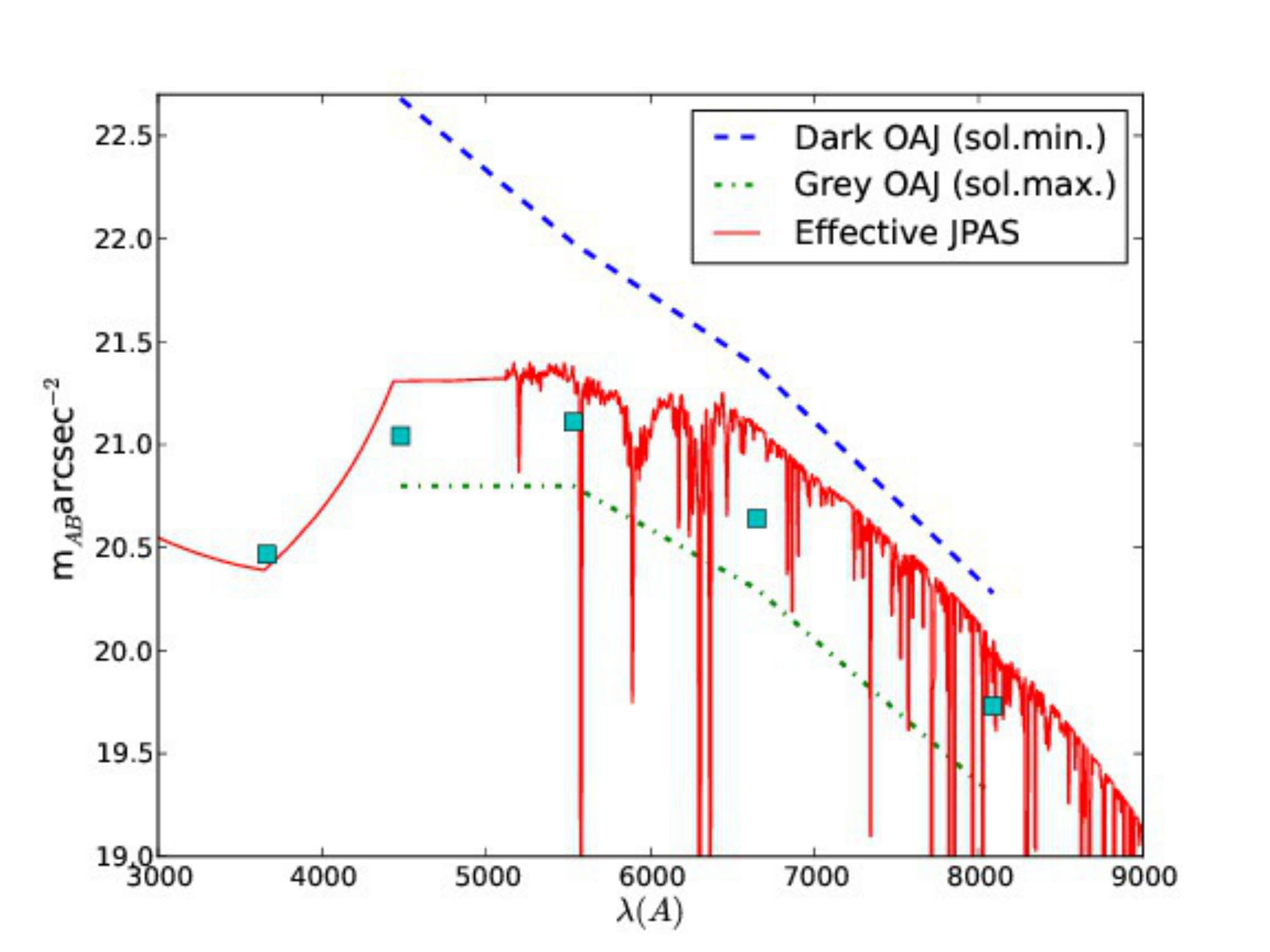}}
\caption{Sky background used to generate our mocks, calculated averaging over the solar (2015.5-2021) and moon cycles (26 darkest nights) for a 1.2 airmass. For reference we include the airmass-corrected measurement of the sky background at the OAJ (dashed line) and the sky that our model would predict at 7 nights from the dark moon at the solar maximum (dash-dot line)} 
\label{fig:sky}
\end{figure}

 To account for the moon phase, we use \citet{walker} and calculate the proper average for each band within the 26 darkest nights in the moon cycle (the brightest two nights will be devoted to other projects). The resulting average sky brightness per broad band filter is listed in Table \ref{sky}. 
We then use the sky spectrum of Puxley et al. (http://www.gemini.edu) and normalize it to the required value in each band. The normalized sky spectrum is plotted in Figure \ref{fig:sky}, together with the published measurements of Javalambre dark sky at the zenith. Note that this differs, in the sense of being much more conservative, from the sky spectrum in \citet{B2009}, where it was assumed that the timing of each filter observation would be adapted to the moon phase in an optimal way. 

\begin{table}
\centering 
\caption{Assumed sky background per $\sq\arcsec$ for the mocks presented here, calculated by averaging over the solar (2015.5-2021) and moon cycles (26 darkest nights) and assuming an airmass of 1.2. For reference we have included the zenith measurements of the OAJ dark sky background at the solar minimum \citep{Moles2010}, without any correction for airmass. The magnitudes are Vega-based.}
\label{sky}
\begin{tabular}{ccc}
\hline  
Band & $m_{mock}$ & $m_{dark}$\\ 
\hline  
U & 20.82 & \\ 
B & 21.43 & 22.8 \\
V & 21.14 & 22.1 \\
R & 20.47 & 21.5 \\
I & 19.24 & 20.4 \\
\hline
\end{tabular}
\end{table}

 We have tested our ETC with the observations of Taniguchi et al. (2007), which are described in enough detail to simulate accurately and provide empirical $S/N$ measurements. We find the agreement excellent, with an average offset of $0.04$ and a scatter of $0.2$.  

\begin{figure}
\centering
\includegraphics[width=0.8\textwidth,keepaspectratio]{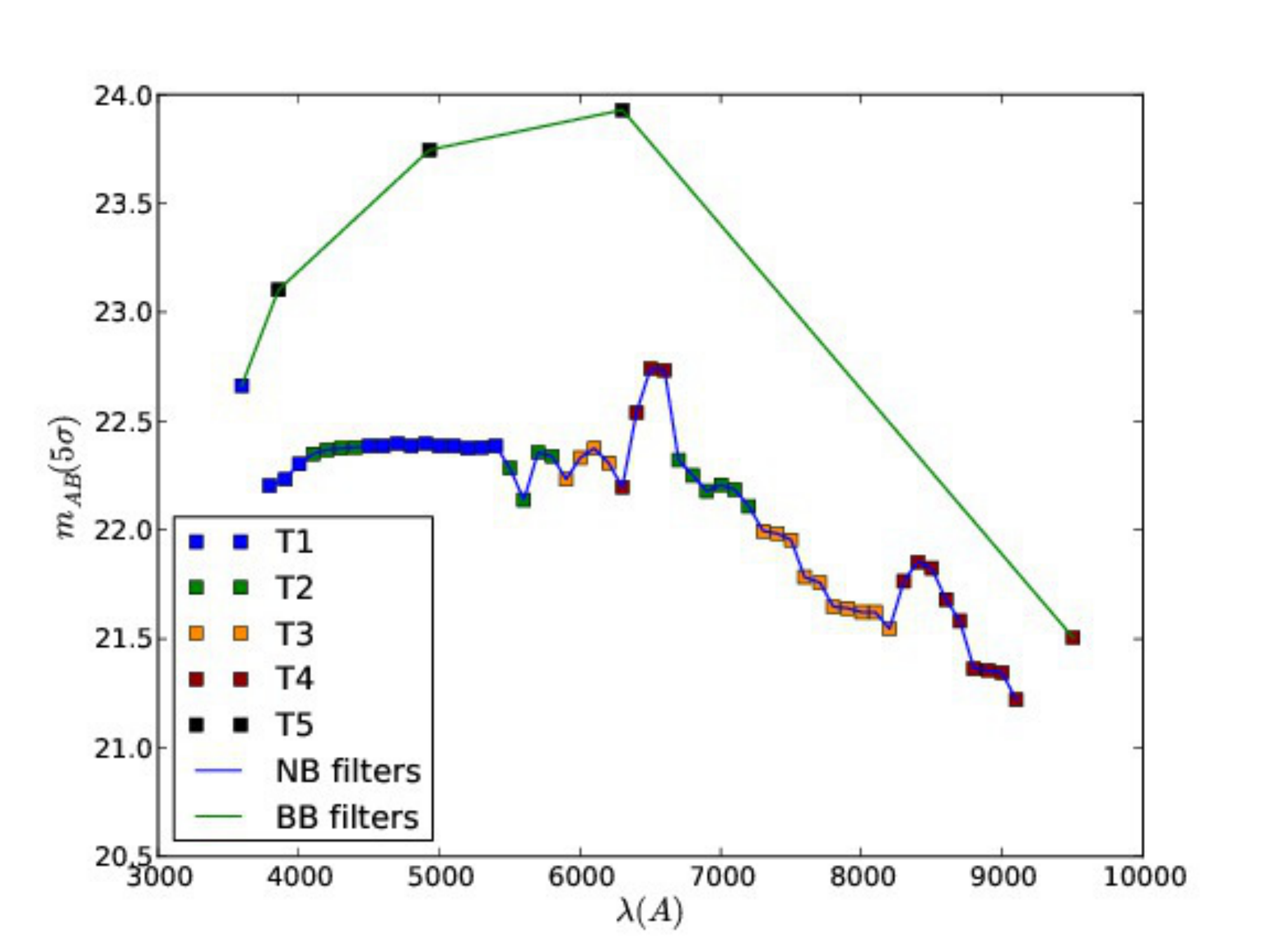}
\caption{Limiting AB magnitudes ($5\sigma$, 3 arcsec aperture) for all the filters in the survey, color coded by their tray distribution}
\label{fig:lim}
\end{figure}

 The final limiting magnitudes and exposure times are plotted in Fig. \ref{fig:lim} and listed in Tables 3., 4. and 5.

\begin{table}[h]
\centering  
\label{t:NBfilters1}
\begin{tabular}{crrrrrrr}
\textbf{Filter}  & \textbf{$\lambda_C$} & \textbf{FWHM($\AA$)} & \textbf{m$_{AB}^{5\sigma}(3\arcsec\diameter)$} & 
\textbf{m$_{AB}^{5\sigma}(/\sq\arcsec)$} &  \textbf{$t_{exp}(s)$} & \textbf{Tray} \\
\hline  
J-PAS3785 & 3791 & 158 & 22.20 & 23.26 & 240 & T1 \\
J-PAS3900 & 3904 & 141 & 22.23 & 23.29 & 240 & T1 \\
J-PAS4000 & 4003 & 141 & 22.30 & 23.36 & 240 & T1 \\
J-PAS4100 & 4102 & 142 & 22.35 & 23.41 & 240 & T2 \\
J-PAS4200 & 4202 & 142 & 22.37 & 23.43 & 240 & T2 \\
J-PAS4300 & 4302 & 142 & 22.37 & 23.44 & 240 & T2 \\
J-PAS4400 & 4402 & 142 & 22.38 & 23.44 & 240 & T2 \\
J-PAS4500 & 4501 & 142 & 22.39 & 23.45 & 240 & T1 \\
J-PAS4600 & 4601 & 143 & 22.39 & 23.45 & 240 & T1 \\
J-PAS4700 & 4701 & 143 & 22.40 & 23.46 & 240 & T1 \\
J-PAS4800 & 4801 & 143 & 22.39 & 23.45 & 240 & T1 \\
J-PAS4900 & 4901 & 143 & 22.40 & 23.46 & 240 & T1 \\
J-PAS5000 & 5001 & 143 & 22.39 & 23.45 & 240 & T1 \\
J-PAS5100 & 5101 & 143 & 22.39 & 23.45 & 240 & T1 \\
J-PAS5200 & 5201 & 143 & 22.37 & 23.44 & 240 & T1 \\
J-PAS5300 & 5301 & 143 & 22.38 & 23.44 & 240 & T1 \\
J-PAS5400 & 5401 & 143 & 22.39 & 23.45 & 240 & T1 \\
J-PAS5500 & 5501 & 143 & 22.28 & 23.35 & 240 & T2 \\
J-PAS5600 & 5601 & 143 & 22.14 & 23.20 & 240 & T2 \\
J-PAS5700 & 5701 & 143 & 22.36 & 23.42 & 240 & T2 \\
J-PAS5800 & 5801 & 143 & 22.34 & 23.40 & 240 & T2 \\
J-PAS5900 & 5901 & 143 & 22.23 & 23.29 & 240 & T3 \\
J-PAS6000 & 6001 & 143 & 22.33 & 23.39 & 240 & T3 \\
J-PAS6100 & 6101 & 143 & 22.37 & 23.44 & 240 & T3 \\
J-PAS6200 & 6201 & 143 & 22.31 & 23.37 & 240 & T3 \\
J-PAS6300 & 6301 & 143 & 22.20 & 23.26 & 480 & T4 \\
J-PAS6400 & 6401 & 143 & 22.54 & 23.60 & 480 & T4 \\
J-PAS6500 & 6501 & 143 & 22.74 & 23.80 & 480 & T4 \\
J-PAS6600 & 6601 & 143 & 22.73 & 23.80 & 480 & T4 \\
J-PAS6700 & 6701 & 143 & 22.32 & 23.38 & 240 & T2 \\
J-PAS6800 & 6800 & 142 & 22.25 & 23.31 & 240 & T2 \\
J-PAS6900 & 6901 & 143 & 22.18 & 23.24 & 240 & T2 \\
J-PAS7000 & 7002 & 142 & 22.21 & 23.27 & 240 & T2 \\
J-PAS7100 & 7100 & 141 & 22.19 & 23.25 & 240 & T2 \\
J-PAS7200 & 7200 & 144 & 22.11 & 23.17 & 240 & T2 \\
J-PAS7300 & 7301 & 143 & 22.00 & 23.06 & 240 & T3 \\
J-PAS7400 & 7401 & 143 & 21.98 & 23.05 & 240 & T3 \\
J-PAS7500 & 7500 & 143 & 21.95 & 23.02 & 240 & T3 \\
J-PAS7600 & 7596 & 142 & 21.78 & 22.84 & 240 & T3 \\
J-PAS7700 & 7705 & 136 & 21.76 & 22.82 & 240 & T3 \\
J-PAS7800 & 7800 & 143 & 21.65 & 22.71 & 240 & T3 \\
J-PAS7900 & 7901 & 143 & 21.64 & 22.70 & 240 & T3 \\
J-PAS8000 & 8000 & 143 & 21.62 & 22.68 & 240 & T3 \\
\hline 
\end{tabular}
\caption{
J-PAS Narrow Band observations. The central wavelengths $\lambda_c$ and filter 
  widths (FWHM) have been calculated taking into account the expected CCD Quantum 
  Efficiency and the Javalambre expected atmosphere at 1.2 airmasses. We also list the $5-\sigma$ detection 
magnitudes in a $3\arcsec$ diameter aperture and per $\sq\arcsec$. 
} 
\end{table}

\begin{table}[h]
\centering  
\label{t:NBfilters2}
\begin{tabular}{crrrrrrr}
\textbf{Filter}  & \textbf{$\lambda_C$} & \textbf{FWHM($\AA$)} & \textbf{m$_{AB}^{5\sigma}(3\arcsec\diameter)$} & 
\textbf{m$_{AB}^{5\sigma}(/\sq\arcsec)$} &  \textbf{$t_{exp}(s)$} & \textbf{Tray} \\
\hline  
J-PAS8100 & 8099 & 142 & 21.62 & 22.68 & 240 & T3 \\
J-PAS8200 & 8200 & 143 & 21.55 & 22.61 & 240 & T3 \\
J-PAS8300 & 8302 & 142 & 21.77 & 22.83 & 480 & T4 \\
J-PAS8400 & 8400 & 143 & 21.85 & 22.91 & 480 & T4 \\
J-PAS8500 & 8500 & 143 & 21.82 & 22.89 & 480 & T4 \\
J-PAS8600 & 8600 & 143 & 21.68 & 22.74 & 480 & T4 \\
J-PAS8700 & 8700 & 143 & 21.58 & 22.64 & 480 & T4 \\
J-PAS8800 & 8800 & 143 & 21.36 & 22.42 & 480 & T4 \\
J-PAS8900 & 8898 & 141 & 21.36 & 22.42 & 480 & T4 \\
J-PAS9000 & 8999 & 143 & 21.34 & 22.41 & 480 & T4 \\
J-PAS9100 & 9100 & 142 & 21.22 & 22.28 & 480 & T4 \\
\hline 
\end{tabular}
\caption{
J-PAS Narrow Band observations. The central wavelengths $\lambda_c$ and filter 
  widths (FWHM) have been calculated taking into account the expected CCD Quantum 
  Efficiency and the Javalambre expected atmosphere at 1.2 airmasses. We also list the $5-\sigma$ detection 
magnitudes in a $3\arcsec$ diameter aperture and per $\sq\arcsec$. 
} 
\end{table}

\begin{table}[h]
\centering
\label{BBfilters}
\begin{tabular}{crrrrrrr}
\hline
\textbf{Filter}  & \textbf{$\lambda_C$} & \textbf{FWHM($\AA$)} & \textbf{m$_{AB}^{5\sigma}(3\arcsec\diameter)$} & 
\textbf{m$_{AB}^{5\sigma}(/\sq\arcsec)$} &  \textbf{$t_{exp}(s)$} & \textbf{Tray} \\
\hline
 J-PAS3518 & 3596 & 261  & 22.66 & 23.73 & 240 & T1 \\
$u_{J-PAS}$ & 3856 & 357  & 23.10 & 24.16 & 225 & T5 \\
$g_{J-PAS}$ & 4931 & 1441 & 23.75 & 24.81 & 225 & T5 \\
$r_{J-PAS}$ & 6301 & 1189 & 23.93 & 24.99 & 600 & T5 \\
J-PAS10069 & 9505 & 618  & 21.51 & 22.57 & 480 & T4 \\
\hline
\end{tabular}
\caption{J-PAS Medium and Broad band observations.
 The central wavelengths $\lambda_c$ and filter 
  widths (FWHM) have been calculated taking into account the expected E2V CCD Quantum 
  Efficiency and the Javalambre expected atmosphere at 1.2 airmasses. We also list the $5-\sigma$ detection 
magnitudes in a $3\arcsec$ diameter aperture and per $\arcsec^2$
}
\end{table}

\FloatBarrier

\subsection{J-PAS Survey Area Definition}
\FloatBarrier

\subsubsection{Area Selection}

In this Section we describe the selection process for the area that will be covered by J-PAS. The survey definition based on the scientific objectives has fixed a minimum area of 8,000$\sq\degr$, selected on the basic criterion of low galactic extinction. Earlier  discussions remarked on the necessity of dividing this area between the Northern and Southern galactic hemispheres, in order to share in an approximately homogeneous manner the area to be observed along the year---or, equivalently, along the different right ascensions. It is also necessary to ensure the compactness of the survey area, in order to optimize the exposure time. 

\subsubsection{Raw observability from Javalambre}

\begin{figure}
\centering 
\includegraphics[width=0.8\textwidth,keepaspectratio]{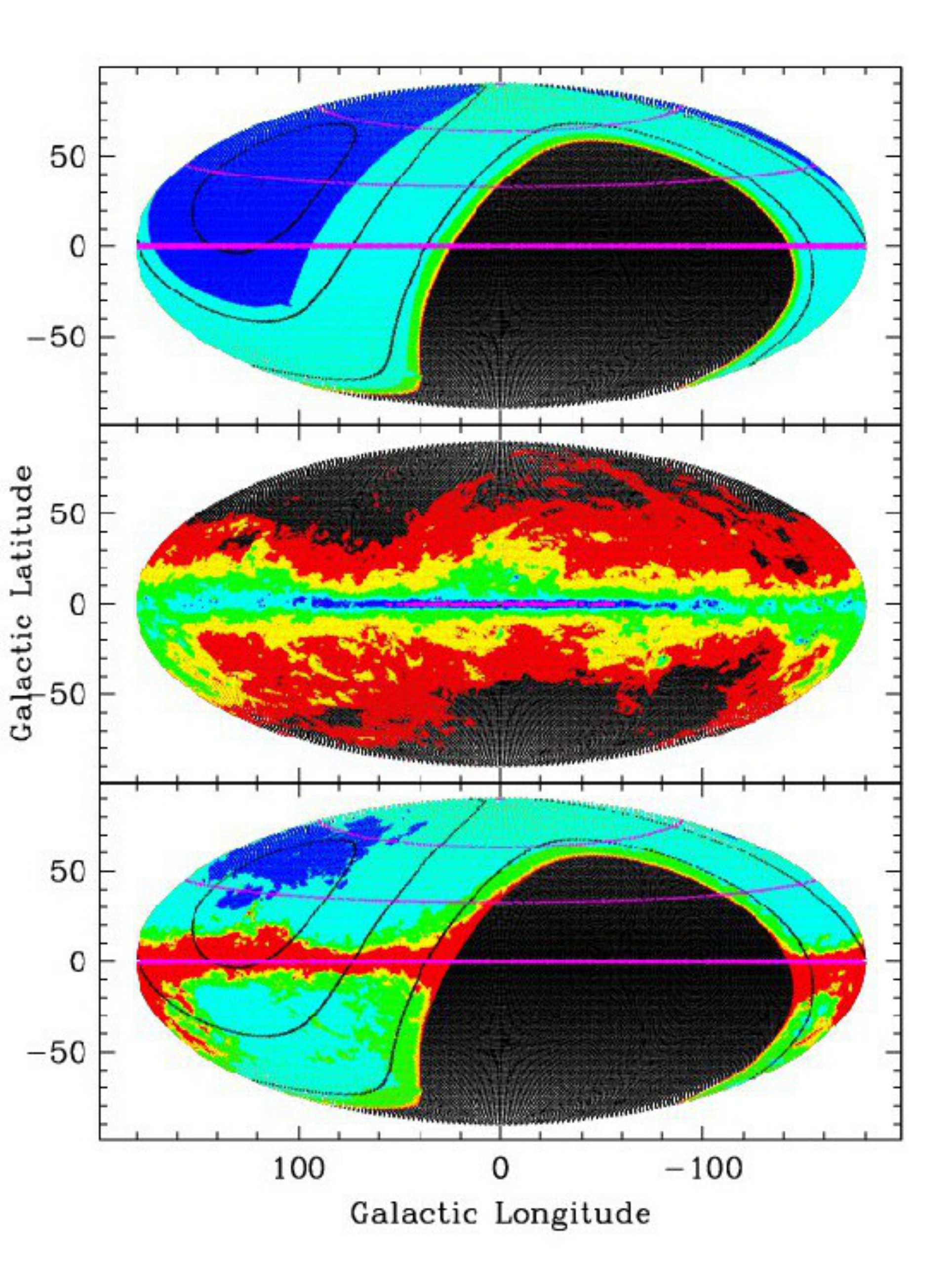}
\caption{(Top) Visibility from OAJ. (Red, yellow, green, cyan, blue) correspond to visibilities greater than (0, 30, 100, 300, 1000) hours/year. Magenta lines represent galactic latitudes $b=(0\degr, 30\degr, 60\degr)$, and black lines represent declinations $\delta=(0\degr, 30\degr, 60\degr)$. (Middle) Dust column in each direction, as measured by Schlegel et al. (1998). The color scale (black, red, yellow, green, cyan, blue, magenta) corresponds to values of E(B-V) $>$ (0.00, 0.03, 0.10, 0.30, 1.00, 3.00, 10.00). (Bottom) Corrected visibility as described in the text. All colors and lines are the same as in the top panel.\label{OAJ-observability}}
\end{figure}

The obvious first point that must be taken into account when selecting the sky area for J-PAS is that the area must be observable from the Javalambre Observatory site (OAJ), for a period of time as long as possible over a natural year. We have measured the total observability for each point in the celestial sphere, defined as the number of night hours\footnote{We define night time using the strict astronomical definition, i.e. time between astronomical twilights.}  when a given point is higher than $40\degr$ above the horizon, as seen from OAJ, in a year. The top panel in Figure \ref{OAJ-observability} shows that information. We have also eliminated (as non-observable) two nights before and after each full Moon and a cone of $30\degr$ around the Moon in grey nights\footnote{Five nights around new Moon are considered dark, the above mentioned two nights around full Moon are completely eliminated, and the rest are considered grey.}. This induces a slight decrease in observability around the ecliptic.

Within these conditions, we see that the area that lies approximately at $\delta  > 40\degr$ is observable for more than 1000 hours per year (dark blue area in Figure 1), and everything at $\delta > 10\degr$ is observable at least for 300 hours per year (cyan area). 

\subsubsection{Correction by dust column extinction}

 In principle one could use an {\it a priori} value of the galactic latitude to define a zone of avoidance in the survey, which has indeed been the method of choice in other surveys. However, because we want to define a large area and the best partition of it in the Northern and Southern galactic hemispheres (NGH and SGH), we decided to use a slightly more sophisticated approach. We have used the DIRBE dust maps published by Schlegel, Finkbeiner \& Davies (1998) to estimate the dust extinction in each direction. The central panel in Figure 1 shows this map, in terms of the values of E(B-V). 

 In order to combine the observability and the dust content in each direction we have defined a corrected observability. For each direction we calculate the extinction at 3800 Angstroms using the standard Milky Way extinction law and the value of E(B-V) given by Schlegel et al. (1998), and estimate the increase in exposure time that would be necessary to reach the nominal depth at 3800 Angstroms taking that extinction into account. We then correct the available time according to that factor, which yields a smaller number of hours of observability in each direction. We must remark several things here. First, this is only a crude approximation, as we have used a very low-resolution dust map and a simple formula for the exposure times. Second, even if the corrections were accurate, they would only apply to the bluest filters in the J-PAS set, and we have applied them to correct all the exposure times. And finally,  the use of this corrected visibility {\it does not mean that we will be correcting the exposure times in the survey}--it is only a convenient way to put together the information about visibility and dust. 

 Taking into account all of these caveats, the map showing the corrected visibility is shown in the bottom panel of Figure 1. As expected, we see the original visibility modulated by the presence of the Milky Way. It is important to point out that there is a relatively large area with visibility $> 300$ hours/year in the SGH. 

\begin{figure}
\centering
\epsscale{.80}
\plottwo{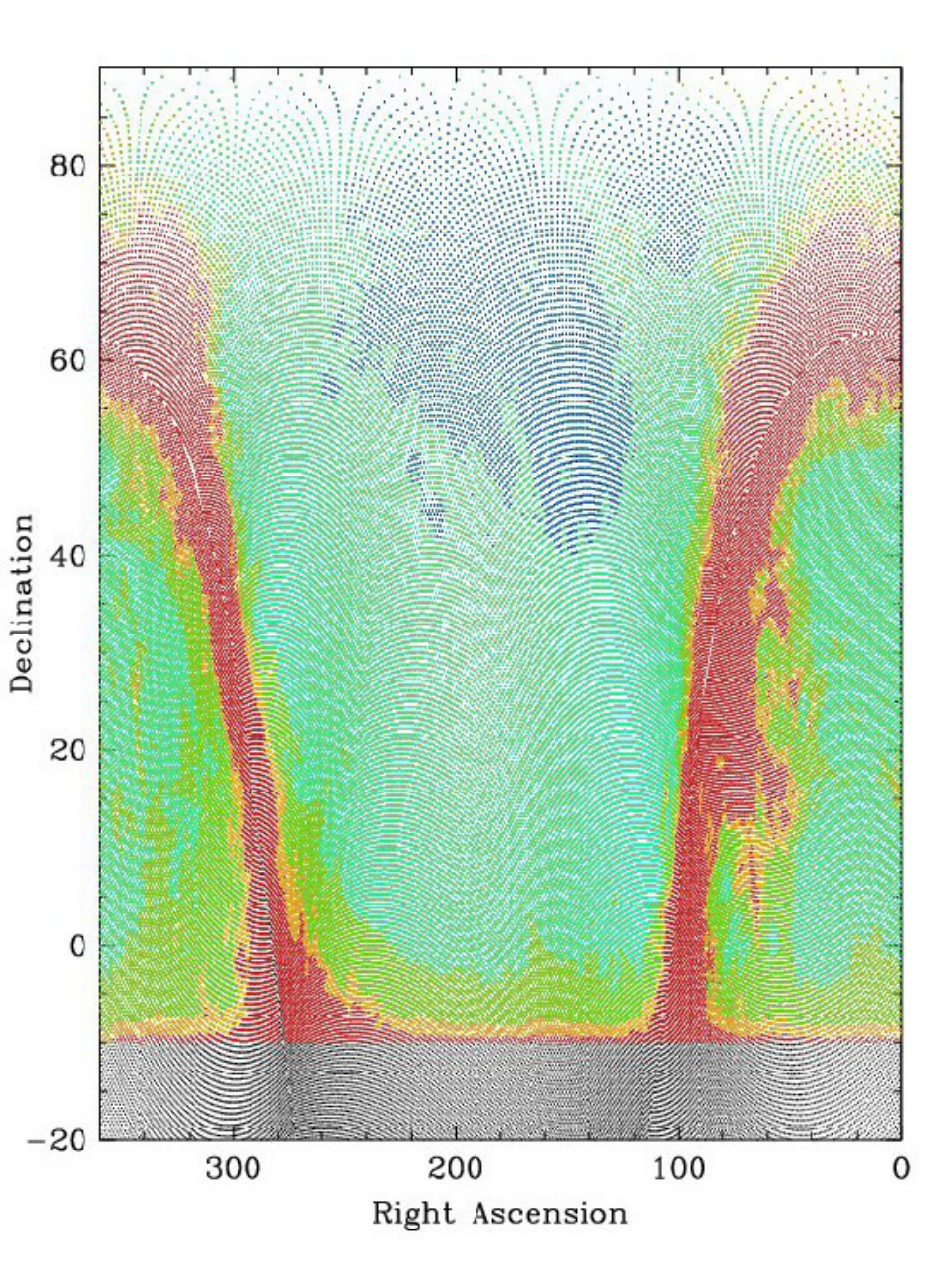}{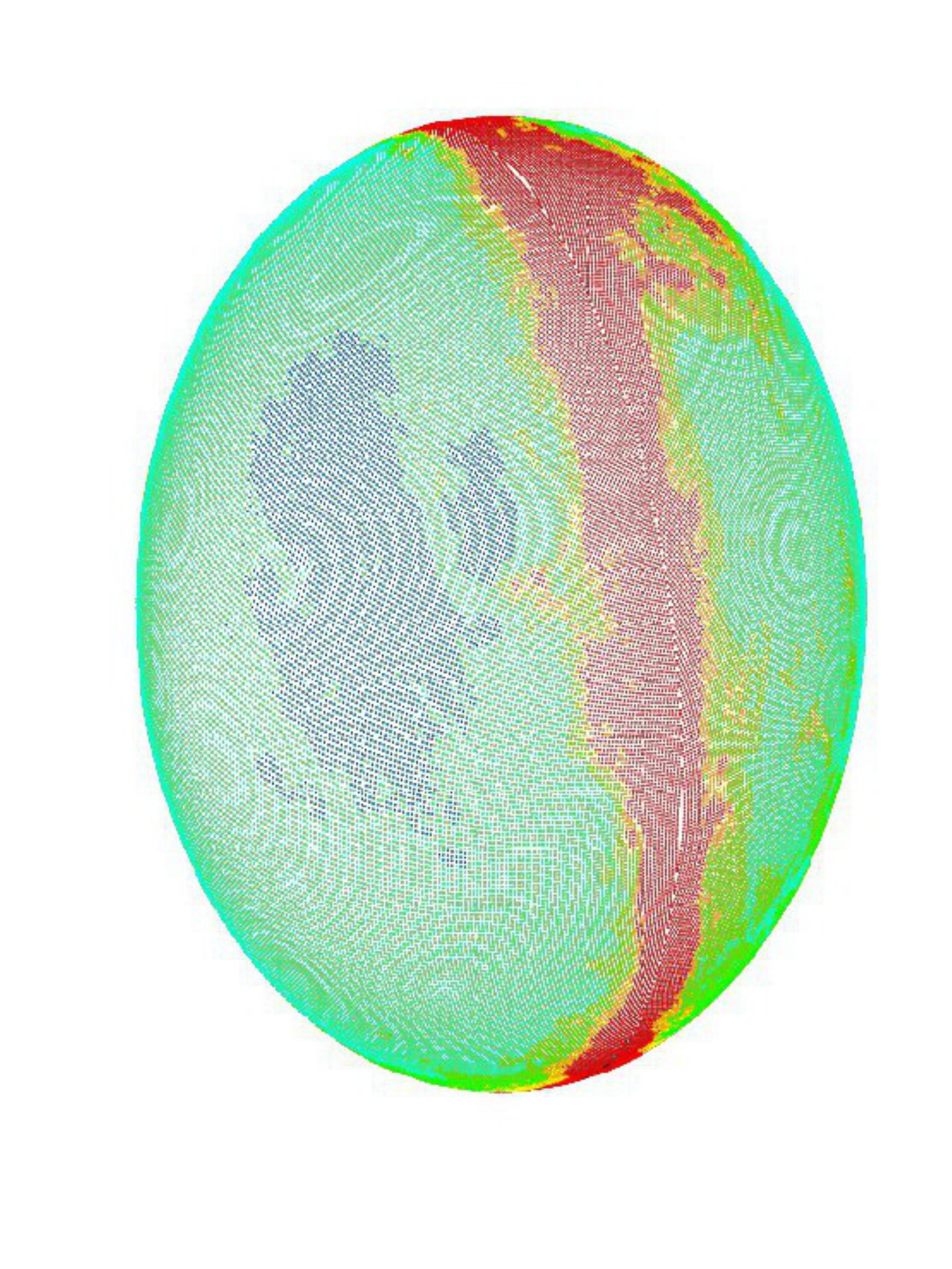}
\caption{(Left) Corrected visibility plotted using equatorial coordinates. Colors are the same as those in Figure \ref{OAJ-observability}. (Right) Corrected visibility projected for the northern celestial hemisphere. Notice that this projection does not show the area below $\delta=0\degr$.\label{OAJ-visibility}}
\end{figure}

 The same data can also be represented in equatorial coordinates, both cartesian and projected, as shown in Figure \ref{OAJ-visibility}. 

\subsubsection{Definition of the survey area}
\begin{figure}
\centering
\epsscale{0.80}
\plottwo{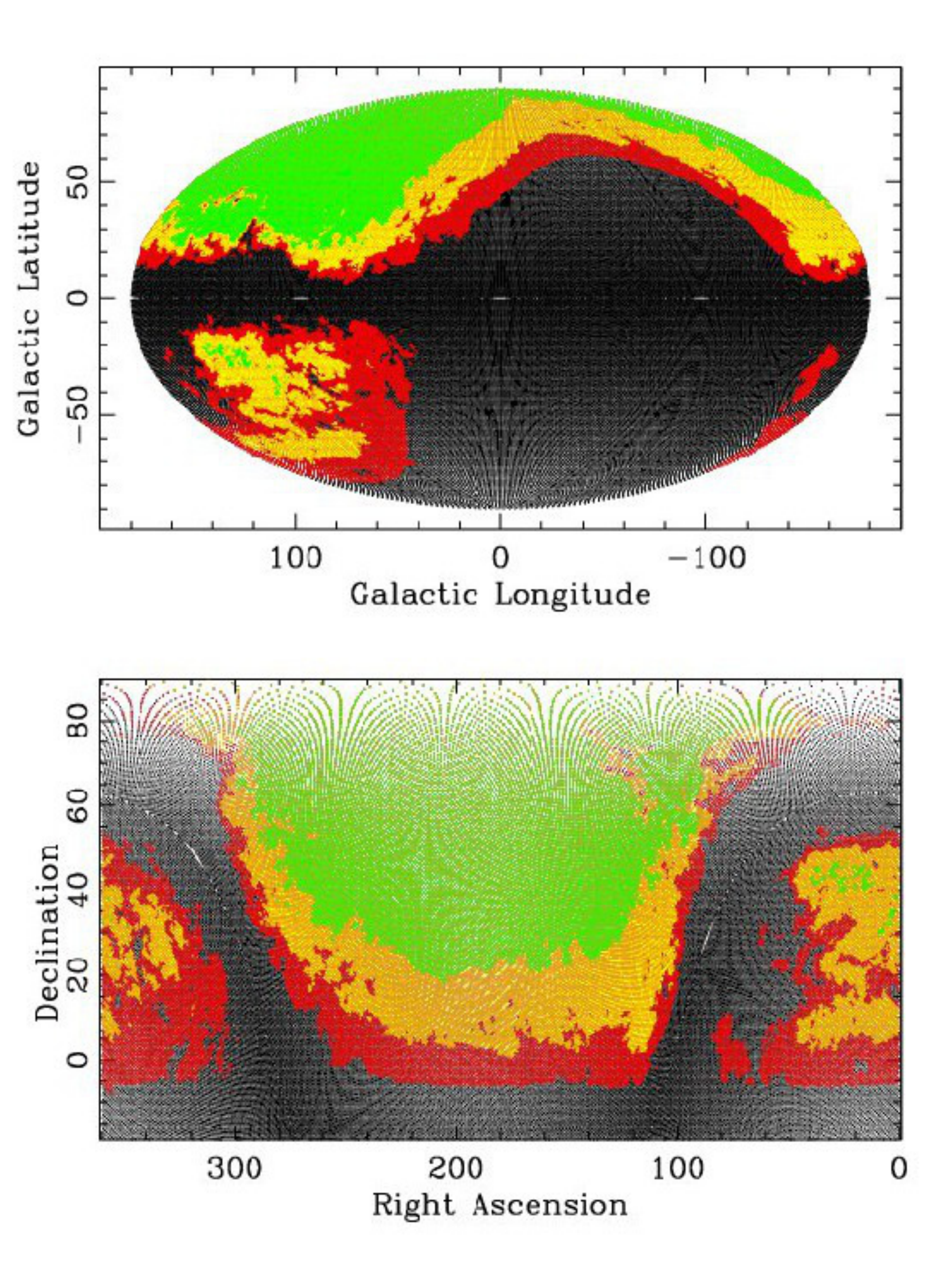}{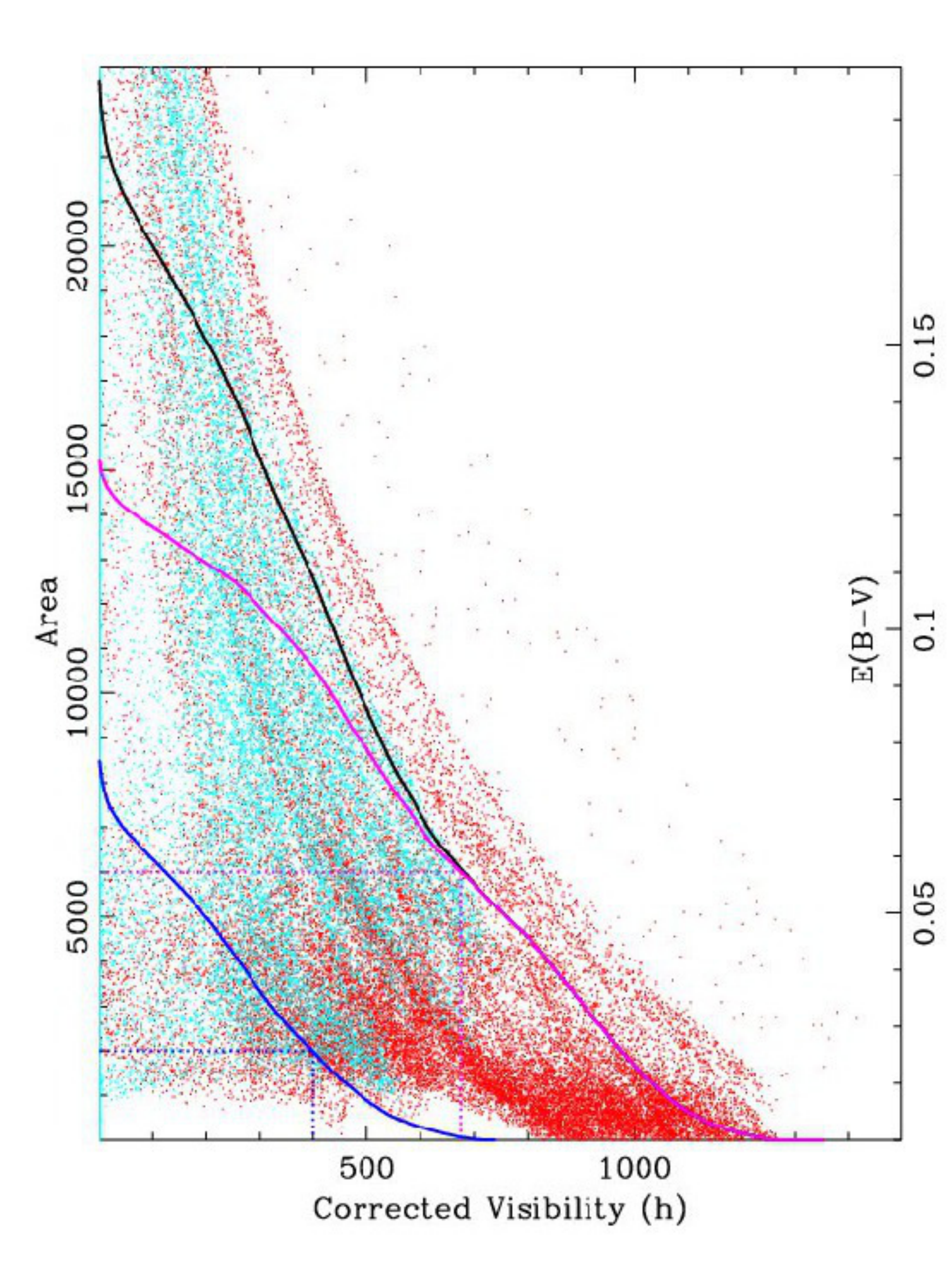}
\caption{(Left) Sky areas with highest corrected visibilities. The
  (6000, 12000, 18000) square degrees with highest corrected
  visibilities are plotted as (green, yellow, red) areas. (Right)
  Cumulative plot of total area vs corrected visibility (black line),
  NGH-only (magenta line and red points) and SGH-only (blue line and
  cyan points). Dots correspond to values of  E(B-V) (see right side
  axis) for each different direction. For the best $6000\sq\degr$ in
  the NGH the visibility is $>$675 hours/year, and $>$400 hours/year 
for the best $2000\sq\degr$ in the SGH.\label{OAJ-jpassky1}}
\end{figure}

Once both the information about visibility and dust are combined, it should not be complicated to select the area to be covered by the survey. In a first approximation, we could simply choose those areas with the highest value of corrected observability. In order to do this, and using this criterion, we choose the best (6000, 12000, 18000) square degrees in the sky and plot them in Figure \ref{OAJ-jpassky1} (left) as the (respectively) green, yellow, and red areas. One problem with this approach is that, by far, most of the best area is necessarily in the NGH, because of the much higher observability from OAJ. Choosing this area without any correction would lead us to a problem during autumn, when only a relatively small area of the NGH is visible, and we may find ourselves without any observable, but not already covered area. This is a well-known issue and, as shown in the left panel of Figure \ref{OAJ-jpassky1}, we need to take it into account unless we want to end up with a survey that only covers  the NGH.

\begin{figure}
\centering
\epsscale{.80}
\plottwo{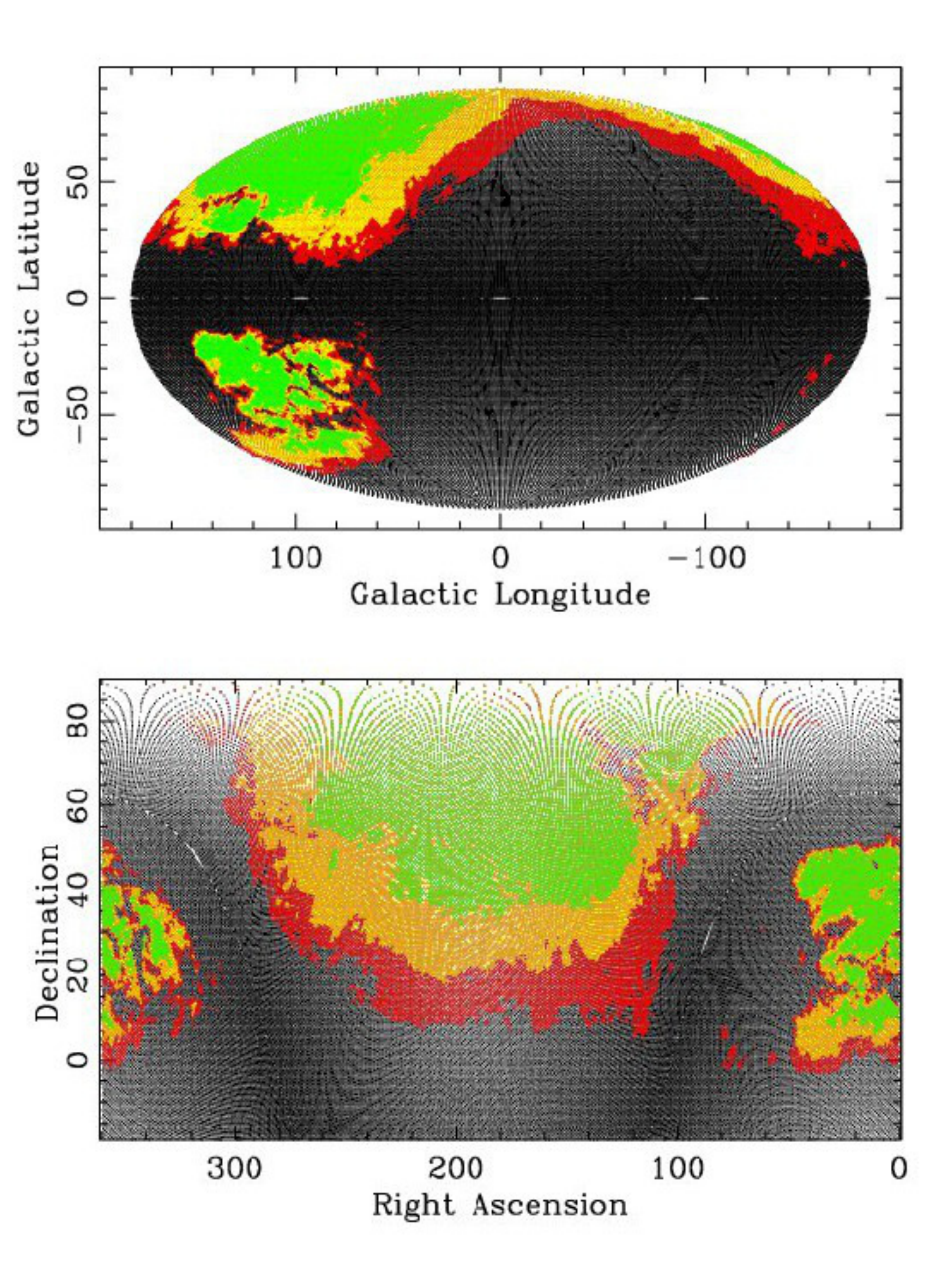}{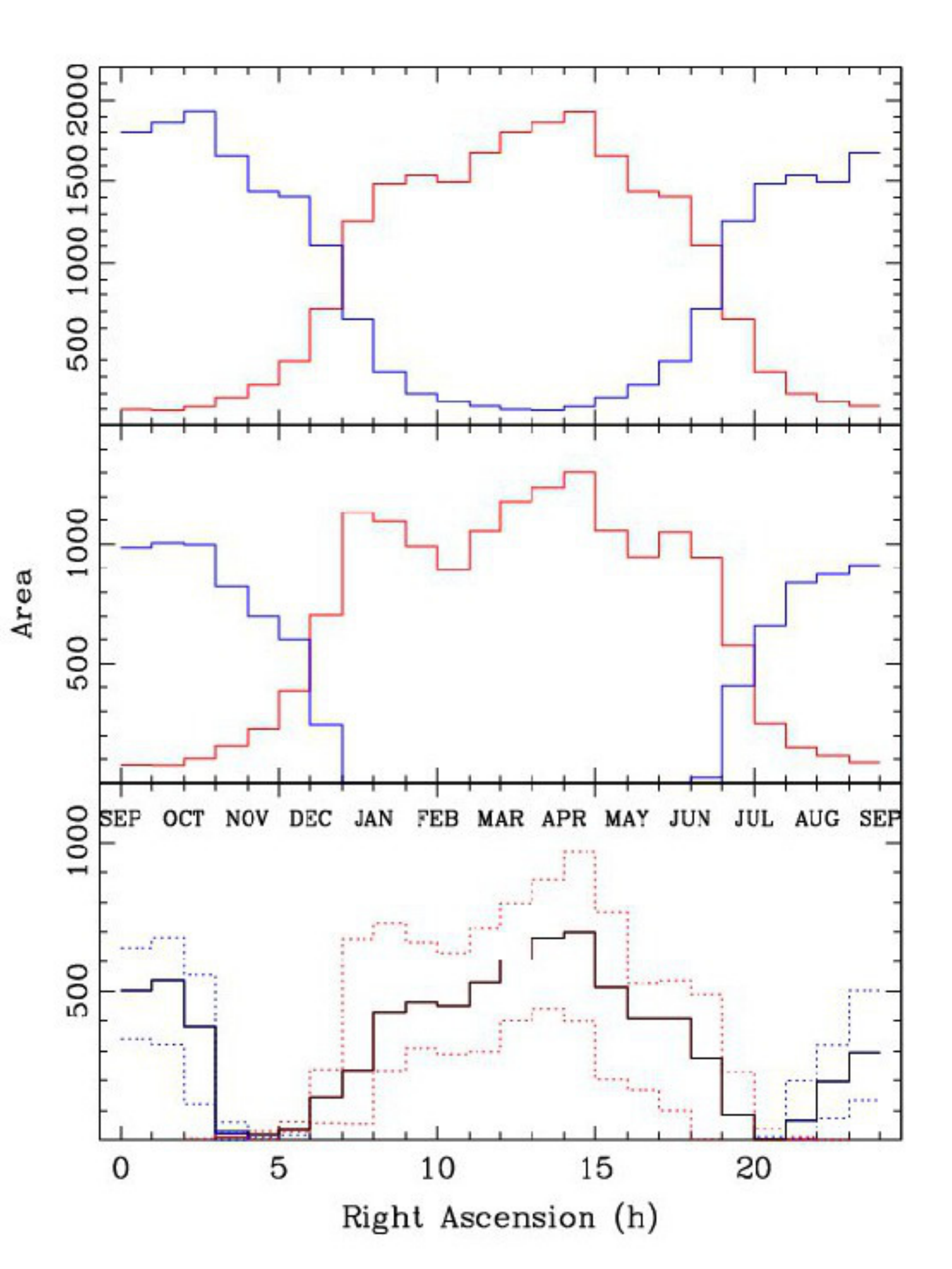}
\caption{Sky areas with highest corrected visibilities, when chosen independently in both galactic hemispheres. The best (3000, 6000, 9000) square degrees are plotted as (green, yellow, red) areas in the NGH, and the best (1000, 2000, 3000) square degrees correspond to the same in the SGH. (Right) Histogram of sky area (red NGH, blue SGH) vs Right Ascension (in hours) for different samples: whole-sky in the top panel, corrected visibility $>0$h in the middle panel, and the three selected areas seen in the left, with the continuous lines corresponding to (6000 + 2000) square degrees, and the dotted lines corresponding to the other two cases. The optimal month for each RA is indicated in the lower panel.\label{OAJ-jpassky2}}
\end{figure}

 In order to avoid this, we take a different approach. We define the best areas independently  for the NGH and SGH, taking the best (3000, 6000, 9000) square degrees from the former, and the best (1000, 2000, 3000) square degrees from the latter. We use these figures because we have from the beginning devised an approximate partition of $6000+2000\sq\degr$ as a reasonable compromise. The result of this exercise is shown in Figure \ref{OAJ-jpassky2} (left). We also show the distribution of those areas in terms of right ascension, to give an indication of the best month when each area can be observed. As a result of those tests, we can see that the transitions between NGH and SGH around June and December are less populated in terms of available area, but this should not pose a serious problem, and may be alleviated with an adequate choice of a deep field.
  
  A possible, perhaps simpler, alternative approach would be to avoid combining both sky quality indicators (visibility and dust column) and just define some limits in each of them. One could, e.g., take the best $8000\sq\degr$  in terms of lowest $E(B-V)$ within some well-defined visibility limit to ensure the feasibility of the observations, for example, areas of the sky with visibility $> 200$ hours. We have repeated all the tests previously described with this criterion, and the result is less satisfactory in terms of the final sample. In particular, when this method is used, the area chosen moves away from the Galaxy in both hemispheres (as it is unmodulated by the visibility), in such a way that the transition from NGH to SGH and back is extremely abrupt, leaving almost no observable area in between.

\begin{figure}
\epsscale{1.20}
\plottwo{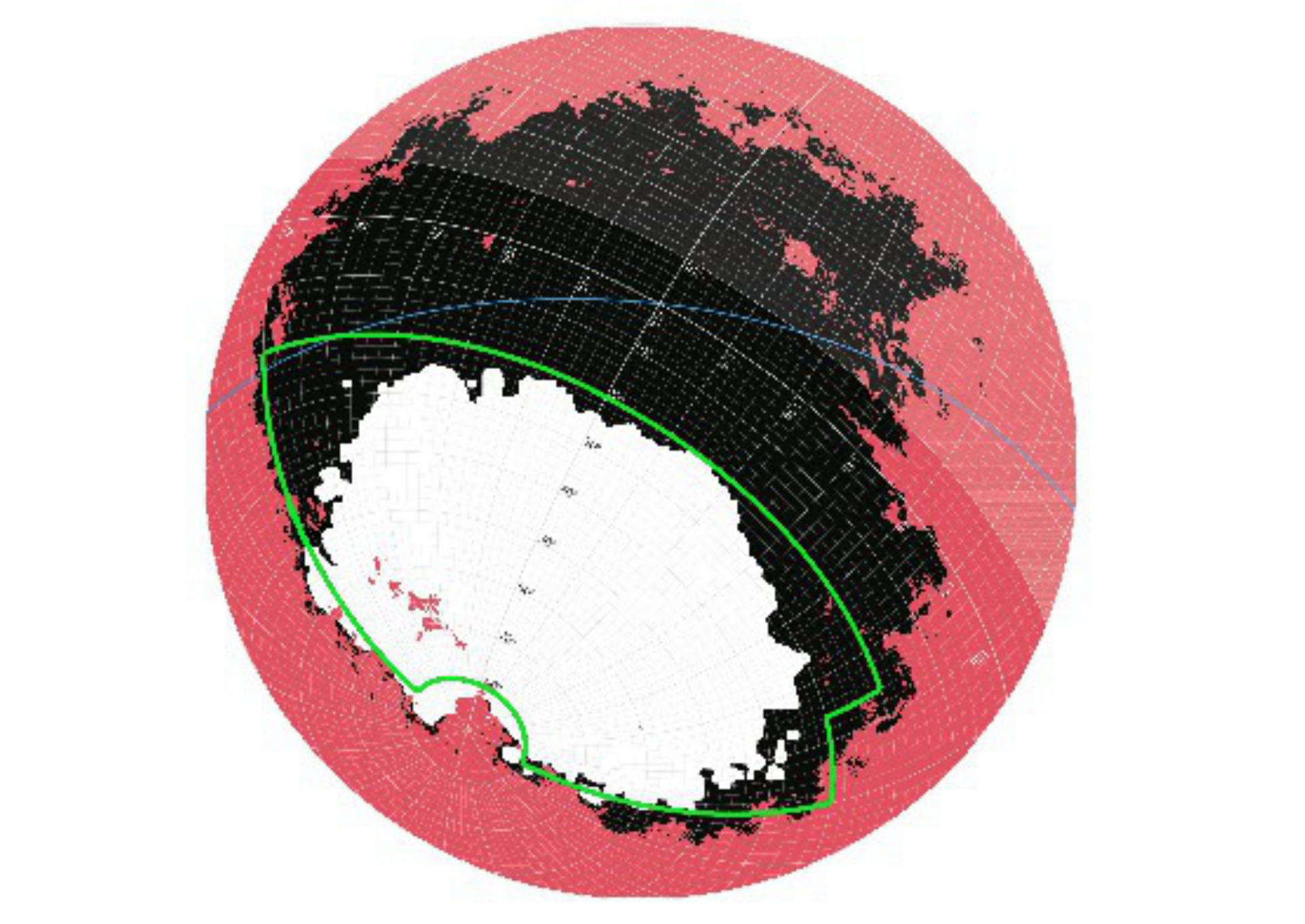}{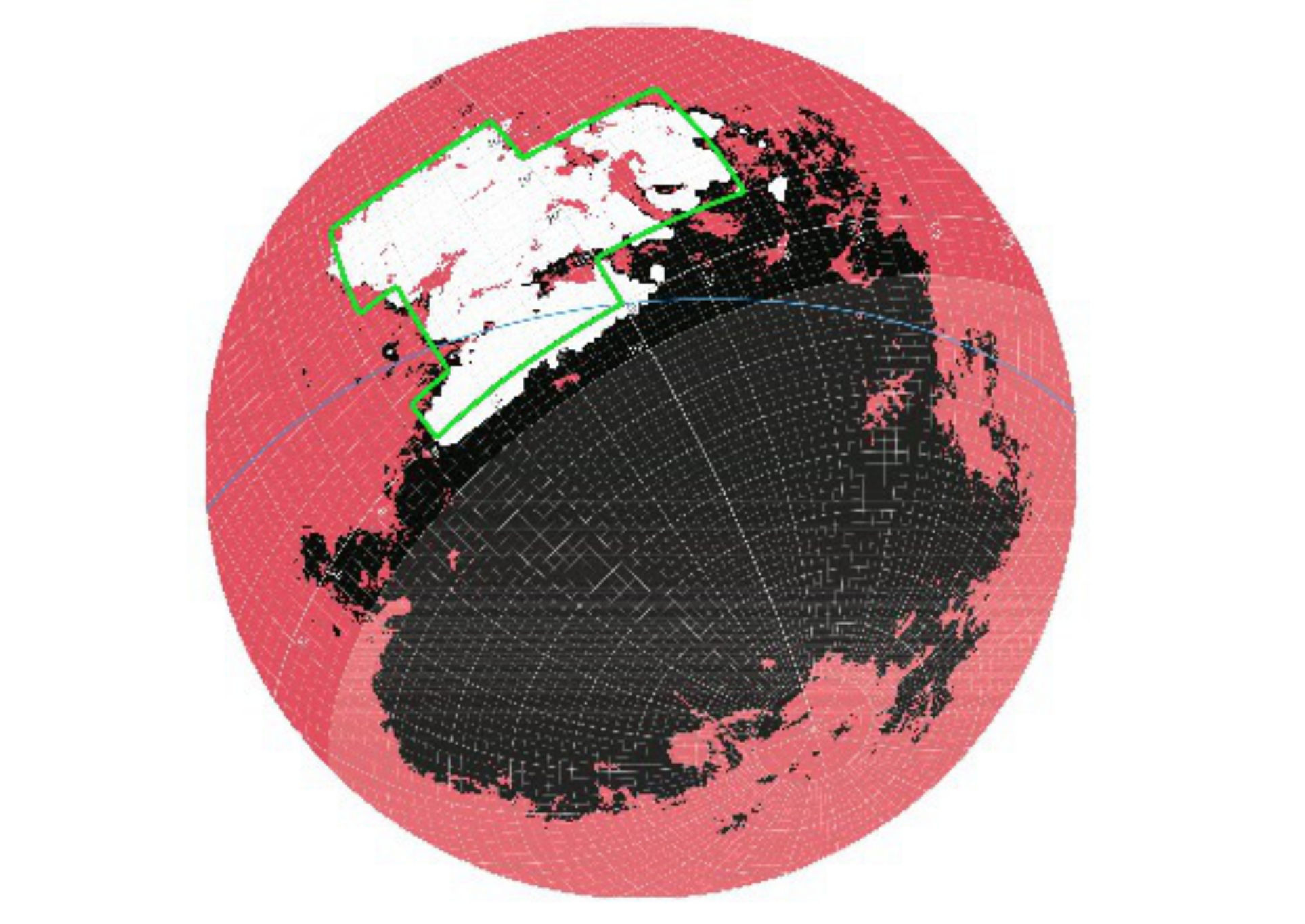}
\caption{Representation in Lambert Projection of the Northern and Southern Galactic Hemispheres and the J-PAS selected areas. Each plot shows in pink the area with relatively high galactic extinction (as given by E(B-V)$>0.1$ in the Schlegel et al. 1998 maps), and in white the area that is selected when taking the best (6000 + 2000) square degrees selected separately in both hemispheres and described in the text. The blue line is the ecliptic. We suggest the areas marked in green as compact versions of the white ones, that will define the J-PAS North and South areas. They cover approximately (6500 + 2250) square degrees.  \label{OAJ-polarmap}}
\end{figure}

  We have thus decided to use the corrected visibility, together with the separation in terms of a northern and southern area, as the main criterion to select the survey area. The characteristics of our survey have also been factored in: our strategy and observational set-up heavily penalizes the use of sparse or irregular areas. With this in mind, we present in Figure \ref{OAJ-polarmap} a possible compact selection of both areas, that covers a total of $\sim 8650\sq\degr$. 

\subsubsection{Overlap with SDSS and SDSS Stripe 82}
\begin{figure}
\epsscale{1.2}
\plottwo{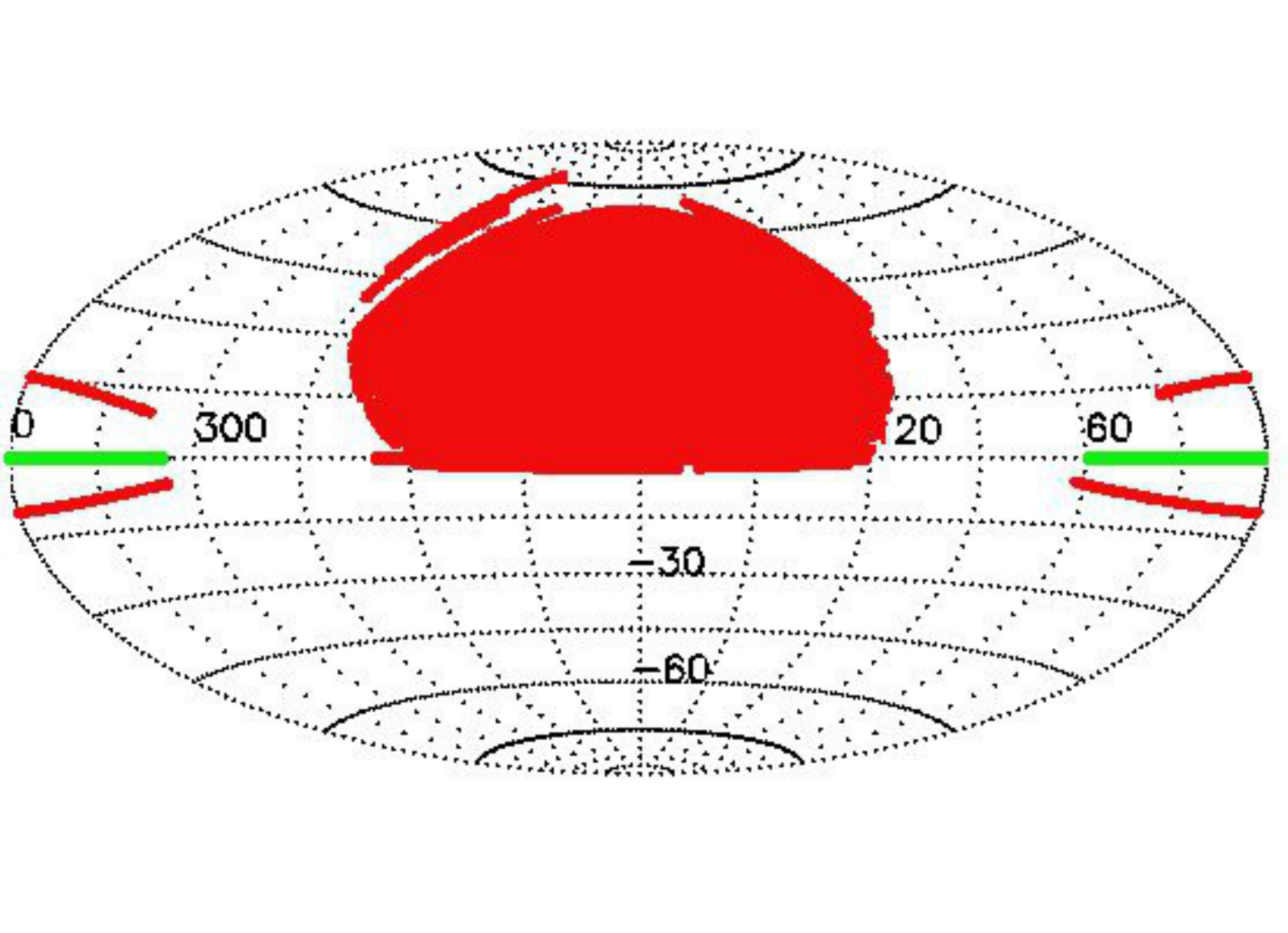}{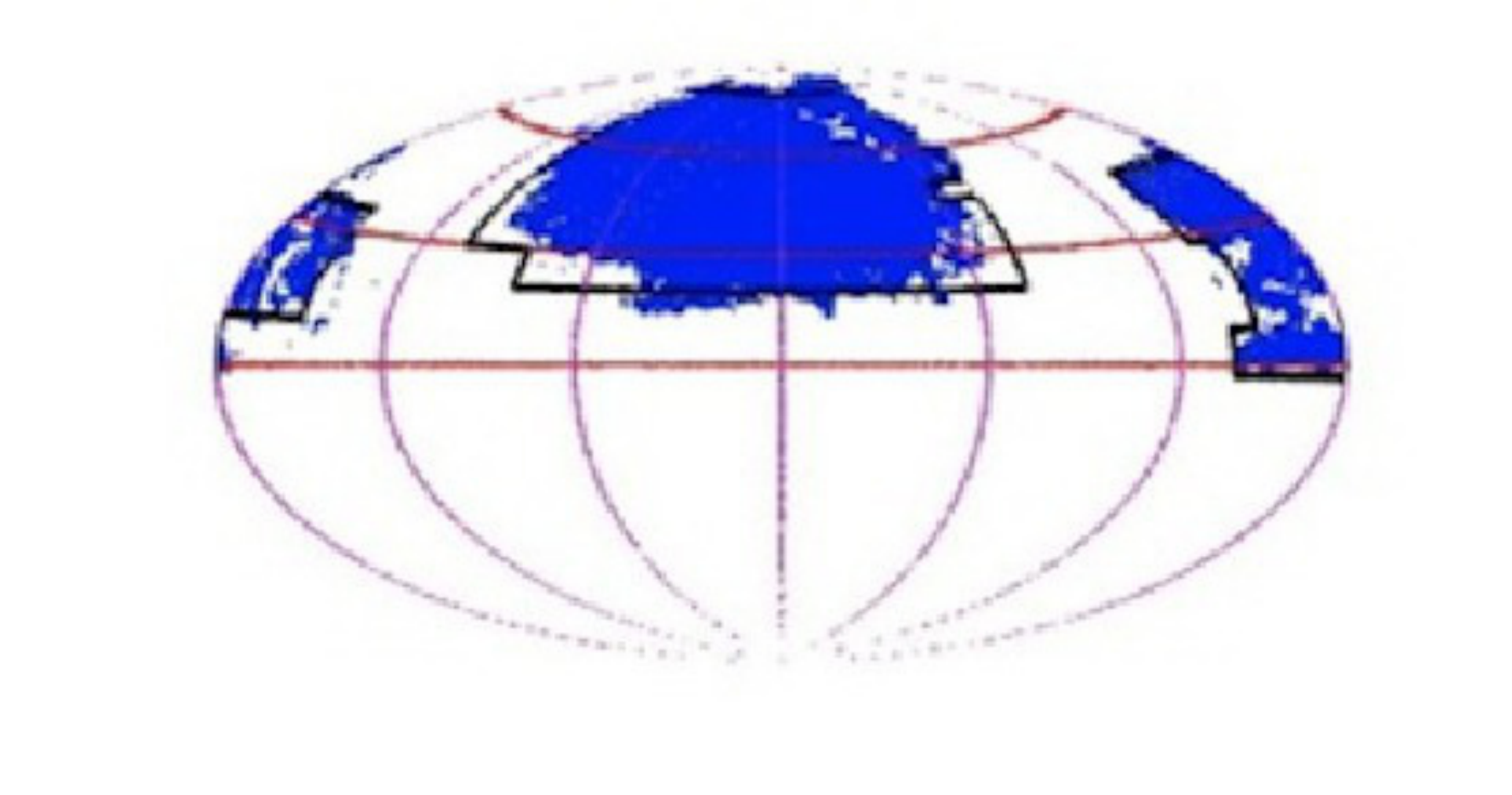}
\caption{Aitoff projection of the full sky in equatorial coordinates, showing (left panel) the area covered by the Sloan Digital Sky Survey, with Stripe 82 highlighted in green. The right panel shows the best 6000+2000 square degrees selected in both galactic hemispheres as described in the text, as well as the J-PAS areas suggested in this document, that amount to ~6500+2250 square degrees. \label{OAJ-sdss}}
\end{figure}

Figure \ref{OAJ-sdss} shows the approximate fingerprint on the sky of both J-PAS and the Sloan Digital Sky Survey. Given that the coordinates of Apache Point are slightly further South than Javalambre, the SDSS coverage reaches further South than J-PAS will. However, we have explicitly included in the Southern Galactic area an equatorial strip that extends down to $\delta=-2.5\degr$, in order to overlap with the deeper Sloan Digital Sky Survey Extension in the area known as Stripe 82. 

\FloatBarrier

\subsection{Expected performance}
\FloatBarrier

\subsubsection{Inputs for the empirical mocks}

\begin{figure}[h]
\centering
\includegraphics[width=0.8\textwidth,keepaspectratio]{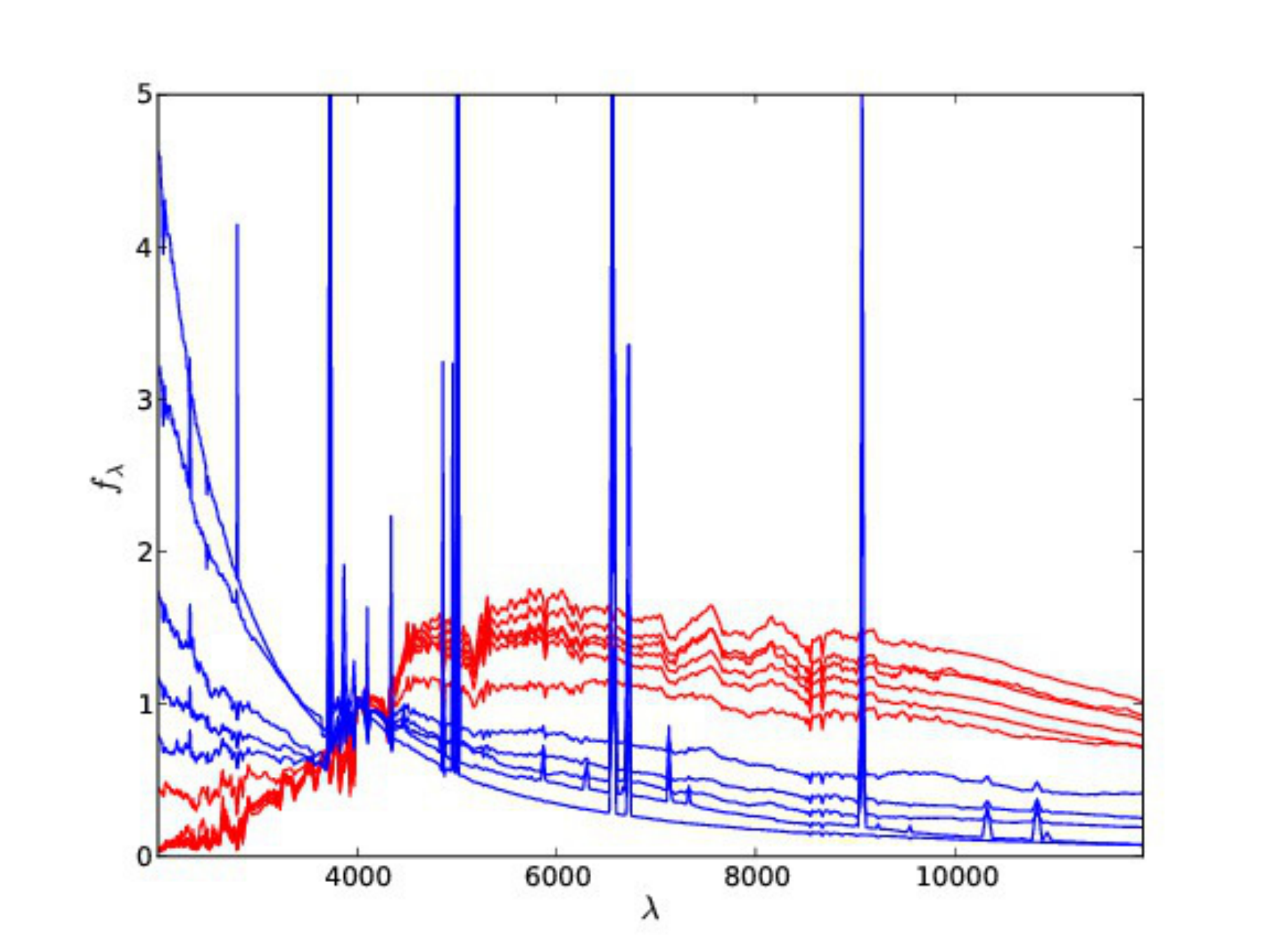}
\caption{BPZ-2 templates, in red, Luminous Red Galaxies (RG), in blue, Emission Line Galaxies (ELG). We interpolate between contiguous templates}
\label{templates}
\end{figure}

\begin{figure}[h]
\centering
\includegraphics[width=0.8\textwidth,keepaspectratio]{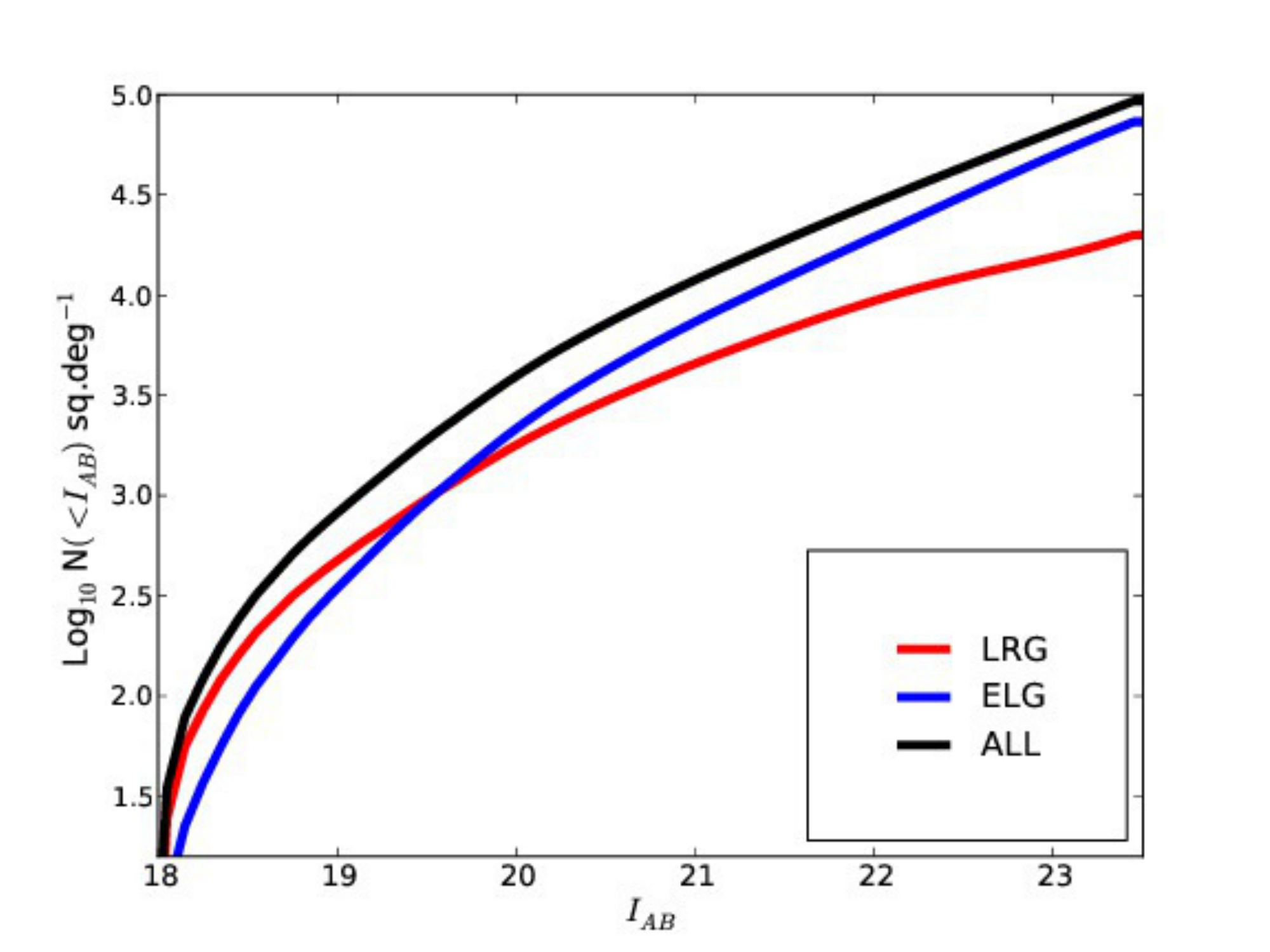}
\caption{Number counts per square degree in our mock catalogs as a function of spectral types (LRG corresponds to early types, ELG to emission line galaxies)}
\label{fig:nmt}
\end{figure}

\begin{figure}[h]
\centering
\includegraphics[width=0.8\textwidth,keepaspectratio]{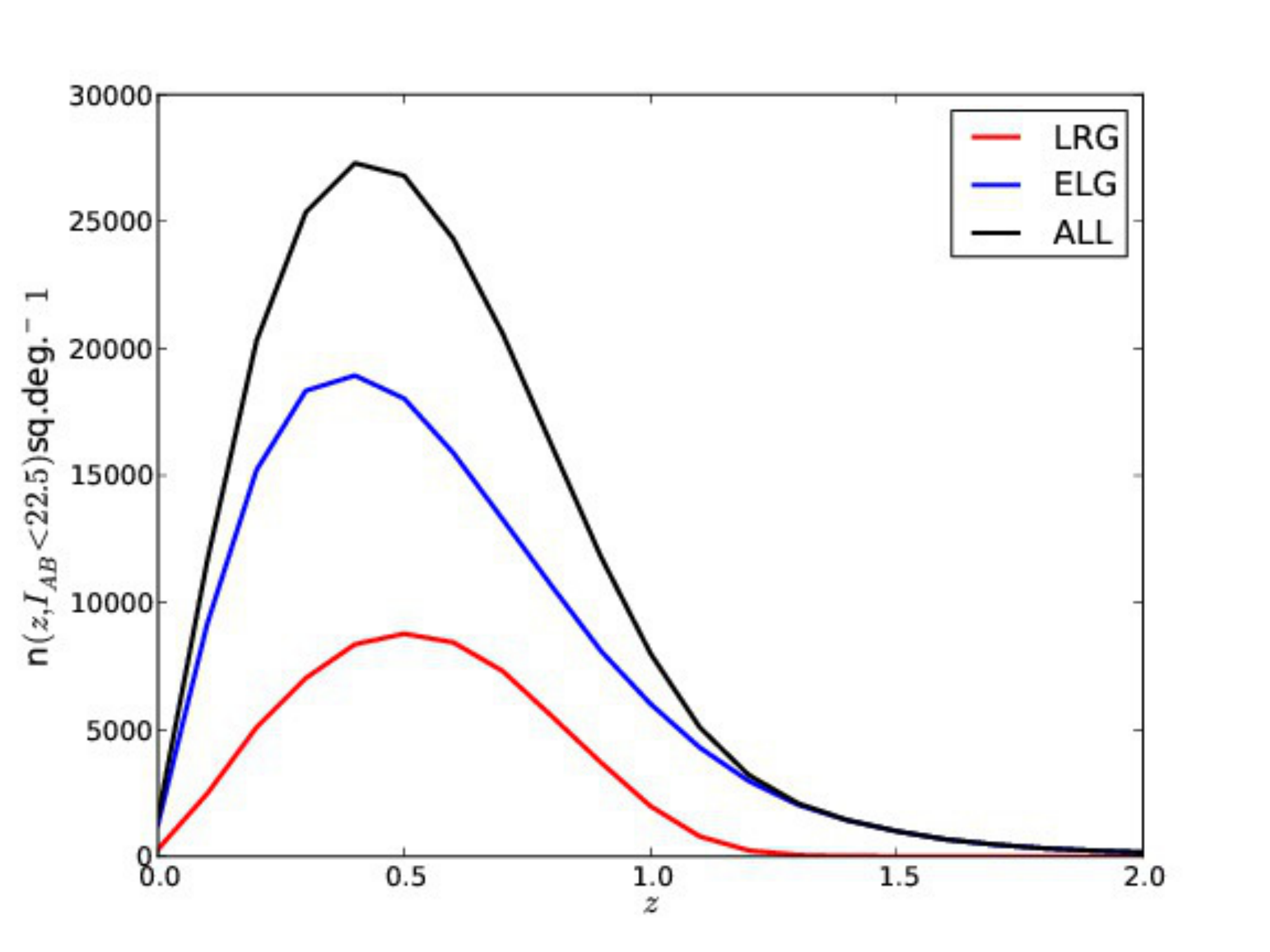}
\caption{Redshift distribution of galaxies in our mock catalogs $I_{AB}<23.5$}
\label{fig:nzt}
\end{figure}

  To generate our mocks we use the prior galaxy distribution of Ben\'\i tez (2014), which produces an accurate distribution of magnitudes, redshifts and spectral types as measured from the COSMOS \citep{ilbert09}, UDF \citep{coe06}, GOODS MUSIC \citep{2006A&A...449..951G} and CFHTLS 
\citep{2009A&A...500..981C} catalogs. The templates are plotted in Fig. \ref{templates}, showing our divide between early type (LRG) templates and emission line galaxy templates (ELG). We repeatedly draw values of $m$, $z$ and $T$ from our prior distributions until we reach the equivalent of an area of $20\sq\degr$, using a fractional step for $T$ of 0.1. The resulting galaxy number counts for different spectral types are shown in Fig. \ref{fig:nmt} and its redshift distribution in Fig. \ref{fig:nzt}. 

 To generate the mocks, we use the ETC described above to calculate realistic instrumental noise within $3\arcsec$ diameter apertures. In addition we add a $0.06-0.08$ systematic noise to the photometry, similar to the one measured in other photometric catalogs with abundant spectroscopy. The addition of this noise, combined with the color ``granularity'' provided by the template interpolation reproduces very well the photometric redshift properties found 
in other real catalogs, both in precision and in number of outliers (see Benitez 2014).   


\subsubsection{Empirical mocks}

  J-PAS will be the first Stage IV project to start operations, and its observing strategy is designed to take advantage of this head-start to produce competitive constraints on dark energy as early as possible after the survey start.  
In Table 6. 
we show the expected schedule of observations. 

\begin{table}[h]
\centering
\begin{tabular}{crcccccc}
\label{trays}
\textbf{Trays} & \textbf{Date} & \textbf{$N_{RG}$} & \textbf{$N_{ELG}$} & \textbf{$V_{eff}$} & \textbf{$N_{RG}^{z>0.7}$} & 
\textbf{$N_{ELG}^{z>0.7}$} & \textbf{$V_{eff}^{z>0.7}$}\\ 
\hline
T543         & Y3 & 4.6 & 33.9 & 9.5 & 0.7 & 9.4 & 5.8 \\
All          & Y6 & 17.6 & 73.1 & 13.9 & 3.7 & 19.7 & 9.9 \\
\hline
\end{tabular}
\caption{J-PAS Observational schedule. The first columns indicates how many trays are expected
  to be completed. The date indicates the number of years after we
  start. $N_{RG}$ and $N_{ELG}$ correspond to the total number of
  respectively, Red and Emission Line galaxies, in $10^6$ units. 
$V_{eff}$ is the effective volume for Power Spectrum measurements}
\end{table}

\begin{figure}[h]
\label{fig:nP}
\centering
\includegraphics[width=0.8\textwidth,keepaspectratio]{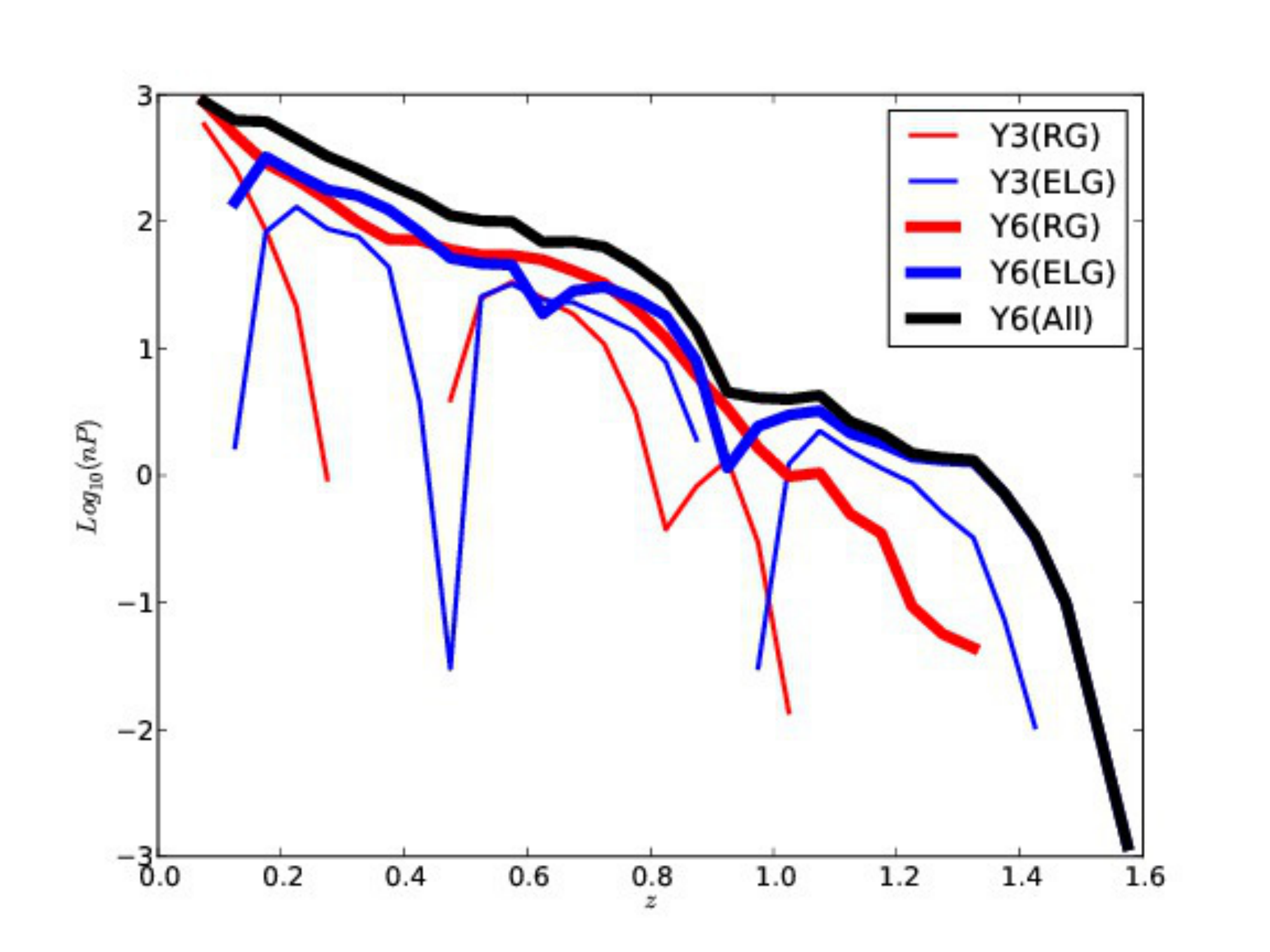}
\caption{Product of the galaxy density for Red Galaxies (RG) and 
Emission Line galaxies (ELG) with $dz/(1+z)<0.003$ 
by the power spectrum (taking into account the corresponding bias) for different stages of completion of J-PAS}
\end{figure}

  For most cosmological applications which rely on a measurement of the Power Spectrum ($P$), a crucial quantity is the number density $N(z)$ 
of different galaxy types as a function of redshift. The effective volume $V_{eff}$ for measuring $P$ increases as 
$nP/(1+nP)^2$, where $n$ is the galaxy number density. Table 6 
lists the expected values of $V_{eff}$, the resulting values of $nP$ are plotted in Fig. 13. 
We use the $P(k)$ of Tegmark et al. (2003).  

\begin{figure}[h]
\centering
\includegraphics[width=0.8\textwidth,keepaspectratio]{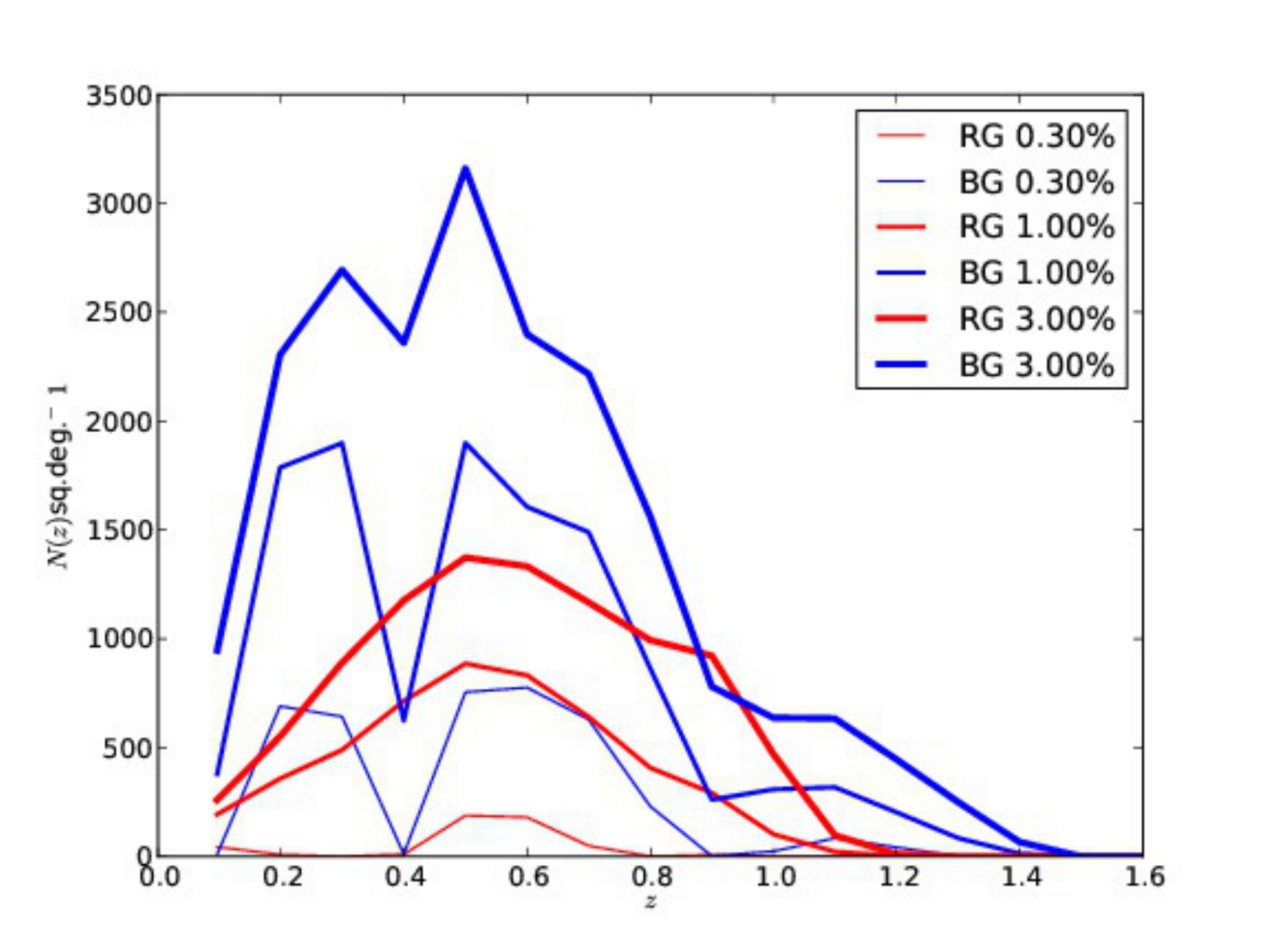}
\caption{Expected surface density of galaxies for different photometric redshift errors at Y3 (half the survey)}
\label{fig:figY3}
\end{figure}

\begin{figure}[h]
\centering
\includegraphics[width=0.8\textwidth,keepaspectratio]{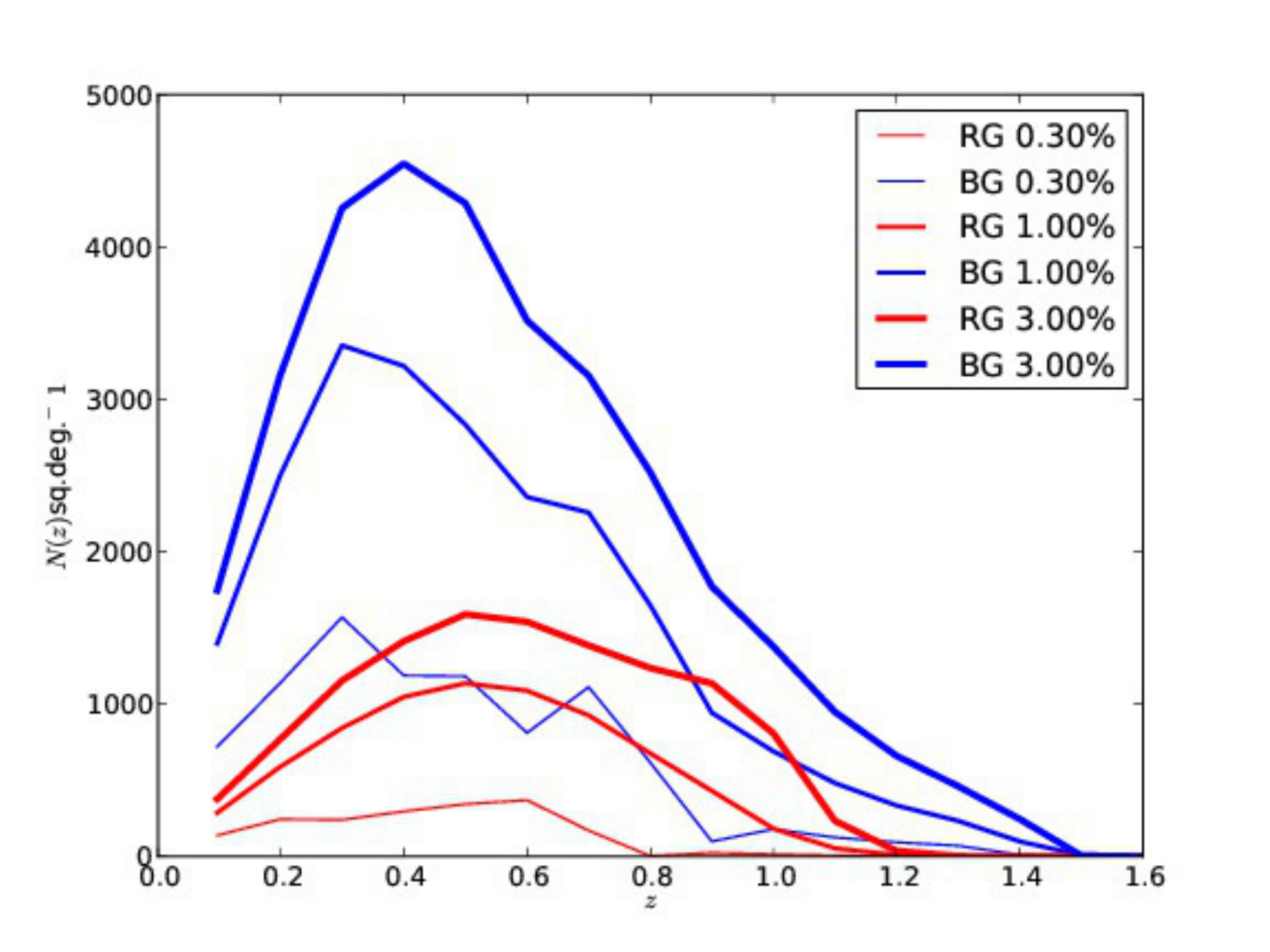}
\caption{Expected surface density of galaxies for different photometric redshift errors at Y6}
\label{fig:figY6}
\end{figure}

 J-PAS will also measure lower precision photometric redshifts for hundreds of millions of galaxies, which can be used for other scientific goals, both in Cosmology and Galaxy Evolution. Figures \ref{fig:figY3} and \ref{fig:figY6} show the expected surface density of galaxies with different photo-z precisions at Y3 (half the survey) and Y6 (end of the survey). Table 7. and 8. 
list the corresponding numbers. 

\begin{table}[h]
\label{NzY3}
\centering
\begin{tabular}{rrrrrrr}
\hline
\textbf{z} & \textbf{$N_{0.3\%}^{RG}$} &  \textbf{$N_{0.3\%}^{ELG}$} &
             \textbf{$N_{1\%}^{RG}$}   &  \textbf{$N_{1\%}^{ELG}$} &
             \textbf{$N_{3\%}^{RG}$}   &  \textbf{$N_{3\%}^{ELG}$} \\
\hline
 0.10 & 43.75 & 10.45 & 193.65 & 379.25 & 260.50 & 946.80 \\
 0.20 & 8.80 & 690.45 & 357.45 & 1786.25 & 550.70 & 2303.90 \\
 0.30 & 0.00 & 642.95 & 489.15 & 1898.45 & 890.20 & 2693.80 \\
 0.40 & 9.80 & 11.35 & 713.95 & 624.15 & 1176.70 & 2359.50 \\
 0.50 & 188.00 & 755.05 & 885.90 & 1899.65 & 1373.10 & 3161.00 \\
 0.60 & 180.50 & 776.05 & 832.30 & 1605.90 & 1332.60 & 2398.10 \\
 0.70 & 49.45 & 629.80 & 641.50 & 1489.70 & 1167.30 & 2216.90 \\
 0.80 & 0.90 & 230.75 & 407.75 & 864.40 & 994.00 & 1561.50 \\
 0.90 & 7.55 & 2.20 & 292.40 & 261.15 & 922.20 & 780.20 \\
 1.00 & 0.10 & 23.80 & 102.55 & 308.45 & 470.80 & 637.40 \\
 1.10 & 0.00 & 84.40 & 21.90 & 318.05 & 97.00 & 633.80 \\
 1.20 & 0.00 & 42.50 & 2.05 & 203.35 & 11.50 & 445.90 \\
 1.30 & 0.00 & 0.95 & 0.20 & 85.20 & 0.50 & 250.50 \\
 1.40 & 0.00 & 0.00 & 0.00 & 13.75 & 0.00 & 66.70 \\
 1.50 & 0.00 & 0.00 & 0.05 & 0.60 & 0.00 & 0.00 \\
 1.60 & 0.00 & 0.00 & 0.00 & 0.05 & 0.00 & 0.00 \\
\hline
\end{tabular}
\caption{Expected Observed Galaxy Density at Y3: 
Expected surface density of galaxies with different photometric redshift errors at Y3 (half the survey)}
\end{table}

\begin{table}[h]
\label{NzY6}
\centering
\begin{tabular}{rrrrrrr}
\hline
\textbf{z} & \textbf{$N_{0.3\%}^{RG}$} &  \textbf{$N_{0.3\%}^{ELG}$} &
             \textbf{$N_{1\%}^{RG}$}   &  \textbf{$N_{1\%}^{ELG}$} &
             \textbf{$N_{3\%}^{RG}$}   &  \textbf{$N_{3\%}^{ELG}$} \\
\hline
 0.10 & 138.80 & 720.30 & 287.65 & 1397.25 & 378.20 & 1746.70 \\
 0.20 & 244.60 & 1136.95 & 588.05 & 2498.50 & 769.10 & 3154.60 \\
 0.30 & 240.90 & 1570.30 & 843.40 & 3353.65 & 1154.30 & 4254.60 \\
 0.40 & 296.45 & 1188.15 & 1047.20 & 3218.35 & 1412.60 & 4546.30 \\
 0.50 & 344.20 & 1182.45 & 1136.95 & 2833.15 & 1589.70 & 4286.00 \\
 0.60 & 370.15 & 811.70 & 1088.45 & 2359.25 & 1539.70 & 3516.60 \\
 0.70 & 170.90 & 1111.90 & 926.35 & 2256.85 & 1385.10 & 3155.00 \\
 0.80 & 3.80 & 617.55 & 677.20 & 1649.75 & 1237.20 & 2521.60 \\
 0.90 & 24.40 & 99.65 & 431.40 & 943.85 & 1136.70 & 1772.10 \\
 1.00 & 14.95 & 179.70 & 183.55 & 690.45 & 809.00 & 1377.30 \\
 1.10 & 6.90 & 125.55 & 54.10 & 482.05 & 234.30 & 948.90 \\
 1.20 & 1.35 & 94.00 & 10.00 & 335.55 & 37.30 & 660.80 \\
 1.30 & 0.40 & 71.10 & 1.50 & 234.20 & 1.70 & 461.40 \\
 1.40 & 0.10 & 11.80 & 0.10 & 99.85 & 0.00 & 246.80 \\
 1.50 & 0.00 & 0.20 & 0.00 & 8.30 & 0.00 & 6.80 \\
 1.60 & 0.00 & 0.00 & 0.00 & 0.05 & 0.00 & 0.00 \\
\hline
\end{tabular}
\caption{Expected surface density of galaxies with different photometric redshift errors at Y6 (end of survey)}
\end{table}

\FloatBarrier

\FloatBarrier 
\section{Scientific Goals I: Cosmology}

 The J-PAS data are so versatile that the survey is, effectively, four different sub-surveys, each of which develops one of the main 
DETF Dark Energy probes. 

  Although J-PAS was initially designed to measure BAOs \citep{B2009}, still one of the main goals of the 
Survey, an imaging instrument with narrow-band filters can do much more than that. 
By tuning the instrument, the filter system and the survey strategy, we 
have been able to construct a tool unique in its capability to detect galaxy groups and clusters. 
Furthermore, the excellent conditions at the site in Pico
del Buitre (with seeing better than 0.7'') have turned an initially modest 
effort on weak lensing into a promising new survey, where not only 
the shapes of several hundred millions of galaxies will be measured, but their
redshifts will be known with high accuracy as well.  

  We will take full advantage of the fact that, by observing the
same areas of the sky hundreds of times (at least three exposures 
in each one of the 56 filters), the survey can have valuable time-domain
information. By tuning the cadence of the observations to coincide with
the typical durations of supernova explosions, we will be able to sample
their spectral surfaces (i.e., the flux as a function of wavelength and time)
in such a way that spectroscopic follow-up is unnecessary for a large 
number of objects. This refined strategy will enable us to conduct one of the most
prolific surveys of Type-Ia supernovas, with thousands of 
science-grade objects up to $z\sim 0.5$.

\subsection{The J-PAS Redshift Survey}
\label{sec:redshift}

The main feature of J-PAS, that distinguishes it from other surveys, 
is that it will achieve very high completeness while still measuring redshifts
with near-spectroscopic accuracies. As will be shown in this Section,
by combining several tracers of large-scale structure such as
luminous red galaxies (LRGs) up to $z\sim1$, emission-line galaxies 
(ELGs) up to $z\sim1.4$, and Ly-$\alpha$ emitters (LAEs) and quasars
up to $z\sim5$, we will be able to assemble a wide \emph{and} deep 
3D map of the Universe over 1/5 of the whole sky ($>8500 \sq\degr$). 

\subsubsection{Cosmology with galaxy surveys}
\label{sec:fisherm}

Galaxy surveys have evolved enormously since late 70's, when the first
maps of the local Universe were laboriously compiled from extremely
scarce resources \citep{1980ApJ...242..448Y,1982ApJ...254..437D}. 
The move from ``retail'' to ``wholesale'' began with the 
IRAS Point Source Catalog Redshift Survey
\citep{1992ApJ...397..395S}, which, despite containing only a few 
thousands of galaxies, was one of the first surveys that enabled 
cosmological applications related to large-scale structure 
\citep{1993ApJ...402...42F}.

It soon became clear that galaxy surveys could be optimized 
for cosmology in general \linebreak \citep{1997MNRAS.290..456H}, and
in particular to obtain information about the cosmological constant 
\linebreak
\citep{1996MNRAS.282..877B,1990Natur.348..705E}.
The 90's saw the first large efforts to collect massive numbers of galaxies
and other extragalactic objects, increasing previous numbers of objects 
with known redshifts by almost two orders of magnitude: 
APM \citep{APM},
the Two-degree Field survey (2dF) 
\citep{1999MNRAS.308..459F,2001MNRAS.328.1039C}
and the Sloan Digital Sky Survey (SDSS) \citep{2000AJ....120.1579Y}.
These precursor surveys were able to measure for the first time 
the clustering of structures over large scales 
\citep{2001Natur.410..169P,2002ApJ...572..140D,2003MNRAS.346...78H,sdss1}, and they allowed for the first measurement of BAOs 
\citep{eisenstein,2df,2010MNRAS.401.2148P}. 
However, these early surveys were unable 
to reach a sufficiently large volume of the cosmos in order to 
allow a measurement of the fine details of the distribution of 
large-scale structure. Hence, despite the ground-breaking checks on
the standard cosmological model that these surveys provided, their 
constraints on dark energy (particularly its equation of state, $w$) 
were not very strong.

More recently, BOSS \citep{2007AAS...21113229S} and WiggleZ
\citep{2007ASPC..379...72G,2011MNRAS.418.1707B}
both achieved high enough densities of galaxies to allow unambiguous 
detections of the BAO features on the power spectrum 
\citep{2011MNRAS.415.2892B,2012MNRAS.427.3435A,2013arXiv1303.4666A}, as well as a vast
array of other applications. The next generation of surveys such as 
DES \citep{2005astro.ph.10346T}, HETDEX \citep{HETDEX},
PAU \citep{PAU}, PFS \citep{2012arXiv1206.0737E}, 
DESI \citep{levi2013}, 4-MOST \citep{4MOST},
LSST \citep{2008arXiv0805.2366I}, and 
Euclid \citep{2011arXiv1110.3193L,2012arXiv1206.1225A},
promises to deliver further leaps in depth, image quality, photometric accuracy,
as well as in the sheer numbers of detected objects. J-PAS, in particular, 
will deliver many millions of galaxies and other extragalactic objects with 
extremely accurate photometric redshifts, over a large fraction of the 
volume of the observable Universe.

The treasure trove of possible applications of these huge datasets is 
immense, and remains mostly untapped \citep{2006astro.ph..9591A}. 
However, a critical gap between the observations and the science
applications is the optimal extraction of information from the catalog.

Given the practical limitations imposed by atmospheric conditions,
intrument performance, surveyed area and galactic contamination, all
instruments end up surveying the cosmos in an uneven way, with some regions 
better observed (and therefore better sampled) than others. Hence, 
when studying large-scale structure through the two-point correlation 
function or its Fourier transform, the power spectrum $P(k)$
\citep{1980lssu.book.....P,1992LNP...408....1P}, we must 
first overcome the angular and radial modulations in the density of 
galaxies that arise not from true fluctuations of the underlying density
field, but from varying observational conditions and instrumental 
performance. Since galaxies can be regarded as (biased) tracers of peaks
of the density field 
\citep{1984ApJ...284L...9K,1986ApJ...304...15B,1999MNRAS.308..119S},
their counts are realizations of random point processes subject to 
shot noise, hence a modulation in the average number of galaxies 
induces modulations in shot noise as well.

A more basic difficulty arises from the fact that one cannot
determine the amplitude of the spectrum with arbitrary precision 
at all scales if observations are limited to a finite volume -- this is known as 
volume sample variance, or cosmic variance.
Finite volumes can also introduce covariances between
power at different scales, and modulations in the
survey's galaxy selection function can generate further 
biases and covariances. When estimating either the two-point
correlation function or the power spectrum from real data, these 
problems should be kept under control  -- see, e.g., 
\citet{1993ApJ...412...64L,1994ApJ...424..569B,1996ApJ...465...34V,
1996ApJ...470..131S,1997MNRAS.289..285H,1998ApJ...499..555T}.

The main problem is how to balance shot noise in light of cosmic variance
in a such a way that we are able to recover the maximal amount of 
information from our catalog -- in other words, how to estimate
the power spectrum while minimizing the total covariance.

A key step forward was obtained by \citet{1994ApJ...426...23F} 
(henceforth FKP), who showed that, under the assumption of Gaussianity, 
there is an optimal weighted
average which minimizes the variance of the amplitude
of the power spectrum averaged over some volume in Fourier space. 
Given fiducial models for the matter power spectrum, 
$P(z;\vk)$, for the average number of
galaxies in our catalog, $\bar{n}(z;\hat{x})$, and for the bias
of the tracer of large-scale structure $b(z)$, the FKP weighted 
average results in an uncertainty for the power spectrum which 
can be expressed as:
\be
\label{Eq:FKP}
\left[ \frac{P}{\sigma_p} \right]^2 = \frac{V_{\vk}}{2} \,
\int d^3 {\vx} \, \left[ 
\frac{\bar{n}(z,\hat{x}) \, b^2(z)  \, P(z;\vk)}{1 \, + \, \bar{n}(z,\hat{x})  \, b^2(z)  \, P(z;\vk)}
\right]^2 \; ,
\ee
where the radial direction $|\vx|=r$ in the volume integral should be expressed 
in terms of $z$, using the fiducial model (e.g., a $\Lambda$CDM FRW model).
In Eq. (\ref{Eq:FKP}), the volume element in Fourier space around the bin $\vk$ 
is defined as $V_{\vk} = \int_{\vk} d^3 \vk /(2\pi)^3$, 
and the integral over volume is known as 
$V_{\rm eff}$, the \emph{effective volume} of the survey
\citep{1997PhRvL..79.3806T,1998ApJ...499..555T}.
Since Eq. (\ref{Eq:FKP}) expresses the inverse of a covariance, it is basically a 
Fisher information matrix \citep{1998ApJ...499..555T}.

Since the constraining power of a galaxy survey is proportional to 
the effective volume, the ideal scenario occurs when the 
product $\bar{n} \, b^2 \, P \gtrsim 1$ for the largest possible fraction
of the survey's volume. When this is the case, shot noise is subdominant, 
and it may become possible to reach the statistical limit set by cosmic 
variance -- in the extreme situation of negligible shot noise, 
$\sigma_P/P \rightarrow \sqrt{2 / V_{\vk} V_{\rm eff} }$.

The formula above can be easily applied to a catalog of galaxies of 
a single type (like luminous red galaxies, LRGs), or to a catalog containing
several different types of galaxies (LRGs, emission-line galaxies, quasars, etc.)
J-PAS is in a unique situation, in the sense that it will be able to
detect galaxies of several different types (\citet{B2009}, see also this paper), 
as well as quasars \citep{2012MNRAS.423.3251A}, in large enough 
numbers to make each one of these types of objects into suitable 
tracers of large-scale structure on their own rights. 

However, it has long been observed that the several distinct types of galaxies, as well as 
quasars, cluster in a different way \citep{dressler80}, which means
that they have different biases \citep{1984ApJ...284L...9K} with respect
to the clustering of the underlying density field. 
This bias is a manifestation of point processes that randomly associate \citep{1999ApJ...520...24D}
peaks of the density contrast $\delta = \delta\rho/\rho$ 
with galaxies of one type or another
\citep{1986ApJ...304...15B,1999MNRAS.308..119S}, in such a way
that more massive objects (with higher biases) tend to form in regions 
of higher density 
\citep{1996MNRAS.282..347M,1998ApJ...503L...9J,2000MNRAS.311..793B}.
Typically, the dependence of bias on mass and other environmental factors
translates into a dependence on the morphology and/or luminosity of the galaxy
\citep{1976ApJ...208...13D,2002MNRAS.332..827N}.

\samepage

Given the mass- or luminosity-dependence of bias, it is clearly sub-optimal to 
simply assume that all galaxies (or quasars) cluster in the same way, and 
then take some averaged bias for the whole catalog -- since this would imply a 
marginalization over the wide variations in bias, which would then lead to a 
degradation in the estimates of the power spectrum. 
When a catalog includes many types of tracers of large-scale
structure, corresponding to halos with different biases, the FKP method can be
generalized in  such a way that each inequivalent
type of tracer is taken into account in an optimal way
\citet{2004MNRAS.347..645P,2012MNRAS.420.2042A}.
By breaking the tracers into subgroups it is not only possible to 
measure the power spectrum and the bias of each individual 
species of tracers to better precision, but we can also obtain dramatic 
improvements on the measurements of redshift-space distortions and 
non-Gaussianities that extrapolate the limits imposed by cosmic variance
\citep{2009PhRvL.102b1302S,2009JCAP...10..007M,2010PhRvD..82d3515H,2011PhRvD..84h3509H,2012PhRvD..86j3513H,2013MNRAS.432..318A}. However, in order to realize
these gains it is necessary to measure these tracers in overlapping volumes, and 
in very high densities. J-PAS will in fact detect millions of different types of galaxies 
(distinguished not only by their types, but also by shapes, spectral types, etc.), as 
well as quasars, hence it will be in a unique position to take advantage of this
exciting new technique.

\subsubsection{The Fisher matrix approach}

Since $\sigma(\log P)=\sigma_p/P$, it is easy to see that Eq. (\ref{Eq:FKP})
gives the Fisher matrix in terms of the power spectrum averaged in 
some bin $V_{\vk}$ around $\vk$. 
The Fisher matrix for some parameters $\theta^i$ ($i=1,\ldots,N_p$)
which we would like to infer from
the power spectrum measured by a galaxy survey can then be written 
using the usual Jacobian for the transformation of a Fisher matrix, and the
result after summing over all the bins in Fourier space, and over the 
volume of the survey, is:
\be
\label{Eq:Fisher}
F_{i j} = 
\frac12 \, \int \frac{ d^3 \vk \, d^3 \vx }{(2 \pi)^3}
\, \frac{\partial \log P}{\partial \theta^i} \,
\left[ 
\frac{\bar{n} \, b^2  \, P}{1 \, + \, \bar{n}  \, b^2  \, P}
\right]^2
\, \frac{\partial \log P}{\partial \theta^j} 
\; .
\ee
It can in fact be shown that this result also follows directly from the 
statistics of counts-in-cells, i.e., from the Fisher matrix in ``pixel space'', where
each cell in position space is regarded as a pixel -- even in the case of multiple
species of tracers \citep{2012MNRAS.420.2042A}.

It is important to notice that the measured power spectrum has both 
angular and redshift dependence.
The power spectrum is the amplitude of the auto-correlation
of the density contrast in Fourier space, which can be expressed as
$\langle \delta(z;{\vk}) \delta^*(z;{{\vk} '}) \rangle = (2\pi)^3 \, P(z;\vk) \,
\delta_D(\vk - {\vk}')$, where $\delta_D$ is the Dirac delta-function.
However, galaxy surveys map our past light-cone in redshift space, 
so in order to infer anything from them we must 
be able to account for this redshift dependence.
 
The most obvious way in which redshift affects the power spectrum is 
through the matter growth function $D(z)$, where in linear perturbation theory 
$\delta_l(z;{\vk}) = D(z) \delta_l(0;{\vk})$, 
from which it follows that $P_l(z;\vk) = D^2(z) P_{l}(0;\vk)$.
Non-linear structure formation introduces a much more complex
dependence of the \linebreak spectrum with redshift,
\citep{1994ApJ...431..495J,1996MNRAS.280L..19P,1998MNRAS.301..797H,2000MNRAS.318..203S,2000ApJ...543..503M,2002PhR...367....1B,2006PhRvD..73f3519C,2008MNRAS.383..755A,2008PhRvD..77f3530M}.
The growth of matter fluctuations depends on the cosmological parameters,
and it could even bear \linebreak imprints of theories of modified gravity
that attempt to explain cosmic acceleration
\citep{2003PhLB..573....1D,2004PhRvD..70d3528C,2007PhRvD..75l4014C,2010RvMP...82..451S}, 
both in the linear \citep{2003MNRAS.346..573L,2005PhRvD..72d3529L,2006ApJ...648..797B,2007PhRvD..75b3519H,2007PhRvD..76j4043H}
and non-linear regimes \linebreak  \citep{2006PhRvD..74h4007S,2007PhRvD..75h4040K,2008PhRvD..77b4048L,2008PhRvD..78l3523O,2008PhRvD..78l3524O}. 

Since there are considerable uncertainties
in the exact form of the galaxy power spectrum when 
non-linearities are present, we
will impose a phenomenological cut-off in the maximal wavenumbers 
that are taken into account for the purposes of forecasting constraints on 
cosmological parameter -- typically, these non-linear scales are believed
to lie near $k_{nl} \sim 0.1 \, h$ Mpc$^{-1}$ at $z=0$. 
There is evidence that the non-linear scale is weakly dependent on redshift 
(since non-linear effects become more pronounced with time)
and on halo bias as well \citep{2007PhRvD..75f3512S}, but we will take a
more conservative approach and fix that scale to that which applies at $z=0$. 
Hence, in order to
protect our forecasts from the unknown effects of non-linear structure formation,
we will damp the effective volume (and, therefore, the Fisher matrix) by a 
an exponential factor, whose effect is to suppress the information
coming from those modes \citep{2007ApJ...665...14S,2007ApJ...664..660E}.

Since one cannot separate, in principle, cosmological redshifts from
peculiar velocities, the clustering of matter in redshift space
introduces an anisotropy in the two-point correlation function, 
$\xi(r) \rightarrow \xi_s (r_\perp,r_{||})$, where $r_\perp$ denotes 
angular distances -- across the line of sight -- and $r_{||}$ denotes
distances along the line of sight $\hat{x}$ -- which are inferred from the redshifts.
These anisotropies are also present in the power spectrum, 
$P(k) \rightarrow P_s (k,\mu)$ \citep{1987MNRAS.227....1K,1998ASSL..231..185H},
where $\mu = \hat{k} \cdot \hat{x}$ is the cosine of the angle between the 
wavenumber and the line of sight. In the
linear regime, the redshift-space power spectrum of some galaxy type
$g$ is given by:
\be
\label{Eq:RedDist}
P_{g \, , \, s} (z; k, \mu) = \left[ b_g + f(z) \, \mu^2 \right]^2 P (z;k) \; ,
\ee
where $f(z) = - d\log D(z)/d \log(1+z)$,
and $P(z;k)$ is the real-space mass power spectrum, which is assumed
isotropic. There are additional distortions arising from the 
quasi-linear and nonlinear regimes of structure formation, 
such as peculiar velocities of galaxies inside clusters. 
These can lead, e.g., to smearing of the clustering on intermediate scales
\citep{1987MNRAS.227....1K,2004PhRvD..70h3007S,2007ApJ...664..660E,2009MNRAS.396.1119C}. 
Finally, the statistics of density peaks of Gaussian fields can also 
lead to a smearing of the redshift-space clustering, even in linear perturbation 
theory \citep{2010PhRvD..81b3526D}.

The effects of nonlinear structure formation on the power spectrum,
especially at or near the BAO scale, can complicate the cosmological
exploitation of the data, however, these effects can be mitigated by
relating the velocity field to the gradient of the density field. 
Up to now, these ``reconstruction'' schemes have relied on spectroscopic
redshifts \citep{2007ApJ...664..675E,2012MNRAS.427.2132P,2012MNRAS.427.2146X,2012MNRAS.427.2168M}, 
and with the superb photo-z's of
J-PAS, we should be able to exploit the benefits of this method.

Perturbation theory and N-body simulations can help to take 
into account, or even parametrize, 
most of these effects \citep{2004PhRvD..70h3007S,2009PhRvD..80l3503T}, 
but despite recent progress in this area 
\citep{2011MNRAS.417.1913R,2011MNRAS.410.2081J}, 
many questions about the proper modeling and interpretation of RSDs 
on small ($r \lesssim 5$ Mpc) and even intermediate 
($r \lesssim 50$ Mpc) scales remain open. 
Here we assume that we will be able to model adequately the
the modes up to $ k\gtrsim 0.1 h $ Mpc$^{-1}$.

Another factor that can erase information contained in the power
spectrum is errors in photometric redshifts, which tend to smear
information along the radial (line of sight) direction 
\citep{2005MNRAS.363.1329B}. The narrow-band filter system of
J-PAS was in fact designed to measure distances down to 
$\sim 20 \, h^{-1}$ Mpc in the radial direction, and hence to detect 
features up to $k_{||} \sim 0.2 \, h$ Mpc$^{-1}$ at $z\sim1$ -- see also 
\citet{B2009}. Since the errors in the radial
direction and in redshift are related by $\sigma_{r_{||}} = c H^{-1}(z)\,  \sigma_z$,
we can factor these uncertainties into the Fisher matrix by 
damping the modes along the line of sight which are smaller than
this uncertainty. In order to take these uncertainties into account, 
we multiply the effective volume by 
$\exp \left[ - k^2_{||} \, c^2 \, H^{-2}(z) \, \sigma^2_{z, \alpha} \right]$
-- and, as discussed in Sec. \ref{survey}, the estimated redshift precision 
is $\sigma_{z,LRG} \simeq 0.003(1+z)$ for LRGs,
$\sigma_{z,ELG} \simeq 0.0025(1+z)$ for ELGs, and
$\sigma_{z,QSO} \simeq 0.0025(1+z)$ for quasars.

It should be noted that the Fisher matrix suffers from well-known 
limitations: first, it relies on a quadratic approximation to the 
likelihood function around its maximum, and on the hypothesis
that the variables of interest obey a Gaussian random distribution.
Under these assumptions, the Cram\'er-Rao theorem assures us 
that the set of constraints derived from the Fisher matrix are in fact
an upper bound -- a best-case scenario \citep{2010bmic.book...99T}.
Although non-Gaussian features in the likelihood are probably
subdominant for the sake of our analysis [at least if we manage to avoid 
the highly non-linear scales, see \citet{2011ApJ...726....7T}],
the Fisher matrix is not the appropriate tool for assessing the 
skewness of joint probabilities. This can be particularly problematic 
for parameters which are not well constrained, such as the equation 
of state of dark energy
and its time dependence \citep{2006astro.ph..9591A}. In extreme 
cases, the Fisher matrix may lead to a significant underestimation 
of the uncertainties \citep{2012JCAP...09..009W} compared to 
methods that are able to sample and to integrate the likelihood
function directly, such as Markov Chain Monte Carlo
\citep{1953JChPh..21.1087M,2002PhRvD..66j3511L}.
However, the influence of priors can have an even larger
impact on forecasts, hence for the purposes of the constraints shown 
in this paper we have chosen to employ the Fisher matrix, 
but to be conservative about priors.

Whenever necessary, we have used a fiducial model specified
in Table \ref{Table:Fiducial}.

\begin{table}
\caption{Fiducial values of basic cosmological parameters}
\begin{center}
\begin{tabular}{@{}lccccccc}
\hline
Parameter & $h$ & $\Omega_m$ & $\Omega_b h^2$ 
& $\Omega_k$ & $n_s$ & $w_0$ & $w_a$ \\
\hline
Value & 0.7 & 0.27 & 0.0223 & 0 & 0.963 & -1 & 0 \\ 
\hline
\end{tabular}
\end{center}
\medskip
We assume a flat $\Lambda$CDM FRW model consistent with the
maximum likelihood set of parameters found by the joint analysis of 
WMAP and other datasets \citep{2011ApJS..192...18K}.
For the fiducial model we also take the neutrino masses, the 
running  ($\alpha$) of the scalar spectral index, as well as 
the non-Gaussian parameter $f_{NL}$, to vanish.
\label{Table:Fiducial}
\end{table}%

\subsubsection{Baryonic Acoustic Oscillations}
\label{sec:baos}

The dynamics of dark matter, baryons and photons in the early Universe
are well understood: the theory 
\citep{2003moco.book.....D,2005pfc..book.....M,peter2009primordial}
is described by a set of Einstein-Boltzmann equations well inside the 
linear regime,
and observations of the CMB have overwhelmingly confirmed this picture 
\linebreak  \citep{1992ApJ...396L...1S,2003ApJS..148....1B,2003ApJS..148..175S,2011ApJS..192...18K,2013arXiv1303.5062P}. Before recombination ($z\sim 1100$) atoms were still 
ionized, so photons were able to transfer some of their pressure to the 
baryon fluid through scatterings with the free electrons in the plasma.
Dark matter, on the other hand, provided most of the 
gravitational drag which pulled matter into the over-dense regions, 
causing the baryons and radiation to heat up inside the gravitational wells.
This competition between gravity and radiation pressure led to acoustic
waves, which are manifested in the photons after decoupling as a series 
of peaks in the angular spectrum of CMB anisotropies 
\citep{2003ApJS..148..175S,2011ApJS..192...18K,2013arXiv1303.5062P}.

From the point of view of the baryonic matter, after recombination these waves 
freeze in, since baryons decouple from radiation at that time (in fact,
decoupling takes place slightly after recombination).
The characteristic scale of these baryonic waves is
given by the sound horizon at decoupling \citep{2003moco.book.....D}, 
and after correcting for the damped peculiar velocities of 
baryons subsequent to decoupling, the scale of frozen baryon acoustic 
oscillations is predicted to be around 150 Mpc (comoving) for the concordance 
$\Lambda$CDM model \citep{2011ApJS..192...18K}. 
After decoupling, gravity takes care of propagating this 
correlation length in the baryonic matter to dark matter, and therefore 
to the full matter transfer function \linebreak 
\citep{1998ApJ...496..605E,1998ApJ...504L..57E,1999MNRAS.304..851M,2007ApJ...664..660E}. The predicted ``wiggles'' in the matter power spectrum are in
excellent agreement with observations of large-scale structure \linebreak 
\citep{2005ApJ...633..560E,2010MNRAS.401.2148P,2011MNRAS.418.1707B,2012MNRAS.427.3435A}.

The acoustic (BAO) scale is a soft feature in the two-point correlation
function and the power spectrum:
baryons are, after all, subdominant with respect to dark matter, which
was only indirectly touched by the acoustic oscillations.
Moreover, the BAO scale is broadened since recombination due to the velocity
dispersion of baryons at the surface of last scattering, and more recently
due to mode-coupling from non-linear structure formation, such that the 
dispersion (or intrinsic smearing) of that scale is around 6\% at $z=3$, 
and about 10\% at $z=1$ \citep{2007ApJ...664..660E}.
Even with these effects factored in, the baryon acoustic scale should 
still be a bump on the two-point correlation function
at scales $80 - 120 \, h^{-1}$ Mpc -- a feature strong enough 
that it can be measured to exquisite precision in the next few years.

BAOs are therefore manifested in the matter distribution as a signature 
correlation length that can be accurately and confidently predicted on
the basis of known physics in the linear regime, and given parameters
that are measured in a completely independent way, by, e.g, the CMB. 
This length scale constitutes a {\em statistical standard ruler} which 
can be directly measured by mapping galaxies in the Universe,
and computing their two-point correlation function
\citep{2003ApJ...594..665B,seo}.

Distances between pairs of galaxies can occur in the radial 
direction ($r_{||}$), or in the angular direction ($r_\perp$), and 
each one corresponds to a different cosmological distance.
In the angular direction, two galaxies at a redshift $z$ that are 
separated by an angle $\delta\theta$ are a distance $d_a(z) \, \delta \theta$
apart. On the radial (line-of-sight) direction, two galaxies at the same
angular position in the sky, separated by a redshift $\delta z$, are
a distance $c \, H^{-1} \, \delta z$ apart. In a flat FRW model where
the equation of state of dark energy is parametrized as 
$w(z) = w_0 + w_a z/(1+z)$
\citep{2001IJMPD..10..213C,2003PhRvL..90i1301L},
the angular distances are given by:
\be
\label{Eq:da}
d_a(z) = \frac{c}{1+z} \int_0^z \frac{dz'}{H(z')} \; .
\ee
where the Hubble parameter at any redshift is (neglecting
the contribution from relativistic particles):
\be
\label{Eq:Hz}
H(z) = H_0 (1+z)^{3/2}
\sqrt{ \Omega_m + (1-\Omega_m) (1+z)^{3(w_0+w_a)}
e^{-3 \,w_a \, z/(1+z)} } \; .
\ee
Clearly, this direct dependence of the radial and angular 
distances on the equation of state and its time variation 
implies that BAOs are a superb tool to study dark energy
\citep{seo}. In fact, radial BAOs are
slightly superior to angular BAOs, since the latter involve
one further integration over redshift.

The scale of BAOs can be measured in the angular and 
radial directions, and that scale should be the same either way
(in position space), hence a direct comparison of the two constitutes a cross-check
that can lead to further constraints on the parameters 
\citep{1979Natur.281..358A}. We can easily include this additional information 
in our Fisher matrix, however, this presupposes that we can model 
RSDs accurately, without introducing systematic
errors, biases, etc. \citep{2003PhRvD..68f3004H}. N-body simulations 
seem to indicate that it is indeed possible to separate, at the level of
the Fisher matrix, the measurements of the scale of BAOs from the information
contained in the RSDs \citep{2007ApJ...665...14S}. 
One could even use the consistency condition that the peculiar velocity field is
determined by the gradient of the gravitational potential, and partially
``reconstruct'' the linear correlation function 
\citep{2007ApJ...664..675E}. This reconstruction scheme can lead to 
better constraints not only on the BAO scale 
\citep{2012MNRAS.427.2132P,2012MNRAS.427.2146X,2012MNRAS.427.2168M}.

The concrete method that we employ
to compute the Fisher matrix is that of \citet{2010MNRAS.409..737W}, 
which is itself an adaptation of the approach of 
\citet{seo,2007ApJ...665...14S}. 
The basic idea is to write the linear theory, redshift-space power 
spectrum explicitly in terms of the distances measured in the angular and radial 
directions, and then use the property that, for cosmologies near the
fiducial one, the power spectrum transforms as the Jacobian between the 
volumes elements for the two cosmologies, 
$d\bar{V}/dV = (\bar{d}_a/d_a)^2 H/\bar{H}$, where we indicate quantities
evaluated at their fiducial values with a bar. After including
an extra shot noise term \cite{2000MNRAS.318..203S}, we have:
\be
\label{Eq:ObsPk}
P_{obs}(z;k,\mu) = \left[ \frac{d_a(z)}{\bar{d}_a(z)} \right]^2
\frac{\bar{H}(z)}{H(z)} \, P_{g}(z;k,\mu) + P_{shot}(z) \; ,
\ee
where:
\begin{eqnarray}
\label{Eq:Plin}
P_{g}(z;k,\mu) 
&=& D^2(z) \left[ b_g(z) + f (z) \, \mu^2 \right]^2 \, P_{l}(0;k)
\\ \nonumber
&=& \left[ b_{s}(z) + f_s (z) \, \mu^2 \right]^2 \, \frac{P_{l}(0;k)}{\sigma_8^2} \; .
\end{eqnarray}
We have absorbed the growth factor, as well as the normalization of the
spectrum (expressed through $\sigma_8$) into the effective bias,
$b_s(z) = b_g(z) D(z) \sigma_8$, and into an effective RSD parameter,
$f_s(z) = f(z) D(z) \sigma_8$. Here 
$P_l(0,k)$ denotes the linear theory, position 
space power spectrum at $z=0$, which can be computed for almost 
any type of cosmological model with the help of the available 
CMB Einstein-Boltzmann codes \citep{Seljak:1996is,Lewis:1999bs}.

In Eq.s (\ref{Eq:ObsPk})-(\ref{Eq:Plin}) the wavenumbers in the fiducial model 
are related to those in a general cosmology, by:
\be
\label{Eq:ks}
k_{||} = \bar{k}_{||} \frac{H(z)}{\bar{H}(z)} \quad , \quad
k_{\perp} = \bar{k}_{\perp} \frac{\bar{d}_a(z)}{d_a(z)} \; ,
\ee
and the properties of $k$ and $\mu$ under changes in the cosmological model 
follow from their definitions,
$k=\sqrt{k_{||}^2 + k_{\perp}^2}$, and $\mu^2 = k_{||}^2/k^2$.

In order to take into account the smearing due to non-linear structure formation
we adopt the procedure of \citet{2007ApJ...664..660E}, which distinguishes
between the radial modes (which inherit additional non-linear effects through
the redshift-space distortions) and the angular modes. The non-linear scales 
in the angular direction are given by $\Sigma_\perp = D(z) \Sigma_0$, and 
those on the radial direction by $\Sigma_{||}=D(z) ( 1+f ) \Sigma_0$, where 
$\Sigma_0$ is the baseline non-linear scale, which we assume to be
$\Sigma_0 = 10$ h$^{-1}$ Mpc.
With these definitions, the Fisher matrix, Eq. (\ref{Eq:Fisher}), becomes:
\beq
\label{Eq:Fishij}
F_{ij} &=& \frac12 
\int_{k_{min}} \frac{d^3 {\vk}}{(2\pi)^3}
\int d^3 {\vx}
\, \frac{\partial \log P_{obs} }{\partial \theta^i} 
\, \frac{\partial \log P_{obs} }{\partial \theta^j}
\\ \nonumber
& \times &  
\left[ 
\frac{\bar{n} \, P_{obs}}{1 \, + \, \bar{n} \, P_{obs}}
\right]^2 \, e^{-k^2 \Sigma_\perp^2 - k^2 \mu^2 (\Sigma_{||}^2-\Sigma_\perp^2) } 
\, e^{-k^2 \mu^2 c^2 H^{-2} \sigma_z^2} \; ,
\eeq
where $\sigma_z$ is the photometric redshift error. 
Notice that, to a good approximation, the smearing due to
photometric redshift errors only affects the radial modes.
The lower limit in the integration over $k$ in Eq. (\ref{Eq:Fishij}) is defined
as $k_{min}=(V_0)^{-1/3}$, where $V_0$ is the volume of the smallest
redshift slice (see below), and its role is to ensure that we do not integrate 
over modes which correspond to scales larger than the typical size of our redshift 
slices.

In order to minimize the effects of systematic errors in the distance
measurements, we adopt the procedure first suggested by 
\citet{seo}, and separate our survey into several redshift
bins, where on each bin we regard $d_a$ and $H$ as free variables
(to be determined from the data). Moreover, we also regard the bias, $b$, 
the redshift distortion parameter, $f$, and the unknown extra shot noise, 
$P_{shot}$, as independent parameters on each redshift slice. 
This means that, besides the
``global'' cosmological parameters such as $\Omega_k$, $h$, etc., we
have another $5 N_{bin}$ free parameters, where $N_{bin}$ is the number 
of redshift bins. Notice that, in the volume integration of Eq. (\ref{Eq:Fishij}),
each slice only contributes to the slice-dependent parameters that belong to 
that same slice, since $d \log P_{obs}[z_n]/d\theta[z_{n'}]=0$ 
when $n\neq n'$.

Our basic set of parameters is, therefore:
\be
\label{Eq:ParFish}
\theta = \left\{ d_a^n  \, , \,  H^n  \, , \, b^n  \, , \, f^n  \, , \, P_{shot}^n \right\}
+ \theta_g \; ,
\ee
where the superscript $n$ refers to the redshift bins, and $\theta_g$ represents
the set of $N_g$ global cosmological parameters which determine the shape of
the power spectrum. In full generality, these global
parameters include the physical densities of cold dark matter ($\omega_c=\Omega_c h^2$)
and baryons ($\omega_b=\Omega_b h^2$), the Hubble parameter ($h$),
the spatial curvature ($\Omega_k$)
scalar spectral index ($n_s$), the amplitude of the spectrum ($A$), the sum of neutrino 
masses ($m_\nu$), and the non-Gaussianity parameter ($f_{NL}$, which enters
as a scale-dependent bias):
\be
\label{Eq:CosmPar}
\theta_g = \left\{ h \, , \, \omega_c \, , \, \omega_c , \, \Omega_k \, ,
n_s \, ,\, A \, , \, m_\nu \, , \, f_{NL} \right\} \; .
\ee
Hence, the total number of parameters is $5N_{bin}+8$.

Notice that the parameters that describe the dark energy equation of state,
$w_0$ and $w_a$, are conspicuously absent from this list -- as are also 
missing any parameters which could hint at modified gravity models, like 
the phenomenological growth parameter 
$\gamma$ in $f(z) \simeq \Omega_m^\gamma(z)$, 
with $\gamma \simeq 0.55$ in General Relativity 
\citep{1980lssu.book.....P,1998ApJ...508..483W,2007APh....28..481L}. 
This is because we are assuming that the equation of state and its time 
variation will be inferred only from the measurements of angular and 
radial distances (i.e., from the BAO scale measured at each redshift). 
Likewise, parameters such as $\gamma$ will be measured only using
information from the shape of RSDs and the BAOs.

We employ bins of $\Delta z =0.2$, 
starting from $z=0.2$, so the bin $n$-th is centered on the
redshift $z_n = 0.1(2n+1)$. Given the expected number of objects
forecasted in Sec. \ref{survey}, the last bin is $n=5$ for the case of LRGs, 
$n=6$ in the case of ELGs, and $n=20$ in the case of quasars.
Table \ref{Table:zbins} presents the redshift bins, and their 
corresponding physical dimensions for our fiducial model. It is clear that the typical length 
scales of each redshift bin are much larger than the BAO scale.

\begin{table}
\caption{First 10 redshift bins, their central redshifts, angular distances, 
values of the Hubble radius, and comoving volumes per unit area, 
for the fiducial cosmology 
(a flat $\Lambda$CDM model with $\Omega_m=0.27$).}
\begin{center}
\begin{tabular}{@{}lcccccc}
\hline
Bin & $z$ & $d_a$ ($h^{-1}$ Gpc) & $\sigma(d_a)/d_a$ (ELGs) 
& $cH^{-1}$ ($h^{-1}$ Gpc) & $\sigma(H)/H$ (ELGs)
& V ($h^{-3}$ Gpc$^3/\sq\degr$)\\
\hline
1 & 0.3 & 0.647 & 0.03  &  0.384 &  0.058 & $1.14 \times 10^{-4}$ \\
2 & 0.5 & 0.891 & 0.019  & 0.427 & 0.035 & $2.55 \times 10^{-4}$ \\
3 & 0.7 & 1.046 & 0.015  &  0.478 & 0.026 & $4.76 \times 10^{-4}$ \\
4 & 0.9 & 1.144 & 0.016  & 0.540 & 0.025 & $5.37 \times 10^{-4}$ \\
5 & 1.1 & 1.203 & 0.017  & 0.600 & 0.024 & $6.49 \times 10^{-4}$ \\
6 & 1.3 & 1.236 &  0.021 &  0.668 &  0.03  & $7.37 \times 10^{-4}$ \\
7 & 1.5 & 1.251 &   -- & 0.742 &   -- & $8.03 \times 10^{-4}$ \\
8 & 1.7 & 1.253 &   -- & 0.820 &  -- & $8.51 \times 10^{-4}$ \\
9 & 1.9 & 1.247 &    -- & 0.902 &  -- & $8.83 \times 10^{-4}$ \\
10 & 2.1 & 1.234 &   -- & 0.988 &  -- & $9.04 \times 10^{-4}$ \\
\hline
\end{tabular}
\end{center}
\medskip
{}
\label{Table:zbins}
\end{table}%

Our fiducial model for the bias is as follows: for luminous red galaxies (LRG), we adopt
$b_{LRG} = 1.8$; for emission-line galaxies (ELG) we assume
that $b_{ELG}=0.9+0.4 \, z$; and for quasars (QSO) we assume that
$b_{QSO} = 0.5 + 0.3(1+z)^2$ -- see, e.g., \citet{2009ApJ...697.1634R}.

The series of steps leading to constraints which employ only information 
from BAOs, with or without including RSDs, 
has been described in detail by \citet{seo} and by 
\citet{2006ApJ...647....1W,2010MNRAS.409..737W}. 
In addition to the Fisher matrix above, we also include priors, such as 
the constraints forecasted for the \Planck satellite arising from temperature anisotropies 
\linebreak  \citet{2008PhRvD..78h3529M}.
In particular, the \Planck Fisher matrix, $F^{(Planck)}_{ij}$, serves to calibrate the 
absolute scale of BAOs, as well as to limit the allowed ranges for the parameters
that affect the shape of the power spectrum.

Our procedure is as follows:

\begin{enumerate}

\item{Add \Planck priors for the BAO scale to the full Fisher matrix of Eq. (\ref{Eq:Fishij}), 
to obtain the survey's Fisher matrix with priors, $F^p_{ij} =  F_{ij} + F^{(Planck)}_{ij}$.}

\item{Marginalize\footnote{By marginalization we mean, concretely:
(a) invert the full Fisher matrix to obtain the covariance matrix; 
(b) eliminate the lines and columns from the covariance matrix that 
correspond to the marginalized parameters; (c) invert the reduced 
covariance matrix to obtain the marginalized Fisher matrix.}
the bias on each slice, $b^n$, the extra shot noise term, $P_{shot}^n$, as well
as the global cosmological parameters, from the Fisher matrix $F^p_{ij}$.
This intermediate marginalized Fisher matrix, $F^m_{ij}$, has $3 N_{bin}$ parameters.}

\item{Project the marginalized Fisher matrix  $F^m_{ij}$
into the final set of cosmological parameters -- which includes, 
naturally, the dark energy parameters $w_0$ and $w_a$.}

\end{enumerate}

For a more conservative approach, we can also marginalize against the RSD 
parameters $f^n$ on each slice, which corresponds to assuming that the form and
redshift dependence of these distortions are completely unknown. In that case, the final
step is identical to $(iii)$ above, except that now the marginalized 
Fisher matrix has $2N_{bin}$ parameters.

\subsubsection{Forecasts for constraints from BAOs}

\begin{figure}
\centering
{\includegraphics[width=10cm]{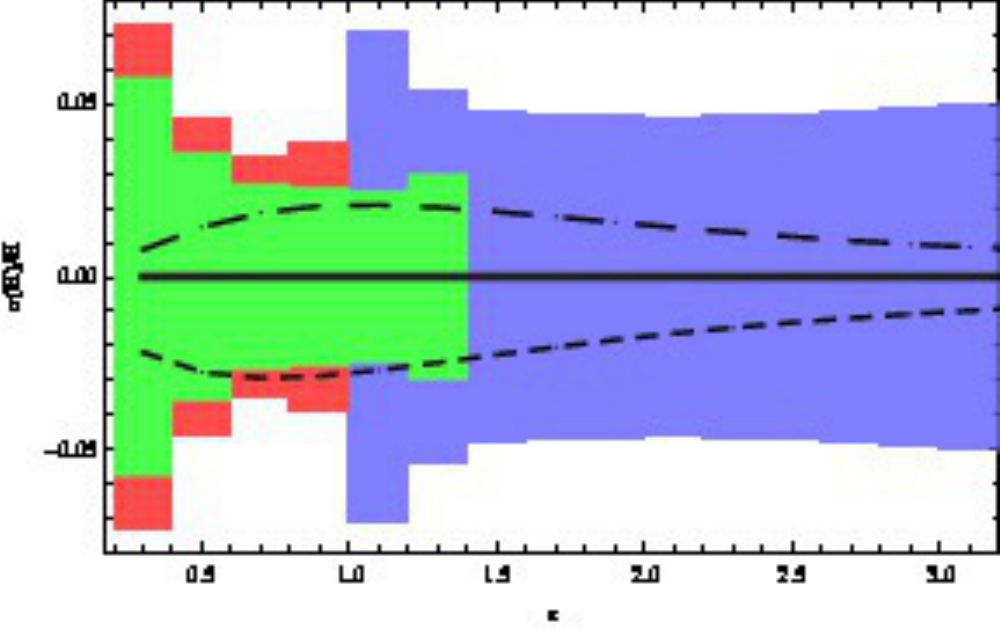}\hfill
\includegraphics[width=10cm]{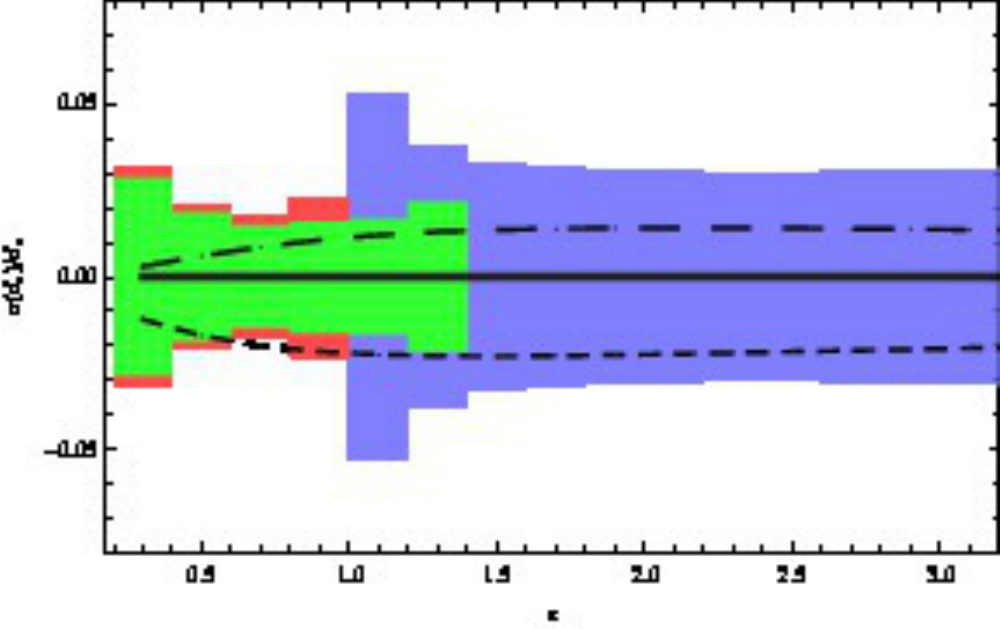}}
  \caption{\label{fig:distance_measurements}
Relative uncertainties in the determination of the Hubble parameter $H(z)$ and
angular-diameter distance $d_a(z)$ on each redshift slice, relative to our
fiducial $\Lambda$CDM model. The red boxes 
(which reach up to $z\sim1$) correspond to the constraints imposed by red galaxies 
alone; the green boxes, to the constraints from emission-line galaxies; and the blue 
boxes, which extend to higher redshifts, correspond to the constraints from quasars.
Also indicated are two alternative dark energy models, one with $w_0=-0.9$ and
$w_a=0$ (upper, long-dashed lines), and another with $w_0=-1.0$ and
$w_a=-0.3$ (lower, short-dashed lines)
}
\end{figure}

The BAO scale provides a statistical standard ruler. This means that
the power spectrum (or, equivalently, the 2-pt correlation function) have a 
characteristic scale. Eq. (\ref{Eq:ObsPk}) shows that measurements of
that scale in the power spectrum inferred from data on a particular redshift 
can be translated into estimates of the Hubble parameter, $H$ and of
the angular-diameter distance, $d_a$, for that redshift.
In Fig.  \ref{fig:distance_measurements} we show how J-PAS can constrain
the radial distance and the angular distance to a given redshift, using three
different types of tracers: red galaxies (RG), emission-line galaxies (ELG), and
quasars (QSO).

\begin{figure}
\centering
\includegraphics[width=0.8\textwidth]{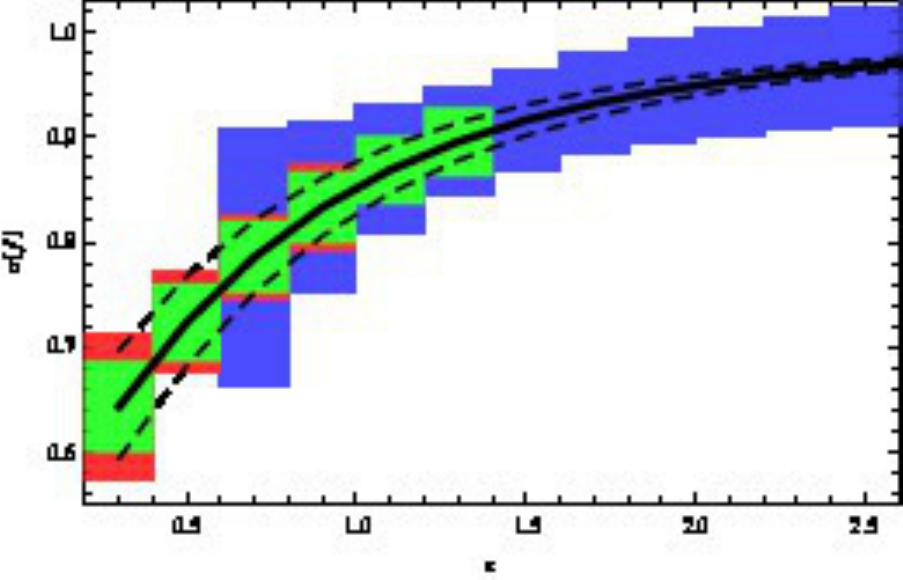}
  \caption{\label{fig:RSDs}
Redshift-distortion (RSD) parameter $f(z)$ for the fiducial $\Lambda$CDM + GR model, and
uncertainties in its measurement on each redshift slice. For GR, the RSD
parameter is very well fitted by $f = \Omega_m^\gamma(z)$, with $\gamma=0.55$.
Also indicated are two hypothetical modified gravity models, with $\gamma_0=0.45$ 
(lower dashed line) and with $\gamma=0.65$ (upper dashed line). Legends are the same 
as for Fig. \ref{fig:distance_measurements}
}
\end{figure}

In Fig.  \ref{fig:RSDs} we present the expected RSD parameter for
our fiducial model, as a function of redshift, and the 
uncertainties estimated for J-PAS. 
It is useful to employ a fit for the RSD that allows us to explore
modified gravity models, and a popular parametrization is $f = \Omega_m^\gamma(z)$, 
where $\gamma=0.55$ for General Relativity (GR).

\paragraph{Neutrinos}

Since neutrinos are only weakly interacting, they can stream freely away from
hot and dense regions much before the time of recombination, which
means that they can carry away some of the structure on small scales. 
Since the effect of neutrinos is most pronounced on scales 
$k \gtrsim 0.02$ $h$ Mpc$ {-1}$, the constraints are particularly sensitive to 
non-linear effects and non-linear bias. 

Assuming that these uncertainties from
non-linear effects only spoil the scales $k > 0.1$ $h$ Mpc$ {-1}$, we 
forecast that J-PAS will be able to constrain the total neutrino mass, $M_\nu = \sum m_{\nu}$, at
the level of $M_\nu \lesssim 0.3$ eV ($1 \sigma$) with ELGs alone. 
This result was obtained using a fiducial value of
$M_\nu \lesssim 0.05$ eV. By combining information from RGs and QSOs
as well, we estimate that J-PAS will be able to improve upon these constraints by 
$\sim 30$ \%.

\paragraph{Spatial curvature}

\begin{figure}
\centering 
\includegraphics[width=0.8\textwidth]{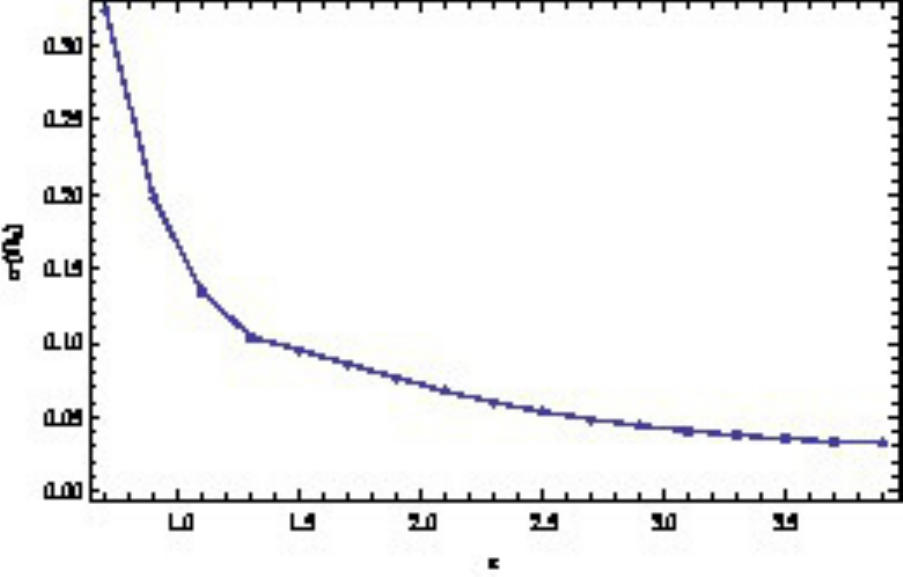}
  \caption{\label{fig:Omegak}
Cumulative constraints on the spatial curvature, as expressed by the
parameter $\Omega_k$. 
}
\end{figure}

\Planck can already constrain the spatial curvature of the Universe to better
than 1\%. However, J-PAS will be able to limit the spatial curvature 
independently from CMB data, to a precision of a few percent. This comes 
basically from comparing clustering in the radial and in the angular directions:
whereas the radial distances are barely influenced by spatial
curvature, the angular distances are strongly affected by the geometry, and comparing
the two allows us to extract the spatial curvature. In Fig. \ref{fig:Omegak} we show 
how the constraint on $\Omega_k$ evolves as we include more redshift slices.
The information from high-redshift quasars is particularly important to break low-redshift
degeneracies, and improve the constraint to a level of $\sim 5 \%$.

\paragraph{Figure of merit}

The distance measurements can be converted into constraints on the cosmological
parameters. This is achieved by projecting the Fisher information 
matrix from all redshift slices, with parameters 
$\theta^i = \{ H(z), d_a(z) \}$, into the set of cosmological parameters, which in our case is
$\tilde{\theta}^a = \{ \omega_0, \omega_a , \Omega_k, \Omega_\lambda\}$. The Fisher
matrix for this final set of parameters is then:
\be
\label{Eq:Proj_distances}
F_{ab} = \sum_{ij} \frac{\partial \theta^i}{\partial \tilde\theta^a} F_{ij}
\frac{\partial \theta^j}{\partial \tilde\theta^b} \; ,
\ee
where the summation above includes all redshift slices which have any information
about distances.

Since $w_0$, the value of the equation of state at $z=0$, is highly
correlated with the value $w_a$ of its time derivative in the CPL parametrization,
it is useful to find the redshift for which the value of the equation of state is
independent of its time derivative. The error in the pivot, $w_p$, is therefore
independent of the error in $w_a$. In Table \ref{Table:FoMs} we present the 
uncertainties in the equation of state expressed in terms of the pivot, its time
derivative, and, for completeness, that of $w_0$ as well. In order to help break some degeneracies,
we have also used \Planck priors and Stage-II priors, as defined in \citet{2006astro.ph..9591A}.
In the last column the dark energy figure of merit, defined
as FoM $= \sigma^{-1}(w_p) \sigma^{-1}(w_a) = \det^{-1} F(w_0,w_a)$.
As a comparison, the latest constraints from BOSS/CMASS \citep{2013MNRAS.tmp.1476S},
and including information from CMB, supernovas and other existing BAO surveys, 
has a FoM $\simeq 25$. J-PAS is the only survey which has the capability to
deliver a FoM $\simeq 100-200$ over the next 5-8 years.

\begin{table}
\caption{Summary of constraints}
\begin{center}
\begin{tabular}{@{}lccccc}
\hline
Tracer & LRGs & ELGs & QSOs & All \\
\hline
\hline
$\sigma(w_p)$ & 0.030 & 0.026 & 0.027 & 0.023 \\
$\sigma(w_a)$ & 0.38 & 0.31 & 0.37 & 0.26 \\
$\sigma(w_0)$ & 0.08 & 0.07 & 0.09 & 0.06 \\
FoM & 87 & 121 & 100 & 164 \\
\hline
$\sigma(M_\nu)$ & 0.39 & 0.28 & 0.29 & 0.2 \\
$\sigma(f_{NL})$ & 6.9 & 10.3 & 3.1 & 1.9 \\
$\sigma(\gamma)$ & 0.1 & 0.06 & 0.11 & 0.05 \\
$\sigma(\Omega_k)$ & 0.28 & 0.12 & 0.05 & 0.03 \\
\hline
\end{tabular}
\end{center}
\medskip
{}
\label{Table:FoMs}
\end{table}%

\pagebreak 
\subsubsection{Field reconstructions from the redshift survey}

Reconstructions of the underlying smooth matter field,
such as, e.g., techniques based on efficient Bayesian inference methods 
\citep[][]{JascheKitaura2010,Kitauraetal2012a},
are useful for a variety of reasons.
(1) It has recently been shown that the linearized cosmic density field recovers information about the primordial fluctuations, leading to better constraints on cosmological parameters 
\citep[see e. g.][]{Neyrinck2009,Joachimi2011,Kitauraetal2012b}.  
(2) It can be used to obtain estimates of the primordial fluctuations by undoing the effects of 
gravity, by such methods 
as inverting the Lagrangian to Eulerian mapping, can be effective in order to reduce the errors in the 
measurements of the location of the BAO peak from spectroscopic redshift surveys \citep[see e. g.][]{2007ApJ...664..660E,Seo2010,Padmanabhan2012,Mehta2012}.
(3) These approaches also allow us to find the cosmic web corresponding to a distribution of matter 
tracers down to an accuracy of a few Mpc \citep[see][]{Kitauraetal2012d,Hess2013}. 
(4) Another interesting application is to perform constrained simulations of the observed 
Universe \citep[see e. g.][]{Hess2013,Wang2013,Dolag2005,Klypin2003,Mathis2002}.
(5) Finally, under certain assumptions, these methods can be used to reduce the impact of photometric 
redshift uncertainties in the estimation of galaxy clustering \citep[][]{Kitaura2008,Jasche2012}. 
J-PAS, with its high density of tracers, should be an ideal dataset where these
methods could be applied.

The number of studies one can perform with the new level of precision in the reconstruction of 
both Eulerian and Lagrangian space go beyond BAO measurements and cosmological 
parameter estimation. One can generate templates for the detection of weak signals, such as 
the WHIM \citep[see e. g.][]{Suarez2013}, the ISW effect \citep[see e. g.][]{Granettetal2009}, 
the kSZ effect  \citep[see e. g.][]{Dedeoetal2005,Hoetal2009}, the cosmic ray signal 
\citep[][]{Dolag2005}  or the dark matter annihilation signal \citep[][]{Cuesta2011}. 

The reconstruction of the cosmic web can help to understand the process of structure formation 
and the importance of the environment in the formation of clusters and galaxies 
\citep[see e. g.][]{Aragon-Calvo2007,Hahn2007,Forero-Romero2009,Libeskind2011,Tempel2011,Tempel2013,Benitez-Llambay2013}.

\subsubsection{Morphology of the Cosmic Web}

 The redshift space distribution of galaxies reveals that galaxies are preferentially distributed 
 in a network frequently referred to as the cosmic web. Ever since the discovery 
 of the CfA wall it had been acknowledged that sheets, filaments and voids are the key 
 components of the cosmic web. This point of view has been supported by the discovery of 
 the Great Wall at $z \simeq 0.08$ in the SDSS survey. The Great Wall is the largest 
contiguous distribution of matter in the currently
observable universe and its discovery leads to the tantalizing issue of whether the wall is the densest large structure in the universe, or whether similar-size structures would be abundant in larger galaxy surveys. The issue of whether or not the Great Wall is an unusual object is important for cosmology, since models of a 
highly inhomogeneous universe have been advanced arguing that, since the dimming of high redshift supernovae can be explained in such models, the presence of a smooth DE component may be rendered unecessary~\cite{celerier,tomita,hunt,foreman,nadathur}.

 A large and deep redshift survey such as J-PAS should be able to shed some useful light on this issue by allowing one to compile an enormous data base of superclusters and voids, which would supplement the one currently available from SDSS. This would help quantify a key property of the supercluster-void network: its morphology. It is well known that standard statistical tools such as the two-point correlation function cannot 
 reveal information about the connectedness of large-scale structure and must therefore be supplemented by geometrical indicators such as percolation analysis, the genus curve, minimal spanning trees and Minkowski functionals~\cite{2}. Ratios of Minkowski functionals, known as Shapefinders, can enable one to answer the question as to whether a given supercluster or void is filamentary, planar or spherical~\cite{3}. 
 Two recent papers have applied the Shapefinders to a catalogue of superclusters and voids in the SDSS survey including a detailed analysis of the morphology of the Great Wall~\cite{4}. In addition, extreme value statistics has been used to ask whether the Sloan Great Wall is an unusual object -- the answer to which seems to be in the affirmative~\cite{5} . Clearly improved deeper datasets are needed to resolve these important issues and the J-PAS survey could play a key role in quantifying supercluster-voids morphology.

\subsubsection{N-body Mocks} 

\FloatBarrier
The complexity of the data processing and scientific exploitation of J-PAS will
demand realistic synthetic observations. We will construct mock universes that
will help in the testing and development of analysis codes, data reduction
pipelines, and in the verification of the forecasts presented in this document.
In addition, the mock observations will assist the science analysis through
comparison with theoretical models, but also these theoretical models will be
constrained by J-PAS observations.

We plan to create these mock universes in a three-step process. First, the
nonlinear mass content of the universe will be given by following the
gravitational interaction of particles in $N$-body simulations \citep[see the recent review of][]{Kuhlen2012}. Second, the properties of galaxies inside
dark matter haloes will be predicted by using i) semi-analytic models of galaxy
formation \citep[e.g.][]{Lagos2012,Henriques2013}, ii) sub-halo abundance
matching \citep[e.g.][]{Vale2004,Zehavi2011}, or iii) empirical rules based on local background
density \citep[e.g.][]{White1987,Cole1998}. Of particular importance for J-PAS will be
addition realistic photometric redshift estimates for each galaxy in the
simulations, which will help to improve and calibrate the relevant algorithms.
Finally, the J-PAS footprint, selection function, redshift completeness, flux
limit, gravitational lensing effects, among other, will be added in
post-processing to the simulated galaxy population.  As a result, we will
deliver a fake but realistic J-PAS survey prior to the arrival of data.

The characteristics of J-PAS requires to simulations covering a few tens of
cubic Gigaparsecs in volume, and with a mass resolution sufficient to robustly
identify haloes of at least $M \sim 10^{10}\,M_{\odot}$. In addition, an
adequate temporal resolution is needed in order to resolve the mass accretion
and merger history of dark matter structures. These specifications are very
demanding in terms of computational resources, and are not met by any of the
state-of-the-art simulations \cite[e.g.][]{Angulo2012,Watson2013}. Therefore, we
plan carry out a dedicated simulation program targeting the desired features,
and also spanning the range of cosmological parameter space currently allowed
by cosmological datasets, and with varying assumptions about the physical
processes affecting simulated galaxies.

Another important aspect of the J-PAS simulation program will be the
construction of accurate covariance matrices for the cosmological
interpretation of data, in particular of the galaxy clustering,
abundance of clusters and gravitational lensing signal. These goals require
thousands of realizations of the cosmological observation. Since
carrying out a large number of direct N-body simulation is beyond current
computational capabilities, we plan the use of approximate methods
\citep[][]{Angulo2013, Kitaura2013, White2013}, which are built on top of an
ensemble of low-resolution $N$-body simulation. All the aspects discussed will
contribute towards the accuracy and correctness of analyses of J-PAS data.

\begin{figure}[h] 
\centering
\includegraphics[width=0.8\textwidth]{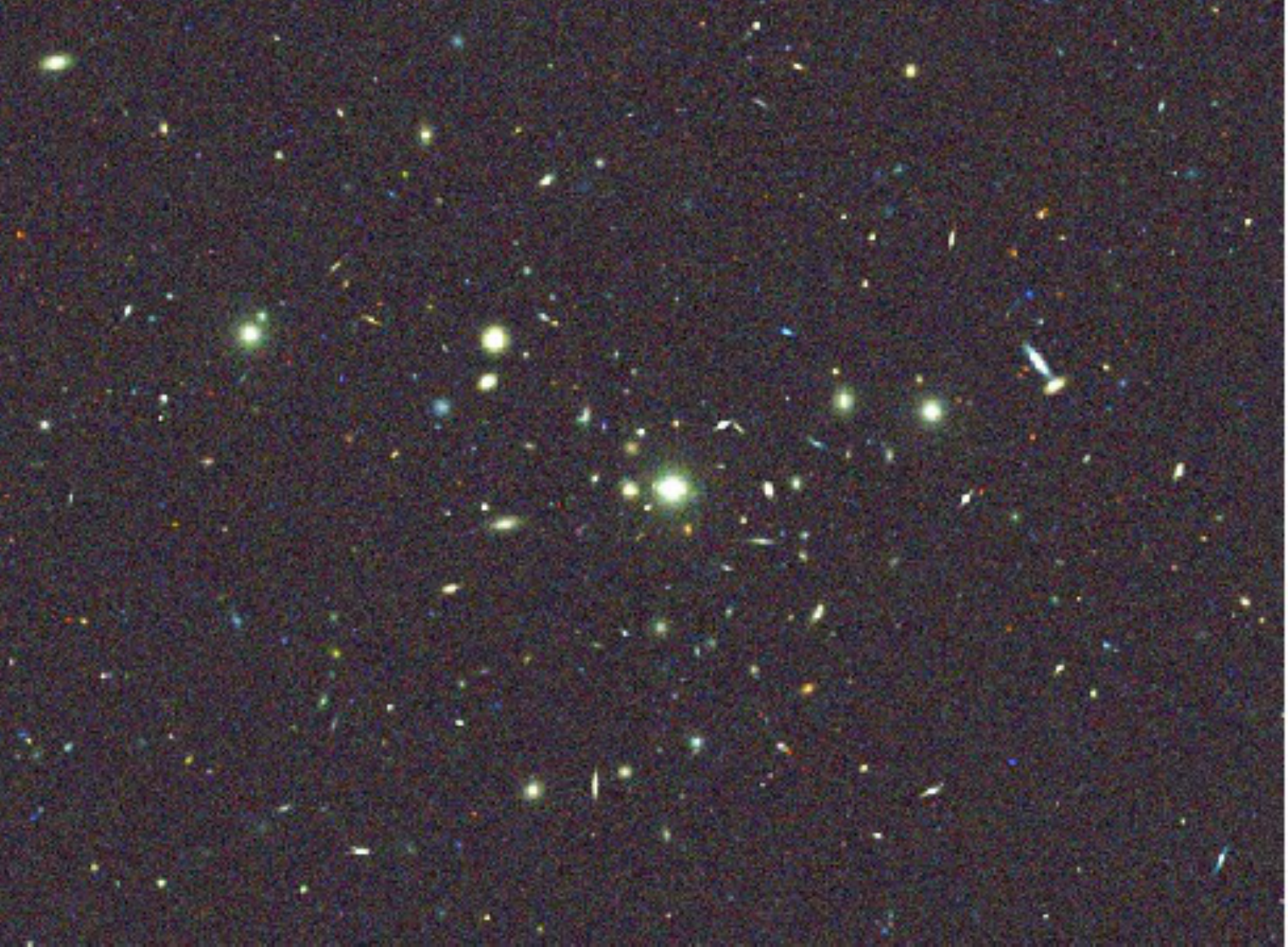}
\caption{A simulated J-PAS image in the filters $g$ (blue), $r$ (green), and
$z$ (red), produced using the Millennium Run Observatory (see text for
details).} 
\label{fig:mrobs}
\end{figure}

The comparison between the observations from J-PAS and cosmological predictions
from $N$-body simulations and semi-analytic models will be extended by creating
simulated data products that more directly correspond to the actual
observations, namely synthetic images and extracted source catalogues. We will
use the Millennium Run Observatory \citep[MRObs][]{overzier13a} to produce
physically-motivated, synthetic images of the night sky by adding the various
observational effects to predictions from cosmological simulations. Halo merger
trees based on the Millennium Run dark matter simulations in WMAP1 and WMAP7
cosmologies form the backbone for the semi-analytic modeling of galaxies
inside haloes. This modeling is based on simple recipes for, e.g., gas
cooling, star formation, supernova and AGN heating, gas stripping, and merging
between galaxies. At each time step of the simulation, the physical properties
of galaxies are translated into theoretical stellar populations in order to
predict the spectra of galaxies. ‘Light-cones’ are constructed that arrange the
simulated galaxies on the sky in a way that is similar to how galaxies would
appear in a galaxy redshift survey. Next, multi-band apparent magnitudes are
calculated, including the effects of absorption by the intergalactic medium.
The light-cone is then projected onto a virtual sky, and the positions, shapes,
sizes and observed-frame apparent magnitudes of the galaxies are used to build
a `perfect’ or `pre-observation’ image. The perfect image is fed into the MRObs
telescope simulator that applies models for the T250+JPCAM system (e.g., pixel
scale, readout noise, dark current, sensitivity and gain), the OAJ site
conditions (e.g., sky background, extinction, point spread function), and the
J-PAS observation strategy (e.g., exposures). The result is a synthetic J-PAS
image of a simulated universe. In Fig. \ref{fig:mrobs} we show an example of a
simulated J-PAS image in the direction of a distant galaxy cluster observed in
the filters $g$ (blue), $r$ (green), and $z$ (red).

\newpage

The simulated J-PAS data set can be used, for example, for testing the data
reduction pipeline, but also for testing many of the J-PAS science projects.
For example, source extraction algorithms are applied to the simulated image,
resulting in a catalogue of the apparent properties of all objects detected in
the image. Then, photometric redshifts, SEDs or morphologies can be determined,
and the results can be tested against the actual physical properties given for
each object by the simulations. The object catalogues can also be cross-matched
with higher level data available from the simulations to find, for example, the
halo masses, the dark matter density field, or to look up progenitors and
descendants.

\FloatBarrier

\subsection{The J-PAS Cluster Survey }
\label{sec:clusters}

The number density of galaxy clusters as a function of mass and redshift can be used to constrain cosmological parameters by measuring the growth of structure in the universe \cite{borgani01,henry00,gladders07,henry09,vikhlinin09,mantz10,rozo10}. 
The cluster mass function $N(M,z)$ probes both the growth factor and the evolution of cosmic volume, and hence can distinguish, a priori,  between a cosmological constant and possible deviations from General Relativity. In particular, 
it depends jointly on the matter density, $\Omega_m$, and on the amplitude of the primordial power spectrum 
through $\sigma_8$. The evolution of the cluster mass function also provides sensitive constraints on the dark energy parameters $(w_0,w_a)$. J-PAS, with its unprecedented volume coverage and redshift accuracy, will be able to map clusters and groups up to very early epochs (z$\sim$1), and down to relatively small masses. Its capability to accurately measure the redshifts of line-emitting galaxies up to z$\sim$1.3 will help avoiding biases in the determination of the galaxy content of each cluster. Finally, we will be able to calibrate optical richness or stellar mass content for the group and cluster candidates with dark matter mass haloes estimated from lensing with a very high precision due to the expected large number of clusters. As a result, J-PAS will produce the most complete and mass-sensitive cluster catalog available for cosmological and galaxy evolution studies.

\paragraph{Detecting galaxy clusters}

The detection of galaxy systems and the completeness of the samples as a function of redshift are crucial for a cluster counting probe. There is a wide range of optical cluster finding algorithms that can be used in the J-PAS data, from methods using positional information of galaxies to detect over-densities to those which include observational properties of the potential member galaxies, like colors and magnitudes. Some of the most relevant methods in the literature are the cluster red sequence methods \citep{gladders00,lopezcruz04,gladders05,wilson08,gilbank11}, the MaxBCG \linebreak 
 \citep{koester07}, the new Gaussian Mixture Brightest Cluster Galaxy (GMBCG) algorithm \citep{hao10}, the cut-and-enhance algorithm \citep{goto02}, the C4 cluster-finding algorithm \citep{miller05}, the counts in cells method \citep{couch91,lidman96}, the Percolation Algorithms \citep{dalton97,botzler04}, the Voronoi Tessellation algorithm \citep{ramella01,kim02,lopes04} or the Friends of Friends  Algorithm \citep{huchra82,ramella02,botzler04,vanbreukelen09} and adapted modifications to photo-z surveys \cite{ariel13}, the Matched Filter technique \citep{postman96,postman02} and later modifications: the Adaptive  Matched Filter \citep{kepner99}, the Hybrid Matched Filter \citep{kim02}, the Simple Smoothing Kernels \citep{shectman85}, the Adaptive Kernel method  \citep{gal00,gal03} the 3D-Matched Filter \citep{milkeraitis10} or the Bayesian Cluster Finder \citep{ascaso12}. For a review on cluster finder techniques, see \cite{ascaso13b} and references herein.

Accurate mass estimates are of utmost importance for studies of galaxy systems, and are absolutely necessary to do precision cosmology. Cluster masses, however, are not trivial to measure precisely. Triaxiality and projection effects can bias the dynamical masses determined by assuming virial relations; masses estimated through gravitational lensing are  sensitive to assumptions about the isotropy of the mass distribution; masses estimated via the Sunyaev-Zel'dovich thermal effect are prone to blending and assymetrizing, potentially biasing the mass function (e.g. \citep{2011ARAA..49..409A}); X-ray-derived masses are sensitive to places where hydrostatic equilibrium breaks down (either near the central AGN activity or even at large scales due to ongoing merging and or residual non-thermal pressure -- \citep{dupke01a}, \citep{dupke01b}, \citep{nagai07}), or because of gas clumping  (\citep{nagai11}). Therefore, masses found using these techniques often disagree at a level that is insufficient to achieve the precision that is desired for maximal cosmological discriminatory power. These are particularly critical issues for photometric surveys, where redshift precision is limited and complementary corroborative data is unevenly found.

In order to circumvent this deficiency a significant amount of effort has been placed into determining the so-called mass proxies, i.e., indicators of the mass of clusters devised by inspecting the least important mass component of clusters -- i.e., galaxies (often of only one Hubble type). These efforts have brought us impressive results, with sophisticated techniques for improving the scatter of the optical mass proxies \citep{andreon12,munari13,rozo09,rozo11}. With its excellent photo-z precision, J-PAS will be able to reduce significantly the galaxy membership noise  for clusters and groups of galaxies. This, combined with weak lensing measurements, will provide much better grounds for richness (of various types) and  mass relations.

\paragraph{The Probability Friend-of-friends (PFoF) cluster set}

By using a mock galaxy catalogue tailored to the J-PAS depth and magnitude limit, a modified version of the Friends-of-Friends algorithm has been used by \cite{ariel13} to detect galaxy groups and to assess the reliability of the algorithm.

In that work, they built a light-cone mock catalogue using synthetic galaxies constructed from the Millennium Run Simulation I \cite{Springel05} combined with a semi-analytical model of galaxy formation developed by Guo et al. 2011. The mock catalogue comprises $\sim$ 800000 galaxies down to an observer frame apparent magnitude of 23 in the SDSS i-band, with a median redshift of $0.72$ and a maximum of $1.5$ within a solid angle of $17.6 \, deg^2$. The solid angle was chosen to avoid repetition of structures down to redshift 1 caused by the limited size of the simulation box. Photometric redshifts were assigned to each galaxy in the mock catalogue in a realistic way by using a technique described in Ascaso et al. 2014 (in prep.) and 
\citet{2013arXiv1311.3280A}.

The identification of groups in the mock catalogue was performed by using the adaptation of the original FoF algorithm to work with photometric redshifts developed by \cite{liu08}. The redshift probability distribution functions for the galaxies were adopted as Lorentzian functions. The sample of photometric groups comprises 15512 groups with four or more members.

The reliability of the finder algorithm as a function of redshift was tested by computing the purity and completeness of the resulting sample. The purity and completeness were defined by a member-to-member comparison of the identified sample to a reference sample. The reference sample was adopted as the sample of groups that have four or more members with $i_{SDSS}  < 23$, where the groups were previously identified in volume-limited catalogue in real space.

\newpage
 The reference sample of groups was obtained from a subsample of the J-PAS mock catalogue defined without introducing any flux limit or redshift space distortions \citep{ariel13}. They used the Friend of Friends (FoF, \citealt{huchra82}) algorithm to detect groups on this ideal subsample, obtaining a sample of 201,032 groups with 4 or more galaxy members, within a solid angle of 17.6 $deg^2$ up to redshift 1.5. This will be considered also the reference sample in this work, in order to compare the detections with other methods. Additionally, we selected from the reference groups those that have 4 or more members with observed-frame magnitude $i_{SDSS}<23$, which refer to those groups that could be identified in the flux limited catalogue. This will be called the restricted-reference sample, and it comprises 11294 groups. 

Considering the fraction of photometric groups that are also in the reference sample and the sample of reference groups that can be recovered with this algorithm, they found that it is possible to identify a sample of photometric groups with less than $40\%$ of completely false groups, while $60\%$ of the underlying true groups are recovered. The purity of the photometric groups can be highly improved if only groups with more than ten members are considered (purity $> 90\%$ in the whole redshift range).

 By applying this algorithm to the future J-PAS, it is expected to find around 700000 photometric groups with more than ten members, among which more than $90\%$ would be related to real groups.

\paragraph{The Bayesian Cluster Finder (BCF) cluster set}

Additionally, we used the mock galaxy catalogues from \citep{merson13}, which were built from a semi-analytical model of galaxy formation, applied to the halo merger trees extracted from a cosmological N-body simulation. The semi-analytical model that they use is the Durham semi-analytical galaxy formation model, GALFORM \citep{cole00}, which models the star formation and merger history for a galaxy. Among other physical processes, this model includes feedback as a result of SNe, active galactic nuclei (AGN) and photo-ionization of the intergalactic medium. The model predicts the star formation history of the galaxy and therefore the spectral energy distribution (SED). The population of dark matter (DM) haloes for the mock catalogue is extracted from the Millennium Simulation \citep{springel05}, a $2160^3$ particle N-body simulation of the $\Lambda$ Cold Dark Matter cosmology starting at $z=127$ and hierarchical growing to the present day. The halo merger trees are constructed using particle and halo data stored at 64 fixed epoch snapshots spaced logarithmically. The minimum halo resolution is 20 particles, corresponding to $1.72 \times 10^{10} h^{-1} M_{\odot}$. Finally, the light-cone was constructed from this simulation by replicating the simulation box and choosing an orientation resulting into a 226.56 $deg^2$ light-cone. In addition, a flux cut in $\sim$24 AB was applied to mimic the condition of the J-PAS survey. All the details can be found in \citep{merson13}. 

In order to obtain realistic J-PAS-like photometric redshifts for this mock catalogue, we follow a similar approach to obtain photometry and photometric redshift as for the ALHAMBRA as in Ascaso et al. 2014 (in prep) and \citet{2013arXiv1311.3280A}. 
We first obtained spectral types from the original rest-frame photometry and spectroscopic redshifts in the mock by running the Bayesian Photometric Redshift package (BPZ, \cite{benitez00,B2014}) with the ONLY\_TYPE yes option. Then, we obtained consistent J-PAS photometry for these spectral types by using the J-PAS filter curve response and adding realistic noise. Finally, we obtained the photometric redshift estimations, together with spectral types and absolute magnitudes associated to the previous photometry by running again BPZ in normal mode. In previous work where we applied this technique (the ALHAMBRA survey), the photometric redshifts that we obtain are found to be very realistic as their performance is very similar to those obtained for real data \citep{molino13}.

In the next step, we detected galaxy clusters and groups in the J-PAS-like mock catalogue by using the Bayesian Cluster Finder (BCF, Ascaso et al. 2012, 2014). We performed a search in twenty-four redshift slices from z = 0.1 to z = 1.2, with redshift bins of z = 0.05. The core radius was selected as 1.5 Mpc, and the Luminosity Function has been chosen to have a value of $M^*(0)=-21.44$ and  $\alpha =-1.05$ (Blanton et al.2003). We calculated the expected $g-i$ and $i-z$ colors from synthetic spectra, and we artificially created these bands by calculating the contribution of each of the J-PAS narrow bands to the new synthetic band (see Molino et al 2014 for details). We also calculated the expected BCG magnitude-redshift relation for the given bands by performing a color transformation to the expected K-band. We merged the galaxy clusters following the same prescription as in Ascaso et al.2014. The expected number of galaxy clusters and groups  per square degree range between $\sim 46$ down to $5 \times10^{13}M_{\odot}$ and $\sim 75$ down to  $3 \times10^{13}M_{\odot}$, obtaining an expected total number of structures between $\sim$ 400.000 and 650.000 for the whole J-PAS down to $5 \times10^{13}M_{\odot}$ and  $3 \times10^{13}M_{\odot}$ respectively.

\paragraph{The selection function}

 In order to assess the completeness and purity of our results, we compared with the initial set of clusters in the simulation. In Fig. 20, we show the completeness and purity results versus Dark Matter halo and its respective stellar mass interval respectively for the output results. According to this results, we are able to detect galaxy clusters with purity and completeness rates $> 80\%$ for clusters and groups down to $M \ge 3 \times  10^13M_{\odot}$ up to redshift 0.8 and down to $M \ge 5 \times  10^13M_{\odot}$ up to redshift 1.2. The extremely good quality of the photometric redshifts in the J-PAS survey make these results comparable to what we would expect for a low-resolution spectroscopic survey. In fact, the photometric redshift resolution becomes directly proportional to the inferior mass limit we can resolve according to our simulations.

\begin{figure}
\centering
\includegraphics[width=0.8\textwidth]{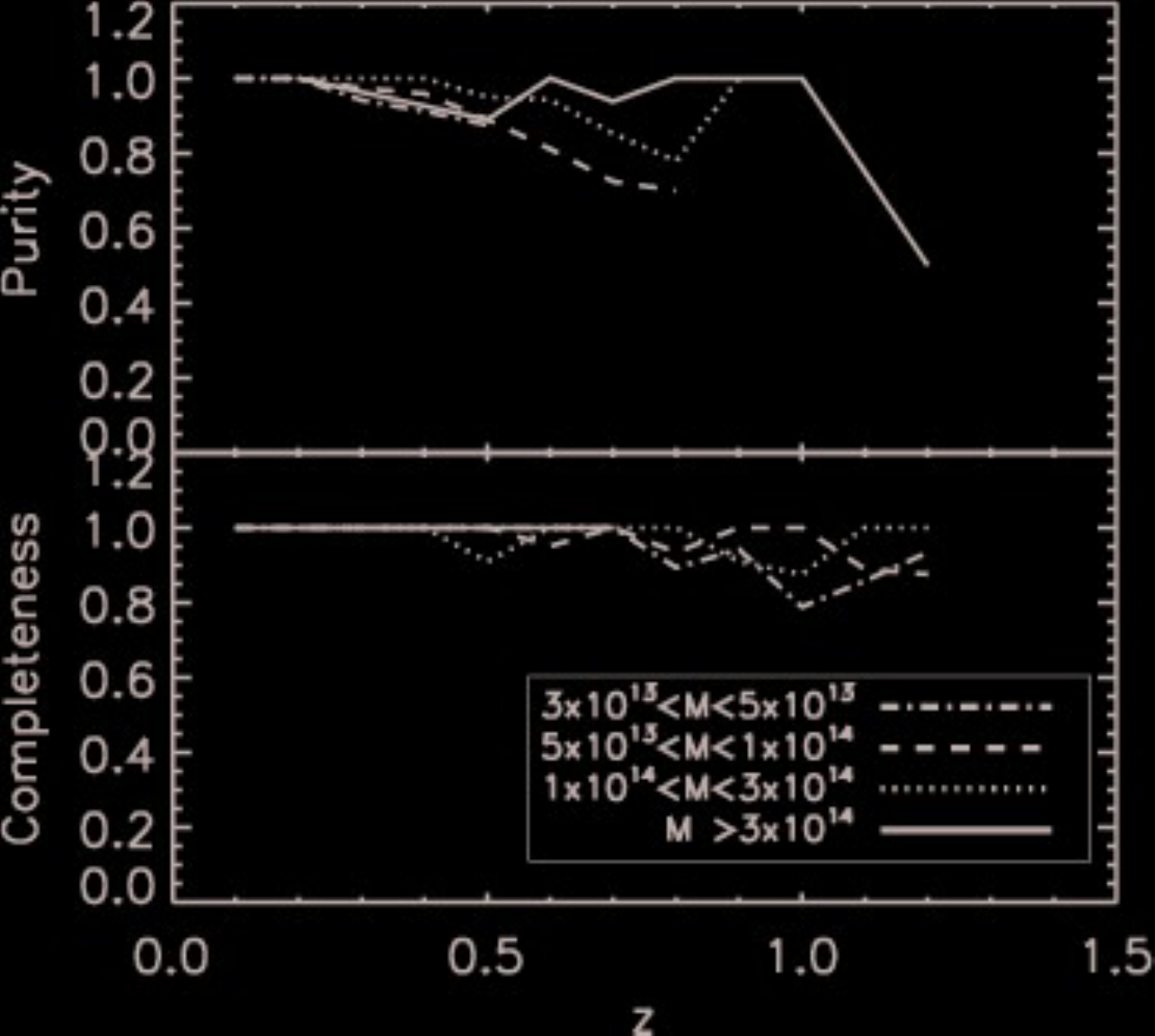} 
\caption{
Completeness and purity rates as a function of redshift for different dark matter halo masses for the J-PAS survey. We see that the purity rates remains constant as a function of redshift as $\sim 80\%$, whereas the completeness rates is always higher than $80\%$ and it starts to decrease at $z\sim 0.8$.}
\label{fig:drate}
\end{figure}

\subsubsection{Self-contained mass calibration}

 The superb seeing conditions ($<0.7^{\prime\prime}$) of which the $r$-band filter is planned to be
conducted will allow to estimate the masses of the clusters using the
weak lensing technique, the only one sensitive to both dark and baryonic matter.

Since many observational properties of clusters correlate well with
mass, we will be able to self-calibrate mass-observable
relations. J-PAS will explore the relation between the mass of galaxy
clusters obtained with weak gravitational lensing and the optical
properties like number of member galaxies, stellar light or total
stellar mass ($N_{\rm gal}$, $L_{\rm tot}$, $M_{\ast}^{\rm tot}$). The
latter showed to be a robust proxy of cluster mass at the same level
of the best X-ray proxies (i.e. $Y_X$, the product of X-ray temperature,
and gas mass) \citep{andreon12}.

The scatter of the mass-observable relation is also an important issue
when using it with cosmological purposes. To bring the errors of the
scatter down to a few percent, J-PAS will adopt a self-calibration
method, binning clusters using mass proxies and the redshift
information to then stack the weak lensing signal of the clusters
belonging to the same bin and measuring averaged masses. Detailed
simulations have proven that averaging out over large number of
clusters/groups is the most robust way to reduce the effects caused by
triaxial of the halos and uncorrelated large scale structure along the
line-of-sight, thus recovering the true value of the averaged mass
within the bin \citep{spinelli12} and decreasing the scatter of
the scaling relations.

\subsubsection{Figure of Merit}

Here we assume the same fiducial cosmology as in previous Sections, i.e.: $h=0.71$, $\Omega_m=0.27$, 
$\Omega_L=0.73$, $\Omega_b=0.024$, $w_0=-1.0$, $w_a=0.$, $\sigma_8=0.8$, $n_s=1.0$, 
$\tau=0.09$, and $\Omega_K=0$.  
Based on our previous predictions, we assume that the cluster catalog will reach down to a mass threshold of $5\times10^{13}$, up to $z=1.2$,  with a photometric redshift precision of $\sigma_z/(1+z)=0.003$ and a 
mass-richness calibration dispersion of $\sigma_{\ln M}=0.25$.

The covariance matrix then, can be expressed as \citep{wang04,lima04,cunha10}:
\begin{equation}
\begin{split}
C=& <(m_{i\mu}-\overline{m}_{i\mu})(m_{j\nu}-\overline{m}_{j\nu})>=\overline{m}_{i\mu}\overline{m}_{j\nu}b_{i\mu}b_{j\nu}\int{\frac{d^3 k}{(2\pi)^3}W_i^{*}(k)W_j(k)P(k,z_{ij})}
\end{split} \, ,
\end{equation}
where $i,j$ refer to bins in redshift and $\mu,\nu$ bins in mass,
\[ 
m_{i\mu}=\overline{m}_{i\mu}(1+b_{i\mu}\delta_i), \qquad \delta(x)=\frac{\rho(x)-\overline{\rho}}{\overline{\rho}} \, ,
\]
where $\delta(x)$ is the dimensionless density perturbation of the underlying matter distribution.

For a detailed window treatment depending on the survey, see \cite{hu03}. According to their Eq. (6), 
the windows can be divided up into slices in redshift. If we consider a series of slices in redshift at 
comoving distances $r_i$ and widths $\delta r_i$, with a field of radius  $\Theta_s$ in radians 
and a flat spatial geometry, then the window function are written as:
\[ 
W_i(k)= 2\exp^{ik_{\parallel}r_i}\frac{sin (k_{\parallel}\delta r_i/2)}{k_{\parallel} \delta r_i/2}\frac{J_1(k_{\perp}r_i\Theta_s)}{k_{\perp}r_i\Theta_s} \; ,
\]
where $k^2=k_{\perp}^2+k_{\parallel}^2$. 

Now, following \cite{wang04,cunha10}, we compute the mean number of counts and the bias term as: 
\begin{equation}
\overline{m}_{i\mu}=\Delta \Omega \int \frac{dV(z_i)}{dz}p_i(z^p_i|z)dz  \int_{M_{obs}^{\mu}}^{M_{obs}^{\mu+1}} \frac{dM_{obs}}{M_{obs}} \int{d \ln M p_{i\mu}(M_{obs}|M)\frac{dn}{d \ln M}} \; ,
\end{equation}
\begin{equation}
b_{i\mu}=\frac{1}{\overline{m}_{i\mu}}\Delta \Omega \int \frac{dV(z_i)}{dz}p_i(z^p_i|z)dz  \int_{M_{obs}^{\mu}}^{M_{obs}^{\mu+1}} \frac{dM_{obs}}{M_{obs}} \int{d \ln M p_{i\mu}(M_{obs}|M)b(M)\frac{dn}{d \ln M}} \; ,
\end{equation}
where $\Delta \Omega$ is the solid angle of the survey,  and
$\frac{dV}{dz}$ is the comoving volume unit at redshift $z_i$:
\[ 
\frac{d^2V}{d\Omega dz}=D_H\frac{(1+z)^2 D_A^2}{E(z)} \; ,
\]
Here, $p_i(z^p_i|z)$ and $p_{i\mu}(M_{obs}|M)$ are the probabilities of measuring a photometric redshift $z^p_i$, given the true cluster redshift z and $M_{obs}$, or given the true mass M respectively.  $\frac{dn}{d \ln M}$ is the halo density distribution and $b(M)$ is the bias function. The two latter ones are extracted from simulations. The bias parameter of halos of a fixed mass $M$ is assumed to be scale independent \citep{1999MNRAS.308..119S}:
\begin{equation}
b(M)=1+ \frac{(a\delta_C^2/\sigma^2)-1}{\delta_c}+\frac{2p}{\delta_c(1+(a\delta_c^2/\sigma^2)^p)} \; ,
\end{equation}   
where $a=0.75$, $p=0.3$ and $\delta_c=1.686$ is the threshold linear
overdensity corresponding to spherical collapse in an Einstein-de
Sitter universe. 
\newpage  
The differential comoving number density of clusters is given by \cite{jenkins01} (although there are other models that one could use for this purpose, 
such as \cite{tinker08}):
\begin{equation}
\frac{dn}{d\ln M}=0.3 \frac{\rho_m}{M}\frac{d \ln \sigma^{-1}}{d\ln M} exp(-|\ln \sigma^{-1}+0.64|^{3.82})   \; ,
\label{eq:equajenkins}
\end{equation}
where $\rho_m= \rho_0 \Omega_m$ is the mean matter density, with $\rho_0=3 H_0^2/(8\pi G)$. 
The variance $\sigma$ is the rms amplitude of mass fluctuation inside a particular spherically 
symmetric window, defined as:
\[     \sigma^2= \int {\frac{d^3k}{(2\pi)^2} k^2 P(k) | W(kR)|^2}  \; .
\]
$P(k)$ is the linear power spectrum, and $W(kR)$ is the Fourier transform of the real-space window function 
$W(x)$, which we have assumed, as usual, to be given by real-space spherical top hat of radius R, so:
\[     \tilde{W}_R(k)=\frac{3}{(kR)^3}(\sin(kR)-kR\cos(kR)) \; ,
\]
where M is the mass included in the window:
\[     M=\frac{4 \pi \rho_m R^3}{3}  \; .
\]

Finally, the total covariance matrix is given by \citep{lima05,cunha10}:
\[Cij:= Cij + m_{i}\delta_{ij} \; ,
\]
where the last term refers to the (shot) noise matrix, and $m_i$ are the cluster counts in each bin. 
In order to make a realistic forecast of the optical cluster constraints, one must marginalize over 
nuisance parameters which are introduced to account for the mass and redshift uncertainties.

\paragraph{The Mass-Observable relation}   
   
The mass selection function can be written as \citep{lima07,cunha10}:
\begin{equation}    
p(M_{obs}|M)=\frac{1}{\sqrt{2\pi}\sigma_{\ln M}} exp(-\chi^2(M_{obs})) \; ,
\end{equation}    
where:
\begin{equation}    
\chi(M_{obs})=\frac{\ln M_{obs} -\ln M - \ln M^{bias}(M_{obs},z)}{\sqrt{2}\sigma_{\ln M}} \; .
\end{equation}

Then, we introduce a series of nuisance parameters as in \cite{lima07}, although more complicated forms 
can also be modified to be dependent also on mass as in  \cite{cunha10}:
\begin{equation}
\ln M^{bias}(M_{obs},z)=\ln M^{bias}_0+ a_1 \ln(1+z) \; , 
\end{equation}
and
\begin{equation}
\sigma_{\ln M}^2= \sigma_0^2+ \sum_{i=1}^3 b_i z^i \; ,
\end{equation}

Hence, we have six nuisance parameters whose fiducial values can be chosen, 
according to \cite{lima07}, as $(\ln M^{bias}_0=0, a_1=0,  \sigma_0=0.25, b_1=0, b_2=0, b_3=0)$.

\newpage 

\paragraph{Photometric redshift true cluster redshift relation}   

The probability of measuring a photometric redshift, $z^p$ given the true cluster redshift $z$, 
can be parametrized as \citep{lima07,cunha10}:
\begin{equation}    
p(z^p|z)=\frac{1}{\sqrt{2\pi}\sigma_{z}} exp(-y^2(z^p)) \; ,
\end{equation}    
where:
\begin{equation}    
y(z^p)=\frac{z^p -z -z^{bias}}{\sqrt{2}\sigma_{z}} \; .
\end{equation}    
Here $z^{bias}$ is the photometric redshift bias, and $\sigma_z$ is the scatter in the photo-z's. We add two more nuisance parameters as $(z^{bias}_0=0., \sigma_z=0.003)$, accounting for a total of eight nuisance parameters to marginalize over before we can say anything about cosmological parameters.

\paragraph{Fisher matrices and priors}  
Finally, we compute the Fisher matrix as:
\begin{equation}
F_{lm}=\sum_{ij\mu\nu}\frac{\partial \overline{m}_{i\mu}}{\partial p_{l}} (C^{-1})_{ij\mu\nu}\frac{\partial \overline{m}_{j\nu}}{\partial p_{m}}
\end{equation}
where $l,m$ run over the  cosmological parameters. This Fisher matrix can be combined with any other 
Fisher matrix -- as long as the two datasets are uncorrelated. As usual, marginalized constraints for any
parameter (or subsets of parameters) are obtained by summing all the relevant Fisher matrices, then 
inverting the total Fisher matrix.

With regard to the dark energy equation of state, we forecast that J-PAS will be able to 
reach a Figure of Merit (FoM; see the previous Subsection) of approximately 170, when we combine 
cluster counts with Planck and Stage-II experiments, as described in \cite{2006astro.ph..9591A}. We would like
to stress the key role of the nuisance parameters (of which we have eight, as described above),
which parametrize our ignorance about the mass calibration, redshift bias and other uncertainties.

\FloatBarrier
\subsection{Joint constraints on Dark Energy from BAOs and cluster counts}

We can combine the constraints obtained through the measurements of BAOs, 
and those coming from cluster counts. Since the information about the clustering 
of halos was not used in the derivation of cosmological constraints from counting the 
numbers of clusters (see Sec. 3.2), we can simply add the Fisher matrices for the 
two datasets.

\begin{figure}[h]
\centering
\includegraphics[width=0.5\textwidth]{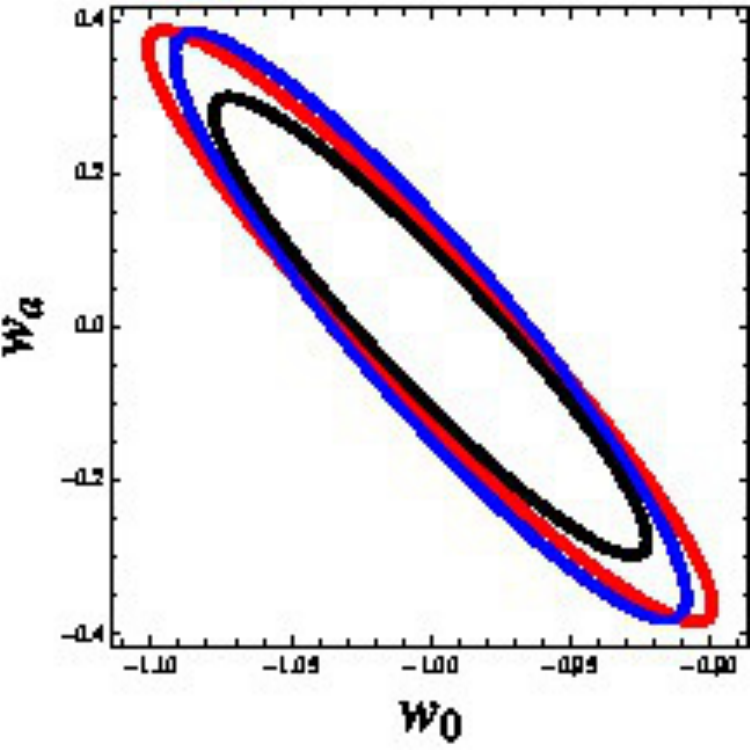}
  \caption{\label{fig:w0_wa_combined}
Constraints for the dark energy equation of state parameters $w_0$ and $w_a$
from cluster counts alone (outer, red contour), from BAOs alone (blue contour), 
and BAOs combined with cluster counts (inner, black contour). We used \Planck and Stage-II 
priors.
}
\end{figure}

In Fig. \ref{fig:w0_wa_combined} we show the constraints for the equation of state
of dark energy for cluster counts alone (outer, red contour), for BAOs alone (blue contour), 
and BAOs combined with cluster counts (inner, black contour). In all cases \Planck and Stage-II
priors were employed \citep{2006astro.ph..9591A}. We can see that the constraints from BAOs
and cluster counts are comparable, but there is a substantial complementarity between the two.

The combined power of cluster counts and BAOs can be gleaned from the DETF Figure of Merit (FoM).
The FoM of cluster counts, combined with \Planck and Stage-II priors, is approximately 175.
In Table \ref{Table:FoMs2} we show the forecasted FoMs for several combinations of the datasets.

\begin{table}[h]
\caption{Figure of merit.}
\begin{center}
\begin{tabular}{@{}lcccc}
\hline
Test & LRGs & ELGs & QSOs & All \\
\hline
BAOs + \Planck + Stage II & 87 & 121 & 100 & 163 \\
BAOs + Clusters + \Planck + Stage II & 195 & 222 & 201 & 256 \\
\hline
\end{tabular}
\end{center}
\medskip
{}
\label{Table:FoMs2}
\end{table}%

\FloatBarrier

\newpage 

\subsection{The J-PAS SN Cosmological Survey }
\label{sec:SN}

Type-Ia supernovae (SNeIa) comprise one of the four main observables that will be the keys to understand the origin of the recent acceleration in the expansion of the Universe. However, there are many aspects of SNeIa properties and their relation to their environments that remain poorly understood \citep{Conley11mn,Smith12mn,Kessler13mn}. For instance, recent studies have pointed out that passive galaxies host faster-declining SNeIa that follow a different color-luminosity relation and that are more luminous after corrections based on light-curve stretch and color \citep{Sullivan06mn,Lampeitl10mn}. Moreover, older passive galaxies -- as those found in galaxy clusters -- tend to host dimmer and even briefer SNeIa \citep{Gallagher08mn,Xavier13mn}. Some of these characteristics were already shown to introduce more scatter and also systematic biases to SNeIa distance determinations and to the cosmological parameters derived from them \citep{Kelly10mn,Lampeitl10mn,Sullivan10mn}, but these effects and relations are still not fully understood. This is in part due to the coarse characterization of the supernovae host galaxies, to the sizes of SNe Ia sub-samples in each environment and to the small number of low-redshift SNeIa that have been studied. The J-PAS Supernova Survey will be a massive, low and intermediate redshift ($z<0.4$) supernova survey (the only one, to our knowledge, being planned at the moment) that will serve to fill that gap.

Future SNeIa experiments will no longer be sample-size limited. In order to achieve a precision of $\sim$1\% in the dark energy equation of state parameter $w=P/\rho$, it will be crucial to control systematic uncertainties. Many of these uncertainties -- like dust extinction, rest-frame ultraviolet variability, intrinsic color variations and correlations between SNe and environment properties -- will need large samples of well observed SNe and host galaxies. A large sample of low redshift ($z<0.1$) objects, with which we can study the properties of SNeIa and their hosts in a cosmology-independent way, is also important.

\begin{figure}
\begin{center}
\includegraphics[width=0.8\textwidth]{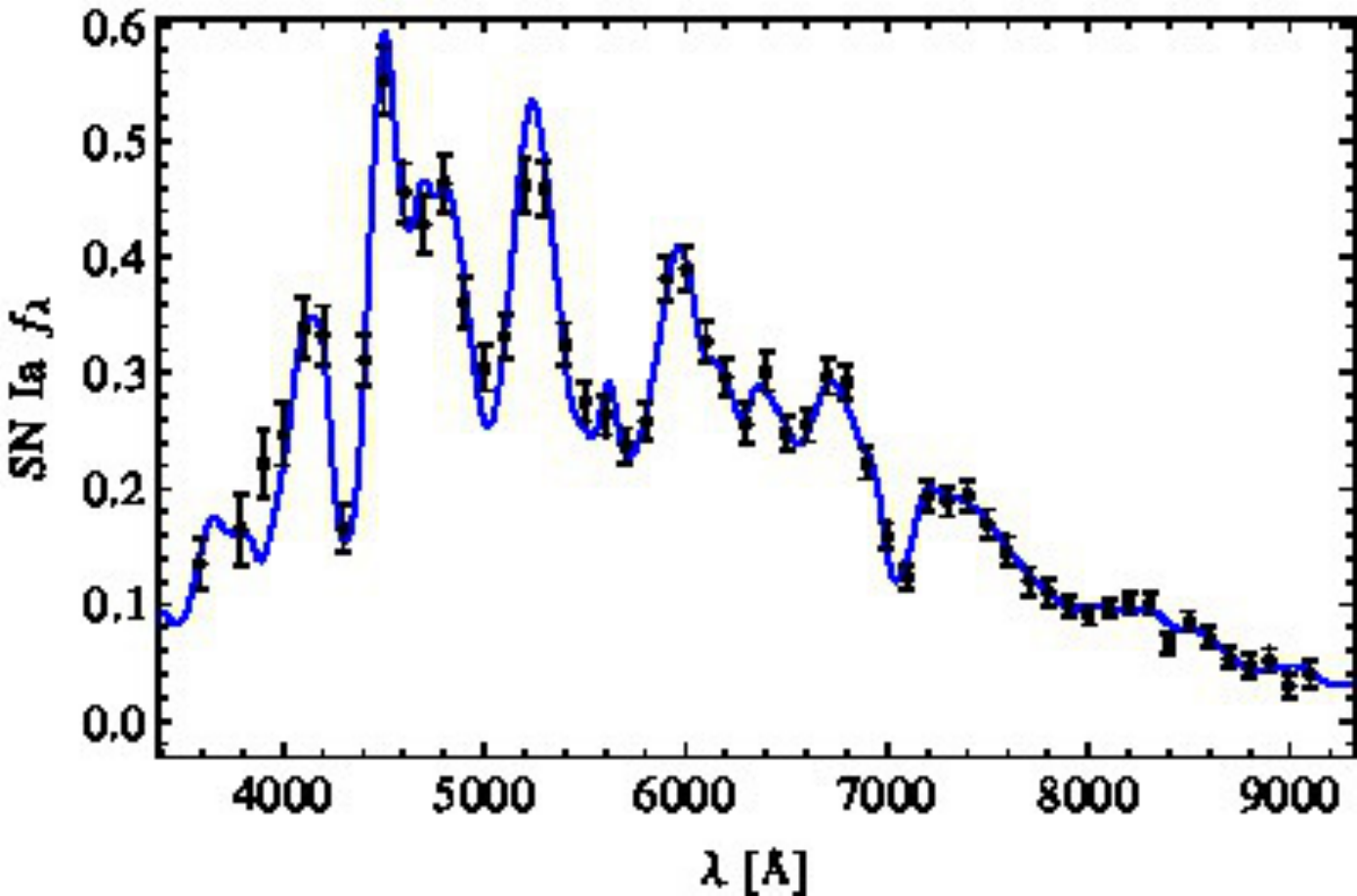}
\caption{Expected J-PAS photometry (black dots) for a SNe Ia at $z=0.148$, seven days after its luminosity peak, and its true spectrum in arbitrary units (blue line). SNe spectral features are broad enough to be detected. To be concise we present the measurements of the whole spectrum on the same day. The reader should keep in mind that, in a given epoch, J-PAS will image the SNe in 14 filters. }
\label{fig:sneia-spectrum} 
\end{center}
\end{figure}

Due to the broad features of the spectra of SNeIa, the filter system of J-PAS makes it an ideal instrument not only to discover them, but also to measure their light curves (albeit often using different filters), to characterize their types (SN Ia/Ib/Ic/II etc.) and to photometrically estimate their redshifts. For a glimpse of J-PAS SNe photometry, see Fig. \ref{fig:sneia-spectrum}. Due to the imaging nature of the survey, the local environments of the supernovae will also be fully characterized -- something that has never been done before in a systematic, massive way. Finally, due to the large area of the survey, the number of SNeIa will be large enough that we can separate them (and their environments) into different subtypes without running into problems related to low statistics. 

Optical surveys are deemed to miss a fraction of exploding supernovae due to host galaxy extinction. This effect is particularly relevant in LIRGs (Luminous Infrared Galaxies) and ULIRGs (Ultraluminous Infrared Galaxies), whose contribution to the SN rate increase with redshift \citep{mannucci07,mattila12}. Since J-PAS will be sensitive to SNe up to $z\sim 0.4$, we will be able to estimate the fraction of missing SNe with unprecedented accuracy (current estimates are between 5\% up to 40\% \citep{mannucci07,mattila12}.

\begin{figure}
\begin{center}
\includegraphics[width=0.8\textwidth]{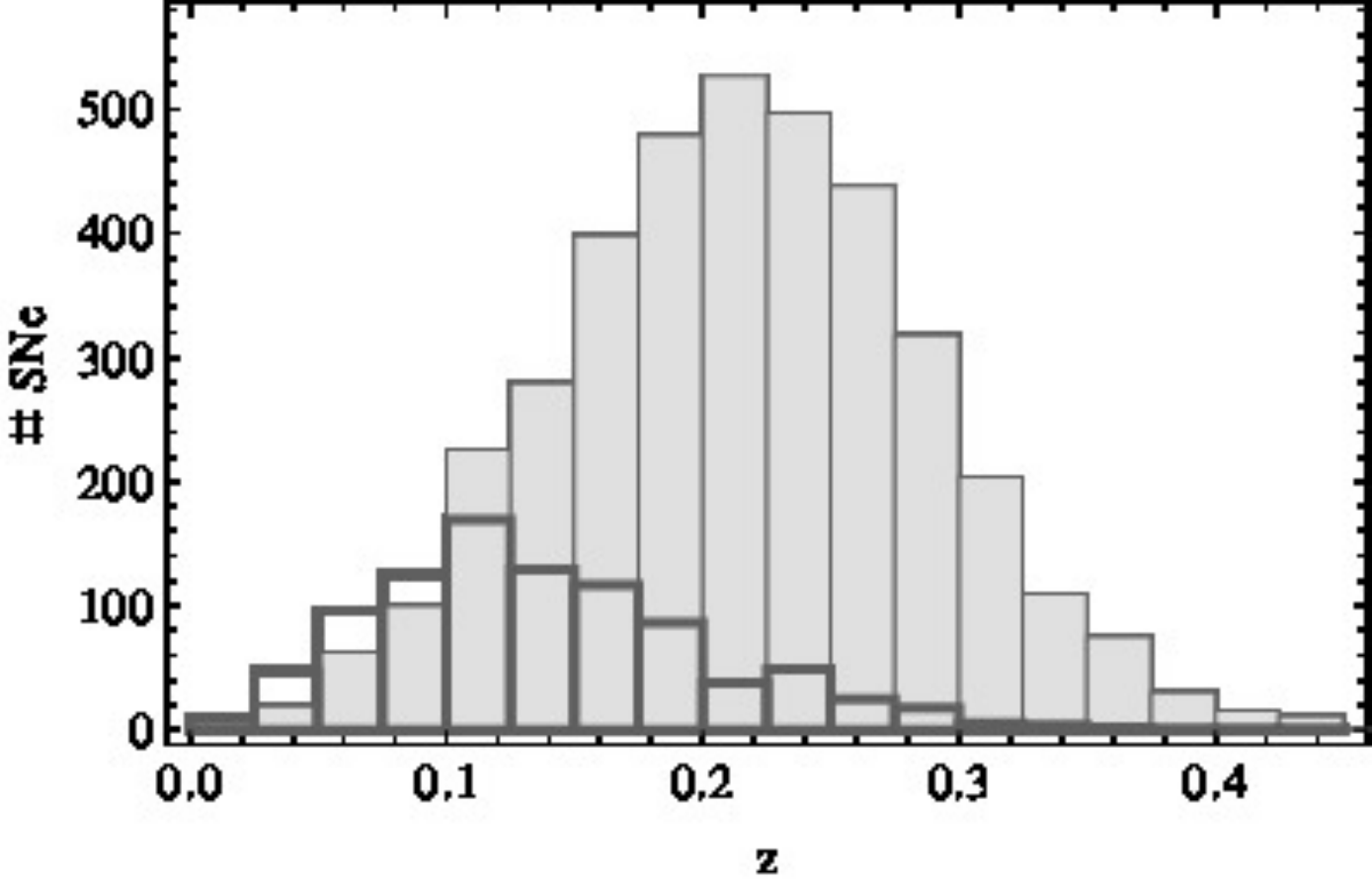}
\caption{Expected SNe Ia (gray filling, thin contours) and CCSNe (no filling, thick contours) redshift distribution for J-PAS. The total number of objects are $\sim 3800$ SNeIa and $\sim 900$ CCSNe. The contamination between these samples should be less than 4\%.}
\label{fig:sne-z-dist} 
\end{center}
\end{figure}

To estimate the J-PAS Supernova Survey performance, we ran detailed simulations with the \texttt{SNANA} software package \citep{Kessler09amn}, Peter Nugent's Core Collapse Supernovae (CCSNe) templates\footnote{http://supernova.lbl.gov/\textasciitilde nugent/nugent\_templates.html} and the SALT2 SNeIa light-curve model \citep{Guy07mn}, assuming that the relation between the distance modulus and the SNe Ia observables, $\mu=m_B-M+\alpha x_1 - \beta c$, has an intrinsic scatter of $\sigma_{\mathrm{int}}=0.14$. 
We set the SNe redshifts to their host galaxies \emph{photo-z}s, which we assumed to have an uncertainty of $\sigma_z=0.005(1+z)$. The SNe typing was performed with the \texttt{psnid} software \citep{Sako11mn} in the \texttt{SNANA} package. Although the results depend on the exact observation schedule which can be affected by other circumstances, our simulations show that, for a conservative scenario, J-PAS will be able to detect and characterize around 3,800 SNeIa and 900 CCSNe up to $z\sim 0.4$ and approximately 190 SNeIa and 280 CCSNe at low redshifts ($z<0.1$). Their redshift distributions are shown in Fig. \ref{fig:sne-z-dist}. 

To classify a particular simulated light-curve as a ``SN type $X$'', we required that its fit by a type $X$ template should have a $\chi^2$ $p$-value of at least 0.01. Also, the probability $P_X$ that the light-curve belongs to the type $X$ SNe, calculated by \texttt{psnid}, should be higher that 0.90. Our simulations show that J-PAS can achieve low contamination rates (less than 4\%) for both SNeIa and CCSNe samples. For SNeIa studies that require a higher purity, further data cuts on the $x_1$--$c$ SALT2 parameters plane like the one described by \citep{Campbell13mn} can lead to samples with $\sim 4000$ objects and less than 1\% contamination. 

To evaluate the SNeIa data quality, we analyzed the average errors on the SALT2 light-curve parameters by calculating the root mean square (rms) of the difference between their fitted and their true values. While purely photometric broad band surveys can detect and measure light-curves of many thousands of SNe, the lack of a good redshift prior undermines its SNe data quality. The use of spectroscopy for constraining the SNe host galaxy's redshift significantly improves the data quality but presents a bottleneck for sample sizes. Table \ref{tab:sne-data-quality} shows the average errors for J-PAS and, as reference points, for simulations of the SDSS Supernova Survey \citep{Frieman08mn}, which have a similar redshift distribution. With the help of its excellent host galaxy \emph{photo-z}, J-PAS can perform as well as broad band surveys backed up by spectroscopy and much better than purely photometric broad band surveys. This advantage (and the full characterization of the SNe's host galaxies) will be beneficial not only for supernovae discovered by J-PAS but for all past and future SNe from other surveys, provided they overlap with J-PAS footprint. For J-PAS in particular, its narrow band filters will allow for the study of correlations between SNe spectral features and broadband properties like light-curve width and color.

\begin{table}
\begin{center}
\begin{tabular}{lcccccccc}
\hline
Survey              & $\sigma_{m_B} $ & $\sigma_{x_1}$ & $\sigma_c$ & $\sigma_{T_0}$ & $\sigma_{\mu}^*$ & $\sigma_{\mu}$ \tabularnewline\hline
\emph{photo-z} SDSS & 0.074          & 0.72          & 0.066     & 0.87          & 0.21            & 0.25          \tabularnewline
\emph{spec-z} SDSS  & 0.069          & 0.69          & 0.043     & 0.77          & 0.13            & 0.19          \tabularnewline
J-PAS               & 0.078          & 0.54          & 0.046     & 0.95          & 0.12            & 0.18          \tabularnewline
\hline
\end{tabular}
\caption{Average errors in the SNeIa SALT2 light-curve parameters (apparent magnitude $m_B$, light-curve width $x_1$, color $c$, epoch of maximum luminosity in days $T_o$ and the distance modulus $\mu$ ignoring and including the intrinsic scatter $\sigma_{\mathrm{int}}$) for J-PAS and for simulations of the SDSS Supernova Survey, both with photometry only (\emph{photo-z} SDSS) and with spectroscopy of all SNe hosts (\emph{spec-z} SDSS). J-PAS have similar data quality to photometric surveys backed up by spectroscopy and much better quality than purely photometric surveys.}
\label{tab:sne-data-quality}
\end{center}
\end{table}

\subsection{The J-PAS Lensing Survey}
\label{sec:lensing}

The combination of two superb characteristics of the OAJ, namely the quality and the time stability of its median 0.71'' seeing (Moles et al. 2010, PASP, 122, 363), with a broad band filter in JPCam, should yield an extremely high quality image of the whole northern sky at a very reasonable cost of observing time and effort. Together with our redshift information, the lensing measurements would produce an outstanding dataset for cosmic lensing studies many years before the arrival of Euclid. 

\subsubsection{Cosmic shear}
 The most obvious application is comic-shear tomography that probes both expansion of the universe and the growth of structures by the variation of the lensing strength between lens and source slices at different redshifts. With the precision of the photometric redshift estimates, we should be able to establish $\sim 10$ non-overlapping  slices which will not only be powerful on its own, but enable thorough systematics tests, crucial for any reliable weak-lensing analysis. We refer to the extensive literature on this topic (see e.g. \cite{Weinberg13.1} for a recent review). 

  Another promising application is the shear-ratio test \citep{Jain03.1, Bernstein04.1}, which probes the geometry of the universe from the scaling of the lensing signal with redshift. Because this application does favor deeper surveys to get a long lever arm, a larger number of non-overlapping slices at redshifts below 0.5 can only yield a significant measurement if the lensing data is truly exquisite. As an alternative route, one could attempt to construct a high-significance lensing analysis  by performing the shape measurement simultaneously across several of the narrow-band filters and the broad-band filter. Modern model-fitting  codes \citep[e.g.][]{Miller13.1, Zuntz13.1} can in principle work in this mode, provided an accurate PSF model can be constructed in each of the filters and that the filters cover a similar wavelength range so that changes to the morphology remain minor.

  This approach would have several advantages: First, it limits the negative influence of pixel noise, which constitutes the most prominent systematic bias in weak-lensing measurements today \citep{Massey07.1, Kitching12.1, Melchior12.1}. Second, it would allow us to extend the magnitude and redshift range, for which we can get reliable shape measurements, critical to both cosmological applications as pointed out above. Third, it would virtually eliminate the chromatic mismatch between the stars, which are used to build the PSF models, and the galaxies \citep{Cypriano10.1}.

 The survey design is advantageous also for treating the most relevant astrophysical systematic: intrinsic alignments. Precise photo-zs will enable us to exclude pairs of galaxies at the same redshift, whose ellipticities are intrinsically coupled. It will furthermore allow a good discrimination of early-type galaxies, for which intrinsic alignments have been confirmed already \citep{Mandelbaum06.1},  from late-type galaxies, for which the current upper limits indicate a much smaller amount of alignment \citep{Hirata07.1, Mandelbaum11.1}.

\subsubsection{Galaxy-galaxy lensing}
Given the relatively shallow depth of J-PAS (at least when using only the broad-band filter), galaxy-galaxy lensing is statistically even more powerful than cosmic-shear measurements. It can be utilized for several kinds of analyses, most prominently constraining the galaxy bias. Because of its multitude of filters, J-PAS will be able to discern several different lens populations, which in turn should allow us to constrain more complex halo-occupation models. This also present a much faster and scalable approach compared to previous work that mostly relied on spectroscopic follow-up to define the lens samples \citep[e.g.][]{Seljak05.1, Hirata07.1, Reyes11.1}. An straightforward extension is the incorporation of galaxy clustering information. For the substantial advantages of this combination, we refer to \cite{Yoo12.1}.

\subsubsection{Cluster Weak Lensing}

Substantial progress has been made through numerical simulations in  understanding the formation and structure of collisionless dark-matter (DM) 
halos in quasi gravitational equilibrium, governed by nonlinear growth 
of cosmic density perturbations.  
In the standard ${\Lambda}$CDM paradigm of hierarchical structure 
formation, galaxy-cluster sized halos form through successive mergers of 
smaller halos, as well as through smooth accretion of matter along 
surrounding filamentary structures \citep{Colberg+2000}. 
Cluster halos are located at dense nodes where the large-scale filaments 
intersect, generally triaxial reflecting the collisionless nature of DM,
and elongated in the preferential infall direction of 
subhalos, namely, along surrounding filaments. 

The internal structure of DM halos constitutes one of the most 
distinct predictions of the CDM paradigm. N-body simulations of 
collisionless CDM established a nearly self-similar form for the 
spherically-averaged density profile $\rho(r)$ of DM halos 
\citep[][hereafter Navarro-Frenk-White, NFW]{1997ApJ...490..493N}
over a wide 
range of halo masses, with some intrinsic variance associated with mass 
assembly histories and dynamical structure of individual halos 
\citep{Jing+Suto2000,2004ApJ...607..125T,Navarro+2010}. 
The degree of mass concentration, $c_{200}=r_{200}/r_{\rm s}$, is 
predicted to correlate with halo mass, since DM halos that are 
more massive collapse later on average, when the mean background density 
of the universe is correspondingly lower 
\citep{2001MNRAS.321..559B,2007MNRAS.381.1450N}. 
Accordingly, cluster-sized halos are predicted to be less concentrated 
than less massive systems, and to have concentrations of $c_{200}\sim 3-4$
\citep{Duffy+2008,Bhatt+2013}. 

Massive clusters serve as powerful gravitational lenses, producing 
various detectable effects, including deflection, magnifying and 
shearing of the images of distant background sources 
\citep{BartelmannSchneider2001}. 
Importantly, there is a weak-lensing regime where lensing effects can be 
linearly related to the mass distribution, which allows us to 
reconstruct the cluster mass distribution in a model-independent 
way. Weak-lensing shear offers a direct means of probing the total 
matter distribution of clusters \citep{1993ApJ...404..441K}
irrespective of the physical nature, composition, and state of lensing 
matter \citep{Okabe+Umetsu08},
providing a direct probe for testing well-defined predictions 
\citep{Oguri+Takada2011}. 

Lensing magnification provides complementary observational alternatives 
to gravitational shear 
\citep{1995ApJ...438...49B,UB2008,Umetsu+2011,Hildebrandt+2011,Ford+2012,Umetsu2013,Coupon+2013}. 
Magnification can influence the observed surface density of background 
sources, expanding the area of sky, and enhancing the observed flux of 
background sources \citep{1995ApJ...438...49B}. 
The former effect reduces the effective observing area in the source 
plane, decreasing the source counts per solid angle. The latter effect 
increases the number of sources above the limiting flux because the 
limiting luminosity at any background redshift lies effectively at a 
fainter limit.  The net effect is known as magnification bias and 
depends on the steepness of the source number counts. 

Magnification bias can be combined with shear to obtain a model-free 
determination of the projected mass profiles of clusters 
\citep{Schneider+2000,UB2008,Umetsu+2011,Umetsu2013},
effectively breaking the mass-sheet degeneracy inherent in 
a standard weak-lensing analysis based on shape information alone 
\citep{1995A&A...294..411S}. 
Recent Subaru weak-lensing work established that deep multicolor 
imaging allows us to simultaneously detect the observationally 
independent shear and magnification signals. 
The combination of shear and magnification allows us not only to 
perform consistency tests of observational systematics but also to 
significantly enhance the precision and accuracy of cluster mass 
estimates \citep{Rozo+Schmidt2010,Umetsu+2012,Umetsu2013}. 

Unlike the shearing effect, magnification is sensitive to the sheet-like 
structure, so that making accurate magnification 
measurements is crucial for a robust statistical detection of the 
two-halo term contribution due to large-scale structure associated 
with the central clusters (Umetsu et al. 2014, in preparation). 

In the J-PAS survey, we will couple our high-precision multi-band 
photometry and deep broadband imaging 
with cluster weak gravitational lensing 
to test fundamental predictions from structure formation models 
with unprecedented precision.

 The J-PAS survey will allow us to measure simultaneously the 
weak-lensing shear and magnification effects from well-defined samples 
of background galaxies, free from significant contamination of unlensed 
cluster member and foreground galaxies. 
Specifically, the main scientific objectives that we will address are 
the following:
\begin{enumerate}
 \item Halo density profile and mass-concentration relation:
  The stacked tangential-shear signal $\Delta\Sigma(R)=\Sigma(<R)-\Sigma(R)$ around a 
       statistical sample of clusters is a sensitive probe of the 
       internal structure of halos within the virial region, where the 
       predicted two-halo contribution $\Delta\Sigma_{\rm 2h}$
       is one-order smaller than that of the one-halo component $\Delta\Sigma_{\rm 1h}$ 
       \citep[e.g.,][]{Oguri+Hamana2011}.  With the J-PAS 
       survey, we will define homogeneous samples of groups and clusters, and obtain the 
       ensemble-averaged halo mass profiles, to compare with a family of 
       standard density profiles, 
       such as the NFW, truncated variant of NFW, and Einasto profiles,
       predicted for CDM halos in gravitational equilibrium. 
       We will establish the halo $c$--$M$ relation as a 
       function of halo mass and redshift, which can be 
       self-consistently obtained from J-PAS data alone.  We will also 
       constrain the mass dependence of the Einasto shape parameter to 
       compare with predictions from numerical simulations \citep{Gao+2008}. 
 \item Halo mass-bias relation, $b_{\rm h}(M,z)$:  The stacked weak-lensing signals on 
       sufficiently large scales $R$ can be used to determine 
       the clustering strength of the halos, which is proportional to 
       $b_{\rm h}\sigma_8^2$
       \citep{Johnston+2007a,Covone+2014}.  We 
       will measure this clustering strength as a function of halo mass 
       and redshift, by combining the observationally-independent shear 
       and magnification effects for greater sensitivity. 
 \item Shear-ratio geometric tests:  
       The amplitude of weak lensing should increase with source distance, rising steeply behind a 
       lens and saturating at high redshift 
       \citep{2007MNRAS.374.1377T,Medezinski+2011}.  Such a 
       characteristic 
       geometric dependence of the lensing strength can be examined in a 
       model-independent matter by using unbiased shape and photo-$z$
       measurements from the J-PAS survey.  We will measure the relative 
       lensing strength of source galaxies behind cluster 
       samples as a function of redshift, for providing model-free 
       constraints on the cosmological parameters.

\end{enumerate}
\subsubsection{Cluster strong lensing}
  
  In the center of the cluster (up to few hundred kpc) , where the
  surface mass density is high enough, often multiple images of
  background sources are seen
  (e.g. \citealt{Kneib1993,Broadhurst2005,Zitrin2012b,Richard2010,Limousin2010},
  see also a review by \citealt{KneibNatarajan2011}). As multiple
  images should be mapped back to the same single source, these are
  used then to place tighter (and high-resolution) constraints on the
  inner mass distribution, which can then be importantly combined with
  the independent WL measurements. We plan to incorporate well-tested
  and commonly-used methods for SL analyses and mass modeling, in
  various different parametrizations
  (e.g. \citealt{Broadhurst2005,Coe2008,Zitrin2012b,Jullo2007}).

  The identification of multiple images, however, is usually a very
  time consuming task, and often requires very high resolution space
  imaging. As a response, we had developed and implemented in recent
  years a unique modeling method which is guided primarily by the
  cluster member luminosity distribution in the cluster. The success
  of this method is remarkable - so that unprecedented numbers of
  multiple images can be in fact be found automatically be the
  luminosity-guided model itself without using any images a priori as
  constraints
  (e.g. \citealt{Broadhurst2005,Zitrin2009a,Zitrin2009b,Zitrin2013a,Zitrin2013b}). Following
  the success of this "Light-Traces-Mass" method in identifying
  multiple-images simply by following the light distribution, we have
  generalized it to \emph{automatically} map the matter in galaxy
  cluster cores, particularly useful for large sky surveys, by scaling
  their light distribution using extrapolations from clusters where
  multiple images are already known. The success of this method and
  its implementation in 10,000 SDSS clusters were shown in
  \citet{Zitrin2012a}. In addition, this method can help traces cosmic
  or structure evolution, as we showed in the works mentioned
  above. We plan to perform the same automated procedure also here, so
  that cluster maps can be reproduced rapidly and automatically.

In addition, recent efforts have proven larger success in identifying arcs in large sky surveys (e.g. \citealt{Maturi2013,Bayliss2011}). Combined with our automated lens model, this will reveal giant arcs (highly stretched and distorted multiply imaged galaxies), which can then be used to refine the lens model. Independently, the number counts of giant arcs was claimed to add constraints on cosmology (e.g. \citealt{Bartelmann1998,Horesh2011}).

\subsubsection{Cross-correlation with Herschel}

The magnification bias due to weak lensing modifies the galaxy angular correlation function because the observed images do not coincide with true source locations \citep{Gunn1967,Kaiser1992,Moessner1998,Loverde2008}, but the effect is generally minor and difficult to single out. A unambiguous manifestation of weak lensing is the cross-correlation between two source samples with non-overlapping redshift distributions. The occurrence of such correlations has been tested and established in several contexts \citep[see, e.g.][and references therein]{2005ApJ...633..560E,Menard2010,BartelmannSchneider2001}.

Since the gravitational magnification decreases the effective detection limit it is obvious that the amplitude of the magnification bias increases with increasing steepness of the number counts of background sources and is then particularly large at sub-mm wavelengths where the counts are extremely steep \citep{Clements10,Oliver2010}. At the same time, for a survey covering a sufficiently large area the counteractive effect on the solid angle is small \citep{JainLima2011}. A substantial fraction of galaxies detected by deep large area \textit{Herschel} surveys at 250, 350 and $500\,\mu$m with the Spectral and Photometric Imaging Receiver \citep[SPIRE;][]{Griffin10} reside at $z\gtrsim 1.5$ \citep{Amblard2010,Lapi11} and therefore constitute an excellent background sample for the J-PAS galaxies, which are located at $z\lesssim 1.4$ (with a peak n the redshift distribution at $z\ll 1$). In particular, two of the largest area extragalactic surveys carried out by the Herschel space observatory \citep{Pilbratt10}, the \textsl{Herschel} Multitiered Extragalactic Survey \citep[HerMES;][]{Oliver2012} and the \textsl{Herschel} Astrophysical Terahertz Large Area Survey \citep[H-ATLAS;][]{Eales10} cover $\gtrsim 200\,\hbox{deg}^2$ in common with the J-PAS survey.

 A first attempt at measuring lensing-induced cross-correlations
 between \textsl{Herschel}/SPIRE galaxies and low-$z$ galaxies was
 carried out by \citet{Wang2011}. Later on, \cite{2014arXiv1401.4094G}
 report a highly significant spatial correlation between galaxies with
 $S_{350\mu\rm m}\ge 30\,$mJy detected in the equatorial fields of
 H-ATLAS ($\backsimeq 161\,\hbox{deg}^2$) with estimated redshift
 $\gtrsim 1.5$ (26,630 sources) and SDSS galaxies at $0.2\le z\le 0.6$
 (686,333 sources). The significance of the measured cross-correlation
 is much higher than those reported so far for samples with
 non-overlapping redshift distributions selected in other wavebands. 

 These works demonstrated that it possible to achieve similar, or 
even better, measured cross-correlation signal significance, compared with the QSO case, with a reduce number of foreground sources. These results open the possibility to extend the analysis on the cross-correlation function to different redshift bins and therefore, to study the evolution of quantities as the typical halo mass, the number of halo satellites or the lensing optical depth. On this respect J-PAS will provide the required large foreground sample with accurate enough photometric redshifts in order to split the cross-correlation analysis in at least $\sim 3-4$ redshift bins between $z=0.2-1.0$.

\pagebreak

\subsection{Correlations with the Cosmic Microwave Background Anisotropies (CMB)}
\label{sec:CMB}

Apart from the blue/red-shift on CMB photons induced by the ISW in over/under-dense regions, the cross-correlation of a J-PAS like survey with observations in the millimeter like those from WMAP\footnote{URL site: {\tt http://map.gsfc.nasa.gov}}, {\it Planck}\footnote{URL site {\tt http://www.rssd.esa.int/index.php?project=planck}}, ACT\footnote{URL site: {\tt http://act.princeton.edu}} or SPT\footnote{URL site: {\tt http://pole.uchicago.edu}} offer a wealth of cosmological tests, related to the physics of galaxy formation, the motion of matter and bulk flows, lensing of the CMB and the search for the missing baryons. In this section we will briefly address foreseen ISW analyses together with all those new approaches, leaving detailed forecasts for future studies.

\subsubsection{Integrated Sachs Wolfe effect}
\label{sec:isw}

The late integrated Sachs-Wolfe effect (ISW) describes the gravitational blue/red-shift imprinted on photons of the Cosmic Microwave Background (CMB) radiation as they travel through large scale, time dependent gravitational potentials at low redshifts ($z< 2$). In an accelerating universe, large scale gravitational potentials shrink and CMB photons leave wells that have become {\em shallower}, hence experiencing a gravitational blue-shift. The opposite mechanism works for large scale voids. This mechanism of gravitational blue/red-shift of CMB photons was first described by \citep{sachs1967}, referring to either non-linear structures (like galaxy clusters, aka {\it Rees-Sciama} effect) or at very early times, during the epoch of recombination (Sachs-Wolfe effect or early Sachs-Wolfe effect). 

The late ISW arises at late epochs and is a distinct signature of DE on the CMB. Since it arises at late epochs and gravitational potentials involve large scale interactions, the ISW contribution is more important on the largest angular scales, which are however dominated by the intrinsic Sachs-Wolfe anisotropies generated at the surface of Last Scattering, ($z\sim 1,050$).

\paragraph{Correlation with the LSS fluctuations}

In order to distinguish late ISW fluctuations introduced in the low redshift universe from those Sachs-Wolfe anisotropies generated during recombination, it was first suggested by ~\citep{crittenden1996} to use galaxies as probes of potential fields in order to detect the presence of ISW via a cross-correlation analysis. 
Galaxies at the appropriate redshift range should spatially sample the same large scale gravitational potential wells giving rise to the ISW, and hence the large scale galaxy angular distribution should be correlated to the ISW component that is embedded in the  CMB temperature anisotropy field. 

After the first attempts on COBE CMB data \citep{bough2002}, analysis with the higher quality CMB data from WMAP were conducted right after temperature maps were publicly released \linebreak   ~\citep{boughn2004,fosalba2003,vielva2006,pietrobon2006,ho2008,
giannantonio2008,mcewen2008,dupe2011,schiavon2012}. Those works claimed detections of the ISW in the 2 -- 4.5\,$\sigma$ range, although a number of other works either found lower statistically significant results and/or warned about the presence of systematics associated to point source emission and abnormal power on the large scales \citep{Hernandezetal2006,Rassatetal2007,Bielbyetal2010,LopezCorredoiraetal2010,HernandezMonteagudo2010,PeacockandFrancis2010,Sawangwitetal2010,Hernandezetal2013}. Recent results from the {\it Planck} collaboration, with better control of foregrounds and systematics, larger sky coverage and lensing information, provide evidence for the ISW at the $\sim$ 2--3$\,\sigma$ level \citep{planckisw}.

There are several estimators to determine the cross-correlation between the ISW fluctuations and a galaxy density field. The most used one is the cross-correlation function~\citep{giannantonio2008}, which, although it is suitable for analysis of partial/small sky coverage surveys, it is a relatively slow technique. Alternatives to this estimator are the covariance of the wavelet coefficients~\cite[CWC; e.g.,][]{vielva2006} or the cross-angular power spectrum~\cite[CAPS; e.g.,][]{HernandezMonteagudo2008}. These two approaches are typically faster than the CCF, although their handling of incomplete skies is less intuitive. There exist also optimal implementations of the cross-angular power spectrum ~\cite{schiavon2012} working on the base of a quadratic maximum-likelihood estimator~\cite[e.g.,][]{Tegmark1997}. These methods, on the one hand, provide an optimal handling of the statistical problem but, on the other hand, are the most CPU expensive (and hence slowest) of the statistical approaches considered so far. In practice, when dealing with a real data set, it is important to check the results with all the approaches, since systematics typically affect the cross-correlation estimators in different manners. For forecast purposes, working with the harmonic space is the most natural option, since, at a first approach, the CAPS can be seen as a quantity of uncorrelated components.

\begin{figure}
\centering
\includegraphics[width=0.8\textwidth]{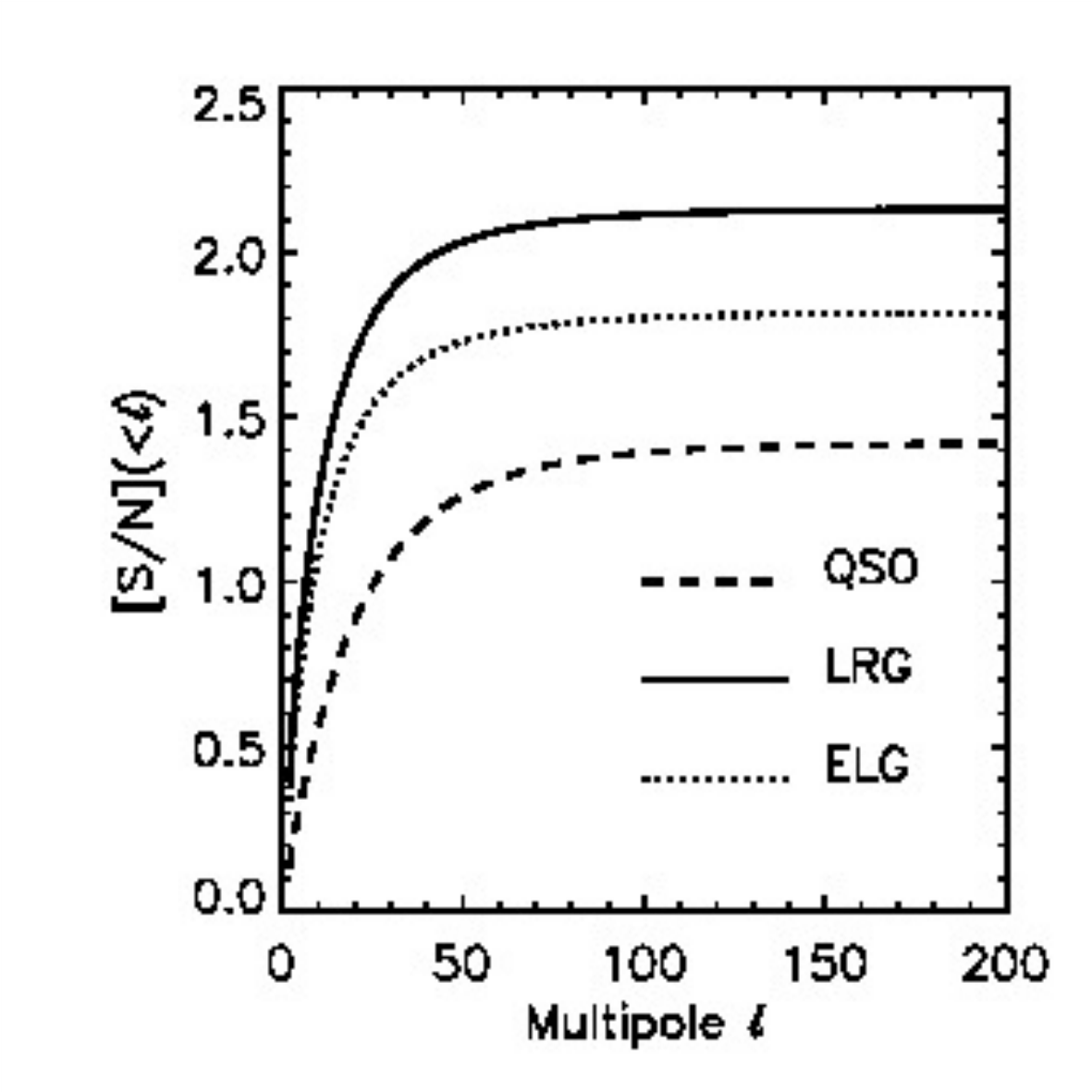}
  \caption{\label{fig:isw_s2n}
  Signal-to-noise ratio for the ISW effect, as a function of the maximum multipole considered in the analysis, for three different probes of J-PAS: LRGs (solid line), ELGs (dotted line) and QSOs (dashed line). After adding the signal from all these probes, we foresee a statistical significance of the ISW via a cross-correlation analysis at the level of $2.6\,\sigma$. 
}
\end{figure}

Both the sky coverage and the redshift depth are critical aspects of a galaxy survey to serve as a dark matter tracer to detect the ISW effect~\cite[e.g.,][]{HernandezMonteagudo2008,douspis2008}. At this respect, J-PAS offers an excellent opportunity to alternatively probe DE through the ISW effect. In Figure~\ref{fig:isw_s2n} we display the signal-to-noise to be obtained, below a given multipole $\ell$, after cross-correlating the angular distribution of LRGs, ELGs and QSOs from J-PAS with CMB maps. The LRGs should provide ISW evidence at the $\sim\,2.1\,\sigma$ level, higher than ELGs ($\sim 1.8\,\sigma$) and QSOs ($\sim 1.4\,\sigma$). When combining these three different probes and after accounting for their correlation, the total foreseen statistical significance for the ISW detection (in the standard $\Lambda$CDM scenario) amounts to $2.6\,\sigma$. This remains at the same level of evidence claimed by the {\it Planck} team, \cite{planckisw}.

\paragraph{Constraints on cosmological parameters}

Although the ISW signal is subdominant with respect to the intrinsic anisotropies in the CMB, it is a complementary probe for the dark energy properties. In particular, the ISW can help to constrain the time evolution of the equation of state of the dark energy fluid. 

\begin{figure}
\centering
\includegraphics[width=0.8\textwidth]{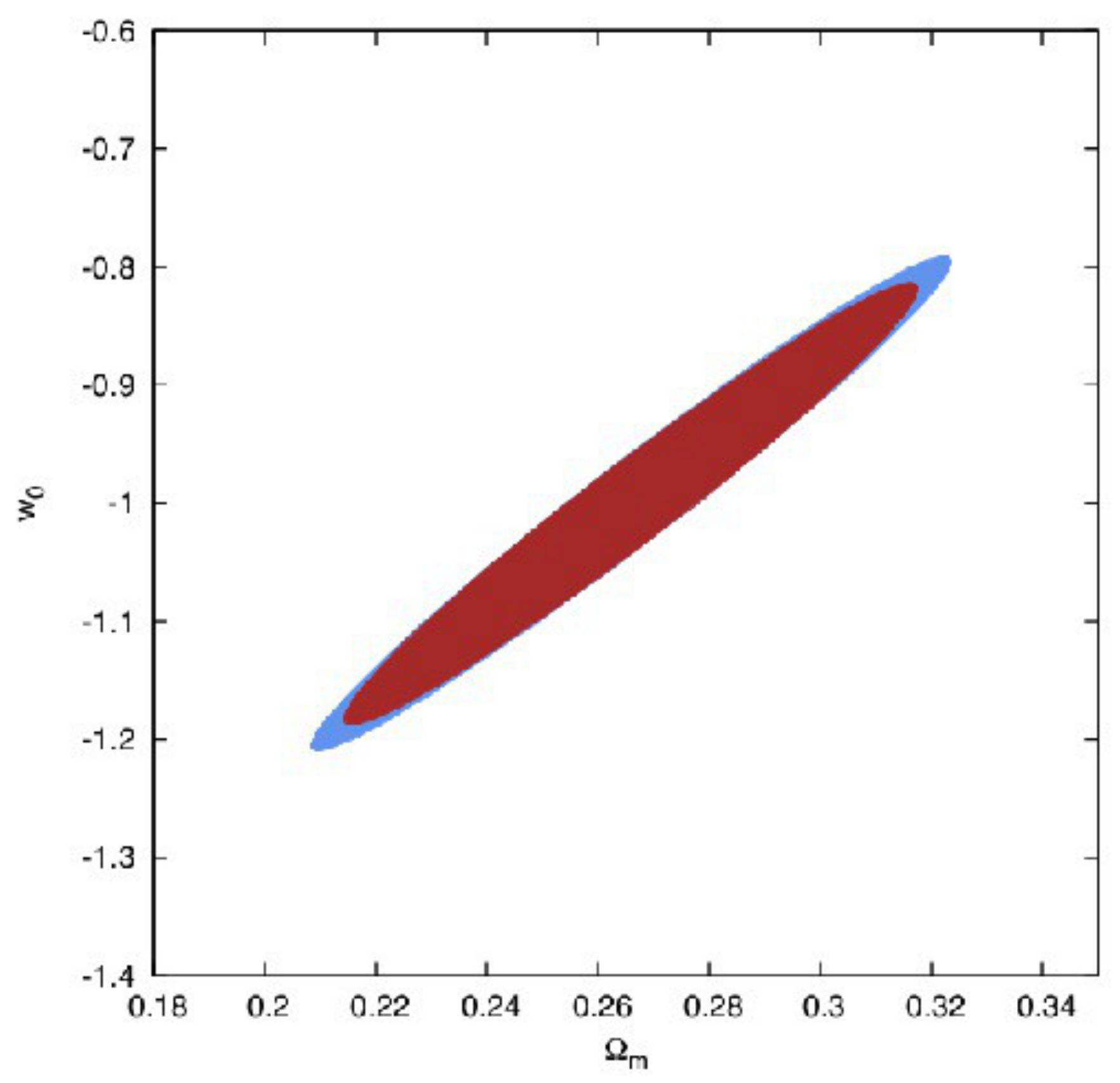}
  \caption{\label{fig:isw_eos_w0}
  Constraints of $\Omega_m$ and $w_0$ from CMB {\it Planck} alone (blue) and CMB {\it Planck} + ISW (red) at $2\sigma$ CL, for a $w_0 \ne 1$ dark energy model ($p = w_0 \rho$).}
\end{figure}

\begin{figure}
\centering
\includegraphics[width=0.8\textwidth]{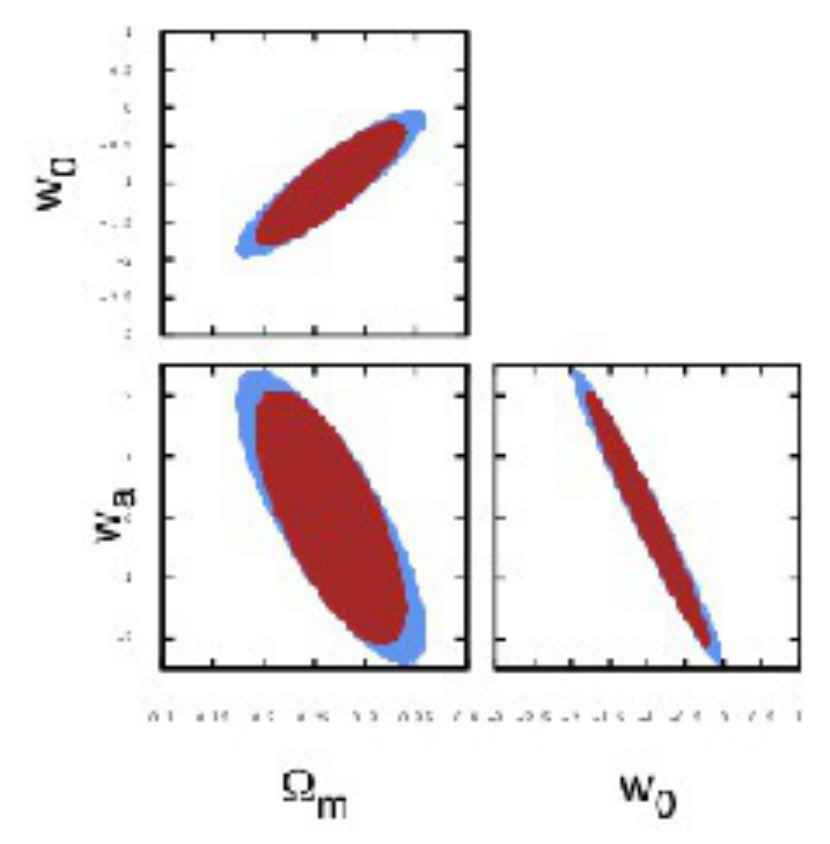}
  \caption{\label{fig:isw_eos_w0_wa}
  Constraints of $\Omega_m$, $w_0$ and $w_a$ from CMB {\it Planck} alone (blue) and CMB {\it Planck} + ISW (red) at $2\sigma$ CL, for a $w(z)$ dark energy model ($w(z) = w_0 + w_a(1+a)$, being $a$ the scale factor).}
\end{figure}

In Figure~\ref{fig:isw_eos_w0} a forecast for the a model of Dark Energy with constant equation of state $w_0 \ne 1$ ($p = w_0 \rho$) is given. The blue area represents the $2\,\sigma$ confidence level imposed by the CMB angular power spectra (from {\it Planck}), whereas the red ones correspond to including the ISW effect into the likelihood function (assumed Gaussian in this forecast). The underlying fiducial model is the WMAP 7yr best-fit~\citep{komatsu2011}. Similarly, in Figure~\ref{fig:isw_eos_w0_wa} we present the forecast for a $w(z)$ dark energy model ($w(z) = w_0 + w_a(1+a)$, being $a$ the scale factor).

\paragraph{Recovery of the ISW fluctuations}
One of the most novel analyses related to the ISW effect is the recovery of the actual ISW fluctuations produced by the gravitational potentials. One can distinguish two different types of approaches in this problem.  \citep{barreiro2008,barreiro2012} propose using 2D information of the CMB and projected galaxy density field in order to yield a minimum variance ISW map estimate. On the other hand, provided the exquisite redshift information to be provided by J-PAS, it is also possible to produce 3D density and gravitational potential maps, which can be then projected along the line of sight to generate ISW shells centered at any arbitrary redshift probed by J-PAS \cite[e.g.,][]{jasche2010}.

These analysis would probe the large angle domain of the CMB, where discussion about possible anomalies challenging the LCDM scenario are ongoing, \cite[e.g.,][]{bennett2011,planckisotropy}.
Consequently, tests of the Cosmological Principle and universal homogeneity would naturally follow.

\subsubsection{The thermal history of the Universe}

The common understanding of the process of galaxy formation pictures baryons cooling down after falling in potential wells seeded by dark matter. In order to avoid the over-cooling problem by which too massive galaxies are generated \citep{LinMohr2004}, additional gas heating mechanisms must be invoked, e.g.,\citep{Borganietal2004,McNamaraNulsen2007} and references therein. How exactly this process proceeds is a matter of active investigation currently. Measurements of the thermal Sunyaev-Zel'dovich effect \cite[hereafter tSZ,][]{tSZ} have recently been used to shed additional light on this problem. The tSZ describes the distortion that the black body spectrum of the CMB undergoes when it Compton scatters off hot electrons in collapsed structures like galaxies and groups and clusters of galaxies. Measurements from CMB experiments like ACT \citep{Handetal2012} and {\it Planck} \citep{PEPXII,PIPXI} have shown that by looking at the amplitude of the tSZ in halos it is possible to put constraints on the amount of baryonic mass residing in halos of different total mass, and hence gain insight on the feedback processes involving baryonic physics in those structures.

The unprecedented depth and volume of group and cluster catalogs to be obtained from J-PAS will constitute a very important contribution to these studies. The photometric depth of J-PAS should allow to identify $\sim 5\times 10^5$  groups down to $\sim 5\times 10^{13}$\,M$_{\odot}$ in the local universe, improving enormously current statistics. Likewise, and by first time, J-PAS should enable extend this study to earlier cosmological epochs and provide alternative constraints of the history of galaxy formation.

\subsubsection{Bulk flows, missing baryons and redshift space distortions}

About half the baryons in the local universe remain hidden to direct observations, \citep{CenOstriker1999,CenOstriker2006}. These {\it missing} baryons are expected to be in an ionized, diffuse phase also known as {\it Warm-Hot Intergalactic Medium} (WHIM). These baryons should be part of comoving flows of matter (also known as {\it bulk flows}) triggered by gravity. It turns out that the moving baryons also leave an imprint on the CMB by means of the kinetic Sunyaev-Zel'dovich effect \citep{kSZ}(hereafter kSZ). The kSZ describes the brightness anisotropies induced on the CMB by moving electrons by means of Thomson scattering: it is sensitive to radial component of the electron peculiar velocity with respect to the CMB, and its spectral dependence is identical to that of the intrinsic CMB anisotropies, making its detection difficult.

With the advent of last generation CMB experiments, the levels of angular resolution and sensitivity are approaching the ballpark required by the kSZ. Indeed, the ACT experiment provided recently a promising claim of kSZ detection \citep{Handetal2012}, while the {\it Planck} surveyor has used the limits on the kSZ to set strong constraints on the homogeneity of the universe on Gpc scales, \citep{PIPXIII}. J-PAS will provide an exquisite mapping of the large scale structure up to $z\sim 1$, with very accurate photo-$z$'s for groups and clusters. By inverting the galaxy density field into the underlying dark matter density and peculiar velocity fields \citep{Hoetal2009,Kitauraetal2012d}, it is possible to search for kSZ signatures in CMB maps by means of cross-correlation studies. This combination of J-PAS data with CMB observations would hence provide the first view of the evolution of peculiar velocity fields at different cosmological epochs.

\subsubsection{CMB lensing maps}

J-PAS will map hundreds of thousands of Quasi Stellar Objects (QSOs) in the redshift range $z\in [1.5,3]$ (see Sect.\ref{sec:redshift}), hence providing an estimate of the density map of the universe at those epochs. It is roughly in this redshift range where CMB photons are more efficiently deflected by gravitational lensing following the inhomogeneities in the distribution of matter. The presence of lensing in CMB maps has been first detected in terms of the convergence field, \citep{actlensing,sptlensing}, although the highest signal-to-noise ratio of the detection is owed to {\it Planck} \citep{planckXlensing}. A further confirmation of this effect can be obtained by cross-correlating tracers of the matter distribution at those epochs with CMB lensing convergence maps, \citep{actlenxqso,planckXlensing}. 

The J-PAS will allow to explore this CMB convergence -- matter correlation by combining its QSO catalog with lensing measurements provided by the all sky {\it Planck} mission. The huge common cosmological volume sampled in this analysis should further improve our constraints on the QSO bias evolution, and will test the model predictions on lensing on scales practically unexplored yet. As shown by, e.g., \citep{Acquavivaetal08}, it is on these larger scales where one can look at a scale dependence of the density linear growth factor to set constraints on alternative gravity theories (like $f(R)$) and further test General Relativity. For that, it is required combining RSD measurements on the larger scales with CMB lensing cross  correlations, so that the bias degeneracy may be avoided with CMB data and direct constraints on the growth factor history may be set.

\subsection{Alternative Cosmologies and Theories of gravity}

The absence of guidance from fundamental physics about the mechanism behind cosmic acceleration has given rise to a number of non-standard cosmologies. These are based either on the existence of new fields in Nature, the role of large-scale inhomogeneities or on modifications of general relativistic gravitation theory on large scales.  Combining  the expansion history measured geometrically with growth of structure data from J-PAS, it will be possible to distinguish among several of these scenarios. In what follows, we briefly discuss some of the most popular alternative models.

\subsubsection{Quintessence}

The simplest approach toward constructing a model for an accelerating universe is to work with the idea that the unknown, un-clumped dark energy component is due exclusively to a minimally coupled scalar field $\phi$ (quintessence field) which has not yet reached its ground state and whose current dynamics is basically determined by its potential energy $V(\phi)$~\cite{ac0}. The dynamics of quintessence in the presence of non-relativistic matter has been studied in detail for many different potentials (see, e.g., \cite{ac1}) and can be broadly classed into three groups: thawing, freezing and hybrid models~\cite{ac2}.

The dynamics of quintessence or $\phi$CDM models is obtained by solving the equations
\begin{equation}
3M_{Pl}^2H^2 = \frac{\dot\phi^2}{2} + V(\phi) + \rho_m\;,
\end{equation}
\begin{equation}
2M_{Pl}^2\dot{H}= -(\dot\phi^2 + \rho_m)
\end{equation}
where the pressure and the energy density of the quintessence field are given, respectively, by $p_{\phi} = \dot\phi^2/2  - V(\phi)$ and $\rho_{\phi} = \dot\phi^2/2 + V(\phi)$ and a dot represents a derivative with respect to $t$. The scalar field satisfies the continuity equation $\dot\rho_{\phi} + 3H\rho_{\phi}(1+w) = 0$, i.e.,
\begin{equation}
\ddot{\phi} + 3H\dot{\phi} + dV(\phi)/d\phi = 0\;.
\end{equation}
where $w = p_{\phi}/\rho_{\phi}$ is the equation-of-state parameter of the dark energy.

In order to realize current cosmic acceleration, the mass of the quintessence field, $m_{\phi} = \sqrt{d^2V(\phi)/d\phi^2}$ should be extremely small, i.e., $|m_{\phi}|\lesssim H_0 \simeq 10^{-33}$ eV. Although being difficulty to reconcile such a ultra light mass with the energy scales appearing in particle physics, there has been some attempts to construct realistic  quintessence models in the framework of fundamental physics~\cite{ac3}. From the observational viewpoint, tight constraints can be paced on the equation-of-state parameter combining geometric probes with the growth rate of matter perturbations $\delta_{m}$, which depends explicitly on $w$.

\subsubsection{Interaction in the dark sector}

Unless some unknown symmetry in Nature prevents or suppresses a non-minimal coupling in the dark sector, the dark energy field may interact with the pressureless component of dark matter. In recent papers, cosmological models with interaction in the dark sector were shown to be a possible alternative to the standard cosmology \citep{c1,ac4_10,Amendola2000,ac4_12,Alcaniz2005,Costa2010}. Among some possibilities, a model with constant-rate particle creation from the vacuum has the same number of parameters as the spatially flat standard model and seems to be able to alleviate some observational/theoretical tensions appearing in the latter scenario~\citep{Borges,PLB}. 

In this class of models, the dimensionless Hubble function is given by \citep{Borges}
\begin{equation} \label{E}
E_{\rm{I}}(z) \equiv \frac{H(z)}{H_0} = 1 - \Omega_{m0} + \Omega_{m0}(1+z)^{\frac{3}{2}}\;,
\end{equation}
whereas for the spatially flat $\Lambda$CDM scenario the well-known expression is written as
\begin{equation} \label{standard}
E_{\rm{\Lambda CDM}}(z) = \left(1 - \Omega_{m0} + \Omega_{m0}(1+z)^{3} \right)^{\frac{1}{2}}.
\end{equation}
Although at low-$z$, the difference between the predicted expansion rate from both models is very small, at higher $z$, e.g., $z = 1$, they provide very distinct results. Assuming $\Omega_{m0} = 0.45$ in Eq. (\ref{E}), as the best-fit concordance value for the present matter density\footnote{If the creation of particles from vacuum is important during the late times of universe expansion, the present matter density is higher than in the standard model, provided it has the standard value at early times \cite{jailson2011}.}, and $\Omega_{m0} = 0.3$ in Eq. (\ref{standard}), we find $E_{\rm{I}}(1)=1.82$ and $E_{\rm{\Lambda CDM}}(1)=1.76$. This amounts to say that the relative difference is
\begin{equation} \label{deltaE}
\frac{E_{\rm{I}}(1) - E_{\rm{\Lambda CDM}}(1)} {E_{\rm{\Lambda CDM}}(1)} = 3.4\%,
\end{equation}
which is slightly larger than the expected uncertainties in J-PAS BAO data at this redshift (see Sec. 3.2). For the interaction models described in \cite{Wang:2006qw} the relative differences are even higher, above $6\%$ for $z = 1$.

Besides the phenomenological models, interacting models based on field theory have also been discussed in the literature~\cite{fieldtheory}. In the particular case of an interaction described by a coupling $\xi$, such that~\cite{He:2010ta}
\begin{equation}
\dot{\rho}+H\left[\left(2-\frac{\xi}{3}\right)\rho_{K}+\left(1-\frac{\xi}{3}\right)\rho_{W}\right] =0\;,\label{alpha}
\end{equation}
constraints on the interacting parameter $\xi$ can be investigated from different sets of observations~\cite{fengetal}. In this regard, an interesting possibility comes from galaxy clusters measurements since close to balance ($\dot{\rho}=0$) the virial ratio should be approximated by 
\begin{equation}
\frac{\rho_{K}}{\rho_{W}}\simeq -\frac{1-\frac{\xi}{3}}{2-\frac{\xi}{3}}\;.\label{eq:KsW}
\end{equation}
Clusters, therefore, turn out to be good probes for these models. Indeed, the very possibility of an interaction of clumping matter with an external object, here dark energy, leads to consequences for the virial condition. Thus, the virial condition is a good test for the dynamics of the dark sector. This has been performed with a small sample of clusters \cite{Abdalla:2007rd,Abdalla:2009mt} and a wider set should provide further restrictions.

\subsubsection{Unified models of dark matter and dark energy}

From the cosmological viewpoint, the main distinction between
pressureless CDM and dark energy is that the former agglomerates at
small length scales whereas the latter is a smooth component on these
scales. Recently, the idea of a unified description for CDM and dark
energy has received much attention
\citep{ac4_1,ac4_2,ac4_3,ac4_4,ac4_5,ac4_6,ac4_7,ac4_8,ac4_9}.
An interesting attempt in this direction was suggested in \citet{ac5}
and further developed in \citet{ac7}.
 It uses to an exotic fluid, the so-called Chaplygin gas (Cg), whose equation of state is given by
\begin{equation}
p_{Cg} = -\frac{A}{\rho_{Cg}^\alpha}\;.
\end{equation}
Inserting the above equation into the energy conservation equation gives the expression for the Cg energy density
\begin{equation}
\rho_{Cg} = \rho_{Cg,0}\left[A_s + (1+A_s)(1+z)^{3(1+\alpha)}\right]^{1/1+\alpha}\;,
\end{equation}
where $A_s = A/\rho_{Cg_0}^{1+\alpha}$ is a quantity related to the sound speed of the Chaplygin gas today. From the above equations, it is clear that the Chaplygin gas interpolates between epochs dominated by non-relativistic matter [$\rho_{Cg}(z >> 1) \propto z^3$] and by a negative-pressure time-independent dark energy [$\rho_{Cg}(z \sim 0) = \rm{const.}$]. Observationally, one of the major difficulties of these models concern the predicted oscillations or instabilities in the matter power spectrum. In this regard, J-PAS data can tightly constrain the idea of unified models of the dark sector and verify if it may or not constitute a viable alternative to the standard model.

\newpage

\subsubsection{The Lema\^itre-Tolman-Bondi Models}

Recently, inhomogeneous cosmologies have gathered considerable interest as a possible explanation for current cosmological observations without invoking a dark energy field. In the simplest class of such models our location is close to the center of a large, spherically symmetric void described by the Lema\^ire-Tolman-Bondi (LTB) metric \cite{ac6} 
\begin{equation} \label{ltb}
ds^2 = dt^2 - \frac{A'^2(r,t)}{1-k(r)}dr^2 - A(r,t)d\Omega^2\;,
\end{equation}
where $d\Omega^2 = d\theta^2 + \sin^2\theta d\phi^2$, $k(r)$ is the radial position-dependent curvature function and $A(r, t)/r$ a position-dependent scale factor. Plugging Eq. (\ref{ltb}) into the Einstein equations, one finds that the two independent equations are
\begin{equation}
\frac{\dot{A} + k(r)}{A^2} + \frac{2\dot{A}\dot{A}' + k'(r)}{AA'} = 8\pi G(\rho_m + \rho_\Lambda)\;,
\end{equation}
\begin{equation}
{\dot{A}^2} + 2{A}\ddot{A} + k(r) = 8\pi G \rho_\Lambda A^2\;,
\end{equation}
which provide the following generalized acceleration equation
\begin{equation}
\frac{2}{3}\frac{\ddot{A}}{A} + \frac{1}{3}\frac{\ddot{A}'}{A'} = -\frac{4\pi G}{3}(\rho_m - 2\rho_\Lambda)\;.
\end{equation}
Clearly, cosmic acceleration is possible in these models even for $\rho_\Lambda = 0$ if the angular or radial scale factor is decelerating fast enough. Since the LTB metric allows for different rates of expansion in the longitudinal and transverse directions, the combination of data constraining the transverse Hubble rate (e.g., SNe Ia observations), together with J-PAS measurements of the radial BAO scale in many redshift slices, will be able to constrain this class of models as well as the hypothesis of large-scale homogeneity and isotropy (see, e.g. \cite{bellido}).

\subsubsection{Most general scalar-tensor theories}

Modified theories of gravity have recently been applied to cosmology as a realistic alternative approach to the late-time cosmic acceleration. There are many modified gravitational theories proposed in literature. Most of them, however, belong to a general class of scalar-tensor theories dubbed Horndeski theories \cite{Horndeski}. The Horndeski theories are constructed to keep the space-time derivatives of the field equations of motion 
up to second order, whose Lagrangian is given by \cite{DGSZ}
\begin{equation}
{\cal L}=\sum_{i=2}^{5}{\cal L}_{i}\,,
\label{Lagsum}
\end{equation}
where 
\begin{eqnarray}
{\cal L}_{2} & = & K(\phi,X),\label{eachlag2}\\
{\cal L}_{3} & = & -G_{3}(\phi,X)\Box\phi,\\
{\cal L}_{4} & = & G_{4}(\phi,X)\, R+G_{4,X}\,
[(\Box\phi)^{2}-(\nabla_{\mu}\nabla_{\nu}\phi)\,(\nabla^{\mu}\nabla^{\nu}\phi)]\,,\\
{\cal L}_{5} & = & G_{5}(\phi,X)\,
G_{\mu\nu}\,(\nabla^{\mu}\nabla^{\nu}\phi) \nonumber\\
& & -\frac{1}{6}\, G_{5,X}\,[(\Box\phi)^{3}-3(\Box\phi)\,(\nabla_{\mu}\nabla_{\nu}\phi)\,(\nabla^{\mu}\nabla^{\nu}\phi)+2(\nabla^{\mu}\nabla_{\alpha}\phi)\,(\nabla^{\alpha}\nabla_{\beta}\phi)\,(\nabla^{\beta}\nabla_{\mu}\phi)]\,.\label{eachlag5}
\end{eqnarray}
\newpage
$K$ and $G_{i}$ ($i=3,4,5$) are functions in terms of a scalar
field $\phi$ and its kinetic energy $X=-\partial^{\mu}\phi\partial_{\mu}\phi/2$
with the partial derivatives $G_{i,X}\equiv\partial G_{i}/\partial X$,
$R$ is the Ricci scalar, and $G_{\mu\nu}$ is the Einstein tensor. 
The Lagrangian (\ref{Lagsum}) involves only one scalar degree of freedom.

Quintessence and k-essence are described by the functions $G_3=0$, 
$G_4=M_{\rm pl}^2/2$, and $G_5=0$, where $M_{\rm pl}$ is 
the reduced Planck mass whereas the Brans-Dicke (BD) theory \cite{BD} corresponds to 
$K=\omega_{\rm BD}X/\phi-V(\phi)$, $G_3=0$, $G_4=\phi/2$, and $G_5=0$, 
where $\omega_{\rm BD}$ is a constant.
The $f(R)$-gravity in the metric and Palatini formalisms 
are the special cases of BD theory with 
$\omega_{\rm BD}=0$ and $\omega_{\rm BD}=-3/2$, respectively \cite{fr}.
The covariant Galileon \cite{Galileon} corresponds to the choice 
$K=X-c_2\phi$, $G_3=c_3X$, $G_4=M_{\rm pl}^2/2+c_4X^2$, 
and $G_5=c_5X^2$, where $c_i$'s are constants.

In the following we also take into account a barotropic perfect fluid of 
non-relativistic matter (cold dark matter and baryons) 
minimally coupled to the field $\phi$.
Then the total 4-dimensional action is given by 
\begin{equation}
S=\int d^{4}x\sqrt{-g}\left({\cal L}+{\cal L}_{m}\right)\,,
\label{action4}
\end{equation}
where $g$ is a determinant of the metric $g_{\mu\nu}$, 
and ${\cal L}_{m}$ is the Lagrangian of 
non-relativistic matter with the energy density $\rho_m$.

\paragraph{Background equations}

Assuming a flat FLRW space-time, the background equations of 
motion following from the action (\ref{action4}) read \cite{DKT}
\begin{eqnarray}
 &  & 3H^2 M_{\rm pl}^2=\rho_{\rm DE}+\rho_m\,,\label{be1}\\
 &  & 2\dot{H}M_{\rm pl}^2=-(\rho_{\rm DE}+P_{\rm DE})-\rho_m \,,
 \label{be2} \\
 &  & \dot{\rho}_m+3H \rho_m =0\,,
 \label{be3}
\end{eqnarray}
where $H=\dot{a}/a$ is the Hubble parameter, a dot represents a derivative with respect to the cosmic time  $t$, and the energy density and pressure , written in terms of derivatives of the scalar field $\phi$ (see, e.g., ~\cite{DKT}) satisfy the usual continuity equation. From the above equations, we can also define the equation-of-state parameter, $w_{\rm DE}=P_{\rm DE}/\rho_{\rm DE}$. For a given model, the evolution of $w_{\rm DE}$ is known by solving Eqs.~(\ref{be1})-(\ref{be3}).
\paragraph{Cosmological perturbations}

We consider the scalar metric perturbations $\Psi$ and $\Phi$
in the longitudinal gauge about the flat FLRW background.
The perturbed line element is then given by
\begin{equation}
ds^{2}=-(1+2\Psi)\, dt^{2}+a^{2}(t)(1+2\Phi) d {\bm x}^2\,.
\label{permet}
\end{equation}
We decompose the scalar field and the non-relativistic 
matter density into the background and inhomogeneous 
parts, as $\phi(t)+\delta\phi(t, \bm{x})$ and $\rho_m(t)
+\delta \rho_m (t, \bm{x})$, respectively. 
The four velocity of non-relativistic matter can be written 
in the form $u^{\mu}=(1-\Psi, \nabla^i v)$, where $v$ is the 
rotational-free velocity potential. 
We also introduce the following quantities
\begin{equation}
\delta \equiv \delta \rho_m/\rho_m\,,\qquad
\theta \equiv \nabla^2 v\,.
\end{equation}
\newpage
In Fourier space the matter perturbation obeys the 
following equations of motion 
\begin{equation}
\dot{\delta}+\theta/a+3\dot{\Phi}=0\,,
\qquad
\dot{\theta}+H \theta-(k^2/a)\Psi=0\,,
\label{per1}
\end{equation}
where $k$ is a comoving wavenumber.
Introducing the gauge-invariant density contrast
$\delta_m \equiv \delta+(3aH/k^2) \theta$, 
it follows that 
\begin{equation}
\ddot{\delta}_m+2H \dot{\delta}_m
+(k^2/a^2) \Psi=3 ( \ddot{I}
+2H \dot{I})\,,\qquad {\rm where} \qquad
I \equiv (aH/k^2) \theta-\Phi\,.
\label{delmeq0}
\end{equation}

The full linear perturbation equations for the action (\ref{action4})
have been derived in Ref.~\cite{DKT}.
For the scales relevant to the large-scale structure
one can employ the quasi-static approximation on 
sub-horizon scales, under which the dominant contributions 
to the perturbation equations are those including the terms
$k^2/a^2$, $\delta$, and the mass $M$ of a scalar 
degree of freedom. Under this approximation we obtain the modified
Poisson equation \cite{DKT}
\begin{equation}
\frac{k^2}{a^2} \Psi \simeq 
-4\pi G_{\rm eff} \rho_m \delta\,,
\label{Poisson}
\end{equation}
where $G_{\rm eff}$ is the effective gravitational coupling 
defined by 
\begin{equation}
G_{\rm eff}=\frac{2M_{\rm pl}^2 [(\C_2 \C_5-\C_3^2)
(k/a)^2-\C_2 M^2]}{(\C_1^2 \C_2+\C_4^2 \C_5-2\C_1
\C_3 \C_4)(k/a)^2-\C_4^2 M^2} G\,.
\end{equation}
Here $G$ is the bare gravitational constant related to $M_{\rm pl}$ 
via $G=1/(8\pi M_{\rm pl}^2)$. 
The coefficients $\C_i$ ($i=1, \cdots, 5$) are 
\begin{eqnarray}
\C_1 & \equiv & -2XG_{{3,X}}-4H\left(G_{{4,X}}+2XG_{{4,{\it XX}}}\right)\dot{\phi}+2G_{{4,\phi}}+4XG_{{4,\phi X}}\nonumber \\
 &  & +4H\left(G_{{5,\phi}}+XG_{{5,\phi X}}\right)\dot{\phi}-2{H}^{2}X\left(3G_{{5,X}}+2XG_{{5,{\it XX}}}\right)\,,\\
\C_2 & \equiv & 4[G_{4}-X(\ddot{\phi}\, G_{5,X}+G_{5,\phi})]\,,\\
\C_3 & \equiv & -4G_{{4,X}}H\dot{\phi}-4(G_{{4,X}}+2XG_{{4,{\it XX}}})\ddot{\phi}+4\, G_{{4,\phi}}-8XG_{{4,\phi X}}\nonumber \\
 &  & +4(G_{{5,\phi}}+XG_{{5,\phi
     X}})\ddot{\phi}-4H[(G_{{5,X}}+XG_{{5,{\it
       XX}}})\ddot{\phi}-G_{{5,\phi}}+XG_{{5,\phi X}}]\dot{\phi}\nonumber\\
& & +4X[G_{{5,\phi\phi}}-(H^{2}+\dot{H})G_{{5,X}}],\\
\C_4 & \equiv & 4[G_{4}-2XG_{4,X}-X(H\dot{\phi}\, G_{5,X}-G_{5,\phi})]\,,\\
\C_5 & \equiv & -K_{,X}-2\left(G_{{3,X}}+XG_{{3,{\it XX}}}\right)\ddot{\phi}-4HG_{{3,X}}\dot{\phi}+2G_{{3,\phi}}-2XG_{{3,\phi X}}\nonumber \\
 &  & +[-4\, H(3\, G_{{4,{\it XX}}}+2XG_{{4,{\it XXX}}})\ddot{\phi}+4H(3G_{{4,\phi X}}-2XG_{{4,\phi{\it XX}}})]\dot{\phi}+(6\, G_{{4,\phi X}}+4XG_{{4,\phi{\it XX}}})\ddot{\phi}\nonumber \\
 &  & -20{H}^{2}XG_{{4,{\it XX}}}+4XG_{{4,\phi\phi X}}-4\dot{H}(G_{4,X}+2XG_{{4,{\it XX}}})-6{H}^{2}G_{4,X}\nonumber \\
 &  & +\{4H(2G_{{5,\phi X}}+XG_{{5,\phi{\it XX}}})\ddot{\phi}-4H[(H^{2}+\dot{H})(G_{{5,X}}+XG_{{5,{\it XX}}})-XG_{{5,\phi\phi X}}]\}\dot{\phi}-4H^{2}X^{2}G_{{5,\phi{\it XX}}}\nonumber \\
 &  & -2H^{2}(G_{{5,X}}+5XG_{{5,{\it XX}}}+2{X}^{2}G_{{5,{\it XXX}}})\ddot{\phi}+2(3H^{2}+2\dot{H})G_{5,\phi}+4\dot{H}XG_{{5,\phi X}}+10{H}^{2}XG_{{5,\phi X}}\,.
\end{eqnarray}
The explicit form of the mass term $M$ can be found 
in Refs.~\cite{DKT}.

Under the quasi-static approximation on sub-horizon scales 
the r.h.s. of Eq.~(\ref{delmeq0}) can be neglected relative 
to the l.h.s. of it.
Since $\delta_m \simeq \delta$,
the matter perturbation obeys the following equation 
\begin{equation}
\delta_{m}''+\left(2+\frac{H'}{H}\right)\delta_{m}'
-\frac{3}{2}\frac{G_{{\rm eff}}}{G}\Omega_{m}\delta_{m}\simeq0\,,
\label{delmeq}
\end{equation}
where $\Omega_{m}\equiv\rho_{m}/(3\Mpl^{2}H^{2})$, and a prime represents
a derivative with respect to $N=\ln a$.
In order to quantify the difference between the two gravitational potentials
we introduce the following anisotropic parameter
\begin{equation}
\eta \equiv -\Phi/\Psi\,.
\label{etadef}
\end{equation}
On sub-horizon scales this is approximately given by 
\begin{equation}
\eta \simeq \frac{(\C_4 \C_5 -\C_1 \C_3)(k/a)^2-\C_4 M^2}
{(\C_2 \C_5-\C_3^2)(k/a)^2-\C_2 M^2}\,.
\end{equation}
We introduce the effective gravitational potential $\Phi_{\rm eff}$
associated with the deviation of the light rays in CMB and 
weak lensing observations, 
$\Phi_{\rm eff} \equiv (\Psi-\Phi)/2$ \cite{Amendola}.
Using Eqs.~(\ref{Poisson}) and (\ref{etadef}), we obtain
\begin{equation}
\Phi_{\rm eff} \simeq -\frac{3}{2} \frac{G_{\rm eff}}{G}
\frac{1+\eta}{2} \left( \frac{aH}{k} \right)^2 
\Omega_m \delta_m\,.
\label{Phieff}
\end{equation}
We also define
\begin{equation}
f_m \equiv \frac{\dot{\delta}_m}{H \delta_m}=
\frac{\delta_m'}{\delta_m}\,.
\end{equation}

The galaxy perturbation $\delta_g$ is related 
with $\delta_m$ via the bias factor $b$, i.e.
$\delta_g=b \delta_m$.
The galaxy power spectrum ${\cal P}_g^{s} ({\bm k})$ in 
the redshift space can be modeled as \cite{1987MNRAS.227....1K} (see also Sec. 3.1.1)
\begin{equation}
{\cal P}_g^{s} ({\bm k})={\cal P}_{gg} ({\bm k})
+2\mu^2 {\cal P}_{g \theta} ({\bm k})
+\mu^4 {\cal P}_{\theta \theta} ({\bm k})\,,
\label{Pred}
\end{equation}
where $\mu={\bm k} \cdot {\bm r}/(kr)$ is the cosine 
of the angle of the ${\bm k}$ vector to the line 
of sight (vector ${\bm r}$).
${\cal P}_{gg} ({\bm k})$ and ${\cal P}_{\theta \theta} ({\bm k})$
are the real space power spectra 
of galaxies and $\theta$, respectively, 
and ${\cal P}_{g \theta} ({\bm k})$ is the cross 
power spectrum of galaxy-$\theta$ fluctuations
in real space.
In Eq.~(\ref{Pred}) we have not taken into account 
the non-linear effect coming from the velocity distribution 
of galaxies in collapsed structures.

For the linearly evolving perturbations, the first of Eq.~(\ref{per1}) shows that 
$\theta$ is related with the growth rate of matter perturbations, i.e. 
\begin{equation}
\theta/(aH) \simeq -f_m \delta_m\,,
\label{continuity}
\end{equation}
where we neglected the $\dot{\Phi}$ term.
In this case the three power spectra on the r.h.s. of Eq.~(\ref{Pred})
have the same shape, leading to \cite{1987MNRAS.227....1K}
\begin{equation}
{\cal P}_g^{s} ({\bm k})={\cal P}_{g} ({\bm k})
\left(1+2 \mu^2 \beta+\mu^4 \beta^2 \right)\,,
\label{Pred2}
\end{equation}
where $\beta=f_m/b$ and ${\cal P}_{g} ({\bm k})$ is the real space galaxy spectrum. Using Eq.~(\ref{Pred2}), one can constrain 
$\beta$ and $b \sigma_{8}$ from observations.
Provided that the continuity equation (\ref{continuity}) 
holds, the normalizations of ${\cal P}_{gg}$, 
${\cal P}_{g \theta}$, and ${\cal P}_{\theta \theta}$ in 
Eq.~(\ref{Pred}) depend on $(b \sigma_8)^2$, 
$(b \sigma_8)(f_m \sigma_8)$, and 
$(f_m \sigma_8)^2$, respectively.
Then the redshift space distortions (RSD) 
can be also modeled as an additive component by observing 
$b \sigma_8$ and $f_m \sigma_8$ \cite{Song09}. 
The quantity $f_m \sigma_8$ has an advantage over $\beta$
in that it can be measured without knowing the bias factor $b$.

\newpage

\subsubsection{$f(R)-Gravity$}

Among several possibilities (see, e.g., \cite{mg}), the simplest extension of Einstein's general relativity is the so-called $f(R)$-gravity~\cite{review2}. In the metric formalism, this theory is characterized by the functions $K=-(M_{\rm pl}^2/2)(Rf_{,R}-f)$, $G_3=0$, $G_4=M_{\rm pl}\phi/2$,  and $G_5=0$ with $\phi=M_{\rm pl}f_{,R}$ in the Horndeski action. The field equations are given by
\begin{equation}
f_{,R}R_{\mu \nu}(g) - \frac{1}{2}f(R)g_{\mu \nu} -  \nabla_{\mu}\nabla_{\nu}f_{,R} + g_{\mu \nu} \square f_{,R} = \kappa^2 T_{\mu \nu}\;,
\end{equation}
whose trace is written as
\begin{equation}
3\square f_{,R} + f_{,R}R - 2f(R) = \kappa^2 T\;,
\end{equation}
where $f_{,R}$ denotes partial derivative with respect to $R$ and $T_{\mu \nu}$ is the usual energy-momentum tensor of matter fields.

In what follows, we consider a metric $f(R)$-gravity model described by the Lagrangian ${\cal L}=(M_{\rm pl}^2/2)f(R)$ with \cite{Humodel}
\begin{equation}
f(R)=R-\lambda R_{c}\frac{(R/R_{c})^{2n}}{(R/R_{c})^{2n}+1}\,,
\label{Hu}
\end{equation}
where $n$, $\lambda$, and $R_c$ are positive constants. In the early cosmological epoch ($R \gg R_c$) the model is close to the $\Lambda$CDM model ($f(R) \simeq R-\lambda R_c$), but there is the deviation from the standard scenario at late times. Substituting these functions into Eqs.~(\ref{be1})-(\ref{be3}) and solving 
them numerically, the dark energy equation of state $w_{\rm DE}$ for the model (\ref{Hu}) starts to evolve from the value $-1$ and then it typically enters the 
phantom region $w_{\rm DE}<-1$ by today \cite{Humodel}.

In $f(R)$ gravity the scalar mass $M$ is approximately given by $M^2 \simeq 1/(3f_{,RR})$ for $M^2 \gg H^2$ \cite{Staro}. When $M^2 \gg k^2/a^2$ the perturbations are in the GR regime  where $G_{\rm eff} \simeq G$ and $\eta \simeq 1$. At late times there is the transition to the ``scalar-tensor'' regime ($M^2 \ll k^2/a^2$) in which $G_{\rm eff} \simeq 4G/3$ and $\eta \simeq 1/2$. For larger $k$ the transition from the GR regime to the scalar-tensor regime occurs earlier \cite{Staro}. The epoch of transition also depends on the model parameters $n$ and $\lambda$. If all the perturbation modes relevant to large-scale structures are in the scalar-tensor regime today, they show at present a $\gamma$ index in the range $[0.40, 0.43]$ \cite{dispersion}. If some of the modes are in the GR regime today, the $\gamma$ values at $z=0$ should range from 0.40 up to 0.55.  Furthermore, in the scenario where all modes are in the scalar-tensor regime, then at higher redshifts the growth index should generally 
decrease with increasing redshift, reaching values as low as $\gamma \simeq 0.1$ at $z\sim 1$ \cite{dispersion}. 

\begin{figure}
\centering
\includegraphics[width=0.8\textwidth]{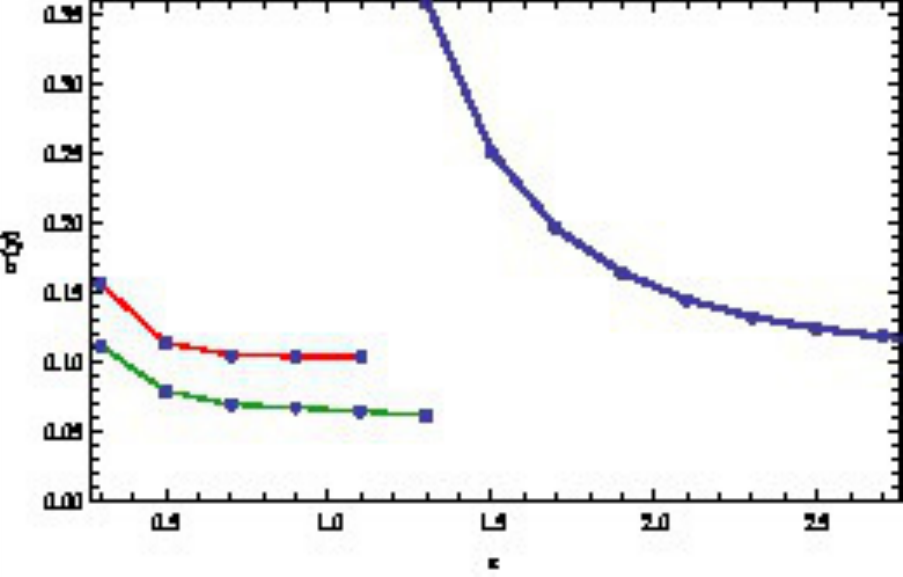}
  \caption{\label{fig:gamma}
Cumulative constraints on the modified gravity parameter $\gamma$, as a function of redshift. 
This plot shows how the constraints improve as we include each additional redshift slice.
The red line, leading up to $z=1.1$, denotes the constraints from RGs; 
the green line, up to $z=1.3$, denotes the constraints from ELGs;
and the blue line, which extends to $z>2.7$, denotes the constraints from QSOs.
}
\end{figure}

In Fig. \ref{fig:gamma} we show the aggregate uncertainties on $\gamma$ from each type of 
tracer detected by J-PAS, as a function of $z$. 
The uncertainties in $\gamma$ displayed in this figure should also allow discerning among different 
modified gravity models for which the growth  index $\gamma$ depends on the  wave number 
$k$ and the redshift $z$, as motivated by the discussion above. 
Provided the foreseen errors on $\gamma$ shown in this figure are roughly at the level of 
$\sigma_{\gamma}\simeq 0.07$--$0.12$ for LRGs and ELGs, respectively, such scenarios should be easily distinguishable from GR. 

We also note that, in the so-called covariant Galileon model \cite{Kase}, the growth rate of matter perturbations and the variation of $\gamma$ are generally larger than those in $f(R)$ gravity. Therefore, we expect this kind of modified theory to be even more tightly constrained by the J-PAS data.

\subsubsection{Vector-tensor theories of gravity}
Modified gravities involving new vector degrees of freedom have received much attention
in recent years motivated in part by the problem of the large angle 
anomalies observed in the CMB temperature maps
which could suggest the existence of preferred spatial directions. Thus, the most general 
action for a vector-tensor theory without any restriction but having linear second order equations
of motion reads: 
\begin{eqnarray}
S=\int d^4x \sqrt{-g}\left(-\frac{R}{16\pi G}+\omega R A_\mu A^\mu
+\sigma R_{\mu\nu}A^\mu A^\nu+\lambda(\nabla_\mu A^\mu)^2+\epsilon F_{\mu\nu}
F^{\mu\nu}-V(A_\mu A^\mu)
\right).
\label{action}
\end{eqnarray}
In the so-called Einstein-Aether theories  \cite{Zlosnik}, the norm of the field is fixed by means of 
a Lagrange multiplier $\lambda(A_\mu A^\mu\pm m^2)$ so that $A_\mu$ can be constrained 
to be either time-like or space-like. In the time-like case, it has been shown that this kind of
fields can act as dark matter.  As a matter of fact this kind of Einstein-Aether theories can 
be understood as relativistic versions of the MOdified Newtonian Dynamics (MOND) theory 
\cite{Milgrom}  proposed to explain galactic rotation curves from a modification
of Newton second law at low accelerations. This kind of theories can mimic some of the 
properties of cosmological dark matter
but the predictions are in tension with CMB and LSS observations. Further developments of this
kind
of theories are the Tensor-Vector-Scalar (TeVeS) \cite{teves} theories which could provide accelerated
expansion solutions \cite{Diaz-Rivera}.

In the general case in which the norm is not fixed,  it is possible to construct dark energy models 
by choosing appropriate
potential terms \cite{Armendariz,Kiselev}.
Nevertheless, one of the most appealing properties of vector-tensor theories is that unlike scalar ones, they can generically give rise to periods of acceleration even in the absence of potential terms \cite{evolution}.
Thus, it is possible to show that in the $V=0$ case, there are six models whose PPN parameters
are exactly the same as in General Relativity and therefore do not suffer from inconsistencies with
local gravity tests \cite{viability}, namely: $\sigma=-4\lambda=-4\epsilon$, $\sigma=-3\lambda=-2\epsilon$, $\sigma=0$ and $\sigma=m\epsilon$ with $m=0,-2,-4$, 
all of them with $\omega=0$. However, in general these models exhibit classical or quantum
instabilities in certain regions of the parameter space. There is however a particular case
which is stable both at the classical and at the quantum level corresponding to $\sigma=\omega=0$ \cite{EM}, and behaves at the background level exactly as $\Lambda CDM$. The corresponding perturbations
have speed of sound $c_s^2=1$ and vanishing anisotropic stress $\pi=0$ i.e. $\Phi=-\Psi$ and therefore the model behaves as a
quintessence theory without potential term. Notice that unlike $\Lambda CDM$ or scalar-tensor
theories, this model does not include  dimensional parameters in the action apart from Newton constant.

This is an example of one of the main difficulties when trying to determine the nature of 
dark energy from observations which is the degeneracy problem \cite{Kunz}, i.e. different dark energy models
or modified gravities can give rise to the same background evolution. This degeneracy can
be broken in certain cases at the level of perturbations. Thus for example, modified gravity theories
involving geometric degrees of freedom  generically  predict non-vanishing anisotropic stress 
$\pi\neq 0$ unlike standard dark energy models. 

\subsubsection{Higher dimensions and massive gravity}
 Modifications of gravity resorting to extra dimensions were proposed by Dvali-Gabadadze-Porrati
(DGP) \cite{DGP}. In these models our universe is understood as a 3-brane embedded in a
five-dimensional bulk space. The corresponding gravitational action reads:
\begin{eqnarray}
S=-\frac{M_5^3}{2}\int d^5x \sqrt{-g}R_5-\frac{M_p^2}{2}\int d^4x \sqrt{-h}R_4+S_{GH}
\end{eqnarray}
with $M_5$ the 5-dimensional Planck scale, $R_5$ is the Ricci scalar in five dimensions and
$S_{GH}$ a boundary term. 
In this model, gravity behaves as ordinary four-dimensional General Relativity on small 
scales, whereas on large scales the gravitational interaction leaks into the bulk. The corresponding 
cross-over scale is: $r_c=M_p^2/(2M_5^3)$. In a flat Robertson-Walker background, the DGP 
model predicts a modified Friedmann equation given by: 
\begin{eqnarray}
\left(1-\frac{\epsilon}{H r_c}\right)H^2=\frac{8\pi G}{3}\rho
\end{eqnarray}
   with $\epsilon=\pm 1$. When $Hr_c\gg 1$, i.e.  the Hubble radius is much smaller than the cross-over 
scale, we recover the standard Friedmann equation. However at late times, the modification 
implies that in a matter dominated universe in the so-called self-accelerating
branch $\epsilon=+1$,  the scale factor accelerates towards a de Sitter regime. Unfortunately 
this branch has a ghost-like instability.  Despite this fact, the DGP model 
provides the first example of degravitation, i.e. the possibility of modifying General Relativity 
in the infrared in such a way that gravity weakens on large scales. This is a generic 
feature of massive gravity theories and has been proposed as a way to weaken the 
effects of vacuum energy on the geometry.  

In massive gravities, the graviton contains five degrees of freedom, namely the two standard
helicity-2 modes, two additional helicity-1 mode and one helicity-0. The standard Fierz-Pauli 
Lagrangian for massive gravity is known to exhibit certain ghosts instabilities associated to 
non-linearities containing higher derivatives in the helicity-0 sector. Recently a new massive gravity theory has been proposed  \cite{deRham} in 
which all the nonlinearities containing higher than second derivatives are eliminated. This model has been shown \cite{Heisenberg}
to exhibit self-accelerated solutions without the instability problems of the original DGP model.

\subsection{Inflation}

 Inflation represents a period of accelerated expansion of the universe at very early times that provides the appropriate conditions to give rise 
to some of the present properties of the universe such as homogeneity, isotropy and flatness. In addition, quantum fluctuations of the inflaton 
field that dominates the dynamics of the universe at those early times are the seeds that evolved via gravitational instability to the present 
large-scale structure of the universe. 

The simplest models of inflation produce a nearly scale invariant primordial power spectrum with a spectral index $n_s$ close to one. 
Beyond the power spectrum, inflation also predicts density perturbations with a distribution very close to that of a homogeneous and isotropic Gaussian random field. The degree of deviation from Gaussianity depends on the specific model and is characterized by the so-called non-linear coupling parameter $f_{NL}$ \citep{2000MNRAS.313..141V,2007PhRvD..76h3004S,2008PhRvD..77l3514D}.

Large-scale galaxy surveys are among the best cosmological observations to provide information about the physics of inflation. J-PAS will be able to measure with high sensitivity the parameters characterizing the primordial power spectrum, $n_s$ and its $running$, as well as possible departures from Gaussianity, $f_{NL}$. This information, in combination with CMB measurements, will help in understanding the properties of the inflationary potential.

Beyond these parameters, J-PAS will also serve to probe fundamental implications of standard inflation as homogeneity, isotropy and Gaussianity. The large-scale anomalies recently confirmed by Planck ~\cite{planck23}, indicates some hints of statistical anisotropy at scales above several degrees. Although the sky coverage of J-PAS is not wide enough to study in detail these anomalies, it is sufficient to explore whether homogeneity and isotropy are hold.

The capabilities of J-PAS to constrain deviations from Gaussianity are much higher. In particular, primordial non-Gaussianity produces a non-linear bias in the galaxy clustering that will be measured by J-PAS, providing constraints on the local shape of $f_{NL}<8.2, 4.7, 1.8$ for ELGs, RGs, QSOs, respectively. The expected combined sensitivity is $f_{NL} < 1$. 

The initial perturbations generated during the inflationary epoch leave their
imprint in the galaxy formation process. In particular, the halo two-point correlation
function contains information about all the higher moments of the
matter distribution (Matarrese 1986). The halo bias is modified 
in the presence of primordial non-Gaussianity that could be originated in different ways in the context of the inflationary 
theory. J-PAS represents an excellent survey to study primordial non-Gaussianity due to its depth and area of sky to be covered.

The Bardeen potential
$\Phi(\mathbf{x})$ in the case of local non-Gaussianity is
\begin{equation}
\Phi(\mathbf{x}) = \phi(\mathbf{x}) + f_{NL} \left( \phi^2(\mathbf{x})
- \langle \phi^2 \rangle \right) \ ,
\end{equation}
where $\phi(\mathbf{x})$ is a Gaussian random field and $f_{NL}$ is
the non-linear coupling parameter. It is possible to relate the matter
overdensity field $\delta_R$, smoothed on a scale $R$, with the
potential through the Poisson equation. In Fourier space this
gives the expression
\begin{equation}
\delta_R(k) = \mathcal{M}_R(k) \Phi(k) \ ,
\end{equation}
where
\begin{equation}
\mathcal{M}_R(k) = \frac{2 c^2 k^2 T(k)}{3\Omega_m H_0^2} W_R(k) \ .
\end{equation}
Here, $T(k)$ is the matter transfer function and $W_R(k)$ is the
Fourier transform of the window function with characteristic radius
$R$. Usually a spherically symmetric top-hat function is assumed
for $W_R(k)$. Primordial non-Gaussianity present in the initial
perturbations modify the bias relation in the following way
\begin{equation}
\label{eqn:ng_bias}
b_R(k,z) = b_g(z) + 2 (b_g(z)-1) \delta_c(z) \frac{\mathcal{F}_R(k)}{\mathcal{M}_R(k)}  ,
\end{equation}
where $b_g(z)$ is the usual Gaussian bias and
\begin{equation}
\mathcal{F}_R(k) = \frac{f_{NL}}{8\pi^2\sigma_R^2} \int_0^\infty
\mathrm{d}k_1 k_1^2 \mathcal{M}_R(k_1) P_\phi(k_1) \int_{-1}^{1}
\mathrm{d}\mu \mathcal{M}_R(k_2) \left(
\frac{P_\phi(k_2)}{P_\phi(k)} + 2 \right) \ .
\end{equation}
In this expression $k_2 = \sqrt{k^2 + k_1^2 + 2 k k_1 \mu}$ and
$P_\phi(k)$ is the power spectrum of the gaussian field $\phi$. In the
large scale limit we have that $\mathcal{F}_R(k) \simeq f_{NL}$ and
the correction to the non-Gaussian bias becomes as in Dalal et
al. (2008).

Equation (\ref{eqn:ng_bias}) depends on the mass $M$ (or equivalent
the radius $R$) of the halo whose distribution is given by the mass
function $n(M,z)$. The total effective bias is a weighted sum of
equation (\ref{eqn:ng_bias}):
\begin{equation}
b(k,z) = \frac{\int_{M_{min}}^\infty \mathrm{d}M \ b_M(k,z)
  n(M,z)}{\int_{M_{min}}^\infty \mathrm{d}M \ n(M,z)} \ .
\end{equation}
The lower limit $M_{min}$ in the integral corresponds to the minimum
mass of the halos present in the survey. The mass $M_{min}$ is a free
parameter depending on the characteristics of the survey. We will
assumed a value of $10^{12 - 13} M_{\odot}$.

Among the catalogues of J-PAS the best to constrain the
primordial non-Gaussianity is the QSO sample. The reason is that this
population has a large bias and also it is deeper in redshift, given a
stronger signal of non-gaussianity (see equation
(\ref{eqn:ng_bias})).

The conditional constraints on $f_{NL}$ obtained from J-PAS are given in Table
\ref{tbl:fnl_constraints}. The error increases as the fiducial value of
$f_{NL}$ is higher. This increment is more pronounced in the case of
the LRG sample. For the QSO catalogue the error in $f_{NL}$ remains almost unaltered
at a value of $\Delta(f_{NL}) \approx 1.5$. 
By combining all the tracers detected by J-PAS, in the sense of 
\citet{2013MNRAS.432..318A}, we expect to achieve a limit $f_{NL} <1$.

\begin{table}

\begin{center}
\begin{tabular}{c|c|c|}
              & QSO  & LRG \\ \hline
 $f_{NL} = 0$  & 1.46 & 3.34 \\ \hline
 $f_{NL} = 10$ & 1.52 & 6.51 \\ \hline
 $f_{NL} = 30$ & 1.67 & 7.46 \\ \hline
\end{tabular}
\end{center}
\caption{
The 1-$\sigma$ errors on $f_{NL}$ obtained from QSO and LRG catalogues derived
from J-PAS. Different fiducials for $f_{NL}$ are considered.
}
\label{tbl:fnl_constraints}
\end{table}

\subsection*{Scale-dependent Non-Gaussianity}

In the case of scale-dependent non-Gaussianity the non-linear coupling
parameter $f_{NL}$ depends on the wavevector $k$. This leads to a
non-local coupling of the fields. The $k$-dependence of $f_{NL}$ can
be parametrized by
\begin{equation}
f_{NL}(k) = f_{NL}(k_p) \left( \frac{k}{k_p} \right)^{n_f} \ ,
\end{equation}
where $k_p$ is the pivot wave vector. This quantity has no physical
meaning and it can be chosen such that the two parameters
$f_{NL}(k_p)$ and $n_f$ have negligible correlation. The index $n_f$
represents the derivative
\begin{equation}
n_f = \left( \frac{\mathrm{d} \log f_{NL}(k)}{\mathrm{d} \log k}
\right)_{k=k_p} \ .
\end{equation}

The modification to the $\mathcal{F}_R(k)$ function when
scale-dependent $f_{NL}$ is taken into account is
\begin{equation}
\mathcal{F}_R(k) = \frac{1}{8\pi^2\sigma_R^2} \int_0^\infty
\mathrm{d}k_1 k_1^2 \mathcal{M}_R(k_1) P_\phi(k_1) \int_{-1}^{1}
\mathrm{d}\mu \mathcal{M}_R(k_2) \left( f_{NL}(k)
\frac{P_\phi(k_2)}{P_\phi(k)} + 2 f_{NL}(k_2) \right) \ .
\end{equation}

The constraints on non-Gaussianity when $f_{NL}$ is
assumed scale-dependent are given in Table
\ref{tbl:fnl_k_constraints}.
In this scenario with scale-dependent non-Gaussianity we have a
two-dimensional parameter space given by the amplitude $f_{NL}(k_p)$
and an index $n_f$. The pivot point is chosen around $k_p = 0.27 \ h
\ \mathrm{Mpc}^{-1}$ in order to cancel the correlation between
them. The fiducial value of the index $n_f$ is zero. Since the correlation between
the amplitude and the index vanishes and the fiducial model is $n_f = 0$,
 then the error in the amplitude $f_{NL}(k_p)$ for this particular pivot scale is the same as for a model with constant $f_{NL}$ (Table \ref{tbl:fnl_constraints}). On other hand the errors in $n_f$ are in
the Table \ref{tbl:fnl_k_constraints}. When the amplitude is
$f_{NL}(k_p) = 0$ there is no information on $n_f$ in the model and
then it is not possible to constrain the index. As the amplitude is
increased the error in $n_f$ decreases because the model is more
sensitive to the tilt.

\begin{table}

\begin{center}
\begin{tabular}{c|c|c|}
                 & QSO  & LRG  \\ \hline
 $f_{NL}(k_p) = 10$ & 3.25 & 15.4 \\ \hline
 $f_{NL}(k_p) = 30$ & 1.13 & 5.56 \\ \hline
\end{tabular}
\end{center}
\caption{
Errors on the index $n_f$ for different values of the amplitude
$f_{NL}(k_p)$. The fiducial value of $n_f$ is assumed to be zero which
corresponds to a scale invariant case.
}
\label{tbl:fnl_k_constraints}
\end{table}

\FloatBarrier 
\newpage
\section{Scientific Goals II: the J-PAS Galaxy Evolution Survey}
\FloatBarrier 

J-PAS will build a formidable legacy data set by delivering low resolution spectroscopy ($R \sim 50$) for every pixel over an SDSS-like area of the sky. A unique characteristic of this type of data is the fact that photo-spectra based on narrow-band imaging, unlike standard spectroscopy, does not suffer from systematic uncertainties in the flux calibration. Every data point of the photo-spectrum -i.e. every filter- is observationally independent, so the resulting SED is not affected by low frequency systematics in the relative flux calibration (or color terms) that can lead to biases in the derived physical properties. Multi-filter spectrophotometry thus provides accurate (low-resolution) SEDs over a wide range in wavelengths and spatial scales.
The four main features of J-PAS that are relevant to the study of galaxy evolution are as follows:\\
\smallskip

(1) a narrow-band filter system providing low resolution spectra ($R \sim 50$) that will result in very high quality photometric redshifts and adequate sampling of galaxy SEDs,\\

(2) a uniform and non-biased spatial sampling allowing environmental studies at small scales, unlike spectroscopic surveys that depend on target selection and are sometimes affected by fiber collision problems,\\

(3) an IFU-like character, allowing a pixel-by-pixel investigation of extended galaxies, and\\

(4) a large survey area and volume which will sample hundreds of millions of galaxies.\\

With these specific capabilities in mind, we have identified five key extra-galactic science drivers for J-PAS. These are the following:\\

{\bf 1. The Nearby Universe:} We will take advantage of the IFU-like capabilities of the J-PAS survey to determine the properties of the spatially resolved components of galaxies in the nearby universe, studying the evolution of galactic disks and spiral structure, bars, satellites, and spheroidal components.\\
 
{\bf 2. Evolution of the galaxy population since $z\sim1$:} Using accurate photometric redshifts and the low-resolution spectra based on the narrow-band filter system we will determine the evolution of the galaxy population from the present-day up to $z\sim1$ when the star formation rate density was an order of magnitude larger, and when most massive galaxies had formed the bulk of their mass. We will study the build-up of the stellar mass function, the evolution of the mass and SFR density, spectral types, and the bimodality of galaxy populations and the transition region (the ``green valley''). Stellar populations will be studied from analysis of the continuum through spectral fitting techniques, spectral indices and emission lines (individual objects and stacked samples).\\

{\bf 3. The High Redshift Universe:} By exploiting data from GALEX in the observed near-ultraviolet as well as several real and synthesized J-PAS broad-band filters we will furthermore be able to select galaxies at $z\approx1-4$ using the Lyman break technique. At $z\sim3$ we will construct the largest sample of LBGs probing the bright end of the UV luminosity function, allowing unprecedented studies of, e.g., their stellar populations and clustering. We will also be able to detect luminous \lya\ emitters in the redshift range $z=2.1-2.5$ using the narrow-band selection technique, and search for giant extended \lya\ (and other) emission line nebulae.\\

{\bf 4. The Growth of the Large-Scale Structure:} J-PAS will allow us to study the build-up of groups and clusters of galaxies, the evolution of the intra-group and intra-cluster light, and the role of environment on galaxy evolution based on an accurate reconstruction of the cosmic density field.\\

{\bf 5. The Build-up of Supermassive Black Holes}: J-PAS will collect large samples of AGN such as Seyfert galaxies, quasars, blazars, and radio galaxies over a very wide redshift range, allowing large statistical studies of the clustering, environments, triggering mechanisms, morphologies of the various AGN populations, and their role in galaxy evolution during the downsizing epoch.\\

In Section \ref{gal:techniques} below we will describe some of the measurement techniques that will be applied to the J-PAS data. Section \ref{gal:science} will describe each of the five key science drivers listed above in more detail. We conclude with a brief overview of other multi-wavelength data that will be available in the J-PAS area in Section \ref{sec:multi}.

\subsection{Measurement Techniques}
\label{gal:techniques}

\subsubsection{Redshifts and Sample size}

The J-PAS narrow-band system will deliver photometric redshifts with a projected accuracy of $\sigma_z<0.003(1+z)$, which is set by the goal of measuring the BAO signal along the line of sight. These photometric redshifts have been estimated for the subset of red $L>L*$ galaxies at $z<0.9$ with good Bayesian redshift ``odds'' using the well-tested Bayesian photometric redshift technique \citep{benitez00,benitez04,coe06,B2009}. 

Summarizing from the data presented in Section 2.4, by year 6 J-PAS will have detected respectively 18 and 73 million red and blue galaxies with 0.3\% redshift errors.  About 20\% of these objects will lie at $z>0.7$. If we relax the 0.3\% redshift accuracy that is required by the cosmology experiment to $\gtrsim 1\%$ errors which 
are more than acceptable for typical measurements of galaxy evolution, J-PAS will detect 64M of red and 200M 
of blue galaxies with $1\%$ redshift error (100M and 286M for $3\%$ redshift error). The redshift distributions peak around $z\sim0.5$, and contain several tens of millions of objects at $z>1$ (primarily blue galaxies).   

\subsubsection{Stellar Population Modeling and Emission Line Measurements}
\label{stelpops}

\paragraph{Spectral Fitting Diagnostics for Old Stellar Populations}

The use of multi-filter photometric surveys to determine SEDs and redshifts with high enough level of accuracy (like SDSS, see also COMBO-17: \cite{wolf08}; COSMOS: \cite{ilbert09}; ALHAMBRA: \cite{moles08}) has opened a new way to analyze the stellar populations of galaxies at different redshifts and in different environments, allowing accurate studies of the evolution of galaxies and cosmology based on very large samples. The combination of the number of filters, the sky coverage and the depth of the survey will make J-PAS an unprecedented experiment for stellar population studies.  

One of the major advantages of a survey like J-PAS is the fact that it
provides low resolution spectroscopy for every pixel of the sky. We
define the term ``J-spectra'' as the low resolution spectra
constructed from the 54 contiguous optical J-PAS NB filters. With low resolution spectrophotometric data, spectral fitting techniques over the full spectral range will allow the maximal exploitation of the information in the data. A unique characteristic of J-PAS is the fact that the J-spectra do not suffer from systematic uncertainties in the flux calibration, unlike standard spectroscopy. Every single point of the J-spectrum -- i.e., every filter -- is observationally independent from the other J-spectrum data points, so the resulting SED is not affected by large scale (low frequency) systematics in the relative flux calibration (hence in the SED colors). This means that the absolute shape of the J-spectrum continuum and its colors have a larger degree of reliability for conducting stellar population studies.  Standard spectroscopy, which is affected by well-known flux calibration issues, is better suited for detailed studies of particular spectral features or line strength indices.

For the proper analysis of the J-spectra of galaxies and stars in
terms of their physical properties it is crucial to use template
stellar libraries with extremely accurate flux calibration, such as
MIUSCAT \citep{vazdekis10,ricciardelli12}. The MIUSCAT SSP models are
perfectly suited for the analysis and interpretation of optical
spectrophotometric data, provided that the flux calibration of the
library stars is accurate over the full optical spectral
coverage. Another important characteristic of the template stellar
libraries required for J-PAS is the spectral coverage. Since J-PAS
will sample the rest frame UV  and optical regions of sources up to
$z\sim 1$, we need synthetic libraries which adequately cover this range.

{\bf Reliability of Stellar Population Studies with J-spectra.} Much of the 
recent progress in our understanding of galaxy evolution has come from the so called fossil-methods, which model the mixture of SSPs of different ages and metallicities to infer the main star formation episodes of galaxies out of medium resolution spectra \citep[][and references therein]{walcher11}. Besides the redshift, stellar masses and the luminosity/mass weighted ages and metallicities of the galaxies, these analysis techniques can recover the mass assembly and even chemical evolution histories, at least in a statistical sense (i.e.., when applied to large samples). The evolutionary information decoded by these methods resides in the continuum shape and stellar absorption features. While the optical SED will be adequately sampled by J-PAS, most absorption features will be very heavily smoothed. This limitation poses the question: what can be learnt about stellar populations from J-spectra alone? From a purely academic perspective, the spectral resolution of the J-spectra (R$\sim$50) is sufficient for identifying and measuring the strongest spectral features of quiescent stellar populations. For instance, the $\lambda4000$\AA\ break, the G-band at $\lambda4300$\AA, the region around the Mg$b$ doublet and the Fe lines at $\sim\lambda5200$\AA, and the strongest TiO bands redwards of $\sim\lambda6000$\AA, are distinguishable in the J-spectra, as illustrated below. It is important to note that the effective resolving power of absorption features is ultimately linked to the S/N ratio of the J-spectra.

\begin{figure*}
\centering
\includegraphics[angle=0,width=0.8\textwidth]{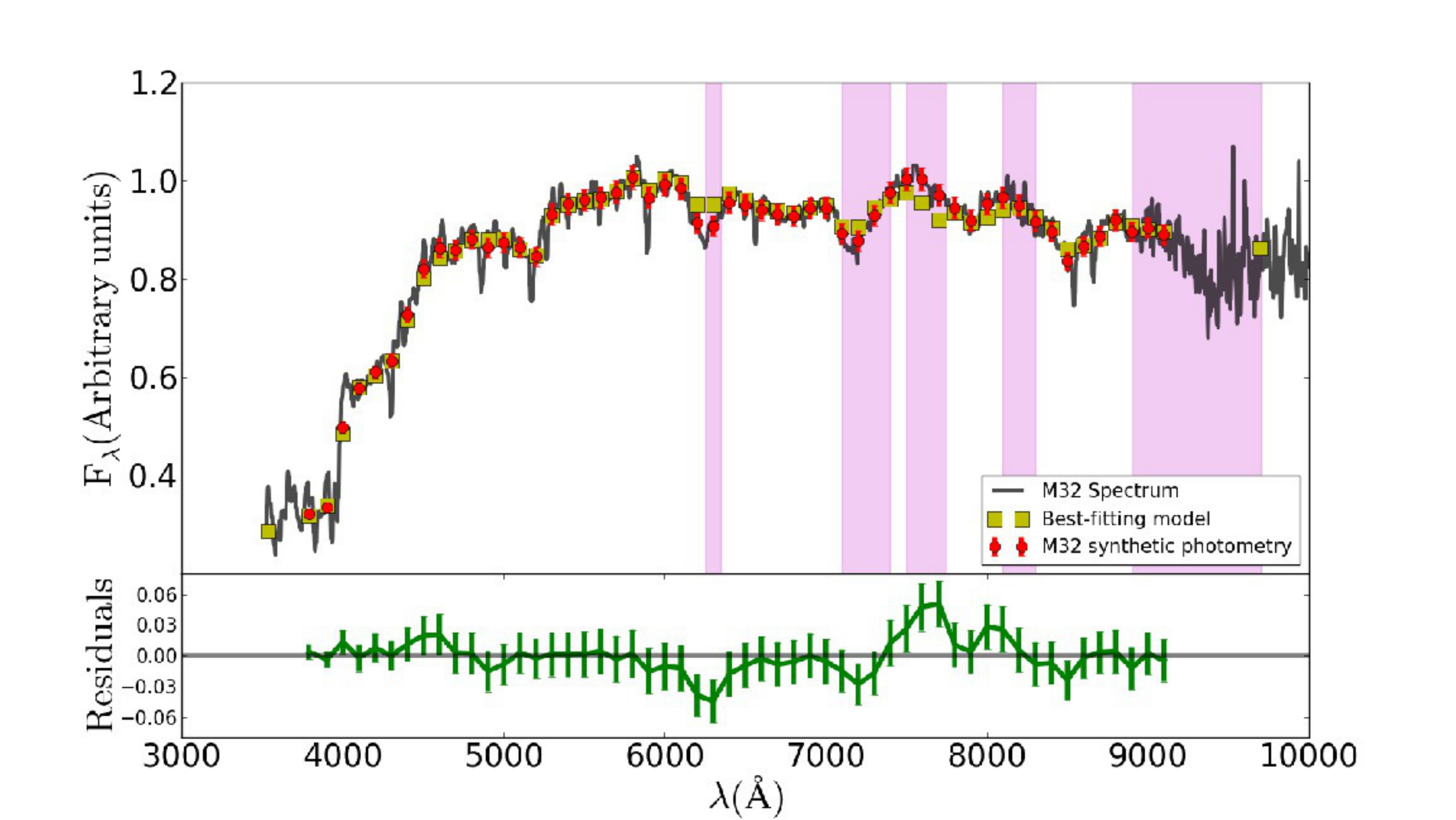}
\caption{Spectral fitting of M32 as seen by J-PAS using the MIUSCAT SSP SEDs as template models. The spectrum of M32 at the nominal spectral resolution is illustrated by the black solid line, whereas the same spectrum at the J-PAS resolution is plotted with red dots. The best fit of a mixture of SSPs to the spectrum of M32, as derived from a standard $\chi^2$ minimization technique, is shown with yellow squares. The residuals are shown in the lower panel in the same scale. Purple bands indicate the locations of potential telluric lines. See the text for more details.
}
\label{M32}
\end{figure*}

\newpage
{\bf The case of M32.} To illustrate this, let us focus on a classic test case of a quiescent stellar population: M32. Figure~\ref{M32} presents the best fit SED derived from the integrated spectrum of M32 from \citet{bica90}, taken from the compilation of \citet{santos02}, and using the MIUSCAT SSP models as input templates. The M32 spectrum and the template spectra have been convolved with the J-PAS filters to simulate a realistic scenario. Errors have been set to 0.025 mag in each filter, which corresponds to a S/N$\sim 43$ in flux units. It is clear from the figure that the best fit, derived from a standard $\chi^2$ minimization technique (D\'{\i}az-Garc\'{\i}a et al., in prep.), reproduces well the observed spectrum at both short and long wavelengths. The residuals are shown in the lower panel. Note the telluric absorption features still present in the spectroscopic data redwards $\lambda6000$\AA. Also, it is worth noting that the MIUSCAT models do not account for different $\alpha$-element abundance ratios. The best fit solution to a single SSP corresponds to a MIUSCAT model of $3.2\pm0.8$ Gyr and a metallicity of around solar (0.11$\pm$0.11 dex). When a  more complex mixture of SSPs is allowed, a luminosity weighted age of $6.5\pm1.5$ Gyr and solar metallicity (0.05$\pm$0.08 dex) is obtained. In both cases, the results are overall in good agreement with those based on much higher resolution spectroscopic data \citep[e.g.][]{vazdekis99,schiavon04,coelho09}, hence illustrating the power of the low resolution data provided by J-PAS for stellar population studies.

\begin{figure*}
\centering
\includegraphics[angle=0,width=1.00\textwidth]{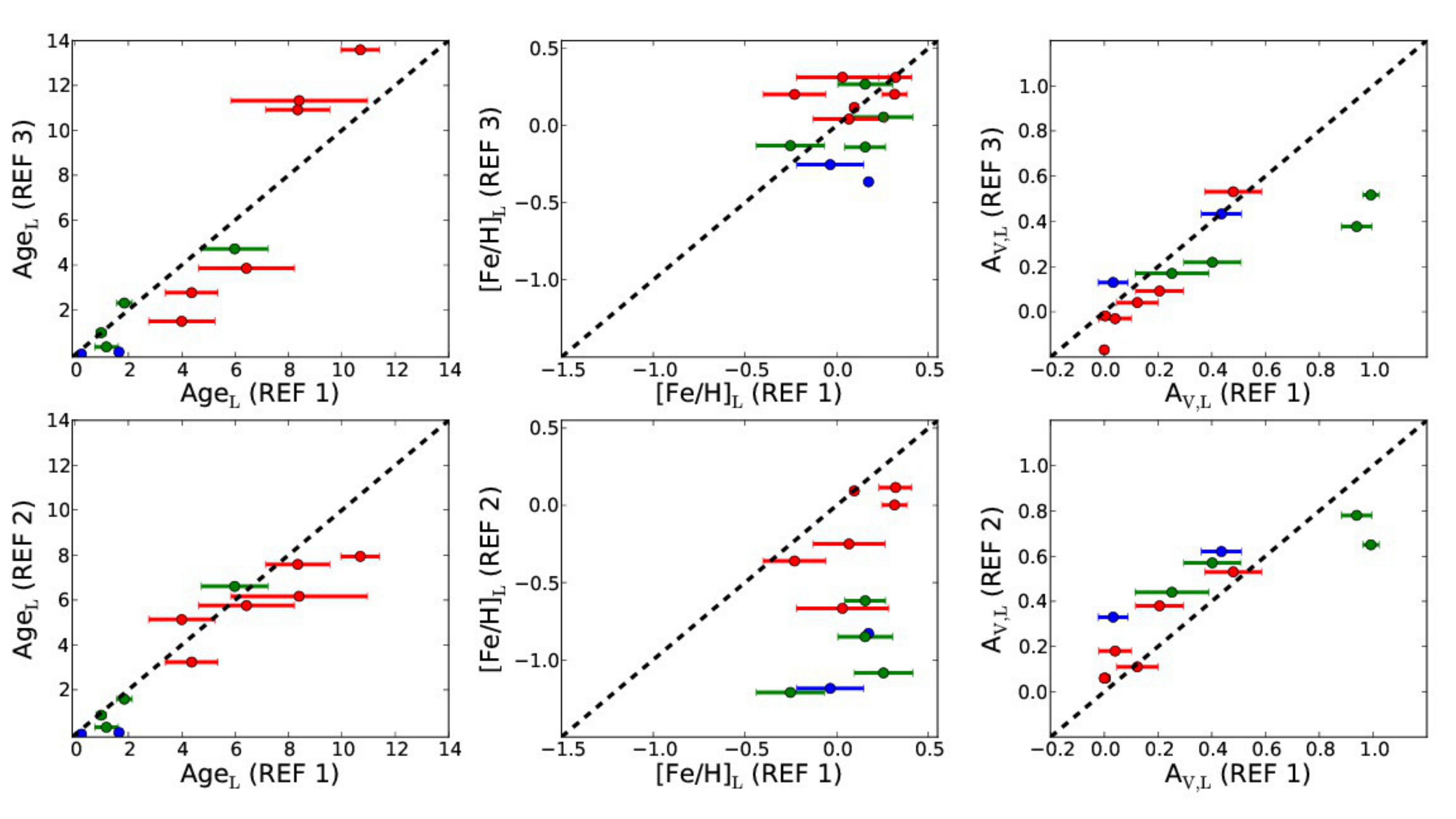}
\caption{Summary of results of the ``J-PAS Stellar Population Challenge'' (see text for details). The figure presents a comparison between the luminosity weighted mean ages, metallicities and extinctions ($A_v$) obtained for 12 SDSS galaxies using three different analysis techniques, namely REF1, REF2 and REF3. Different colours correspond to galaxies with no emission lines and strong stellar continuum (red), weak emission lines and mild stellar continuum (green) and strong emission lines and weak stellar continuum (blue).
}
\label{j-pas_challenge}
\end{figure*}

\begin{figure*}
\centering
\includegraphics[angle=0,width=0.97\textwidth]{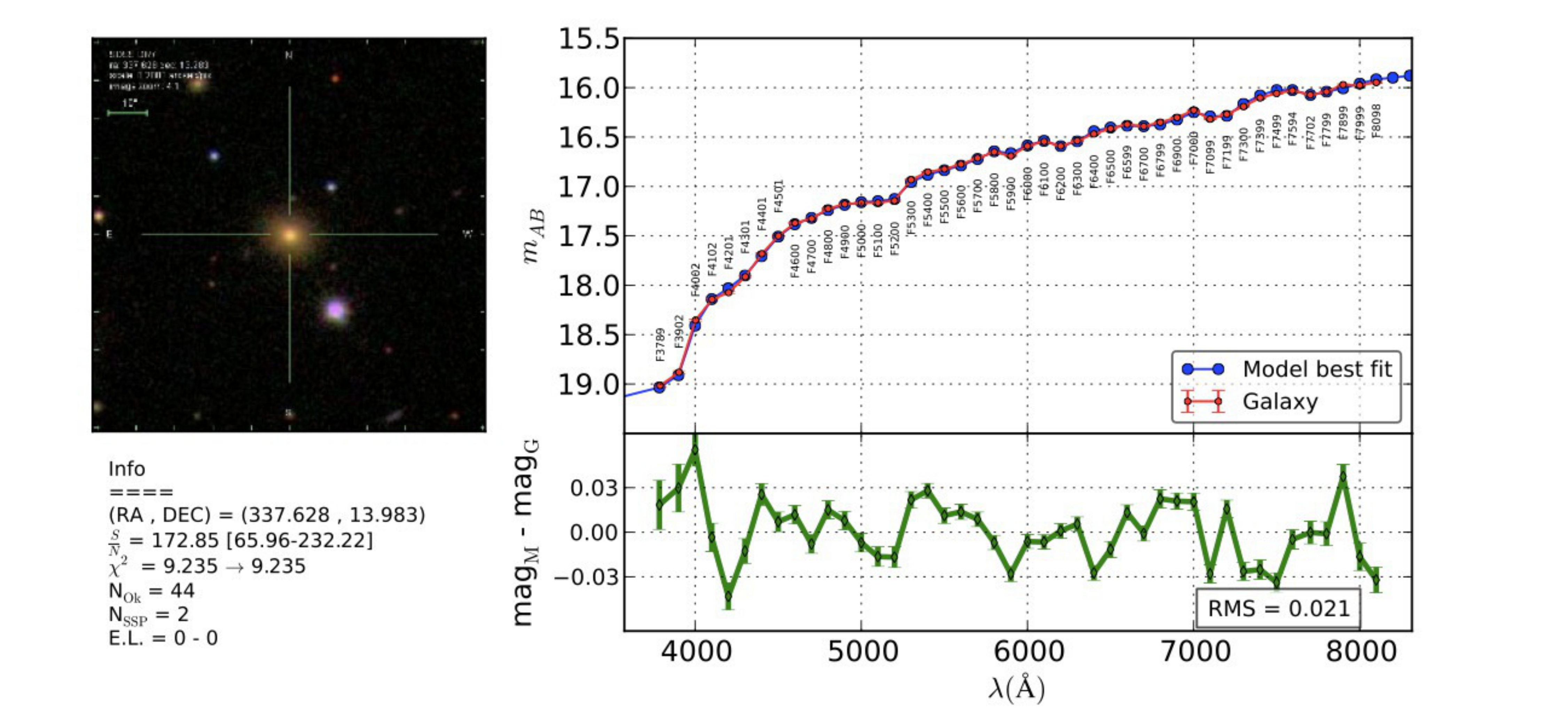}
\includegraphics[angle=0,width=0.99\textwidth]{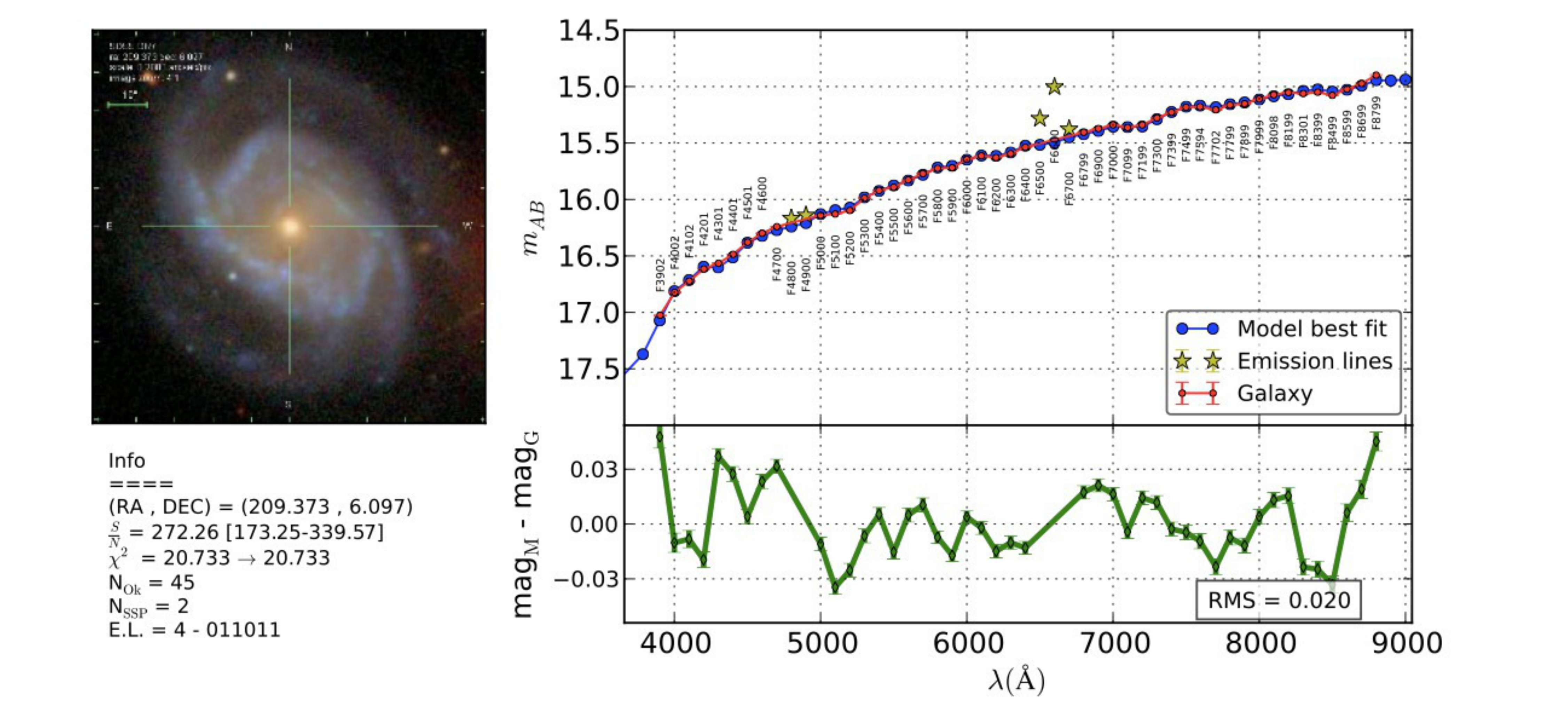}
\caption{Two examples of SDSS galaxies taken from the ``J-PAS Stellar Population Challenge'' (see text for details). The SDSS images are shown on the left. On the right, the constructed J-spectrum (red) and the best fit (blue) are illustrated. Bottom panels show the residuals of the fits. In this particular case, emission lines (yellow stars) are detected but ignored in the overall fit.
}
\label{challenge_galaxies}
\end{figure*}

\newpage
{\bf The J-PAS Stellar Population Challenge.} The reliability of the stellar population parameters derived from the J-spectra depends not only on resolution, quality and spectral type,  but also on the analysis technique employed. To account for this, we have set up an internal test within the J-PAS collaboration: the ``J-PAS Stellar Population Challenge''. A dozen galaxies of various types were retrieved from the SDSS and their spectra were convolved with the J-PAS filters. The corresponding J-spectra were then distributed to the participants, who analyzed them using a common set of ingredients. This is necessary for homogenizing the results and allow for fair comparisons. Input SSP models were set to \citet{bc03}, with the STELIB library, Padova (1994) evolutionary tracks and a Chabrier initial mass function (IMF). Also, a simple foreground dust screen with a \citet{cardelli89} reddening law was assumed. The participants were asked to provide estimates of the luminosity and mass weighted mean ages and metallicities, the $V$-band extinction, the stellar mass. In some cases, filters were masked out because of possible contamination by line emission.
Figure~\ref{j-pas_challenge} presents a comparative view of the luminosity weighted mean ages, metallicities and extinctions retrieved by the three different participants (``REF1'', ``REF2'', and ``REF3''). Suffice it to say that the analysis techniques employed in each case constitute a representative sample of methods employed in the current literature, ranging from full spectral fits, mixed populations fits, as well as a novel technique involving matching with SDSS galaxies. The important point is that the results derived by these different methods show a satisfactory degree of consistency, except for one particular technique that seems to underestimate the metallicities of, mainly, galaxies with emission lines. It is equally important to note that the results agree well with those obtained from full spectral fits of the original SDSS spectra analyzed with the same ingredients and the Starlight code \citep{cidfernandes05}. This test demonstrates that, at least for the global properties analyzed here, the J-spectra do not lead to a substantial loss of information with respect to conventional medium resolution spectroscopy.

For illustration, Figure~\ref{challenge_galaxies} provides two examples of SDSS galaxies employed in the test. Since the SDSS spectra do not cover the whole spectral range of J-PAS and because in some cases certain spectral windows are unusable, these tests used less data (typically 45 out of the 54+5 J-PAS filters) than will actually be available. Notice also that these experiments were carried out in the rest-frame (i.e., at $z=0$), but the overall conclusion should remain valid up to at least $z\sim0.5$. As the redshift increases, more UV light will be sampled, forcing analysis methods to deal with a spectral range not as well consolidated as the optical. On the other hand, the reduced age range should help alleviating some of the main degeneracies which affect population synthesis. In addition to that, the fact that the width of the narrow-band J-PAS filters (defined to be constant with lambda) effectively decreases with redshift by a factor of ($1+z$), leads to a significant improvement in the overall spectral resolution of the J-spectra at high redshift. In fact, as presented later, the higher effective spectral resolution toward higher redshifts helps to reduce the intrinsic uncertainties in the determination of stellar population parameters.

The different techniques employed in this test can (and will) be
improved and fine-tuned to retrieve more robust determinations, making
use of large spectroscopic datasets in the literature as training
sets. On the other hand, by no means we try to convince the reader
that the J-PAS data provide the same information that high resolution
spectroscopy can provide. J-spectra will not be sensitive to weak
absorption lines and will not allow detailed studies of element
abundance ratios based on individual line strengths. What we hope to
illustrate here is that J-spectra, even though they are of much lower
resolution, provide meaningful information for stellar population
studies.  The lack of spectral resolution is partially compensated for by a
  much more reliable continuum determination (as compared to standard
  spectroscopy), so that spectral fitting techniques over the full
  spectral coverage can provide reliable information on, e.g., ages,
  metallicities, extinction, unlike standard spectroscopy. In fact,
  the definition of line strength indices in the 80's and 90's
  \citep[e.g. those in the Lick system][]{gorgas93,worthey94} was
  motivated by the need to overcome the intrinsic uncertainties of
  flux calibration in standard spectroscopy.

 J-PAS will thus offer a new and fresh approach to stellar population studies. It is important to note that the state-of-the-art SSP models aim to provide accurate predictions for the observed SEDs, not only for different ages and metallicities, but also for different $\alpha$-element abundance ratios. Different parameters in the  models do not only affect predictions for the individual line strengths of Mg, Na, Ti, etc., but also the overall shape of the continuum \citep[see, e.g.][]{sansom13}. Therefore, when the proper set of SSP models is employed, the J-spectra may be sufficient for distinguishing between different abundance ratios from the full spectral fitting. 
  
{\bf Random Uncertainties in Stellar Population Determinations.} To have a first estimate of the minimum random uncertainties that we may expect when measuring ages, metallicities and extinctions from the J-spectra, we have performed a Monte Carlo simulation with galaxy templates covering the expected parameter space of age, metallicity, extinction, redshift and S/N. Overabundances or varying IMFs were not considered in this test. The procedure can be summarized as follows: 
\begin{itemize}
\item A set of 9 MIUSCAT SSPs with ages of 0.5, 3, and 10 Gyr, and metallicities of 0.0, $-0.4$, $-0.7$ dex were selected as the model target galaxies. The chosen values are representative of the typical ages and metallicities of red sequence galaxies over a range in mass and up to redshift $\sim 1$ that J-PAS is expected to observe. 
\item The 9 target SSPs were modified to match three different extinction values ($A_v=0.0$, 0.3 and 0.6), six different redshifts ($z=0.0$, 0.2, 0.4, 0.6, 0.8 and 1.0), and four different values of the average S/N per filter (10, 20, 50 and 100). Overall, this translates into 648 target SSPs. 
\item For each target SSP, one thousand simulations were created according to the assumed S/N per filter. For each simulation, a $\chi^2$ minimization fitting technique that mixes two (younger + older) SSPs (see details in D\'{\i}az-Garc\'{\i}a et al.\, in prep.) was performed, resulting in a best fit with its corresponding redshift, luminosity weighted mean age, metallicity and extinction. The whole set of SSPs ($0.06 \leq$ age $\leq 14.1$ Gyr; 48 steps; $-2.3 \leq$ [Fe/H] $\leq +0.2$ dex; 7 steps) of the MIUSCAT database has been employed to perform the fits. 
\item With all the best solutions for each target SSP, the mean values of age and metallicity and the root mean square (rms) of the obtained solutions ($\sigma_{Age}$ and $\sigma_{[Fe/H]}$)  were computed. These values can be considered as a first order estimate of the best-case uncertainties in the parameter estimation.
\end{itemize}

\begin{table}
 \centering{\scriptsize
 \caption{\small Typical uncertainties in the determination of ages and metallicities ($\sigma_{Age}$ in Gyr and $\sigma_{[Fe/H]}$ in dex) for old stellar populations using standard spectral fitting techniques applied to J-spectra. According to the photometric errors in the J-PAS filters given by the different S/N per filter (10, 20, 50 and 100), Monte Carlo simulations have been performed for different redshifts ($z=0.0$, 0.2, 0.4, 0.6, 0.8 and 1.0), extinction ($A_v=0.0$, 0.3 and 0.6), age (0.5, 3 and 10 Gyr) and metallicity ([Fe/H] = $-0.7$, $-0.4$ and 0.0 dex). In each case, $\sigma_{Age}$ and $\sigma_{[Fe/H]}$ represent the rms standard deviation of the 3$\times$1000 best solutions obtained around the nominal age and metallicity input values for the 3 different extinction values. See the text for more details on the procedure.}
 \vspace{4mm}
 \begin{tabular}{@{}cclc@{}c@{}cc@{}c@{}cc@{}c@{}cc@{}c@{}cc@{}c@{}cc@{}c@{}c@{}}
 \hline\hline
& & &\multicolumn{3}{c}{z=0.0} & \multicolumn{3}{c}{z=0.2} & \multicolumn{3}{c}{z=0.4} & \multicolumn{3}{c}{z=0.6} & \multicolumn{3}{c}{z=0.8} & \multicolumn{3}{c}{z=1.0} \\  
\hline
\hline
          & & &\multicolumn{3}{c}{} & \multicolumn{3}{c}{} & \multicolumn{3}{c}{[Fe/H] (dex)} & \multicolumn{3}{c}{} & \multicolumn{3}{c}{} & \multicolumn{3}{c}{} \\  
 S/N   & Age & &$-0.7$ & $-0.4$ & $\ 0.0$ & $-0.7$ & $-0.4$ & $\ 0.0$ & $-0.7$ & $-0.4$ & $\ 0.0$ & $-0.7$ & $-0.4$ & $\ 0.0$ & $-0.7$ & $-0.4$ & $\ 0.0$ & $-0.7$ & $-0.4$ & $\ 0.0$ \\
\hline
\hline

                    & 0.5\,Gyr  & $\sigma_{Age}$     & $0.98$ &\ $1.04$ &\  $1.04$ & $0.84$ &\  $0.93$ &\  $0.69$ & $0.70$ &\  $0.79$ &\  $0.56$ & $0.68$ &\  $0.61$ &\  $0.57$ & $0.52$ &\  $0.48$ &\  $0.49$ & $0.43$ &\  $0.47$ &\  $0.55$ \\
                    &                & $\sigma_{[Fe/H]}$  & $0.35$ &\  $0.26$ &\  $0.23$ & $0.31$ &\  $0.19$ &\  $0.15$ & $0.35$ &\  $0.27$ &\  $0.15$ & $0.35$ &\  $0.29$ &\  $0.17$ & $0.39$ &\  $0.31$ &\  $0.18$ & $0.48$ &\  $0.34$ &\  $0.17$ \\
              10  & 3\,Gyr  & $\sigma_{Age}$    & $2.55$ &\  $2.69$ &\  $2.75$ & $2.57$ &\  $2.68$ &\  $2.71$ & $2.44$ &\  $2.16$ &\  $1.95$ & $2.20$ &\  $1.90$ &\  $1.31$ & $2.01$ &\  $1.86$ &\  $0.91$ & $1.76$ &\  $1.75$ &\  $0.73$ \\
                    &                & $\sigma_{[Fe/H]}$ & $0.29$ &\  $0.32$ &\  $0.26$ & $0.26$ &\  $0.24$ &\  $0.23$ & $0.28$ &\  $0.25$ &\  $0.13$ & $0.27$ &\  $0.20$ &\  $0.06$ & $0.24$ &\  $0.16$ &\  $0.04$ & $0.22$ &\  $0.16$ &\  $0.02$ \\
                    & 10\,Gyr & $\sigma_{Age}$   & $2.97$ &\  $2.86$ &\  $2.62$ & $2.73$ &\  $2.72$ &\  $2.43$ & $2.80$ &\  $2.63$ &\  $1.50$ & $2.60$ &\  $2.31$ &\  $1.47$ & $2.47$ &\  $1.84$ &\  $1.41$ & $2.33$ &\  $1.61$ &\  $1.06$ \\ 
                    &                & $\sigma_{[Fe/H]}$ & $0.26$ &\  $0.22$ &\  $0.14$ & $0.23$ &\  $0.18$ &\  $0.13$ & $0.22$ &\  $0.20$ &\  $0.06$ & $0.18$ &\  $0.18$ &\  $0.04$ & $0.15$ &\  $0.13$ &\  $0.01$ & $0.13$ &\  $0.13$ &\  $0.01$ \\ \hline 
                   & 0.5\,Gyr  & $\sigma_{Age}$   & $0.85$ &\  $0.70$ &\  $0.65$ & $0.63$ &\  $0.39$ &\  $0.50$ & $0.67$ &\  $0.57$ &\  $0.35$ & $0.53$ &\  $0.38$ &\  $0.26$ & $0.38$ &\  $0.31$ &\  $0.27$ & $0.35$ &\  $0.32$ &\  $0.33$ \\
                   &                & $\sigma_{[Fe/H]}$ & $0.25$ &\  $0.13$ &\  $0.13$ & $0.20$ &\  $0.11$ &\  $0.10$ & $0.25$ &\  $0.26$ &\  $0.12$ & $0.29$ &\  $0.25$ &\  $0.12$ & $0.31$ &\  $0.26$ &\  $0.11$ & $0.38$ &\  $0.26$ &\  $0.12$ \\
             20  & 3\,Gyr  & $\sigma_{Age}$    & $1.90$ &\  $1.96$ &\  $1.93$ & $2.00$ &\  $1.79$ &\  $2.11$ & $1.79$ &\  $1.51$ &\  $1.02$ & $1.42$ &\  $1.05$ &\  $0.65$ & $1.21$ &\  $1.17$ &\  $0.62$ & $1.10$ &\  $0.94$ &\  $0.58$ \\
                   &                & $\sigma_{[Fe/H]}$ & $0.19$ &\  $0.16$ &\  $0.15$ & $0.17$ &\  $0.14$ &\  $0.14$ & $0.19$ &\  $0.14$ &\  $0.05$ & $0.16$ &\  $0.09$ &\  $0.02$ & $0.14$ &\  $0.08$ &\  $0.02$ & $0.13$ &\  $0.08$ &\  $0.01$ \\
                   &10\,Gyr & $\sigma_{Age}$   & $2.09$ &\  $1.97$ &\  $2.17$ & $2.19$ &\  $2.10$ &\  $1.99$ & $1.98$ &\  $1.98$ &\  $0.96$ & $1.75$ &\  $1.54$ &\  $0.83$ & $1.62$ &\  $1.41$ &\  $0.75$ & $1.64$ &\  $1.12$ &\  $0.49$ \\ 
                   &                & $\sigma_{[Fe/H]}$ & $0.13$ &\  $0.11$ &\  $0.11$ & $0.12$ &\  $0.12$ &\  $0.09$ & $0.13$ &\  $0.13$ &\  $0.03$ & $0.11$ &\  $0.10$ &\  $0.01$ & $0.09$ &\  $0.08$ &\  $0.01$ & $0.09$ &\  $0.07$ &\  $0.01$ \\ \hline 
                   & 0.5\,Gyr  & $\sigma_{Age}$    & $0.37$ &\  $0.30$ &\  $0.25$ & $0.39$ &\  $0.13$ &\  $0.22$ & $0.43$ &\  $0.29$ &\  $0.09$ & $0.24$ &\  $0.17$ &\  $0.05$ & $0.26$ &\  $0.04$ &\  $0.04$ & $0.23$ &\  $0.06$ &\  $0.05$ \\
                   &                & $\sigma_{[Fe/H]}$ & $0.13$ &\  $0.03$ &\  $0.05$ & $0.12$ &\  $0.02$ &\  $0.05$ & $0.18$ &\  $0.15$ &\  $0.07$ & $0.18$ &\  $0.13$ &\  $0.05$ & $0.17$ &\  $0.09$ &\  $0.05$ & $0.10$ &\  $0.05$ &\  $0.04$ \\
             50 & 3\,Gyr  & $\sigma_{Age}$     & $1.05$ &\  $1.30$ &\  $1.16$ & $1.01$ &\  $1.20$ &\  $1.40$ & $0.97$ &\  $0.72$ &\  $0.53$ & $0.81$ &\  $0.46$ &\  $0.48$ & $0.70$ &\  $0.27$ &\  $0.47$ & $0.73$ &\  $0.27$ &\  $0.38$ \\
                   &                & $\sigma_{[Fe/H]}$ & $0.07$ &\  $0.08$ &\  $0.08$ & $0.07$ &\  $0.07$ &\  $0.08$ & $0.09$ &\  $0.06$ &\  $0.02$ & $0.07$ &\  $0.04$ &\  $0.01$ & $0.04$ &\  $0.03$ &\  $0.00$ & $0.04$ &\  $0.03$ &\  $0.01$ \\
                   & 10\,Gyr & $\sigma_{Age}$    & $1.28$ &\  $1.33$ &\  $1.35$ & $1.26$ &\  $1.37$ &\  $1.17$ & $1.05$ &\  $1.18$ &\  $0.43$ & $0.58$ &\  $0.86$ &\  $0.30$ & $0.53$ &\  $0.58$ &\  $0.27$ & $0.42$ &\  $0.47$ &\  $0.15$ \\
                   &                 & $\sigma_{[Fe/H]}$ & $0.06$ &\  $0.05$ &\  $0.05$ & $0.06$ &\  $0.06$ &\  $0.04$ & $0.06$ &\  $0.06$ &\  $0.01$ & $0.04$ &\  $0.04$ &\  $0.01$ & $0.03$ &\  $0.03$ &\  $0.01$ & $0.02$ &\  $0.03$ &\  $0.01$ \\ \hline 
                   & 0.5\,Gyr  & $\sigma_{Age}$    & $0.20$ &\  $0.19$ &\  $0.16$ & $0.30$ &\  $0.07$ &\  $0.05$ & $0.17$ &\  $0.03$ &\  $0.03$ & $0.12$ &\  $0.02$ &\  $0.04$ & $0.11$ &\  $0.03$ &\  $0.03$ & $0.14$ &\  $0.02$ &\  $0.03$ \\ 
                   &                & $\sigma_{[Fe/H]}$ & $0.02$ &\  $0.01$ &\  $0.01$ & $0.06$ &\  $0.01$ &\  $0.01$ & $0.08$ &\  $0.01$ &\  $0.02$ & $0.07$ &\  $0.01$ &\  $0.02$ & $0.07$ &\  $0.01$ &\  $0.02$ & $0.05$ &\  $0.02$ &\  $0.01$ \\
            100 & 3\,Gyr  & $\sigma_{Age}$    & $0.45$ &\  $0.69$ &\  $0.92$ & $0.49$ &\  $0.70$ &\  $0.83$ & $0.44$ &\  $0.25$ &\  $0.40$ & $0.41$ &\  $0.14$ &\  $0.37$ & $0.50$ &\  $0.07$ &\  $0.29$ & $0.52$ &\  $0.06$ &\  $0.20$ \\ 
                   &                & $\sigma_{[Fe/H]}$ & $0.03$ &\  $0.04$ &\  $0.04$ & $0.04$ &\  $0.03$ &\  $0.03$ & $0.04$ &\  $0.02$ &\  $0.01$ & $0.03$ &\  $0.01$ &\  $0.00$ & $0.02$ &\  $0.01$ &\  $0.01$ & $0.01$ &\  $0.01$ &\  $0.01$ \\ 
                   & 10\,Gyr & $\sigma_{Age}$    & $0.65$ &\  $0.74$ &\  $0.54$ & $0.65$ &\  $0.65$ &\  $0.42$ & $0.36$ &\  $0.65$ &\  $0.11$ & $0.17$ &\  $0.52$ &\  $0.02$ & $0.15$ &\  $0.04$ &\  $0.01$ & $0.11$ &\  $0.06$ &\  $0.01$ \\
                   &                 & $\sigma_{[Fe/H]}$ & $0.02$ &\  $0.02$ &\  $0.03$ & $0.03$ &\  $0.03$ &\  $0.02$ & $0.02$ &\  $0.02$ &\  $0.01$ & $0.01$ &\  $0.01$ &\  $0.01$ & $0.01$ &\  $0.01$ &\  $0.01$ & $0.01$ &\  $0.01$ &\  $0.01$ \\
\hline
\hline
 \label{agemet-uncertainties}
 \end{tabular}
 }
 \end{table}

The results of this test are presented in Table~\ref{agemet-uncertainties}. It is worth noting that the uncertainties presented here illustrate the typical random errors that we may expect just due to the noise in the J-PAS photometry. Systematic effects coming from differences in the spectrophotometric system of real J-spectra and SSP templates, or intrinsic differences between simple template models and more complex real galaxies will add additional uncertainty to the derived values, certainly dominating the final errors for high S/N data. In this sense, the numbers in Table~\ref{agemet-uncertainties} must be considered as a lower limit (best case) to the final errors expected in luminosity-weighted SSP-equivalent ages and metallicities. As expected, the uncertainties in age and metallicity decrease as the S/N per filter increases, also depending on the parameter space region (Age-[Fe/H]). For instance, at $z=0$, $\sigma_{Age}$ and $\sigma_{[Fe/H]}$ vary from $\sim$3 Gyr and $\sim$0.2 dex for S/N$=10$ down to $\sim$0.6 Gyr and $\sim$0.02 dex for S/N$=100$. Interestingly, we also see a trend of smaller errors obtained at higher redshift. This is probably due to the fact that the effective spectral resolution increases with redshift as ($1+z$), which improves the power to disentangle the age-metallicity degeneracy. This is an interesting result that, to some extent, helps to alleviate the effects of a decreasing S/N with increasing redshift when determining the stellar population parameters.

\newpage
\paragraph{Pseudo Line-Strength Indices for Old Stellar Populations}

Spectroscopic absorption line indices, as for instance the Lick system \citep{worthey94}, have been widely used to build diagnostic diagrams, where an age-sensitive index and a metallicity indicator are used to disentangle the age/metallicity degeneracy.  Although this approach can give accurate estimates of age and metallicity, it is very expensive in terms of telescope time, since it requires high S/N spectra.  A similar approach can be applied to narrow-band photometric surveys such as J-PAS, where the galaxy SED sampled at regular and small intervals in wavelength can be considered as a low-resolution spectrum.

We model photometric absorption indices on the basis of the MILES stellar population models \citep{vazdekis10}, and using the J-PAS filter definitions. We focus on two age-sensitive features (D4000 and $H_{\beta}$) and one metallicity indicator (Mg). To measure the photometric indices ($H_{\beta}$ and Mg), we consider three J-PAS filters: one filter containing the feature of interest, and two filters directly on the red and blue side of the feature to measure the continuum. Hence, the index is given by the difference in magnitude between feature and continuum. The 4000\AA\ spectral break is defined by adopting the classical definition of \citet{bruzual83}, which uses the magnitude difference between the bands at [4050, 4250] and [3750, 3950]\AA.

\begin{figure*}
\centering
\includegraphics[angle=0,width=0.45\textwidth]{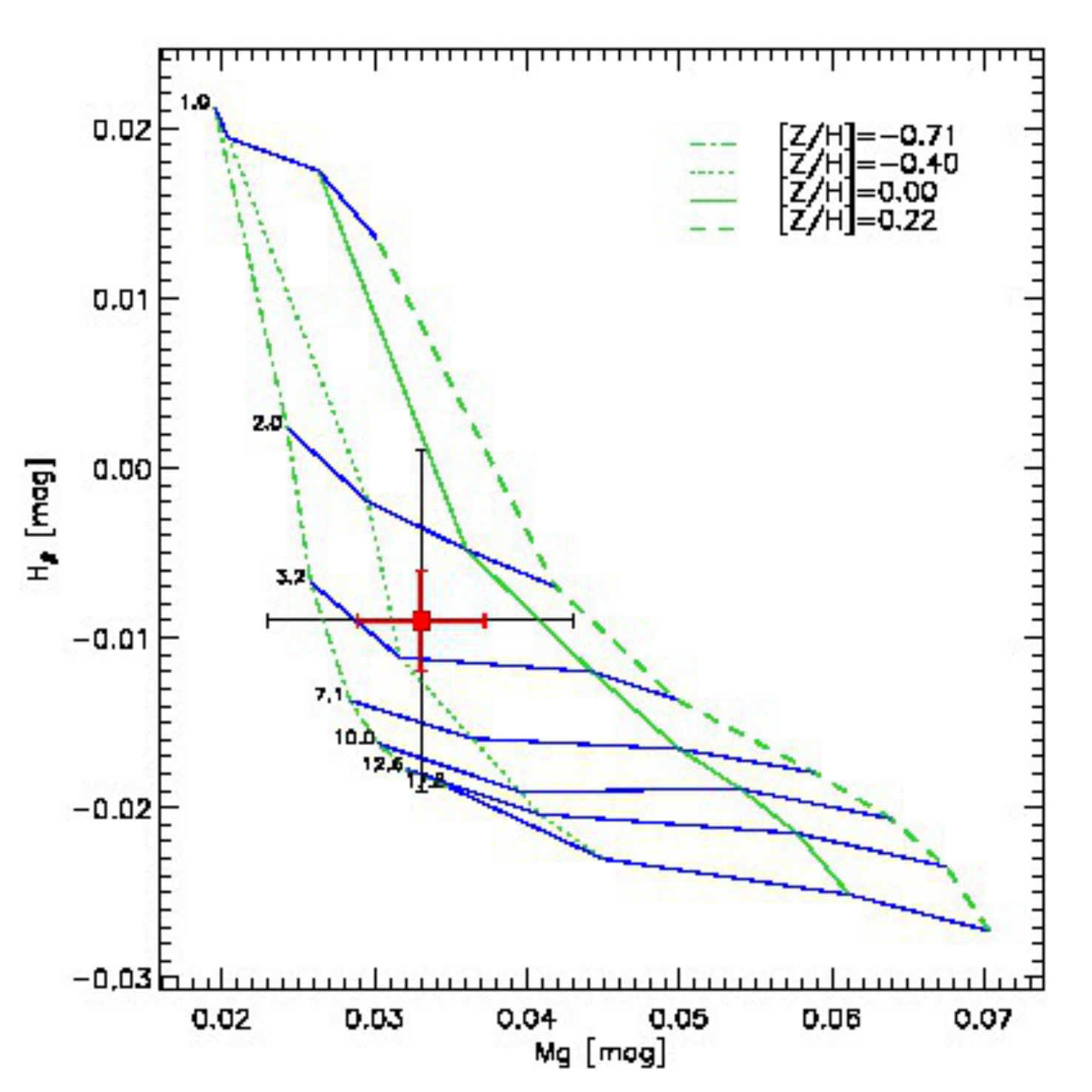}
\includegraphics[angle=0,width=0.45\textwidth]{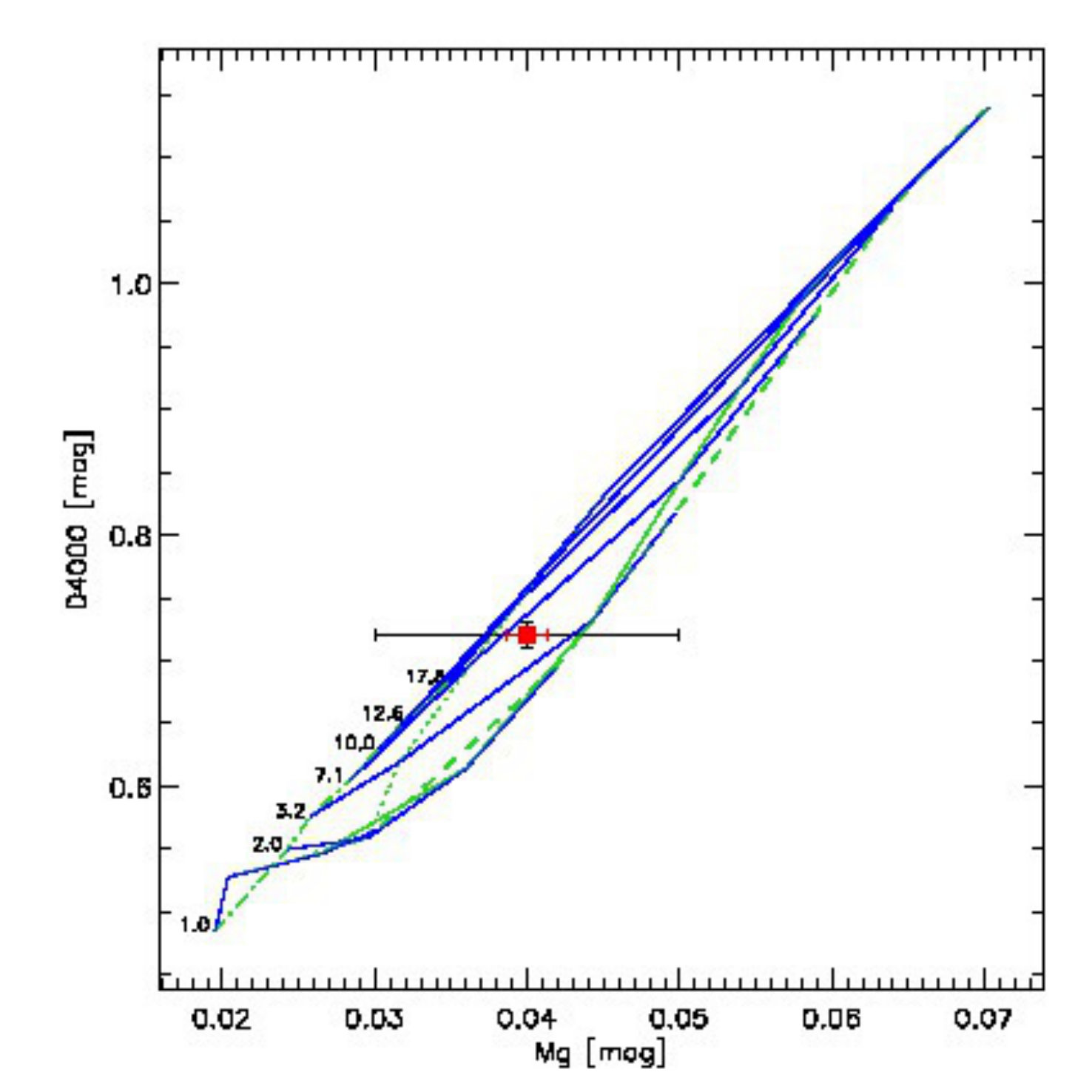}
\caption{Index-index diagrams for  age-sensitive indicators
($H_{\beta}$ in the left-hand panel and D4000 in the right-hand one)
versus the metallicity-sensitive index Mg. The model grids are shown for
ages ranging from 1 to 17.8 Gyr and for metallicities ranging from
$-0.71$ up to $0.22$ dex. The thin black bar shows an index
error of 0.01 mag, equivalent to a S/N ratio of $\sim$150 per filter. The
thick red bar indicates the uncertainty assuming  a redshift
error of 150 km s$^{-1}$.}
\label{indices}
\end{figure*}

Figure \ref{indices} shows the potential of this approach for disentangling age and metallicity in J-PAS.  The orthogonality of the diagram is appreciable mainly when $H_{\beta}$ is used as age indicator.  On the other hand, the D4000 diagnostic is not independent of metallicity, but it is also more sensitive to age variation, showing a higher dynamical range.  We have verified that the accuracy of the J-PAS photometric redshifts is not sufficient for the purpose of measuring indices, as it is large enough to move the filter containing the index by almost one filter-width, hence compromising the measurement.  Thus, we plan to apply this approach to objects with known spectroscopic redshift.  By assuming a redshift error of $\pm$150 km s$^{-1}$, taken as a conservative upper limit for the uncertainty on the galaxy rotational velocity, we obtain the index error indicated in Figure \ref{indices} by the thick red bar. The resulting uncertainties in the stellar population parameters are $<1$ Gyr for the age and $\sim 0.2$ dex for the metallicity.  The thin black bars in the same figure indicate an error in the index of 0.01 mag, which we consider to be the minimum photometric error obtainable. Photometric errors smaller than this will be hard to achieve because of zero-point errors.  Such an error in the index translates into a S/N requirement of $\sim$150 for each filter.  For nearby galaxies, we aim to reach such a S/N ratio by azimuthal integration over rings of increasing radius.  With such an approach we will be able to obtain spatially resolved stellar population analysis for a large number of nearby galaxies, for which spectroscopic redshifts are already available in the SDSS database.

\paragraph{Spectral Diagnostics for Emission Line Galaxies}
\label{sec:lines}

Emission lines carry information about the excitation mechanism (AGN, young or old stars, shocks), the 
chemical abundance of the warm gas, and its dust content. In star-forming systems, the H$\alpha$ 
luminosity is a well-known tracer of the current star formation rate \citep{kennicutt98}, while 
for AGN [OIII]$\lambda$5007 is a useful proxy for the accretion power of their super-massive black 
holes \citep{heckman04}.

\begin{figure}[t]
\begin{center}
\includegraphics[width=0.7\textwidth]{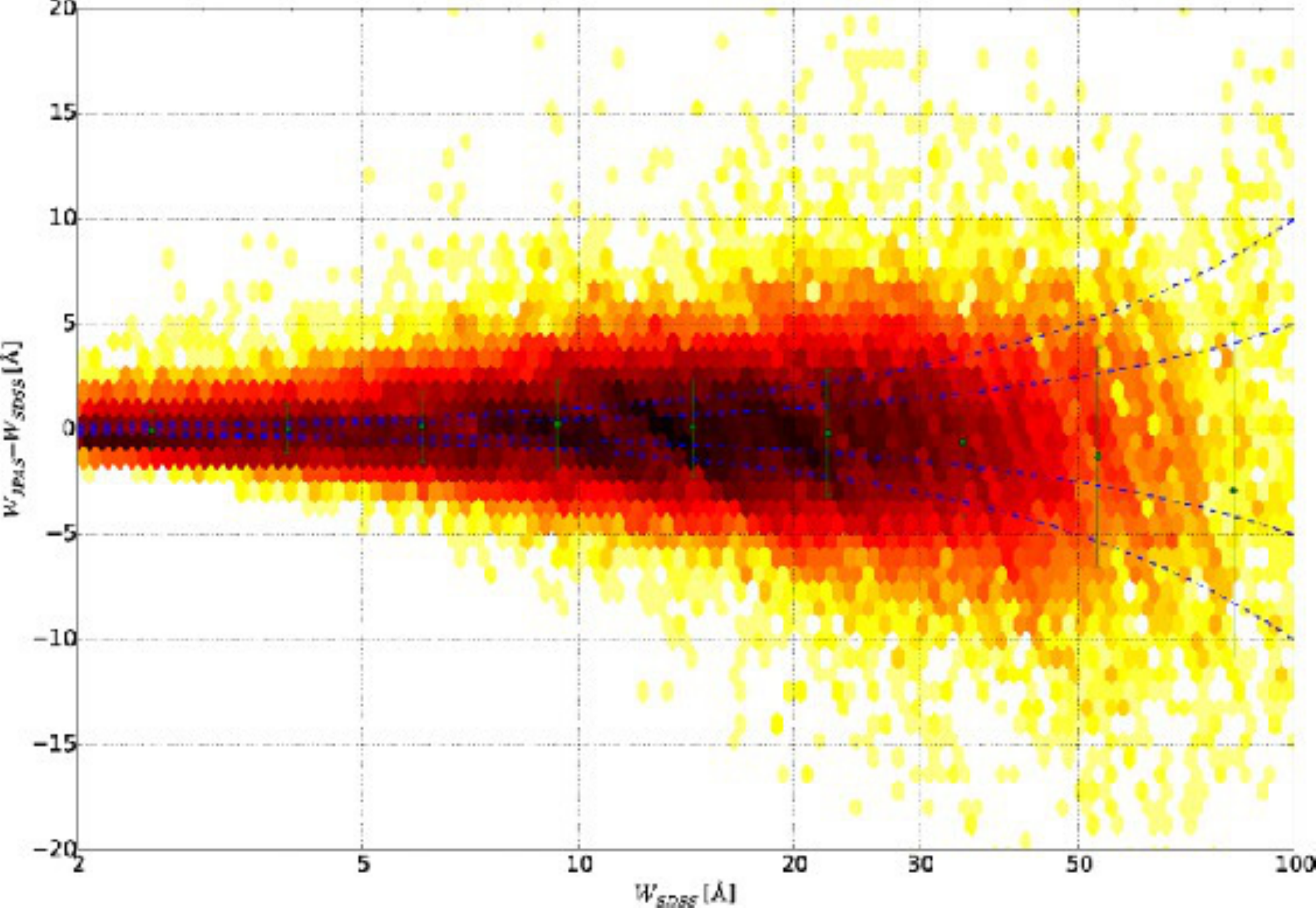}
\end{center}
\caption{Tests of the spectral matching method. On the x axis we have the
$H\alpha$ equivalent width measured on the SDSS spectra ($W_{SDSS}$)
and, on the y axis, we have the difference between the value measured by
the method explained in the text and the spectroscopic one ($W_{JPAS} -
W_{SDSS}$). The color-scale is logarithmic. The two pairs of dotted lines 
indicate the $\Delta W =$ 5 and 10\% of $W_{SDSS}$.}
\label{EmissionLine6563}
\end{figure}

Due to the low spectral resolution of J-spectra, direct measurements of emission line fluxes will be a challenge when using J-PAS data alone. A line of equivalent width $W$ increases the flux in a filter of width $\Delta\lambda$ by a factor $(1 + W/\Delta \lambda)$. For $\Delta\lambda = 100$\AA\ and a photometric accuracy of 2\%, lines stronger than $W \sim 7$\AA\ should be detected with a $S/N>3$ assuming the adjacent filters trace the continuum appropriately and that no other strong line is present within the filter.  Even in the most favorable situation, J-PAS data will not be able to separate H$\alpha$ from [NII]$\lambda\lambda$6548,6584\AA\, precluding the application of traditional SF and AGN classification schemes \citep{baldwin81}. In short, direct emission line flux estimates from J-PAS will be of very limited use, except for the most extreme cases. 

A way to circumvent these problems was devised within our collaboration \citep{schoenell10}.  The idea is to ``borrow'' emission line measurements from galaxies in the SDSS (or any other reference spectroscopic data set) that have approximately (say, in a $\chi^2$ sense) the same J-spectrum as the J-PAS target. The underlying premise is that galaxies which are similar in so many filters should also be similar if observed under higher resolution.  Experiments with this spectral matching scheme produced encouraging results. The method is able to recover the \nii/H$\alpha$ and H$\alpha$/H$\beta$ ratios to within 0.16 dex. Other emission line indices (line ratios and equivalent widths) can be recovered with a similar accuracy. Figure \ref{EmissionLine6563} and Table \ref{tab:ResultadosLinhasCMD} illustrate the application of this method to simulate J-PAS data out of actual SDSS spectra. As shown by \citet{schoenell10,schoenell13}, this method can be easily cast into a fully Bayesian framework, producing posterior probability distributions for any observed quantity.  Conceivably, and in analogy with photo-$z$ methods, even better results could be obtained by using the appropriate priors.

This technique opens up the possibility to use J-PAS to study emission lines at a level of detail much 
beyond initial expectations, substantially enlarging the scope of the project. In fact, 
this indirect (but very efficient) spectral matching trick can be applied to {\em any} 
observed or physical property derived from a SDSS spectrum. For instance, stellar population 
properties such as mean ages, stellar extinction, mass-to-light ratios and velocity dispersion 
derived from a full spectral analysis, such as those obtained by {\sc starlight}\footnote{http://www.starlight.ufsc.br} fits 
\citep{cidfernandes05}, can be estimated through exactly the same formalism. More details on this method are presented in
\citet{schoenell13}

\newpage
\begin{table}
\caption{Emission line accuracy simulation. For each property (which can
be an emission line equivalent width or an emission line ratio) we
measured the  average $\overline{\Delta p}$, median $\tilde{\Delta p}$
and standard deviation {\bf $\sigma (\Delta p )$} of the difference
between our estimation and the value given by the {\sc STARLIGHT}-SDSS database, which is
based on the actual spectra.}
\centering
\begin{tabular}{lcccccc}
\hline
\hline
Property     &       $\overline{\Delta p}$   &       $\tilde{\Delta p}$      &       {\bf    $\sigma ( \Delta p )$    }\\
\hline
$\log W_{ [ OII ] }$    &       0.051   &       0.065   &       {\bf    0.223    }      \\
$\log W_{H\beta}$       &       0.024   &       0.020   &       {\bf    0.145    }      \\
$\log W_{ [ OIII ] }$   &       0.046   &       0.048   &       {\bf    0.245    }      \\
$\log W_{H\alpha}$      &       0.010   &       0.008   &       {\bf    0.160    }      \\
$\log W_{[ \rm {N} II ]}$       &       -0.028  &       -0.024  &       {\bf    0.159    }      \\
$\log [ NII ] / H_\alpha$       &       -0.045  &       -0.042  &       {\bf    0.141    }      \\
$\log [ OIII ] / H_\beta$       &       0.026   &       0.027   &       {\bf    0.250    }      \\
$\log H_\alpha / H_\beta$       &       -0.011  &       -0.013  &       {\bf    0.107    }      \\
$\log [ SII ] / H_\alpha$       &       -0.006  &       0.019   &       {\bf    0.172    }      \\
$\log [ O II ] / H_\beta$       &       0.036   &       0.049   &       {\bf    0.202    }      \\
$\log [ O III ]/ [ N II ]$      &       0.075   &       0.063   &       {\bf    0.265    }      \\      \hline
\hline
\hline
\label{tab:ResultadosLinhasCMD}
\end{tabular}
\end{table}

\paragraph{Confronting the Models}

One key aspect that has per force been excluded from the considerations above is the uncertainty in the stellar population models to which the J-PAS data are to be compared. These models are provided as sets of estimates of the spectral energy distributions (SEDs) of populations of stars of fixed age, metallicity and, in some cases, $\alpha$-element abundance ratios, which are referred to as ``single stellar populations'' (SSPs). It is well known that the various stellar libraries, isochrones and other modeling constituents used in the competing model sets result in significant differences in the flux levels of the SEDs derived. These can be at the 5\% level, and often vary systematically with wavelength. However, while this presents a challenge to the analysis of J-PAS data, it also presents an opportunity. The large quantity of data for nearby galaxies that will be obtained by the J-PAS survey, particularly those for which high resolution spectroscopy is available, will allow the comparison of the results from various model sets. By comparing the best fitting SSPs from both high resolution absorption-line analysis and the low resolution J-spectra in each of the competing model sets, for the first time an analysis of the consistency and quality of fits of the modeling will be possible. This will allow feedback to the stellar synthesis community, hopefully resulting in insights into the wavelength dependent differences between models.


\subsubsection{Density Field Construction}

The estimation of the cosmic density field is of capital importance for large area surveys which are able to cover a wide range of environments, from the low density voids to the high density cores of clusters. In practice, the reconstruction of the galaxy density field reduces to the (weighted) count of objects within some aperture around a set of positions where the density field is to be evaluated. In the general case, the density at an observationally defined position $\mathbf{r} = (RA, DEC, z)$ can be estimated as in \citet{kovac10}:
		\begin{equation}
			\rho(\mathbf{r}) = \sum_i m_i W(|\mathbf{r} - \mathbf{r}_i|; R)\label{rhoeq},
		\end{equation}
\newpage
where the summation is over those galaxies in the sample that have been chosen to define the density field, which we refer to as {\it tracer galaxies}, $m_i$ is the astrophysical weight of the tracer galaxy, and the function $W(|\mathbf{r} - \mathbf{r}_i|; R)$ is the kernel used to weight the tracer galaxies, which is a spatial smoothing function, and $R$ is the smoothing length. The $W$ function is typically chosen such that it weights tracer galaxies depending on their distance $|\mathbf{r} - \mathbf{r}_i|$ from the position where the density field is being reconstructed. \citet{kovac10} show that photometric redshifts can be used in the estimation of the density field by using the probability distribution function (PDF) of the $z_{\rm phot}$. It is common to express the resulting measurement of density as a dimensionless density contrast $\delta(\mathbf{r})$ defined as $\delta(\mathbf{r}) = [\rho(\mathbf{r}) - \overline{\rho}(z)]/\overline{\rho}(z)$, where $\overline{\rho}(z)$ is the mean density at a given redshift. We will test our methodology using {\it mock catalogues} derived from cosmological simulations available to the J-PAS collaboration. These mock catalogues mimic the J-PAS observational strategy, and are essential to assess the reliability and accuracy of the recovered density field. 

Note that we can measure the density field from the galaxy distribution ($\delta^{\rm g}$), while we are ultimately interested in the underlying dark matter (DM) density field ($\delta^{\rm m}$), which defines the structures in the Universe. Both distributions are linked through the bias parameter $b$, with $\delta^{\rm g} = b \times \delta^{\rm m}$. The bias could be a complicated function of redshift, galaxy population, etc. The astrophysical weights $m_i$ in Eq.~(\ref{rhoeq}) can be used to give less importance to more biased populations, thus improving the relation between the measured galaxy density field and $\delta^{\rm m}$. For example, the bias of red massive galaxies is higher than that of blue galaxies. Therefore, the optimum combination of different galaxy populations will enhance the precision and the reliability of our $\delta^{\rm g}$ measurements. To reach this goal, instead of using as the weight some galaxy property, such as luminosity or mass, we can use the real bias of each population, estimated directly through clustering or weak lensing analysis.

J-PAS will allow us to compare and combine numerous estimators of density and environment. For example, the distance to the $n$th nearest neighbor is also a widely used density field estimator \citep[see][]{haas12}. In this case the aperture $R$ varies with the local density, from short lengths in dense environments to large ones in voids. Finally, decomposition of the density field into the main virialised structures such as clusters, groups, filaments, and voids can provide an alternative approach to quantifying the environment. J-PAS will allow us to search for differences between galaxies that are situated in similar local density but in different topological structures, and vice versa. This will allow important tests designed to understand the precise role of environment in galaxy evolution.

\subsubsection{Morphological measurements}

Since the first discovery of galaxies, classifications of their morphologies have been proposed. Hubble established a classification in which galaxies were divided in two main classes according to their global shape: ellipticals (E) and spirals (S). E galaxies were sub-divided in seven groups according to their ellipticity from E0 (round Es) to E7 (the most flattened Es). S galaxies were ordered into three groups depending on the relationship between the bulge and spiral arms (Sa, Sb, and Sc). Hubble established a sequence of shapes from E0 to Sc, with lenticular galaxies forming the bridge between E7 and Sa galaxies. Almost a hundred years after this classification there are still many open questions related to our understanding of the physics behind the formation and evolution of these different morphological types. 

Large-scale imaging surveys such as the SDSS, Pan-STARRS, DES and J-PAS present us with a greatly increased number of galaxies to classify. Moreover, multi-wavelength, spatially-resolved datasets require us to also investigate the galaxy colors and the colours of many different components within those galaxies (i.e. thin and thick disk, bulge, bar, arms). These large data sets make any classification scheme based on visual classifications an enormous challenge, unless a large number of classifiers is involved \citep[see the Galaxy Zoo project;][]{lintott08}. In most instances, therefore, automatic algorithms for galaxy classification are needed. Automatic methods for the classification of galaxies can be divided into two broad groups: parametric and non-parametric methods. Parametric methods measure a set of physical parameters by fitting some parametric laws to the light distribution of galaxies, and attempt to classify them accordingly. In contrast, non-parametric techniques characterize the morphological types of galaxies by translating them into a different mathematical or physical representation and then identify the most significant components. 

In J-PAS different methods will be used for the morphological classification. The first two methods are based on the modeling of the surface brightness distribution of the galaxies. In the first one, a parametric method, we will model the galaxy surface brightness distribution by fitting the traditional parametric laws \citep[see e.g.,][and references therein]{prieto01,aguerri04}. We will use standard codes like GASPH2D \citep{mendez08} or GALFIT \citep{peng02}. The modelization will provide us with an effective radius and surface brightness profile of the main galaxy components \citep[see e.g.,][and references therein]{aguerri04,mendez08}. These can then be used to evaluate the main scaling relations of galaxies and their evolution with time. This modeling will also allow us to perform a broad galaxy classification (early- versus late-type) based on the bulge-to-disc ratio or the Sersic shape parameter. Exploiting more fully the multi-band nature of  J-PAS, we will also apply a recently developed multi-wavelength version of GALFIT named MegaMorph \citep{haussler13}. MegaMorph enables the automated measurement of wavelength-dependent structural parameters for very large samples of galaxies. In fact, fitting galaxy light profiles with multi-wavelength data increases the stability and accuracy of the measured parameters, and hence produces more complete and meaningful multi-wavelength photometry than has been available previously. We will recover the color  and the color gradient of each galaxy  component and we will study how it varies for different Hubble types. We will be able to understand how many components galaxies have, which components formed first, and if it has been rejuvenated by star formation due to recent mergers, or perhaps quenched. 

In the second method, a non-parametric one, the modelization of the galaxy surface brightness distribution will be performed by fitting Chebyshev polynomials \citep[CHEFs,][]{jimenez12}. The CHEF method will not be directly performed on all objects to be classified, but on a visually, well classified, and complete set of galaxies \citep[e.g., the EFIGI catalog;][]{baillard11}. In this way, we will project the CHEF mathematical basis onto a physically meaningful basis composed by the CHEF models of these EFIGI galaxies (after scaling, rotating, and flux normalizing them). Then, we will decompose the J-PAS galaxies using this physical basis providing us with a probability for each source to belong to a certain morphological type (according to the EFIGI precise classification).

\begin{figure}[t]
\begin{center}
\includegraphics[width=0.8\textwidth]{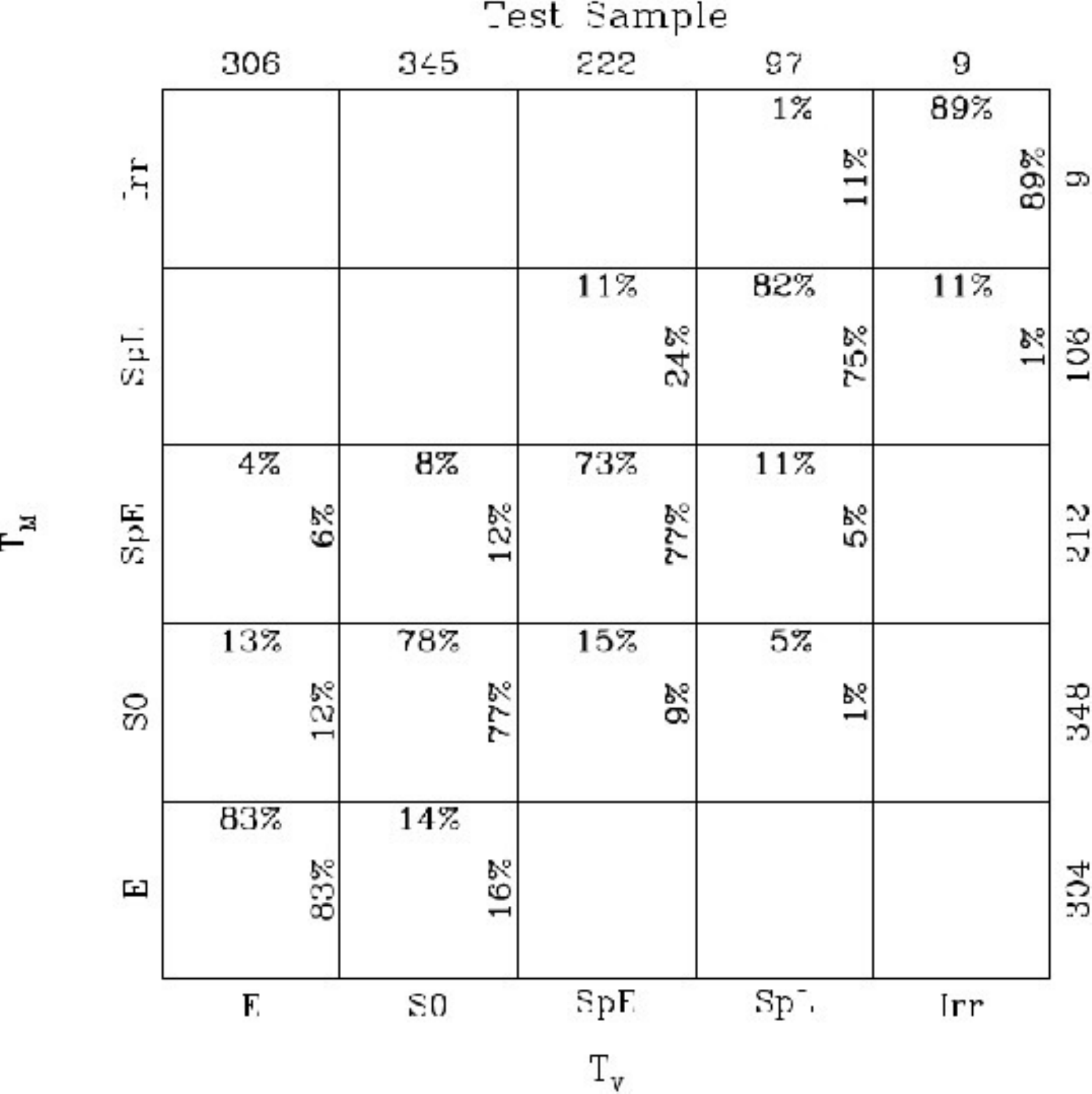}
\end{center}
\caption{Comparison between visual and MORPHOT 'broad' morphological classes
for the galaxies of the MORPHOT test sample. At the top of the
2D-bins the percentages of the visual classes (Es, S0s, early spirals
[SpE], late spirals [SpL] and irregulars) falling in different bins of
the MORPHOT classification are shown. The percentages of
the MORPHOT classes falling in different bins of the visual
classification are shown at the right-hand side of the 2D-bins. Finally,
on the top (columns) and on the right (rows) of the plot, we report
the total number of galaxies in each 'broad' class of the visual and
MORPHOT estimates, respectively.}
\label{broadcomp}
\end{figure}

Other non-parametric classifications will be achieved by using the codes GALSVM and MORPHOT. These two algorithms classify galaxies using a multi-dimensional set of galaxy parameters. The approach of the MORPHOT tool is fully empirical. In particular, MORPHOT exploits 21 morphological diagnostics, directly and easily computable from the galaxy image, to provide two independent classifications: one based on a Maximum Likelihood, semi-analytical technique, the other one using a Neural Network. The technique has been tested on a sample of $\sim$1000 visually classified WINGS galaxies, proving to be almost as effective as 'eyeball' estimates. In particular, at variance with most existing tools for automatic morphological classification of galaxies, MORPHOT has been shown to be able to distinguish between ellipticals and S0 galaxies with unprecedented accuracy (see Figure~\ref{broadcomp}). 
This morphological classification scheme is expected to be most efficient for those galaxies with an area larger than 200 pixels \citep[see][]{fasano12}. 

The second non-parametric algorithm that we will use for the morphological galaxy classification will be GALSVM. This code was developed by \citet{huertas08} and has been applied to several samples of galaxies at different redshifts including galaxies from the ALHAMBRA survey \citep[see][and references therein]{povic13}. The ALHAMBRA images are of similar quality as the expected J-PAS data. The algorithm is a generalization of the non-parametric classifications by using an unlimited number of dimensions. The classification provided by this algorithm is probabilistic following a Bayesian approach \citep{huertas08}. The algorithm is trained with a set of galaxies visually classified. These galaxies are inserted into the real scientific images according to the observed redshift distribution of the galaxies that we wish to classify. For each classified galaxy the algorithm then provides a probability for it to belong to each of the considered morphological classes. For example, the ALHAMBRA galaxies were classified in two groups (early and late types). Thus, each classified galaxy has a probability to belong to these two classes. Due to the similarities between the J-PAS and ALHAMBRA images, we expect to be able to classify all J-PAS galaxies down to 22 AB mag in the $r^\prime$-band filter in at least in two broad groups (early and late). For the ALHAMBRA galaxy survey ($4\sq\degr$) we have obtained a sample of 22\,051 well-classified objects at $z<1.5$ having F613W$<$22 mag \citep{povic13}. This means that we expect to classify several millions of galaxies in the full J-PAS survey. 
Large galaxies (larger than 200 pixels) will be classified using a finer classification in which all the Hubble galaxy types will be considered. 

In summary, J-PAS will deliver the largest sample of galaxies with morphological classifications, bulge-to-disk ratios, and integrated and component colors in the literature, which will be useful for a wide variety of studies of galaxy evolution.

\newpage

\subsection{Science Themes}
\label{gal:science}

\subsubsection{Theme I. The Nearby Universe}

What can J-PAS do for improving our understanding of the formation of galaxies across the local Hubble sequence in general, and for galaxies like our own Milky Way in particular? Although the general picture of disk galaxy formation has more or less been established, the relative importance of the various secular and accretion-driven processes is an ongoing topic of investigation. Fortunately, galaxies possess a long memory in terms of the fossil record in their stars which we can use to trace their evolutionary history. Since the seminal work of \citet{eggen62}  the importance of 
the study of ages and chemical abundances has been recognized. Of particular relevance is the problem of radial migrations of gas and stars within the disk. The realization that stars in galactic disks can migrate radially across significant distances has, in recent years, completely changed the discourse on spiral galaxy evolution. The subject of disk migration has received particular attention in the last few years in light of new astro-archaeology surveys of the Milky Way (e.g., APOGEE, HERMES, and Gaia). Radial migration in the Milky Way brings stars from the inner and the outer disk, where the mean abundances are different, into the solar neighborhood. The result is a change in the age-metallicity relation, and in the relations between ages and metallicities on one hand and velocity dispersion on the other. 
However, radial mixing is very much a theoretical concept, and its relative importance to the evolution of the Galactic disk is still unknown.  We also do not know what the main mechanism is that produces the radial migration. \citet{sellwood02} postulated that resonant scattering of disk stars off of successive, transient spiral density waves can produce significant displacements ($>$4 kpc). \citet{minchev10} further argued that an overlap of bar and spiral arms resonances could drastically enhance the migration efficiency within disks. Lastly, \citet{quillen09} showed that radial migrations of stars to the outskirts of disk galaxies could arise via tidal perturbations during the peri-center passages of dwarf satellites. These different mechanisms are furthermore expected to have different efficiencies in galaxies of different masses, different bar strengths, and different environments. Therefore, studying the properties of the spatially resolved stellar populations in large samples of galaxies covering a large range of masses, structural properties and environments should allow us to constrain the importance of secular evolution/radial migrations and the main physical mechanisms responsible.
 
Another problem that has received considerable attention in the last few years relates to the formation of the thick disk.  Because of its old age and because it constitutes a kinematically and chemically recognizable relic of the early Galaxy, the thick disk is a highly significant component for the study of galaxy formation. How did the thick disk form?  Several mechanisms have been proposed, including (i) gas rich mergers at high redshift \citep[e.g.,][]{brook04}, (ii) accretion debris \citep{abadi03}, (iii) heating of the thin disk via disruption of its early massive clusters \citep[e.g.,][]{kroupa02}, (iv) heating of the thin disk by accretion events, and (v) migration of more energetic orbits from the inner galaxy to larger radii where the potential gradient is weaker \citep{schonrich09}. To test these formation models, detailed comparison of thin and thick disk properties are required across a range of galaxy masses. In particular, the relative ages and chemical enrichment patterns of the thin and thick disks are expected to differ among these different formation models. If the thick disk results from a gradual kinematical heating of the thin disk, there should be a smooth age and enrichment gradient between the two. In contrast, if the thick disk is formed from accreted stars we should expect the ages and metallicities of the thin and thick disk to be only weakly correlated. We may also expect to see variations with the mass of the galaxies, with less massive galaxies being more susceptible to external heating and more massive galaxies being better able to tidally disrupt satellites. Measuring the ages and metallicities of thick disks outside the local group has proved to be challenging \citep[see][for an early attempt]{yoachim08}. 

Large spectroscopic studies of stellar populations across the disks of spiral galaxies have been very scarce \citep{yoachim08,yoachim12,macarthur09,sanchez-blazquez09}. In total, less than $\sim$30 galaxies have been studied and these studies were mostly limited to the inner disk. Furthermore, disk galaxies are intrinsically complex, with multiple structural components (e.g., disks, bulges, bars, and halos). Long-slit spectroscopic studies therefore often loose valuable information or introduce confusion bias. Spectroscopic surveys such as CALIFA \citep{sanchez12}, VENGA \citep{blanc13}, and SAMI \citep{croom12} are using integral field units to perform spatially resolved studies of the stellar populations in nearby disk galaxies. Although these surveys will allow a major step forward, the number of galaxies that they reach is still fairly limited. J-PAS will offer a number of benefits over other studies. First, the large survey area and corresponding large sample size will allow us to isolate statistically the influence of parameters such as mass, morphological type, and environment, on the spatially resolved stellar populations and population gradients across the disks. Second, J-PAS will be able to trace the low surface brightness external parts of disks beyond 3 scale-lengths, which are very difficult to reach for the spectroscopic surveys mentioned above. The spatial resolution offered by J-PAS will allow us to resolve the stellar populations in the different components of galaxies, such as arms, inter-arms, bars, rings, and central components. 

\paragraph{Dwarf Galaxies}

Dwarf elliptical galaxies (dEs) are small, low-luminosity galaxies which
constitute the dominant population of nearby galaxy clusters. Indeed, dEs alone
outnumber high luminosity galaxies by a factor of 6 in the Local Group \citep{mateo98}, 
and they represent more than 50\% of the  galaxies in the Virgo cluster
\citep{sandage85}. As potential building blocks of massive galaxies in
hierarchical frameworks of galaxy formation, dwarf elliptical galaxies may
provide important clues on the main processes involved in galaxy assembly and
evolution. 

With the advent of larger telescopes and more sophisticated instrumentation, we
now know that dEs display a much wider range of properties than originally
thought, opening again the debate about their origin. The three most widely
adopted scenarios are: (1) They might be primordial objects which expelled
their gas in early stages of their evolution because of supernova explosions
\citep[e.g.][]{mori99}, (2) dEs could be the by-product of late-type disky
galaxies that entered clusters $\sim$5 Gyr ago and evolved into a hot
spheroid because of internal dynamical processes \citep{conselice01}. (3)
Tidal harassment within the cluster. Dwarf ellipticals are mostly found in
clusters and groups of galaxies, while star forming dwarfs are predominantly
found in the field \citep{dressler80}. This very pronounced morphology 
density-relation for dwarfs shows that indeed the environment plays a very
important role in  their evolution.

In recent years, a growing number of studies has shown that they are a surprisingly
inhomogenous class: photometric studies of large samples of Virgo dwarf
early-types have revealed the presence of disks, bars, spiral arms, and nuclei
\citep[e.g.][]{lisker07,janz12}. A diversity of properties has
also been found through the analysis of dE stellar populations, such as their ages,
metallicities, and the gradients thereof \citep[e.g.][]{chilingarian09,koleva11}.
Kinematic studies confirm and add to the variety: the degree of rotation is not
correlated with the (projected) flattening, and kinematically-decoupled 
components are found in some early-type field dwarfs \citep{toloba11,rys13}. 
This diversity has made it challenging to both relate the different
subtypes to each other, as well as to place the whole class in the larger
context of galaxy assembly and (trans)formation processes.

Despite their large numbers and paramount importance in our understanding of
galaxy evolution, the low-luminosity character of these systems has always
prevented extensive studies of similar quality as those performed on normal 
galaxies. On one hand, major photometric surveys (e.g. SDSS), while a
good source for identifying candidates, are often too shallow to map the
properties of these galaxies far out in radius. Spectroscopic studies, on the
other hand, are based on a rather limited number of dwarf galaxies and are
typically restricted to a single aperture measurement or a short long-slit
profile. Today, integral-field spectroscopic studies, while providing a
wealth of detailed spectral information, are still scarce \citep[e.g.][]{rys13}. None of
the upcoming major integral-field surveys (e.g. CALIFA, SAMI, ManGA) will
change this picture in the foreseeable future.

The J-PAS Galaxy Evolution Survey presented here opens up a new and important window in this
field. Not only will it allow the identification of many dwarf galaxies in
different environments over the surveyed  area of $8500\sq\degr$, but, most importantly, it
will be deep enough to probe regions well beyond where the surface brightness
profiles of these galaxies are no longer described well by a single exponential
profile. The multi-band observing strategy of the J-PAS survey will
allow us, for the first time, to produce a very detailed study of the stellar
populations of dwarf galaxies well into their outskirts. The analysis of their
star formation histories at different radii will reveal whether star formation 
takes place in an inside-out fashion (i.e. as most ordinary galaxies exhibit) or if
on the contrary secular evolutionary processes dominate their evolution. It
will also reveal the importance of environmental processes in those dwarfs
living in clusters. Combined with the results obtained for ordinary galaxies, the
J-PAS survey has the potential to become the absolute reference in the field of 
stellar populations by providing the complete picture of galaxy evolution as a
function of mass, luminosity, and galacto-centric radius for the largest
set of galaxies ever observed.

\paragraph{Extragalactic Globular Clusters}

The formation of globular clusters (GCs) is thought to be linked to major episodes of star formation in galaxies \citep{larson96,elmegreen97,ashman01}. A key observational result on this topic is the existence of a bimodal color distributions in the GC systems of most galaxies, including our Milky Way. This fact has been widely interpreted as evidence for two distinct GC subpopulations -- metal rich (red) and metal poor (blue) -- and this has been confirmed in many cases by conducting detailed spectroscopic studies of extragalactic GC systems in nearby galaxies. Different galaxy formation scenarios are proposed to explain the existence of the GC subpopulations, involving mergers, in situ formation or accretion processes. In this sense, GCs are relics that provide valuable information on how the main star formation episodes of their host galaxies took place. An interesting review on extragalactic GCS and their capability to shed light on galaxy formation can be found in \citet{brodie06}.

How can J-PAS contribute to our understanding of GC systems and, therefore, galaxy formation and evolution? Extragalactic GCs appear as point-like sources in the outskirts of galaxies. A good characterization of GC subpopulations in terms of ages and metallicities in all kind of galaxies is essential to have robust statistics and put constraints on the complex process of GC and galaxy formation. Detailed studies have been limited to spectroscopic work on $8-10$\,m class telescopes, and are therefore scarce and time-consuming \citep[e.g.][]{strader05,cenarro07}. As a low resolution IFU, J-PAS will constitute a revolution in this topic by providing a massive census of extragalactic GCs for thousands of nearby galaxies in the $8500\sq\degr$ survey area. The multi-filter approach will allow not only to detect GC candidates but also to characterize their stellar populations in the same way as explained in Section~\ref{stelpops}. 

It is well  known that the number of GCs scales with the galaxy luminosity \citep{harris79}. This introduces the definition of the so called {\it specific frequency} of GCs, $S_{\rm N}$, which can be considered approximately as the number of GCs per unit luminosity \citep{harris81,harris91} normalized to $M_{\rm v} = -15$. The specific frequency varies in the range $0.3-1$ for spiral galaxies, $1-15$ for giant ellipticals and $1-30$ for dwarf elliptical galaxies. For instance, the Milky Way has $S_{\rm N} \sim 0.6$ ($\sim 150$ GCs), whereas the giant elliptical M87 has a $S_{\rm N}$ of $14.1 \pm 1.6$ \citep{harris98}, with more than 1000 GCs. Interestingly, GCs are also considered as standard rulers for inferring cosmological distances. GC systems follow a roughly universal, Gaussian-like luminosity function (LF) that peaks at $M_{\rm V} \sim -7.5$ \citep[e.g.][]{harris01,cezario13}. Therefore  the number of GCs that J-PAS will detect depends on the distance to the galaxy and the galaxy type and luminosity. 

To get an idea of the impact that J-PAS will have in this field, let us consider the galaxies in the Virgo Cluster. Assuming a distance of $\sim 17$\,Mpc (m$-$M $\sim 31.15$) the globular cluster LF peaks at V$\sim23.65$, or $g\sim24$ (depending on the GC color). The J-PAS magnitude limit in $g$ band reaches down to 23.75 (S/N$=5$; 3\,arcsec aperture). This means that J-PAS will be able to detect nearly all GCs in the bright half of the globular cluster LF, which amounts to several hundreds of GCs in a typical massive elliptical galaxy. The brightest GCs in Virgo galaxies have $g\sim20-20.5$, depending on the colour. So, making use of the survey broad-band filters, J-PAS will detect all the GCs $2.5-3$\,mag fainter than this value. More interestingly, if we focus on the J-PAS narrow-band filters, at the Virgo cluster distance J-PAS will detect all the GCs down to $2-2.5$\,mag fainter than the brightest GCs. This amounts to from around one hundred GCs per giant elliptical in Virgo to a few ($0-5$) GCs in dwarf ellipticals, as the number of GCs scales with the galaxy luminosity.

Putting all the above numbers in context: integrated over the $8500\sq\degr$ area that J-PAS will cover, it is expected to observe tens of thousands of GCs in nearby galaxies (say $< 20$\,Mpc), with a J-spectrum for each GC. This will provide a first estimate of the GC metallicity and age, allowing to split between metal rich and metal poor GCs, as well as to study the ages of the GCs and infer new clues on the formation epoch depending on the host galaxy type.

\paragraph{Tidal Disruption Events in Globular Clusters}

J-PAS will also provide a highly efficient means of detecting the
aftermath of tidal disruption events within extragalactic GCs caused
by stars being torn apart by tidal forces from intermediate-mass black
holes (IMBHs, masses in the 100--10,000 M$_{\odot}$ range) within the
cluster. The possibility that globular clusters harbor IMBHs has been
a bone of contention for more than 30 years. Demonstrating whether or
not IMBHs exist in globular clusters has important ramifications on
our understanding of not only black hole formation, but also of the
postulated feedback mechanism linking the growth of black holes and
galaxy formation that is believed to cause the well-known $M_{BH} -
\sigma$ relation in massive galaxies. If IMBHs exist in the centers of
globular clusters, they should occasionally disrupt passing stars
\citep{rees88,baumgardt04}. It is predicted that the debris from the
disrupted star forms a precessing, self-interacting stream, which
ultimately forms an accretion disk, an optically-thick envelope, and a
quasi-spherical $\sim10^4$ K diffuse photosphere around the black
hole. 

  This envelope of stellar debris is expected to intercept X-rays from matter in the accretion disk around the black hole and reprocess it in the optical/UV part of the spectrum in the form of emission lines superimposed on the stellar continuum of the stars within the GC. This emission should be detectable for a few hundred years after the disruption event \citep{clausen11,strubbe09}. Such a post-tidal disruption event by a GC black hole is believed to have been observed in a GC near the Fornax Cluster elliptical galaxy NGC~1399 that harbors the luminous X-ray source CXO J033831.8--352604. \citet{irwin10} detected strong [N II] and [O III] emission lines in the optical spectrum of this GC, and \citet{clausen12} argued that the X-ray and optical properties of this cluster are consistent with the tidal disruption of a star by a 100--200 M$_{\odot}$ black hole 100--200 years ago. The tens of thousands of extragalactic GCs that J-PAS will observe within 20 Mpc will provide fertile hunting grounds for further examples of tidal disruption aftermaths. The J-spectra of the brighter ($m_g \sim 20-20.5$ mag) systems should be sufficient for detecting the strong emission lines such as those found in CXO J033831.8--352604.

\subsubsection{Theme II. Evolution of the Galaxy Population since $z\sim1$}

\begin{figure}[t]
\centering
\includegraphics[width=0.8\textwidth]{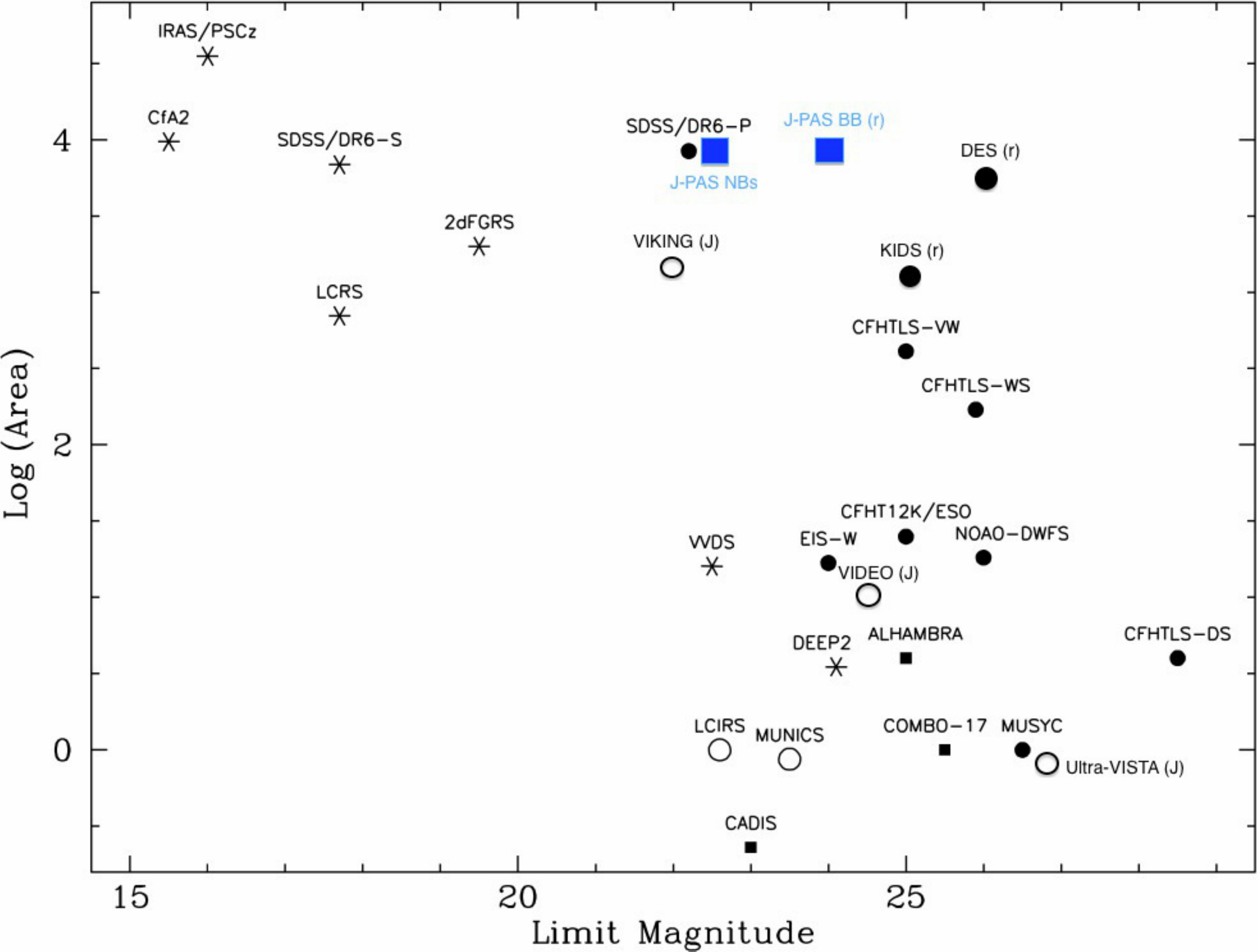} 
\caption{An overview of extra-galactic surveys in the area (square degrees) versus depth (limiting magnitude) plane. Circles correspond to broadband photometric surveys. Stars correspond to spectroscopic surveys. Squares correspond to multi-filter system photometric redshift surveys (i.e. CADIS, COMBO-17, ALHAMBRA and J-PAS). Filled symbols are surveys that are performed (primarily) in the optical, while open circles are surveys performed primarily in the near-infrared. The areas covered by the SDSS, DES and J-PAS are comparable. The J-PAS broad-band filters reach a depth of almost two magnitudes fainter than SDSS, while each of the J-PAS narrow-bands reaches a depth that is comparable to that only achieved by SDSS in the broad-bands. Both DES and KIDS will reach much fainter magnitudes in the optical than J-PAS, but will not have the great leverage in photometric redshifts achieved by the 56 filters of J-PAS. Note that the magnitude limits in general refer to a variety of bandpasses. Figure taken from \citet{moles08}.}
\label{fig:surveys}
\end{figure}

Over the past decade, large-area sky surveys of the relatively nearby universe, such as SDSS \citep{york00}, 2MASS \citep{skrutskie97}, and GALEX \citep{martin05}, have proven the power of large data sets for answering fundamental questions on extragalactic astronomy. These data, combined with the much deeper multi-wavelength pencil-beam surveys of the high-redshift universe (e.g. AEGIS, COMBO-17, GEMS, CANDELS), have improved significantly our understanding of the evolution of galaxies and stellar populations in galaxies over the last 9 billion years and more. Systematic studies of, for example, morphologies, number densities, luminosity and stellar mass functions, stellar populations, and the effect of the environment over a wide range in redshift are required to construct a detailed picture of galaxy evolution. In the relatively nearby universe, the SDSS has been instrumental in constraining many of the relevant parameters through a combination of large-area imaging with targeted medium-resolution spectroscopic follow-up. Large ``value-added'' data sets based on the combination of SDSS optical samples with samples in the UV from GALEX, in the near-infrared from 2MASS and WISE, and in the radio from FIRST have greatly extended the range of extra-galactic science questions that can be addressed with these data. At higher redshifts, multi-filter photometric surveys such as COMBO-17 \citep{wolf08}, COSMOS  \citep{ilbert09}, and ALHAMBRA \citep{moles08} have allowed to determine photometric redshifts while at the same time sample the SEDs of galaxies with an accuracy sufficient for evaluating stellar populations as a function of, e.g., redshift and environment.

For the majority of galaxies that will be unresolved, J-PAS will be used to determine, e.g., the stellar masses, luminosity/mass-weighted ages, dust content, some line-strength indices suited for low-resolution data, current star-formation rates, past star-formation histories, and the presence of AGNs. All these parameters will be studied as a function of, e.g., redshift and environment thereby constraining the main mechanisms responsible for galaxy evolution over the crucial redshift range $0\lesssim z\lesssim1.5$, which can be compared to data from deeper surveys probing higher redshifts. 

In Fig. \ref{fig:surveys} we present an overview of many of the main past and present galaxy surveys in terms of their sky coverage and limiting magnitude. The largest spectroscopic surveys do not typically reach very deep, while the deepest photometric surveys are typically limited in sky coverage. J-PAS will populate a ``sweet-spot'' in the area--depth plane. The sky coverage is comparable to that of the SDSS photometric survey.  Its main power however lies in the fact that it achieves a similar or greater depth compared to SDSS in each of its many narrow-band filters. This adds great leverage to the study of galaxy evolution when spectra are not feasible and a small set of broad band filters offer a spectral characterization that is too coarse for most detailed diagnostic studies. The J-PAS broad-band filters will reach $\sim2$ mag deeper than those of the SDSS, allowing the detection of lower-mass galaxies, higher redshift galaxies, and more low surface brightness details in nearby systems compared to SDSS. Other large optical surveys, such as KIDS and DES will observe $>1500\sq\degr$ of sky, and will go several magnitudes fainter than J-PAS in the optical broad-bands, but will not have the great leverage in photometric redshifts and spectral classifications achieved by the 56 filters of J-PAS. It is therefore expected that these surveys will be highly complementary, rather than repetitive.

\begin{figure}[t]
\centering
\includegraphics[width=0.8\textwidth]{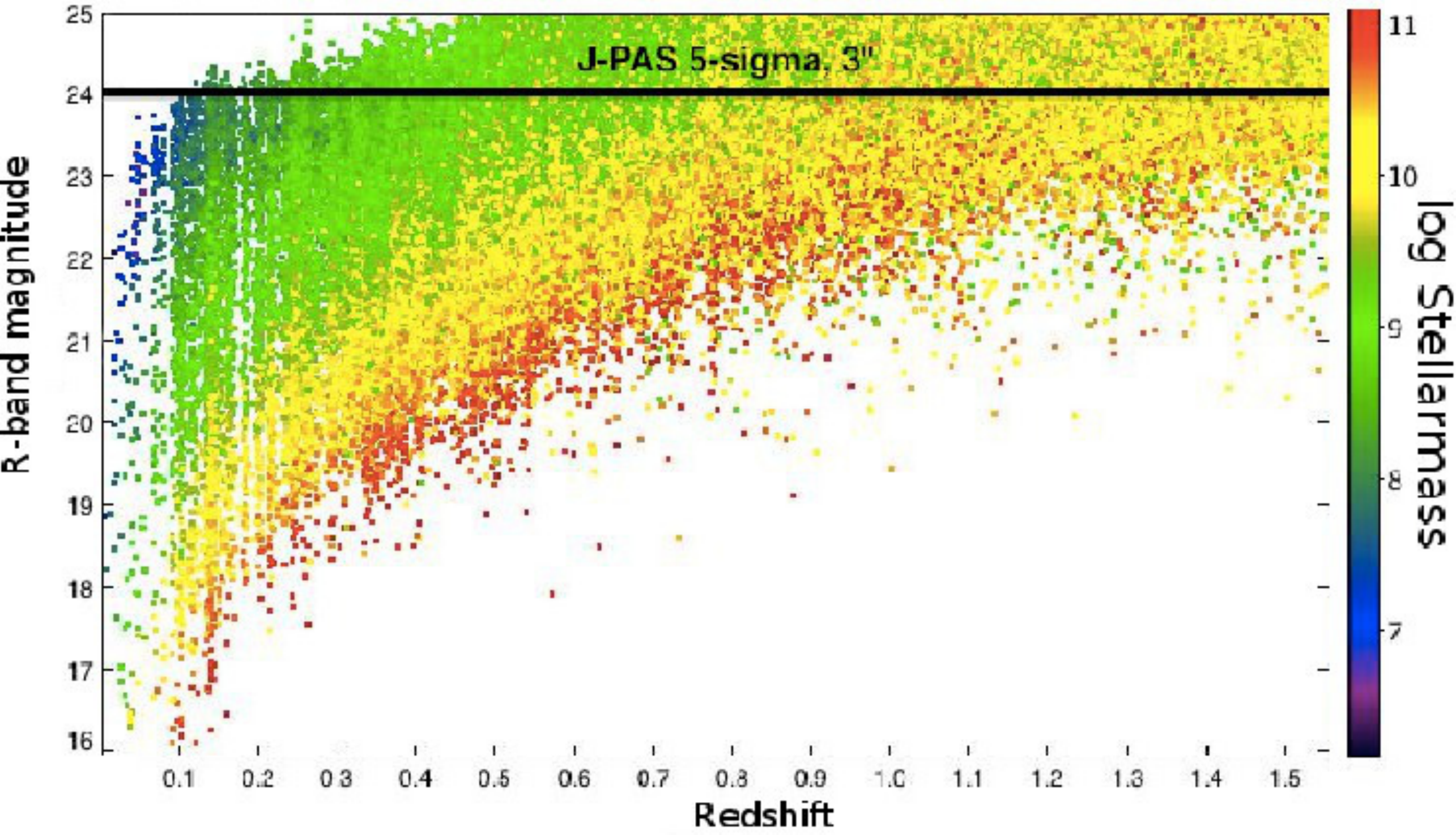} 
\caption{The range in stellar mass probed by J-PAS as a function of redshift and limiting magnitude in the $R$-band. The black horizontal line indicates the approximate limiting magnitude achieved by J-PAS ($5\sigma$, 3\arcsec\ aperture). Galaxies are colour-coded according to their stellar mass. J-PAS should be largely complete for galaxies more massive than $\sim10^{9}$ $M_\odot$ out to $z\sim0.4$ and $\sim10^{10}$ $M_\odot$ out to $z\sim0.8$ This prediction is based on a 2 square degree mock galaxy redshift survey from the Millennium simulations \citep[see][]{henriques12,overzier13a}.}
\label{fig:masslimits}
\end{figure}

What kind of galaxies will J-PAS be able to detect? In Fig. \ref{fig:masslimits} we show a recent simulation of galaxies in an area of about 2 square degrees as a function of redshift and $R$-band magnitude, colour-coded according to their stellar mass. The J-PAS detection limit in the $R$-band is about 24 mag ($5\sigma$, AB) measured inside a 3\arcsec\ diameter circular aperture (horizontal line). In principle, J-PAS will thus be able to detect large numbers of galaxies down to $M_*\approx10^{9}$ $M_\odot$ up to $z\sim0.4$ and $M_*\approx10^{10}$ $M_\odot$ up to $z\sim1.5$. Although the exact amount of information that we will be able to extract from these galaxies based on the 56 J-PAS bands will depend on, for example, the achieved $S/N$ in each filter, the redshift and the spectral type of each galaxy, Fig. \ref{fig:masslimits} illustrates the enormous leverage power in stellar mass and redshift that J-PAS will bring to the field of galaxy evolution. This will allow us to open (or re-open) a large number of parallel investigations in this field. It would be cumbersome to discuss all the possibilities here, but we will list a few. 

In a colour-magnitude diagram (CMD) the distribution of galaxies appears bimodal, and this is true at both low and high redshift. Quiescent early-type galaxies (ETGs) and star-forming late-type galaxies (SFGs) populate preferentially the red sequence and the blue cloud, respectively. The colors of non-star-forming galaxies in the red sequence change with redshift as expected for passively aging stellar populations. In the blue cloud, galaxy colors are determined by recently born stars and vary little with redshift.
The cosmic star formation rate (SFR) density evolved strongly with time, achieving a peak when the universe had about half of its current age, at $z \sim 1.5 - 2$. Galaxies in the blue cloud that see their star formation quenched should move quickly toward the red sequence, traversing the so-called ``green valley'' in the interim. Even though the details are not yet fully understood, the triggering of star formation by post-merger starbursts in blue cloud galaxies, the quenching of star formation by AGN feedback, and the amounts of neutral gas available, are all crucial for regulating the evolution of galaxies. Understanding how and why galaxies traverse the CMD and evolve with time will be one of the main goals pursued by J-PAS. 

If we look at galaxy assembly in detail, the evolutionary scheme is more complicated. The stellar mass in red galaxies increases by a factor of two since $z=1$ \citep{bell04,faber07,cristobal09}, in agreement with hierarchical models of galaxy formation and evolution. However, red sequence galaxies as massive as 3$\times10^{11}$ $M_\odot$ were already in place at $z\sim2$ \citep{nicol11}, and many of these galaxies are very compact \citep{daddi05,trujillo06}. The fact that such massive, dense systems ($M\simeq10^{11} M_\odot$, $R_{e}\sim1.5$ kpc) appear to be scarce in the nearby universe (\citet{trujillo09} find $<0.03\%$ based on the SDSS, see also \citet{taylor10}) implies that the structural properties of these massive objects have evolved strongly between $z\sim2$ and the present \citep{trujillo07,buitrago08}. J-PAS will be able to contribute to this problem in at least two ways. First, it will be possible to sample the evolution from massive galaxies at high redshift to massive galaxies at low redshift in various redshift bins from $z\sim1$ to $z=0$ with  unprecedented statistics. The statistics of this evolving population could then shed light on the main mechanisms that transform the structural properties of these galaxies over time. Second, J-PAS will allow us to search for rare local galaxies that are in a stage analogous to the dense and compact stage as those at high redshift, thereby shedding light on the formation mechanisms of these systems \citep[e.g.][]{overzier09}. It is believed that the formation of (compact) spheroids at high redshift is related to the core--cusp dichotomy observed in local early-type galaxies \citep{kormendy09}, in which the most massive early-types tend to have a deficit of light in their inner regions with respect to their outer Sersic profile, and lower mass early-types tend to have an excess of light in their inner regions. These observations are consistent with the latter being the result of so-called ``wet'' or dissipative mergers at high redshift that form dense mass concentrations in the core, while the former are the result of ``dry'' mergers that lead to a ``cored'' inner profile. Examples of this process can be seen in various classes of nearby galaxies that are good analogues of the compact spheroids at high redshift in various stages of their evolution \citep[e.g. see][]{overzier09,trujillo12,jiang12}. 

Another area in which J-PAS data could excel is that of the population of intensely star-forming (star-bursting) galaxy population. In the very nearby universe, the relatively rare class of starburst galaxies are the only set of galaxies in which we can directly observe the interplay between (massive) star formation and the interstellar medium at high spatial and spectral resolution. Though interesting by itself, these studies are important for providing insight into similar processes that were much more common in galaxies at much earlier times. Despite detailed observations of star-forming populations at high redshift observed with for example the Hubble Space Telescope, we still rely largely on locally determined calibrations and diagnostics when determining their physical properties. J-PAS will allow us to not only establish new large ``training sets'' to aid in the determination of galaxy properties at higher redshifts, it will also allow us to directly compare the main properties of the heavily star-forming population as a function of redshift. At $z\gtrsim1$, the far-UV is directly accessible in the $U$-band, while at lower redshifts a combination between GALEX and J-PAS will allow us to select large numbers of UV-luminous starburst systems that share many similarities with the typical star-forming population at $z\gtrsim2-4$ \citep{heckman05}. 

A related area of research will involve studying the effects of galactic outflows and their importance in galaxy evolution. For systems at $z\gtrsim0.3$, the Fe II$\lambda\lambda2586,2600$ and Mg II$\lambda\lambda2796,2803$ absorption line doublets, which are sensitive probes of the cool galaxy-wide ionized winds, are accessible in the observed optical. J-PAS should deliver the largest sample to date of systems covering a wide range of redshift, stellar mass, and SFR (density) that are well-suited for detailed follow-up spectroscopy of such wind features. This will allow an unprecedented survey of the interplay between gas and stars that is crucial to constrain the importance of stellar winds feedback in galaxy evolution \citep[e.g.][]{tremonti07,rubin10,heckman11,diamond12,rubin13}.

\subsubsection{Theme III. The Growth of Large-scale Structure and Environment}

\paragraph{Galaxy evolution as a function of environment}

\cite{dressler80} showed that the fraction of elliptical and lenticular galaxies increases towards denser regions, while the fraction of spiral galaxies shows the opposite behavior. This so-called morphology -- density relation motivated many studies of the relation between galaxy properties and local density, and its evolution with redshift. 
Some of the main properties that have been found to correlate with local density are the fraction of red or early-type galaxies  \citep[e.g.,][]{dressler97,postdam05,cucciati06,tasca09,iovino10,capellari11}, the galaxy luminosity and stellar mass functions \citep[e.g.,][]{bolzonella10,pozzetti10,vulcani11}, and the merger fraction \citep{lin10,deravel11,pawel12}. We have since learnt that the stellar mass of galaxies is one of the primary drivers responsible for the observed correlations, in the sense that, for example, the red/early-type fraction is higher for more massive galaxies. However, the local density or the environment must still play an important role, particularly in the transformation of low-mass, blue galaxies into red galaxies \citep{cucciati10den,peng10}. However, the precise scale lengths on which the various environmental effects operate on galaxies are still a matter of debate. For example, it has been shown that the colors of galaxies, H$\alpha$ equivalent widths,  and $D_{4000\AA}$ depend on environment on small scales ($\lesssim 1h^{-1}$ Mpc), but that the environmental dependence of these quantities is significantly weaker on larger scales \citep[e.g.,][]{kauffmann04, blanton06, cucciati10den}. Therefore, a robust computation of both the density field as well as the main galaxy properties such as color, morphology, and metallicity are needed in order to better constrain the interplay between galaxies and their environment.

With J-PAS, we will be able to combine detailed studies of the density field or environment with those of the properties of the galaxies that they contain. For example, we plan to extend the calculation of the cross-correlation function between early and late type galaxies to shorter distances over a wide redshift range. The ALHAMBRA survey has provided reliable results for the projected correlation function $w_p$ \citep{davis83} over the range $r_p \in [0.03,10] h^{-1}$ Mpc (Hurtado-Gil et al. in prep.), showing new distinctive features in the clustering of various galaxy populations that are typically not seen on scales above 0.1 $h^{-1}$ Mpc. These results suggest an increase of the galaxy clustering among late-type galaxies for short distances, breaking its power law pattern. This effect is due to interactions between early and late type galaxies with their direct environment. 

With the density field, several projects and synergies will be explored in the J-PAS collaboration, such as {\it (i)} the dependence of the red fraction/SFR/age/metallicity on the local density, {\it (ii)} the study of the luminosity/mass function in different environments, and {\it (iii)} state which environment is more important for galaxy evolution: the large-scale environment (groups, clusters, filaments), or the very small-scale environment. All these studies will provide fundamental clues about the dominant environmental effects involved in galaxy evolution as a function of redshift, and with their corresponding time-scales and the spatial extent of their reach.

\paragraph{Astrophysics of Groups and Clusters of Galaxies}

Clusters of galaxies are not only important cosmological probes (see the discussion in Sect. 3.2 in this document), but they also offer a unique view on many important astrophysical processes that are highly pronounced and sometimes exclusively found in these dense environments. Galaxies in groups and clusters are under the influence of a number of environmental processes (e.g., harassment, strangulation, ram-pressure stripping, dynamical friction, cannibalism) that can have a strong impact on, e.g., their stellar and gas components, morphology, and star formation rate, as well as their spatial distribution within the cluster. However, the specific contribution of each process as a function of environment and redshift is still a matter of great debate. J-PAS, being complete for clusters of mass $>5\times10^{13}$ $M_\odot$ up to $z\sim0.8$ -- a wider extent in both redshift and mass compared to other surveys -- will have the potential to  revolutionize this particular area of research. The power of J-PAS for the study of galaxy properties in groups and clusters resides in its wide-area coverage together with its precise photo-$z$'s, which will yield a map of the (3D) large-scale structure.  This allows not only the identification of the most likely members of groups/clusters in and outside the virial radius from $0 < z < 0.8$, but also yields coarse spectral information on a pixel-to-pixel basis. This will allow us to link internal galaxy properties to the overall cluster/group properties, at least in a large number of relatively nearby groups and clusters in which the galaxies are resolved. We will also study the evolution of the galaxy populations in groups and clusters by determining how the different properties, e.g., the sizes, morphologies, colors, stellar masses, $M/L$, SFRs, SEDs, ages, and metallicities have evolved with redshift.

In addition to the study of galaxy properties as a function of local environment with 'cleaner' samples (i.e., less background contamination) than have so far been available, the real novelty of J-PAS will lie in the fact that we can study the properties of galaxies in the outskirts ($R\gtrsim R_{vir}$) of groups/clusters. We will be able to scrutinize the environments of groups and clusters and the filamentary structure around them.  It may be possible to witness the large numbers of groups and individual galaxies in the process of accretion onto clusters \citep[e.g., see][]{gonzalez05,berrier09,chiang13}. This will allow us to address the importance of the ``pre-processing'' of galaxy properties inside these sub-clumps, and to investigate how these infalling groups relate to the structure of the cosmic web on larger scales. J-PAS will thus also offer important topological information about the cosmic web up to $z\sim1$.

We will be able to check if the so called fossil groups (massive groups formed by one dominant isolated elliptical galaxy) are at the junctions of filaments. We will be able to map the space around the groups and clusters quite accurately, as well as determine the alignment of galaxies with each other as well as other structures on small and large scales. We will be able to study how the magnitude gap between the BCG and the second brightest member, and the luminosity and stellar mass of the BCG change with the density of the environment. The relatively high precision in photometric redshifts will reduce significantly the uncertainties related to halo membership of galaxies of various types, allowing a substantial improvement in the dark matter halo mass determination through the mass-optical richness relation for both groups and clusters. We will use  self-calibration based on weak lensing estimates and with cluster catalogs based on other frequencies such as the X-rays.

We will systematically investigate the deficit of low-mass galaxies in the center of groups and clusters as compared to the outskirts (mass segregation) and correlate the deficit with other parameters of the host groups and clusters.  We will study the distributions of dwarf galaxies in clusters and groups and compare with the expectations from simulations \citep[e.g.][]{weinmann11}.  We will investigate in which cases the dwarf galaxies are associated with the group/cluster as a whole, around a common halo, or to specific galaxies within the groups and clusters, and if this is somehow related to the dynamical stage of the sytem.

At the extremely massive end of the galaxy mass function, the large J-PAS cluster sample will be ideal for the study of the first-ranked galaxies in groups and clusters, the BCGs and BGGs. The present-day structure of these galaxies and their evolution with redshift sets strong constraints on important components of our galaxy evolution models \citep{delucia07,conroy07,hopkins10,dubois13,shankar13}, such as merging processes \citep{bernardi07,vonderlinden07,liu09,liu13,burke13,ascaso13}, AGN feedback \citep{fan08,collins09,stott11,ascaso11}, the formation of the most massive black holes \citep{mcconnell11,hlavacek-larrondo12,postman12,volonteri13}, and the gaseous and stellar components of the intra-cluster medium \citealt{gonzalez07,conroy07,murante07,rudick11}.

Given the large area covered by J-PAS, the survey will also be an exquisite tool to find rare objects. For example, the J-PAS sample of clusters and groups may yield the largest sample of AGNs in dense environments. We also hope to improve our knowledge about shocks in the intracluster/intragroup medium (traced by broad emission line systems), as seen, for example, in Stephan's quintet. Cluster-scale feedback from AGN versus central star formation will be studied in order to determine the important role of feedback in clusters. The low surface brightness limits of J-PAS will furthermore allow an unprecedented study of the intra-group and intra-cluster light components. Similarly, the study of mergers, shell galaxies, and tidal dwarf galaxies in structures ranging from small groups to dense environments will benefit greatly from the large area covered by the survey and the possibility to determine spatially-resolved properties at the low spectral resolution offered by J-PAS. 

In summary, the J-PAS sample of groups and clusters will allow us to investigate in detail the formation of the cluster and cluster galaxy population since $z\sim1.5$, where the results can be connected to dedicated observations of high redshift clusters based on optical-IR detections, X-ray, Sunyaev-Zel'dovich, and lensing techniques that are all sensitive to $z\sim2$ \citep[e.g.][]{blakeslee03,mei06,andreon08,rettura10,rettura11,fassbender11,foley11,menanteau12}, and to the progenitors of clusters (the so-called ``proto-clusters'') based on the identification of large-scale galaxy overdensities detected in galaxy redshift surveys at $z\gtrsim2$ \citep[e.g.][]{chiang13}. 

\paragraph{Close Pairs and Minor/Major Mergers}

In their pioneering study, \citet{toomre72} were able to explain the tails and the distortions of four peculiar galaxies as the intermediate stage of a merger event between two spiral galaxies. Since then, the role of mergers in galaxy evolution has been recognized and studied systematically, both observationally and theoretically. To constrain the role of mergers in galaxy evolution two observational approaches are needed: (i) understand precisely how interactions modify the properties of galaxies and what is the fate of the merger remnants, and (ii) measure the merger history of different populations over cosmic time to estimate the integrated effect of mergers.

Regarding the first approach, it is well-known that ``major mergers'' (i.e., the merger of two galaxies with similar masses, $\mu \equiv M_2/M_1 \geq 1/4$) of two spiral galaxies is an efficient mechanism to create new red sequence galaxies \citep[RSGs; ][]{naab06ss,rothberg06a,rothberg06b,hopkins08ss,rothberg10,bournaud11}, while both major and minor mergers have been proposed in order to explain the observed mass and size evolution for massive RSGs since $z \sim 1$. When the separation $r_{\rm p}$ between two galaxies in a close pair decreases, the star formation rate (SFR) is enhanced \citep{barton00,lambas03,robaina09,knapen09,patton11}, the metallicity decreases \citep{kewley06,ellison08,scudder12} and the AGN fraction increases \citep{ellison11}. 
The 2D photo-spectra provided by  J-PAS will allow us to explore how the SFR in the central and in the external parts of galaxies in close pairs depend on $r_{\rm p}$. We will also be able to study how other geometrical parameters of the pair, e.g., the angle between the semi-major axis of the galaxies or their inclination, affect the spatial distribution of the star formation. In relation to galaxy groups and clusters, we will be able to study the role of mergers in the size evolution of BCGs that appear to have evolved strongly since $z=1$, in contrast to their stellar masses. 
\begin{figure}[H]
\centering
\includegraphics[width=8cm]{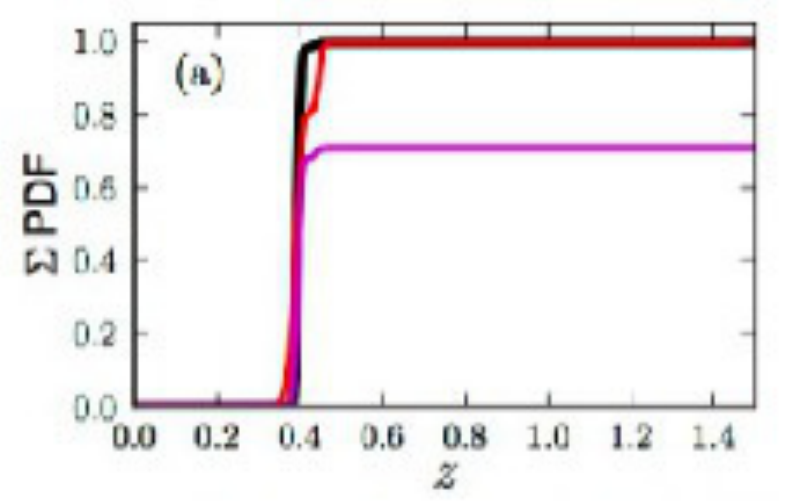}
\includegraphics[width=8cm]{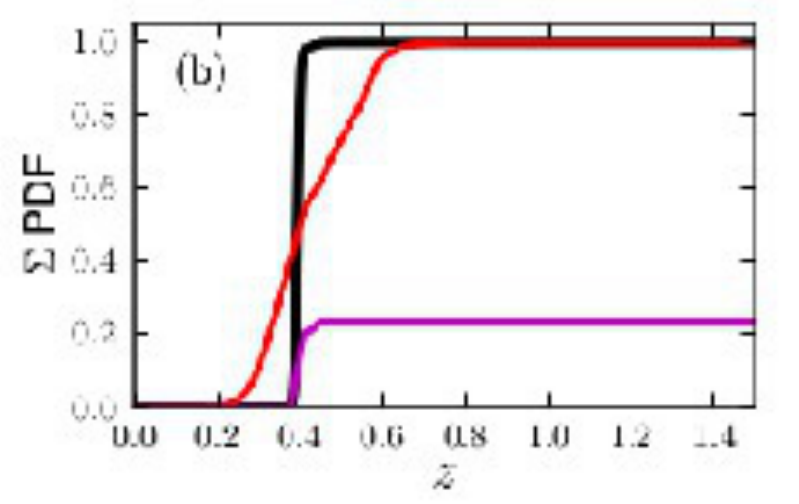}
\includegraphics[width=8cm]{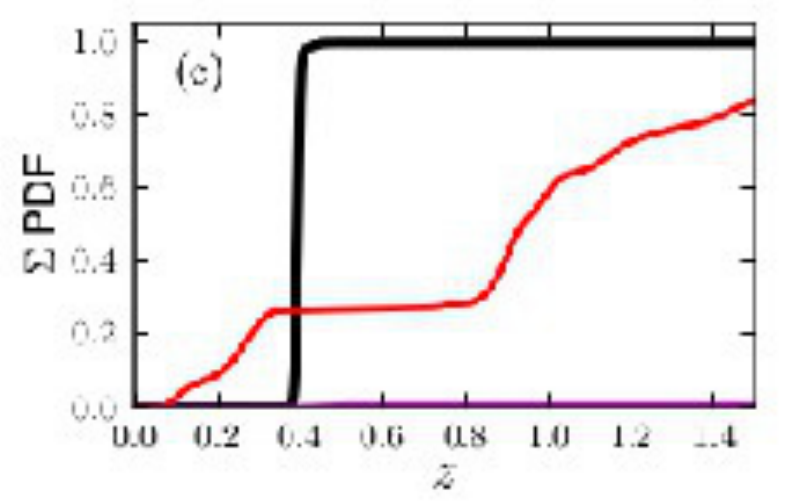}
\includegraphics[width=8cm]{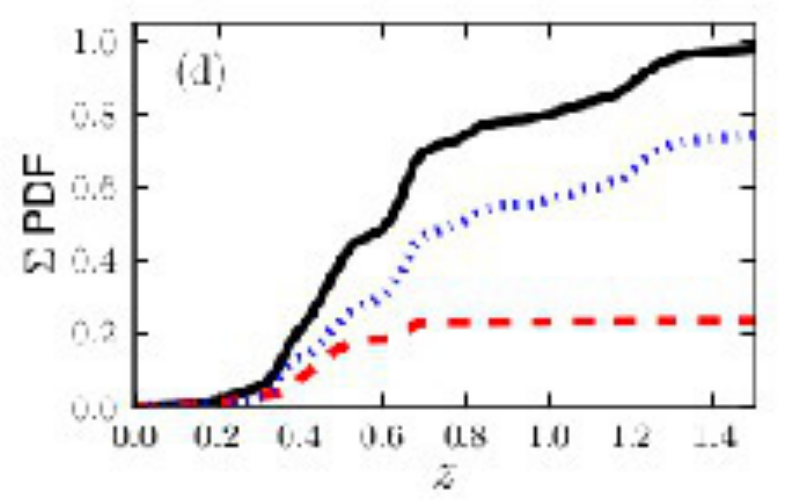}
\caption{Cumulative probability distribution functions in the ALHAMBRA survey. {\it Panels (a)}, {\it (b)}, and {\it (c)} show a principal galaxy (black line) at $z = 0.395$ and a companion galaxy (red line) at $z = 0.395$, $z = 0.400$, and $z = 0.934$, respectively. The purple line shows the probability of both galaxies to be at the same redshift. The cumulative pair probability is $71$\%, $23$\%, and $0$\%, respectively. {\it Panel (d)} shows the PDF of a single source split by spectral types: red types (E+S0, red dashed line, $24$\%), blue types (S + starburst, blue dotted line $76$\%), and all the types (black solid line). We will be able to estimate reliable red and blue merger fractions.}
\label{photpair}
\end{figure}

Regarding the second approach, the merger history of a given galaxy population can be characterized by estimating its merger fraction $f_{\rm m}$, i.e., the fraction of galaxies in a sample undergoing merging. This can be determined either on the basis of morphological information (highly distorted galaxies are merger remnants \citep[e.g.][]{conselice03,conselice08,cassata05,lotz08ff,clsj09ffgs,clsj09ffgoods,jogee09,bridge10}) or based on close-pair statistics (two galaxies close in the sky plane, $r_{\rm p} \leq r_{\rm p}^{\rm max}$, and in redshift space, $\Delta v \leq 500$ km s$^{-1}$, are likely to merge, \citep[e.g.][]{lefevre00,patton00,patton08,lin04,lin08,deravel09,deravel11,clsj10pargoods,clsj13ffmassiv}). With a parametrization of the merger fraction evolution following $f_{\rm m} \propto (1 + z)^{m}$, the {\it major merger} fraction evolution has been shown to depend on the luminosity and the stellar mass of the galaxies. Massive galaxies with $M_{\star} > 10^{11}\ M_{\odot}$ have a higher merger fraction, but with little redshift evolution ($m \sim 0 - 2$), while lower mass galaxies ($M_{\star} = 10^{9}-10^{11}\ M_{\odot}$) have a lower merging fraction but with stronger redshift evolution ($m \sim 3 - 4$). Regarding {\it minor mergers} with $\mu < 1/4$, the observations show a nearly constant evolution ($m \sim 0$) up to $z \sim 2$ \citep{clsj11mmvvds, clsj12sizecos, lotz11, williams11, marmol12}.

Reliable merger fractions and rates can be determined based on photometric redshift surveys like J-PAS, by following the methodology of \citet{clsj10pargoods}. This methodology uses the Probability Distribution Functions (PDF) of the photometric redshifts, $z_{\rm phot}$, to estimate the probability that a galaxy pair with a projected separation $r_{\rm p} \leq r_{\rm p}^{\rm max}$ measured in the sky plane is also a close pair in redshift space (relative velocity $\Delta v \leq 500$ km s$^{-1}$). This methodology has been tested in the Millennium Galaxy Catalogue (MGC, \citet{mgc}) at $z \sim 0.1$ and in the zCOSMOS spectroscopic survey \citep{zcosmos10k} up to $z \sim 1$. The results show that we can recover reliable merger fractions from photometric redshift surveys. We have also applied this methodology to measure successfully the merger fraction in GOODS-South \citep{clsj10pargoods}, COSMOS \citep{clsj12sizecos} and ALHAMBRA (L\'opez-Sanjuan 2013b, in prep.). Thanks to the high accuracy of J-PAS photometric redshifts ($\Delta z/(1+z) \sim 0.3$\%) we will be able to estimate the major merger fraction from close pairs up to $z \sim 1$, and the minor merger fraction up to $z \sim 0.5$. In addition, we will characterize with unprecedented detail the dependence of the merger fraction on stellar mass, color, environment, etc. The methodology developed by \citet{clsj10pargoods} assumes that the
PDFs are Gaussian in redshift space, and we are upgrading their methodology to use the more general PDFs (i.e., asymmetric and with multiple peaks) that are expected to be provided by the J-PAS photometric redshift techniques (see Fig.~\ref{photpair}).
	
In summary, J-PAS will greatly improve our knowledge about the impact of interactions on galaxy properties and the role played by mergers in the evolution of the red sequence and the blue cloud since $z \sim 1$.

\subsubsection{Theme IV. The High Redshift Universe}

The two methods that have proved most effective in recent years for identifying high redshift galaxies are the so called Lyman-break and \lya\ selection techniques. Both of these selections are based on the ultraviolet properties of the galaxy spectrum - redshifted into the optical/IR window at high redshifts - and are hence selecting only the galaxies which are young enough to produce copious amounts of ultraviolet light, and are sufficiently dust-free for a fair amount of this light to escape the galaxy.

The identification of galaxies through the Lyman-break technique is mainly based on two ultraviolet spectral features introduced by the blanketing effect of neutral hydrogen both within the galaxy itself, and by intervening clouds along the observers line-of-sight: the Lyman break at 912\AA\ and the Lyman forest between 912\AA\ and 1216\AA. Traditionally, these galaxies are discovered based on their broad-band colors measuring the drop in brightness due to the Lyman break and/or Lyman forest, and the galaxies hence selected are called Lyman Break Galaxies (LBGs). The second method selects galaxies which are \lya\ emitters. The \lya\ emission line is produced in the interstellar medium of the galaxy where the hydrogen atoms have been excited by the ultraviolet light from young stars. The traditional selection technique involves comparing images taken through a narrow-band filter with that taken through a broad-band (or another narrow-band) at comparable wavelengths. The galaxies selected using this method are generally called Lyman-$\alpha$ Emitters (LAEs).

The differences and similarities between the properties of LBGs and LAEs have been widely discussed in the recent literature. Most likely, however, the reported differences are a consequence of the different selection techniques \citep[e.g. see][]{dunlop13}. 
The main reason why LAEs are often not detected by LBG selections is because they are typically very faint in the UV continuum, beyond the reach of a broad-band selection technique, despite them having similar UV continuum breaks as LBGs. The reason why not all LBGs are detected as LAEs is that while interstellar extinction peaks at the UV range, \lya\ photons are affected by the resonant scattering, being easily scattered and destroyed by the neutral gas in the local and intergalactic medium. This characteristic permits the LBG/LAE ratio to be used to trace the neutral hydrogen fraction of the Universe, giving information about the last epochs of reionization.

We aim at studying this early galaxy population in the huge J-PAS volume avoiding biases due to the cosmic variance, a general problem in most LBG/LAE studies performed to date. Based on the J-PAS SEDs, we will be able to detect the brightest (in the continuum) LBGs in the redshift range $z\sim2-3$. The J-PAS narrow-band filters will also permit us to identify which of these objects are luminous LAEs, thus avoiding the selection biases mentioned above. Combining J-PAS data with the GALEX UV data, we will also be able to identify and study LBGs at redshifts of $z\sim1$. Finally, using the traditional narrow-band selection, we will be able to identify LAEs reaching objects too faint in their continuum to be identified by their SEDs. In addition, as a by-product of this LAE selection, we will also be able to detect the separate rare class of \lya\ blobs. These approaches will be detailed below.

\paragraph{Lyman-Alpha Emitters}

\lya\ emitters (LAEs) are within the more distant baryonic structures so far detected in the universe. As most high-redshift objects, they are classified according to their selection method, the so-called narrow-band technique. It employs a combination of narrow and broad band filters to isolate the \lya\ emission and characterize its energy distribution in the continuum. Although due to the resonant nature of the \lya\ line, a huge fraction ($\sim$90\%) of star-forming galaxies emit insufficient \lya\ photons to be detected by narrow-band surveys \citep{hayes10}, LAEs can be found at almost any redshift from local \citep{ostlin09,deharveng08,cowie10,cowie11} up to $z\sim7$ \citep{iye06} and beyond \citep{sobral09}.  However, some low-redshift LAEs show quite different properties from those at $z > 2$ \citep{finkelstein09a,finkelstein09b,oteo11,oteo12a,oteo12b}. Thus, LAEs are representative of different effects related to galaxy evolution and to the complex resonant scattering mechanisms of the Lyman-$\alpha$ line. 

At the highest redshifts, LAEs are indicative of the stage of reionization of the universe. Early reionization models claim that reionization is nearly complete at $z\sim8$ and ends at around $z\sim6.6$ \citep{choudhury06}. This is supported by the number density evolution of LAEs, that seems to decrease beyond $z\sim6$ \citep{kobayashi07}. Analyzing samples of LAEs between $z\sim3.1$ and $z\sim5.7$, \citet{ouchi08} found that LAEs were more common at earlier epochs. \citet{kovac07} measured the spatial correlation function of a LAE sample at $z\sim4.5$, finding a significant clustering strength consistent with those of the halos of Lyman break galaxies, albeit with a lower occupation number. In contrast, the relatively scarce number of LAEs detected at $z\sim3$ could be consistent with them being the progenitors of present day $L^*$ galaxies \citep{gawiser07,guaita10}. These authors did not find evidence for strong obscuration or a substantial AGN fraction ($\sim$1\%), indicating that the LAEs are young, low stellar mass objects. In contrast, in a large sample at $z\sim2.3$, \citet{nilsson09} detect a significant AGN contribution and red spectral energy distributions (SEDs), implying a contribution from more massive, dustier and older sources than among the LAEs observed at $z>3$ \citep[see also][]{bongiovanni10,oteo12a,oteo12b}.  

Most of the above results were obtained from sky areas no larger than $\sim1 \sq\degr$ and, consequently, the cosmic variance is a significant handicap of these surveys. The substantial area surveyed by J-PAS will allow us to address some unanswered problems. By taking advantage of the current design of the J-PAS filter set, we will perform a systematic search for LAEs at $2\lesssim z \lesssim2.4$ on a (proposed) J-PAS Deep Field (JDF), using the well-known narrow band technique employed to find high-redshift galaxies \citep[e.g.,][]{cowie98,gronwall07,ouchi08}, but combining the 4 bluest filters of J-PAS centered at 360, 379, 390 and 400 nm, alternatively for separated detection and continuum subtraction. 

A similar approach, but using intermediate-band filters of the ALHAMBRA Survey \citep{moles08} was successfully used in \citet{bongiovanni10}, searching for LAEs at $z\sim2.2$ in the GOODS-North field. Nevertheless, in this case, the medium-band filters employed favored the finding of large equivalent width (EW) objects. A JDF would be sensitive to LAEs with a minimum rest-frame $EW_{Ly\alpha}$ of $40\pm10$\AA\ for a $m_{OFF}$-$m_{ON}$ color $>$0.3 mag. For all J-PAS filters involved in the LAE detection, we performed several simulations by convolving their total transmission with real, conveniently redshifted spectra. The rest-frame EW threshold is similar for these filters.

Down to a magnitude limit of $i_{AB} = 23$ ($i_{AB}=24$), we estimate a mean of $7\pm2$ ($37\pm4$) LAEs $/\sq\degr$ with an $EW_{Ly\alpha}>35$\AA\ at a median redshift of $z\sim2.25$ \citep{bongiovanni10,nilsson09}. The AGN fraction among this sample is expected to be about 40\%. Despite the fact that we will only be able to detect the brightest LAEs at this mean redshift (i.e. between 6 and 23\% of the total), the J-PAS data set will allow us to characterize some of the fundamental properties (stellar mass, metallicity, age, SFR, AGN fraction and dust content) by modeling the stellar populations from the low-resolution SED. The large number of LAEs detected over a wide range of spatial scales will furthermore allow an unprecedented study of the clustering of bright LAEs.  
We will search for signs of density evolution and test whether LAEs at $2 < z < 3$ are progenitors of local $L^*$ galaxies.

Furthermore, the wide area covered by J-PAS allows us to search for the most extreme LAEs that are missed by typical deep pencil-beam surveys \citep[e.g.,][]{matsuda04,matsuda11,yang09,bridge13}. In particular, we will be able to perform a systematic search for the highly rare population of bright and extended \lya\ nebulae or ``\lya\ blobs'' (LABs). This will be the subject of the next section. 

\begin{figure*}[t]
\centering
\includegraphics[width=0.85\textwidth]{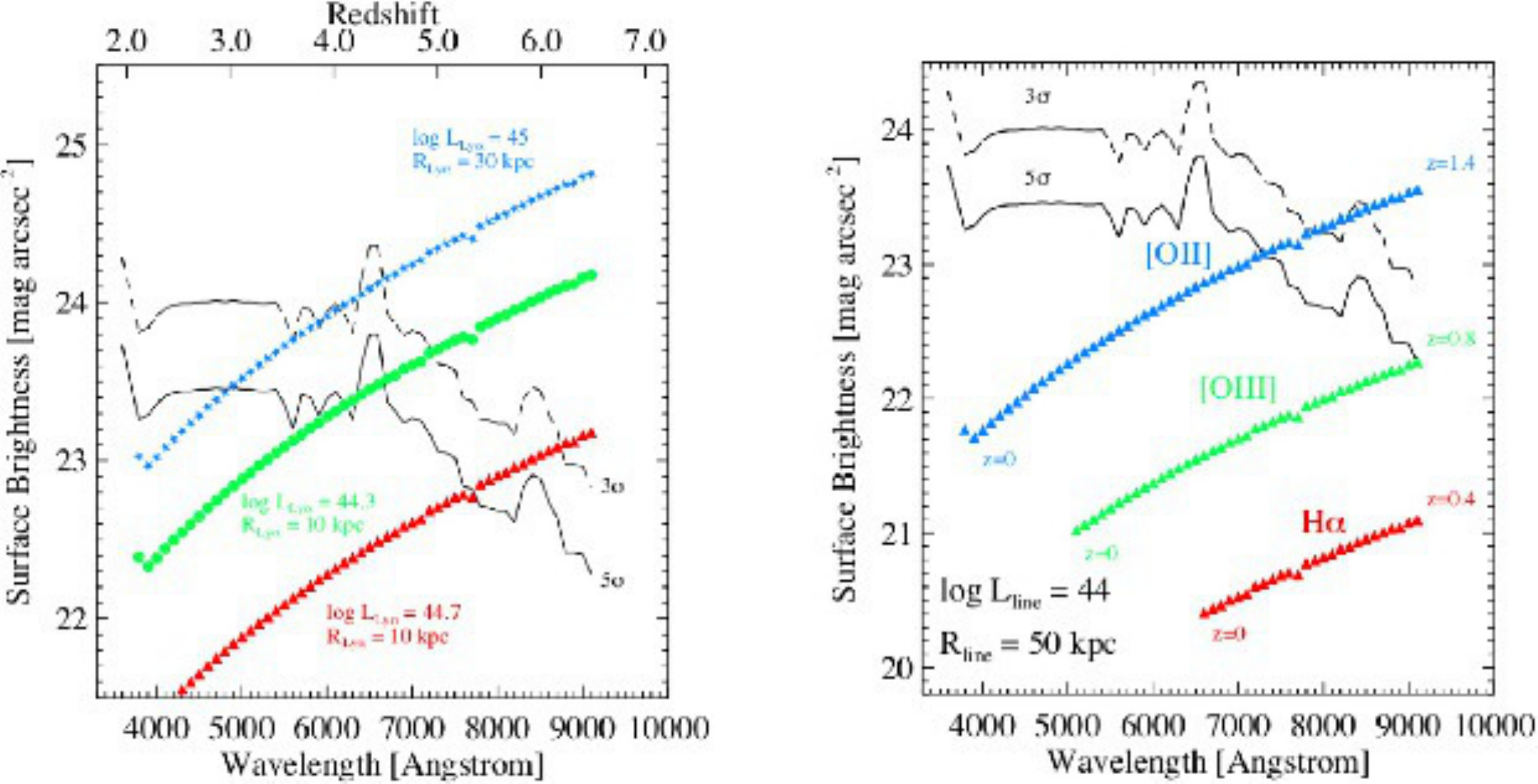} 
\caption{Detectability of LABs and other extended emission line
  objects. Left panel: The expected surface brightness of \lya\ in
  each J-PAS filter for three types of LABs (blue points: log
  $L_{Ly\alpha}=45$, $R_{Ly\alpha}=30$ kpc; green points: log
  $L_{Ly\alpha}=44.3$, $R_{Ly\alpha}=10$ kpc; red points: log
  $L_{Ly\alpha}=44.7$, $R_{Ly\alpha}=10$ kpc). The limiting surface
  brightness achieved by J-PAS is indicated by the solid (5$\sigma$)
  and dashed (3$\sigma$) curves. Right panel: Expected surface
  brightness of EELRs having a line luminosity of 10$^{44}$ erg
  s$^{-1}$ and a radius of 50 kpc. Blue points: \oii\ emission at
  $z=0-1.4$, green points: \oiii\ emission at $z=0-0.8$, red points:
  \ha\ emission at $z=0-0.4$.}
\label{fig:labs}
\end{figure*}

\paragraph{Ly$\alpha$ Blobs and Other Extended Emission Line Objects}
 
A particularly rare subset of objects at high redshift that is also selected through their high equivalent width \lya\ emission involves the population of giant, luminous \lya\ nebulae. These nebulae, also referred to as ``\lya\ blobs" (LABs), have sizes ranging from a few tens to a few hundreds kpc and \lya\ line luminosities ranging from a few times $10^{43}$ to $10^{45}$ erg s$^{-1}$ \cite[e.g.][]{francis96,steidel00,overzier01,matsuda04,venemans07}. Luminous extended \lya\ nebulae have been known to exist around low and high redshift AGN such as radio galaxies, quasars, and Seyfert galaxies for over three decades \cite[e.g.][]{mccarthy90,heckman91,fu09}. The primary energy source powering the \lya\ emission in these type of sources is the ionizing radiation from a central AGN, sometimes with contributions from star formation or radio jet-cloud interactions. The main mechanisms responsible for the spatially extended neutral gas are however still largely unknown. If the gas originates from within the source itself, it could have been driven out by radio jets (in the case of radio galaxies) or quasar and/or starburst superwinds. The gas may also originate externally to the galaxy, perhaps related to the same processes that provide fuel to the central black hole and power the AGN. At high redshifts, the gas could also be related to cold, dense gas recently accreted from the intergalactic medium during structure formation \citep{dijkstra09} or from a reservoir of previously expelled gas.

More recently, LABs have also been found serendipitously in \lya\
surveys
\cite[e.g.][]{francis96,steidel00,matsuda04,matsuda09,matsuda11,prescott13}. Although
they are similar in size and luminosity, these LABs are typically not
associated with any known radio galaxies or Type I (i.e. unobscured)
quasars. Multi-wavelength follow-up observations of these systems have
shown that they frequently host Type II (i.e. obscured) AGN,
starbursts, and/or outflowing superwinds
\cite[e.g.][]{bower04,dey05,geach05,overzier13b}. Nebular line
metallicity measurements further show evidence that the gas is usually
not pristine, indicating that it has previously been processed
\cite[e.g.][]{overzier01,overzier13b}. Furthermore, numerous
observations show that the LABs tend to occur predominantly in
overdense environments, as evidenced by the fact that they frequently
sit in local maxima in the distributions of LAEs or LBGs
\cite[e.g.][]{steidel00,francis01,overzier08,venemans07,erb11}. Recently,
\citet{overzier13b} presented an analysis of a complete sample of LABs
having \lya\ luminosities in excess of $5\times10^{43}$ erg s$^{-1}$
and sizes of $\gtrsim$50 kpc, showing that essentially all of the
luminous LABs harbor obscured AGN. Because AGN typically have short
duty-cycles (10-100 Myr), the fact that we know of almost no LABs
without an AGN suggests that they must be a direct consequence of the
AGN activity. In the AGN scenario, the ionizing luminosity required to
power the \lya\ emission out to $\sim$100 kpc is provided by the AGN,
which can be obscured (e.g. for radio galaxies and other LABs) or
unobscured (e.g. for quasars) along the line of sight. 

This scenario also explains the empirical relation between LABs and environment, because at high redshift the most luminous AGN are also preferentially found in overdense regions.  Alternative energy sources for the LABs have been suggested by invoking massive starbursts (of order 1000 $M_\odot$ yr$^{-1}$) or gravitational cooling radiation related to structure formation. Although these processes appear capable of producing a sufficient number of ionizing photons, the observational evidence appears weak, at least for the brightest LABs. It is possible that cooling radiation and/or star formation play a greater role in \lya\ sources of much more modest sizes and luminosities, such as the population of LAEs. It has been pointed out that LABs become rarer, less luminous, and smaller with decreasing redshift \citep{keel09,zirm09,overzier13b}. It is not yet clear whether this effect, if real, is related to the fact that the most luminous AGN and starbursts have died out following the decline in the cosmic star formation rate density and AGN activity, or whether it reflects the lack of extended reservoirs of dense neutral gas around galaxies at $z\lesssim2$. However, the volumes of existing narrow-band surveys have been quite small, with essentially no constraints at $z<1.6$ where \lya\ is inaccessible from the ground.

While J-PAS does not reach the depths achieved by narrow band surveys performed with 4-8m range telescopes, this is compensated for by its unprecedented combination of survey area and number of narrow band filters. J-PAS will be able to make a significant contribution to the study of LABs. In order to assess the sensitivity to LABs, the left panel of Fig. \ref{fig:labs} shows the J-PAS surface brightness limits as a function of wavelength. We compare these with some expected surface brightnesses for LABs that fall within each narrow-band filter. Assuming circular LABs with a flat surface brightness distribution and no detectable continuum, J-PAS can detect an LAB with a luminosity of $10^{45}$ erg s$^{-1}$ and a radius of 30 kpc at $z\simeq2-3$ at 5$\sigma$. A more typical LAB with a luminosity of a few times $10^{44}$ erg s$^{-1}$ could easily be detected out to $z\sim4$, provided that the bulk of the emission comes from a relatively compact region ($R\sim10$ kpc). In reality, the LABs will likely have rather clumpy morphologies of low and high surface brightness regions, as well as varying levels of continuum. At the very least, J-PAS will generate large numbers of candidates that could be followed up to confirm the redshifts, total sizes, and luminosities.

Furthermore, analogous to the extended emission line regions (EELRs) observed around radio galaxies and quasars, LABs often have luminous extended line emission in lines other than \lya, the brightest of which are \oii, \oiii, and \ha. At a luminosity of $\gtrsim10^{44}$ erg s$^{-1}$ and sizes of a few tens of kpc, there is little chance of confusing these EELRs with the less luminous and more compact line emission from more typical star-forming galaxies and AGN. Although these lines are redshifted into the NIR at $z>2$, at lower redshifts they offer a chance to study LAB-like objects at redshifts inaccessible by \lya\ \citep[e.g.][]{yuma13,brammer13}. J-PAS will be able to perform for the very first time a general census of the population of luminous emission line halos at a wide redshift range. In the right panel of Fig. \ref{fig:labs}, we show the expected surface brightnesses for EELRs with a line luminosity of $10^{44}$ erg s$^{-1}$ and a radius of 50 kpc. J-PAS will be able to detect such EELRs as traced by \ha\ at $z\simeq0-0.4$, by \oiii$\lambda$5008 at $z\simeq0-0.8$, and by \oii$\lambda$3727 at $z\simeq0-1.4$. Because J-PAS is a blind, low-resolution survey we will detect all EELRs above the surface brightness limits of the survey, allowing us to study in detail to which classes of objects they belong and to determine their evolution as a function of redshift.

How many LABs at $z\gtrsim2$ will J-PAS find? We can get a sense of the numbers involved by looking at the different classes of luminous, extended \lya\ emitting objects involved. The first category is that of LABs associated with quasars. With an area of $8650\sq\degr$ J-PAS will detect about two million quasars down to $g_{AB}=24$ in the comoving volume of 100 Gpc$^3$ at $2<z<3$ \citep{2012MNRAS.423.3251A}. We do not currently have good statistics on the fraction of quasars that host \lya\ halos, but there is good evidence that a significant fraction does \citep{heckman91,christensen06}. The second category is that of the LABs associated with radio galaxies and radio-loud quasars.  The number density of radio sources at $z>2$ having a luminosity $L_{2.7GHz}>10^{33}$ erg s$^{-1}$ Hz$^{-1}$ ster$^{-1}$ at $2<z<5$ that is known to host extremely luminous \lya\ halos, is about $4\times10^{-8}$ Mpc$^{-3}$ \citep{willott01}. This implies at least 400 potential radio source halos in J-PAS, but these numbers will increase as $n\propto L_{radio}^{-2}$ as we go down the radio luminosity function. Potentially, J-PAS will thus probe the extended emission line gas around thousands of radio sources at high redshift. The third category is that of LABs hosted by sources that are not quasars or radio galaxies (e.g., Type 2 AGN, starbursts, and other \lya\ emitting sources). The numbers in this category are much harder to estimate. We currently know about 15 luminous LABs of this type, selected from a handful of surveys with different redshifts, different selection techniques, and different survey volumes. Most useful for estimating a number density for these LABs is perhaps the recent study of \citet{prescott13} who found two LABs of $L_{Ly\alpha}>10^{44}$ erg s$^{-1}$ at $2<z<3$ within the $8.5\sq\degr$ NOAO Wide-Field Survey Bo\"otes field, corresponding to a number density of $2\times10^{-8}$ Mpc$^{-3}$. J-PAS should therefore discover hundreds of these kinds of LABs.

In terms of number statistics, J-PAS thus has the potential of being one of the most powerful surveys of LABs to date. With this in mind, we will be able to focus on the following key questions: (1) What fraction of LABs are associated with radio galaxies, quasars, and other sources? (2) What is the main powering mechanism of the luminous line emission?
(3) What determines the kinematics of the gas? (4) What is the origin of the line-emitting gas? (5) What is the fate of the extended gas? (6) What drives their strong number density evolution with redshift? (7) What is the role of environment? (8) What is the nature and evolution of the related class of luminous extended emission line blobs at lower redshifts as traced by, e.g., \ha, \oiii\, and \oii?

\begin{figure}[t]
\centering
\includegraphics[width=0.5\textwidth,height=0.5\textwidth]{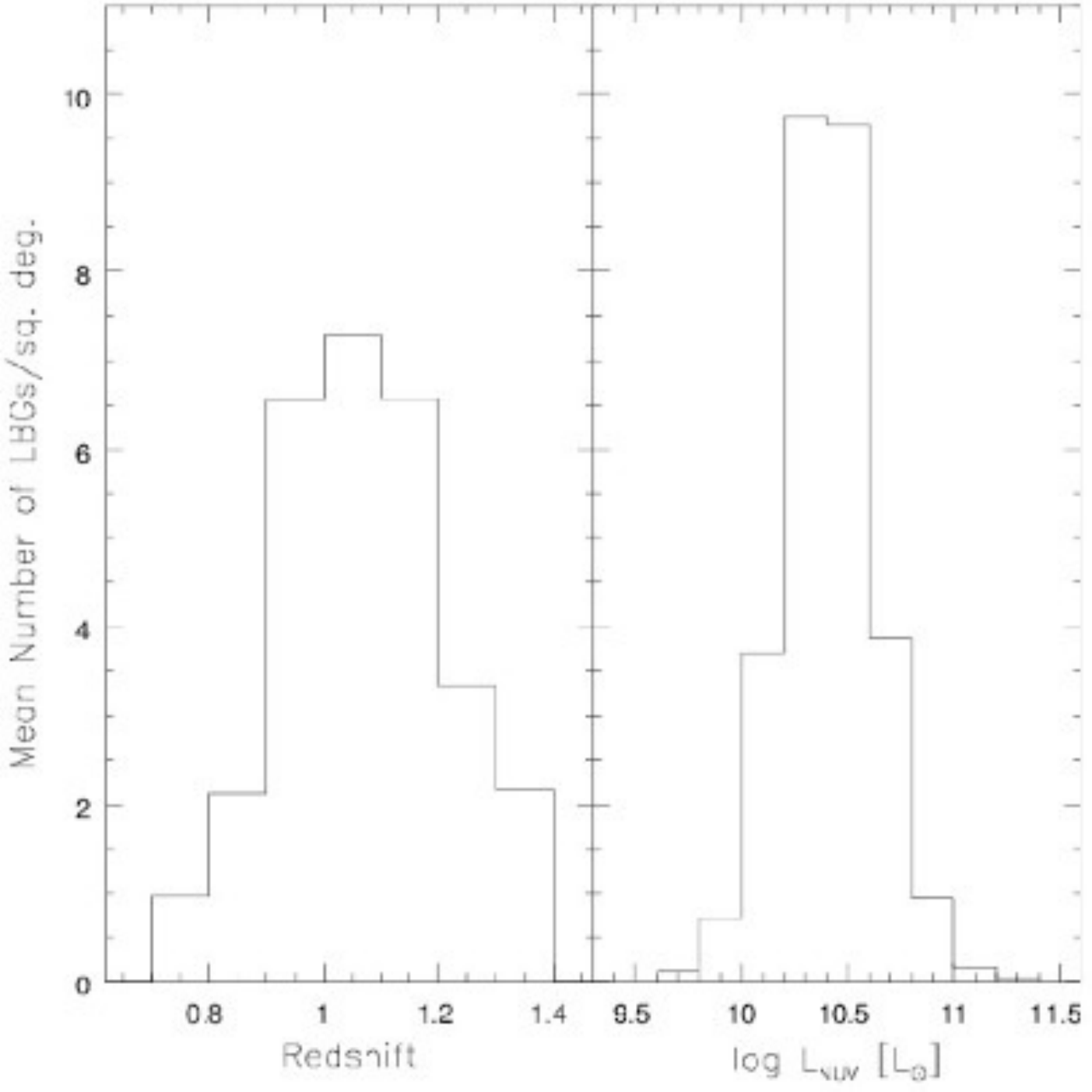} 
\caption{Photometric redshift and UV luminosity mean distributions of
  a GALEX selected LBG sample at the limiting magnitude of J-PAS, from
  a $\sim 125\sq\degr$ region with $55\deg < b_{II} < 60\deg$.}
\label{fig:lbg1}
\end{figure}

\begin{figure}[t]
\centering
\includegraphics[width=0.5\textwidth]{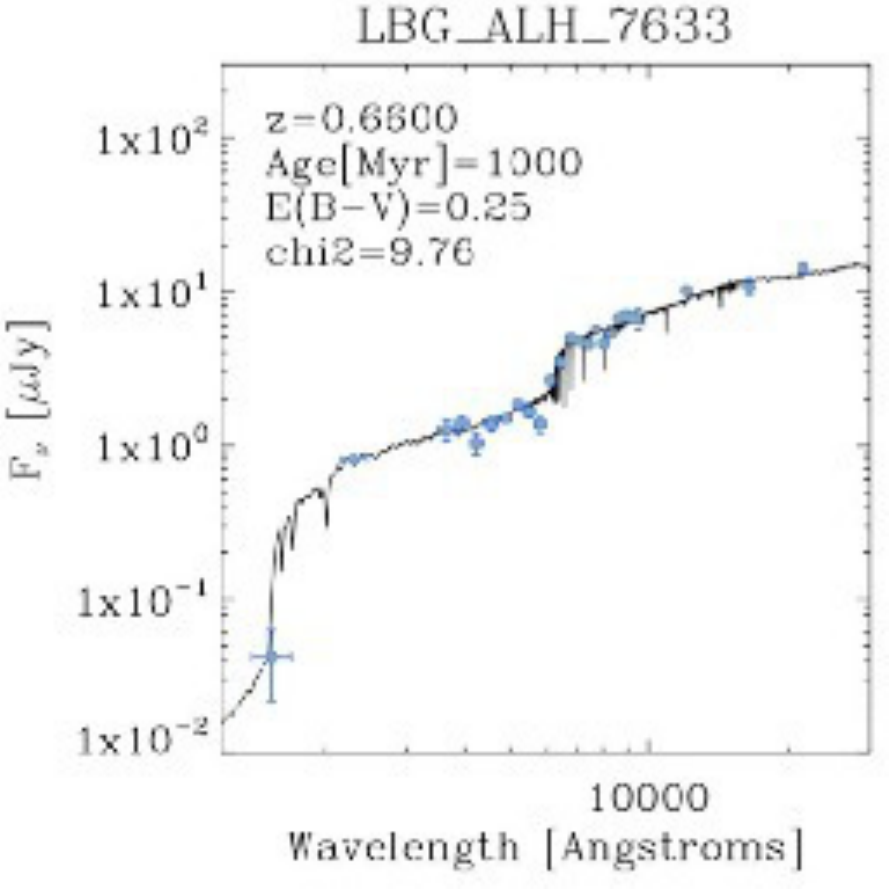} 
\caption{Template fitting (BC03) and derived parameters of a confirmed LBG in the ALHAMBRA survey. The two bluest dots correspond to GALEX photometry and the remaining ones to ALHAMBRA (optical + near-IR).}
\label{fig:lbgfit}
\end{figure}

\paragraph{Lyman-Break Galaxies}

Lyman-break galaxies (LBGs) constitute the dominant star-forming population at high-redshift (Steidel et al. 1995), and are a popular choice for estimating the star formation rate (SFR) density in the early universe (Madau et al. 1998). At $z\gtrsim2$, the Lyman break redshifts into the optical domain where it is accessible with ground-based telescopes, using the so-called dropout technique (see for example Steidel et al. 2003). At lower redshifts of $z\sim1$, LBGs can be selected by applying the dropout technique to deep optical data combined with data in the observed ultraviolet obtained by the Galaxy Evolution Explorer (GALEX). The enormous area of the J-PAS main survey will be exploited to probe the bright end of the LBG luminosity function at $z\approx1-3$ to unprecedented detail.  

{\bf LBGs at ${\mathbf z\sim1}$:} Taking advantage of the around
$700\sq\degr$ of area in common between the footprints of the J-PAS
Main Survey and the GALEX-Medium Imaging Survey (MIS), we will
generate by far the largest known and most robust sample of bright
LBGs at $0.8 < z < 1.2$. Using the 56-band SED sampling, we will then
be able to obtain accurate photometric redshifts. Possible
contaminants (AGNs and stars) will be ruled out based on SED
diagnostics. Based on a preliminary study using GALEX-MIS and SDSS
optical data in a $\sim 125\sq\degr$ region  we can recover $\sim$30 robust LBG candidates at $0.8<z_{phot}<1.2$ per $\sq\degr$ (Fig. 1, left panel) after enforcing a limiting magnitude of 22.4 in $r$ band. The LBG mean UV luminosity distribution obtained is shown in Fig. 1 (right panel) and it is complete above $L_{NUV}=2\times10^{10}L_\odot$ (i.e. corresponding to $\sim0.5L^*_{UV,z=3}$ and the lower limit to define UV luminous galaxies in the nearby universe; see \citet{heckman05}), and accounts for the $\sim$50\% of the total LBG NUV luminosity contribution in the LBG LF \citep{burgarella07}. Therefore, we estimate it will be possible to detect up to $\sim24,000$ LBG candidates in the GALEX-MIS/J-PAS common area above $\delta = -10\deg$. 

\begin{figure}[t]
\centering
\includegraphics[width=0.5\textwidth]{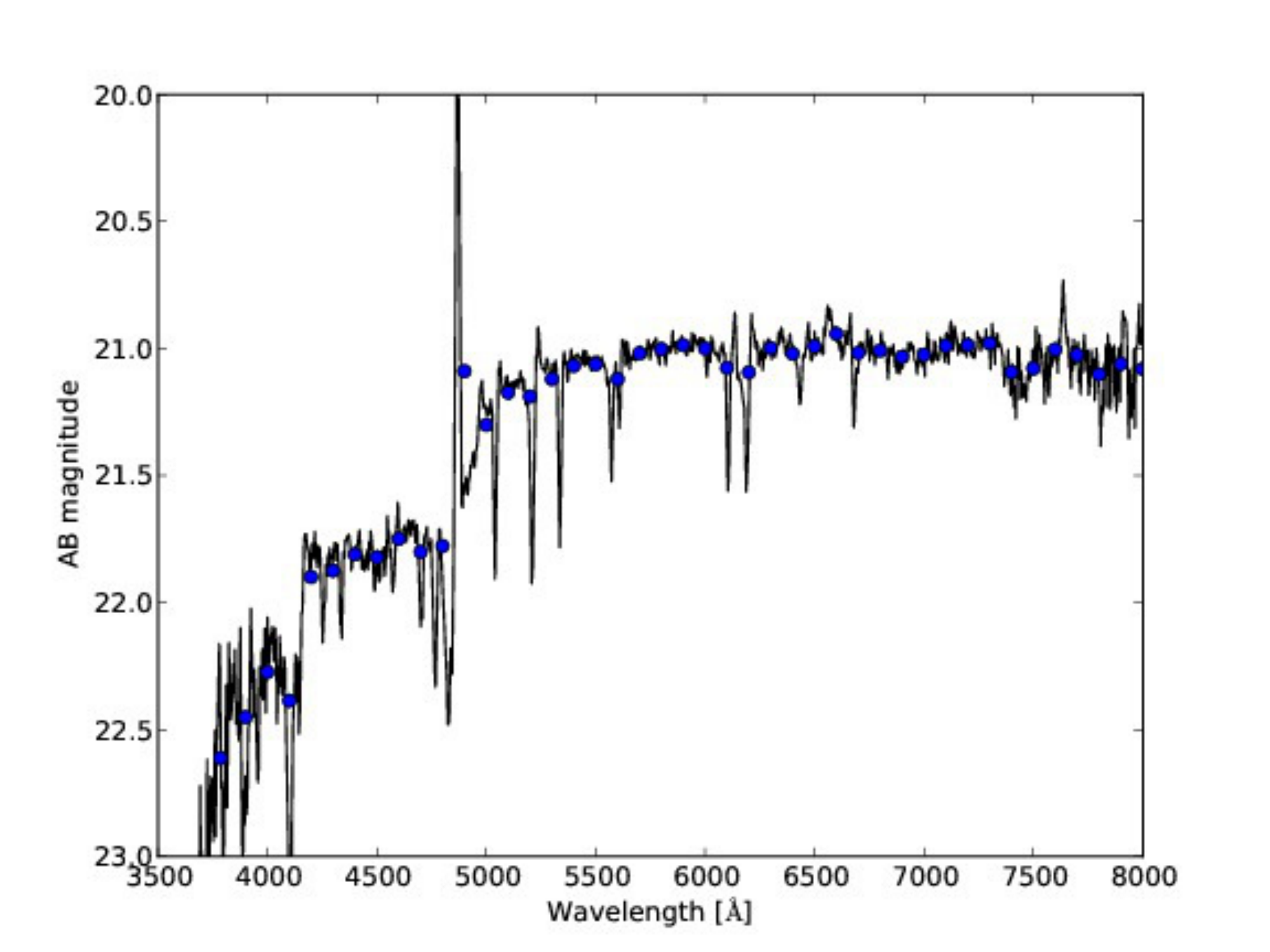} 
\caption{A composite spectrum of 811 LBGs of \citet{shapley03} (black line) and the corresponding synthesized J-spectrum (blue dots).}
\label{fig:lbg4}
\end{figure}

{\bf LBGs at ${\mathbf z\sim 2-3}$:} At redshifts $z\gtrsim2$, the Lyman forest is shifted to the wavelength range covered by J-PAS filters. This permits us to select LBGs/LAEs based directly on their J-PAS SEDs and spectral fitting of model templates, including theoretical LBG spectra, and spectra of other types of galaxies and stars. The spectral fitting also provides us with information about the newly selected objects and permits ruling out contaminants, as mentioned above. In addition, by selecting candidates using spectral fitting instead of color cuts, we will equally select LBGs and LAEs up to the limiting continuum magnitude set up by the bluest J-PAS filters ($m_{AB} \sim 22.3$). This will permit us to study the LBG-LAE connection in a sample free of selection biases. An example of how J-PAS would see a LBG/LAE spectrum at $z\sim3$ is shown in Fig.~\ref{fig:lbg4}, where we have plotted the original, and J-PAS filter convolved, composite spectrum of 811 LBGs of \cite{shapley03}.

\begin{figure}[t]
\centering
\includegraphics[width=0.8\textwidth]{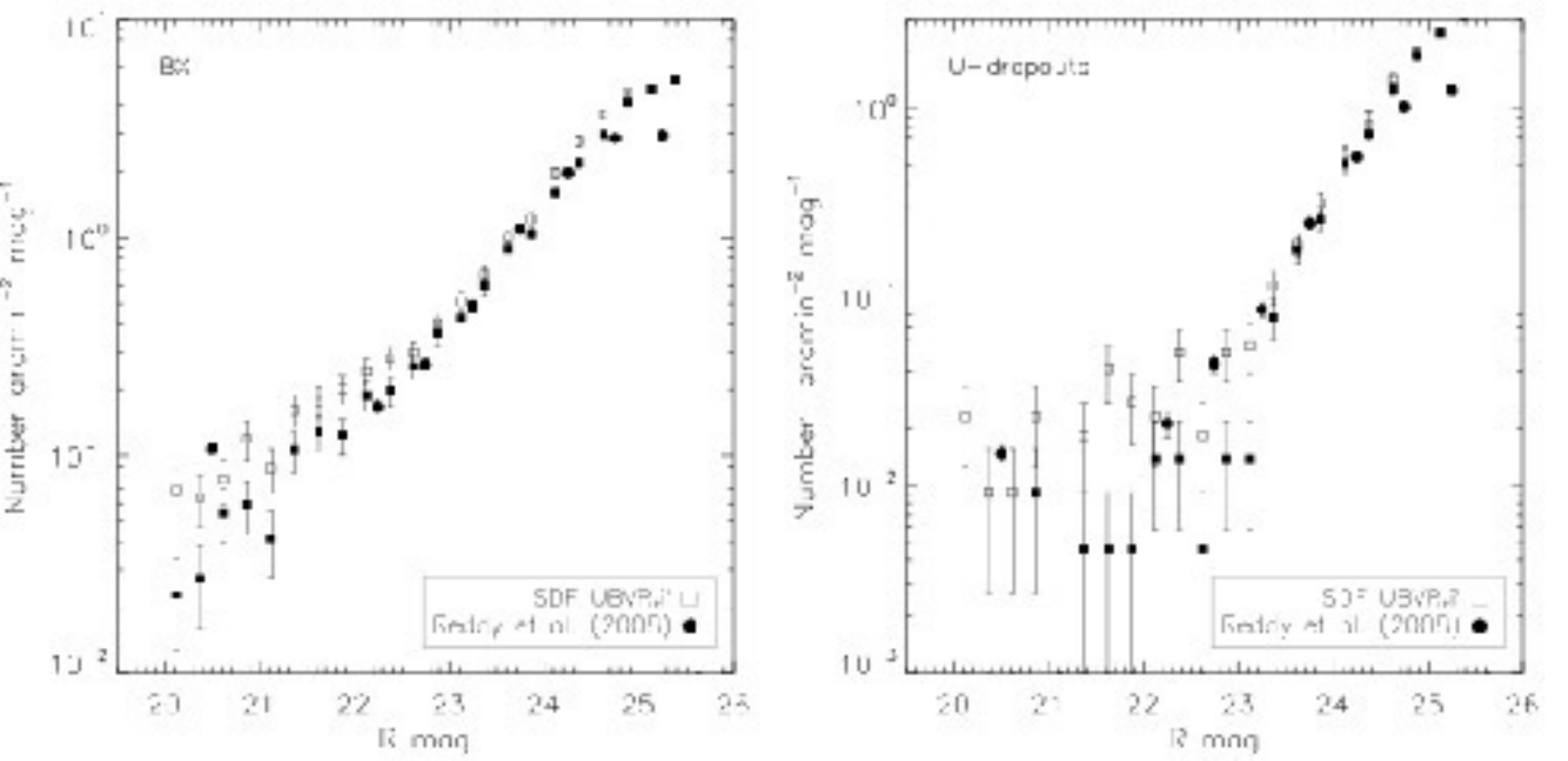} 
\caption{The LBG luminosity functions at $z\sim2$ (left) and at $z\sim3$ (right) from Ly et al. 2011.}
\label{fig:lbg5}
\end{figure}

Fig.~\ref{fig:lbg5} shows the LBG luminosity function (LF) at $z\sim2$ and $z\sim3$ \citep{ly11}. It is clear that both LFs are badly defined at their brightest ends. Considering the J-PAS limiting magnitude in the bluest filters and the survey area, we estimate we will discover 
$> 10^6$ LBGs/LAEs at $z\sim2$ and $>10^5$ at $z\sim3$. This would significantly contribute in constraining the bright end of the LFs at these redshifts. Discovering and studying the brightest LBGs/LAEs is also interesting in their own right to understand the nature and properties of these UV ultraluminous galaxies \citep[see, e.g.][]{bian12}.
\medskip

The resulting catalogues of $z\sim1$ and $z\sim2-3$ LBGs will be used in the following science:
\begin{itemize}
\item
Building the largest sample of LBGs at their corresponding redshifts - virtually free from cosmic variance effects - that represents the bright end of the UV luminosity function of this kind of galaxies. 

\item
A characterization of fundamental properties (redshift, stellar mass, mean age, SFR, and dust content) of these bright LBGs, using stellar population modeling resources under Maximum Likelihood (ML) or Bayesian approaches. An example of a ML-template fitting and derived parameters for a LBG in the COSMOS field using data from ALHAMBRA survey is given in Fig. \ref{fig:lbgfit} from \citet{oteo12b}. 

\item
Cross-correlating the LBG data to be obtained with the QSO sample of J-PAS (see Section \ref{sec:agn}), we will be able to perform an unprecedented analysis of the LBG-AGN spatial correlation. 	 	
\end{itemize}

\paragraph{Damped Ly$\alpha$ systems (DLAs)}

Damped Lyman Alpha systems (DLAs) are high column density neutral hydrogen systems, observed through their characteristic signature (damped \lya\ absorption) in the spectra of high redshift quasars. They are probably a consequence of the gas-rich outer regions of young, forming galaxies along the line of sight of the quasars, and are of significant interest for probing the early stages of galaxy formation at high redshift \citep[see,][for a review]{wolfe05}. Here we will evaluate the detectability of a DLA based on the J-PAS spectra of quasars. For a DLA to be observable, it must cause an absorption trough at a wavelength that is both redwards of the Lyman break of the quasar and the blue limit of the J-PAS filter transmission system. This translates into $\rm max(1.9;912/1215(1+z_{quasar}) < z_{DLA} < z_{quasar}$.  Even though it is clear that very broad DLAs will be easily detected by J-PAS -- see for an example the DLA recently published by \citet{kulkarni12} which has $\sim$200\AA\ (FWHM) -- these systems tend  to be very rare. To what extent will J-PAS be able to detect the more common DLAs that have typical widths that are just a fraction of a single J-PAS narrow-band filter? In order to evaluate this more quantitatively, we built a toy simulator which creates observed quasar spectra with intervening DLAs.
\begin{figure}[t]
\centering
\includegraphics[width=0.9\textwidth]{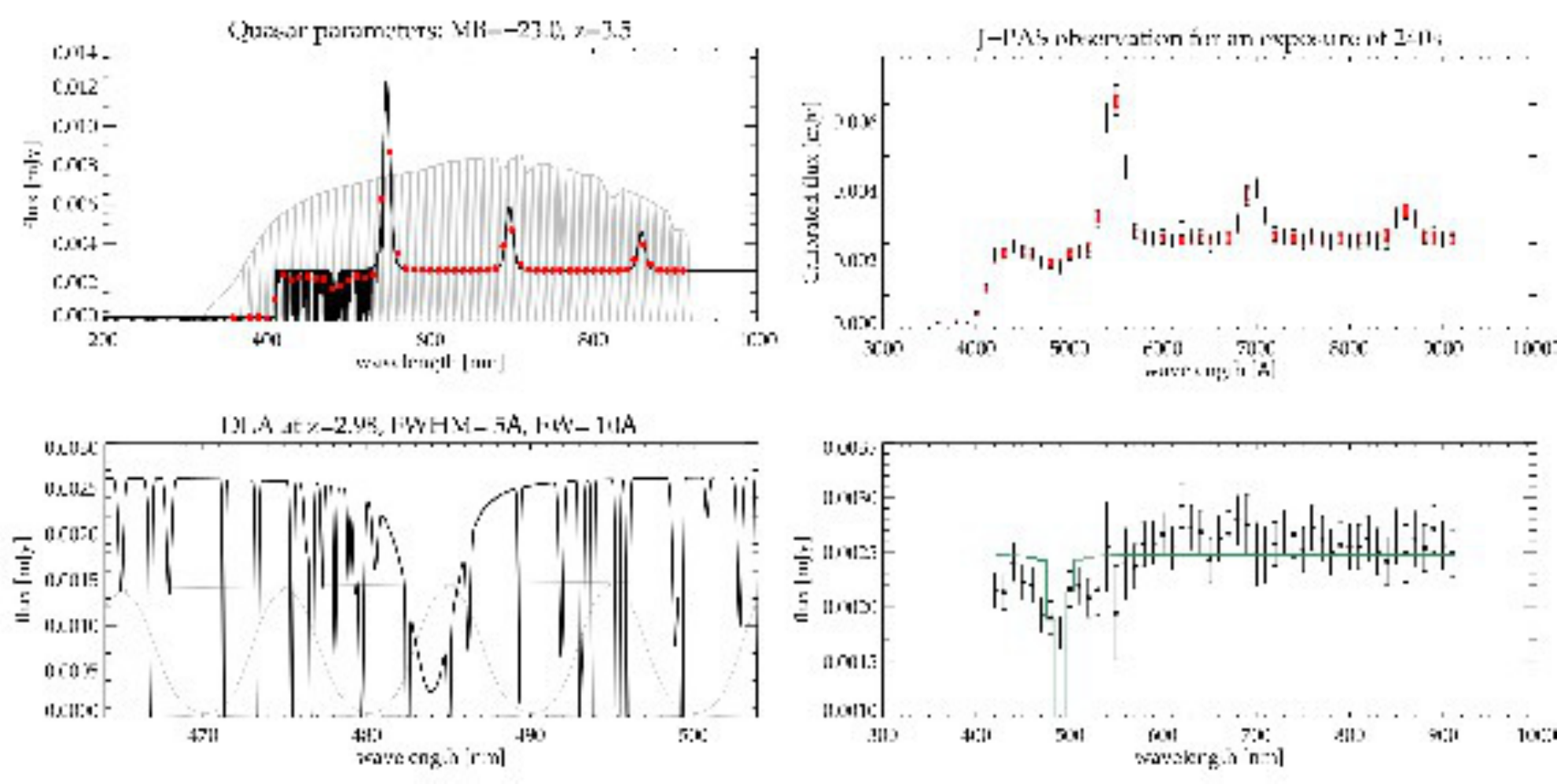} 
\caption{Illustration of the result for one simulation of a DLA observed by
J-PAS. The quasar parameters are: $\rm M_B=-23$ and
$z=3.5$. For the DLA, we chose $z_{DLA}=2.98$, FWHM$=$5\AA\ and an
EW$=$10\AA. Since the profile is a Voigt profile, what we call FWHM is
in fact 2.35 $\times$ $\sigma_{Voigt}$ at rest wavelength. The
observed width is multiplied by a factor ($1+z_{DLA}$). Top to
bottom, left to right: (1) The full quasar simulated spectrum in
black, the red dots mark the flux densities through the filters, and
the grey curves show the filter transmissions. (2) A zoom on the DLA
absorber, with the grey filter transmission curves in the region. (3)
The resulting measurements through J-PAS for an exposure time of
240s. The red and thick bars show the computed values and the poisson
noise, while the thin and black bars show a random draw for the
measurement, like an actual observation, and the full uncertainty
(quadratic sum of the poisson noise and 3\% of the flux). (4)
Residuals after a fit of the spectrum by a simple quasar model made of
a flat continuum and three emission lines with a Voigt profile
shape. A simple Voigt absorption fit is performed on this residual and
is shown with the green line.}
\label{fig:dla}
\end{figure}

\pagebreak 

We construct a simple quasar model spectrum consisting of a flat continuum and three broad emission lines, \lya\ at 1215\AA, CIV at 1549\AA, and CIII] at 1908\AA. These emission lines ensure that we will be able to detect and identify the quasar as well as measure its redshift.  We choose relevant parameters for these lines, which need not be detailed here, and simply note that the model quasar spectrum is realistic. The luminosity of the quasar is set by $\rm M_B$, its absolute magnitude in the B filter. A realistic luminosity range for quasars spans from $\rm M_B=$--23 to $\rm M_B=$--26. From the value of $\rm M_B$, we set the flux density value of the continuum in the \lya\ region assuming that $\rm \nu L_\nu$ is constant over the relevant wavelength range. An artificial and randomly distributed \lya\ forest is added to the spectrum. We use the VLT/UVES spectrum of a $z=2.4$ quasar\footnote{http://www.eso.org/public/images/eso0013f/} to choose realistic parameters for the FWHM and equivalent width (EW) of the ``trees'' populating this forest.  The DLA is modeled by a Voigt absorption profile. The rest-frame FWHM is the input parameter, while the EW is set to twice the FWHM (to reproduce the saturation of the line) and the $\alpha$ parameter of the Voigt profile is set to 2.5. The transformation from luminosity to the local flux density is computed using the luminosity distance $\rm D_L$ for the familiar cosmological parameter values $\rm \Omega_m=0.3$ and $\rm \Omega_{\Lambda}=0.7$. For the resulting spectrum, a J-PAS set of measurements (54 flux densities and their associated uncertainties) is then evaluated. The result is shown in Fig. \ref{fig:dla}, for a quasar of $\rm MB=-23$ at $z=3.5$. For the DLA, we chose $z_{DLA}=2.98$, a FWHM of 5\AA\ and an EW of 10\AA. For this case, the imprint of the DLA on the low-resolution quasar spectrum can be seen quite clearly. We are currently modeling the detectability of DLAs in J-PAS based on a wide range of parameters. A possibly complication in these kind of measurements may be the variability of the quasars. However, quasar variability will only be a source of uncertainty when observations through the different filters will be spread over long periods.

\subsubsection{Theme V. Active Galactic Nuclei}
\label{sec:agn}

\paragraph{The spatially resolved properties of nearby AGN}

For nearby AGN, J-PAS will be able to spatially resolve the stellar populations. While stellar populations in nearby AGN host galaxies have been mapped in detail with CALIFA, J-PAS will allow a far greater number of galaxies to be observed. The resulting low resolution spectra will then be interpreted using spectral synthesis methods. 
Together with a properly constructed control sample, this will allow us to investigate the differences between the stellar populations of active and non-active galaxies, as well as the differences in the stellar populations of Type 1 and Type 2 AGN \citep{storchi01}, and between more luminous and less luminous classes of AGN (e.g. LINERs).

J-PAS will furthermore allow us to map the nebular emission over the AGN host galaxies in the strongest emission lines, by using  filters containing emission lines and adjacent filters to obtain the continuum to be subtracted and isolate the gas emission (see Sect. \ref{sec:lines}). Line ratio maps between two emission line images can be used to map also the excitation of the gas. Again, comparisons between the extended emission of different AGN types can be investigated, such as the spatial extent of the emission and the gas excitation. Other properties that can be investigated are the presence of star-formation across the host galaxy, and the frequency of star formation in the disk for different types of AGNs  \citep[e.g.,][]{storchi95,storchi12}.

\paragraph{Optically selected AGN}

Active galactic nuclei (AGN) play an important role in galaxy
formation and evolution. The growth of central supermassive black holes (SMBHs) is
suggested to be related to the formation of the bulge of the host galaxy. This
scenario is supported by the Magorrian relation \citep{magorrian98}, and by
the fact that both star formation and AGN activity show similar evolutions from 
$z\sim1$ to the current epoch \citep[e.g.][]{silverman08}. 

Quasars are the most luminous sources to redshifts of at least $z\simeq 6.5$, and have been 
key cosmological probes. Observations of high-redshift QSOs have revealed a marked increase 
in the optical depth to neutral hydrogen (H{\sc i}) at redshifts $z\geq 5.7$, which signal 
the end of cosmological reionization. Measurements of the the quasar luminosity 
function (QLF) at high redshifts also constrain the growth of structures and the early 
formation of super-massive black holes in the first billion years of the Universe. QSOs
have been used as tracers of large-scale structure \citep{porciani04,daangela05,shen07,ross09}, and they can even be employed to 
measure baryon acoustic oscillations  (BAOs) at high redshifts \citep{2012MNRAS.423.3251A}. 

\begin{figure}
\centering
\includegraphics[width=0.45\textwidth]{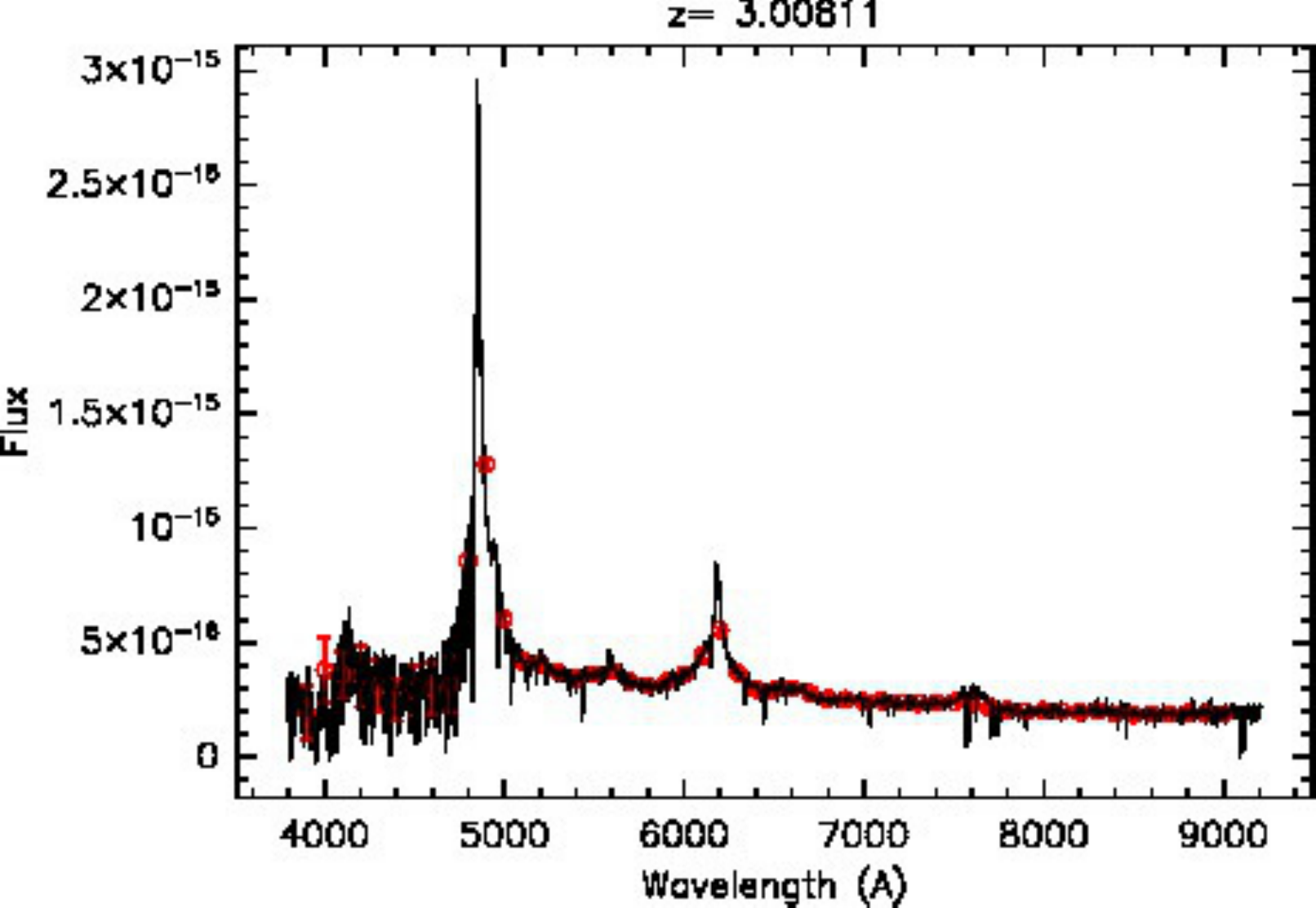}
\includegraphics[width=0.45\textwidth]{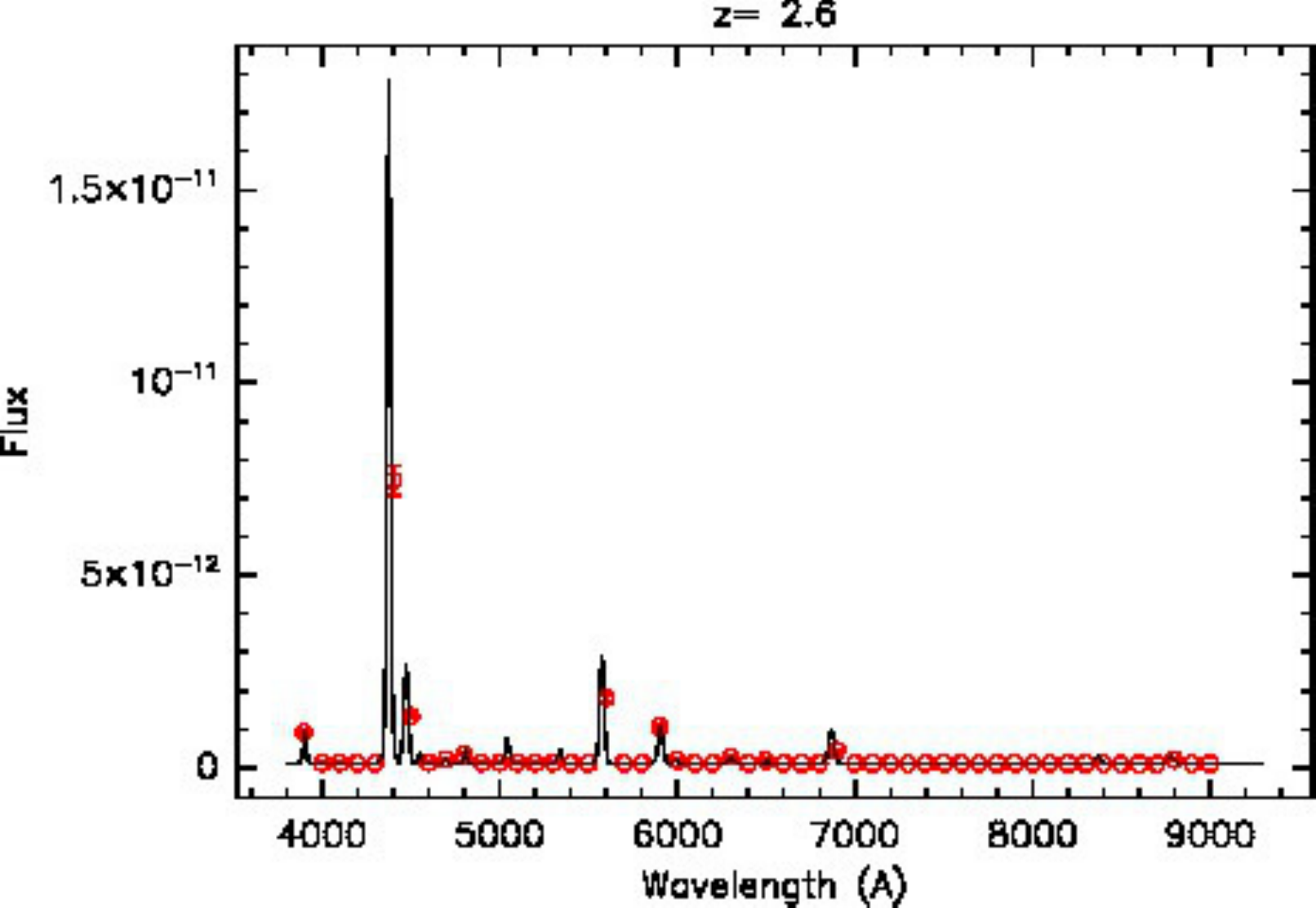}
\caption{A QSO at $z=3$ and a Seyfert-2 galaxy at $z=2.6$ as they would be observed by J-PAS. 
The black line is the original spectrum at SDSS resolution and the red dots, with error bars, 
are the corresponding fluxes for the J-PAS narrow-band filters.}
\label{fig:qso}
\end{figure}

We have explored the feasibility of identifying type-I and type-II AGNs by simulating how 
these objects will be observed with J-PAS. Since type-I AGNs have broad, high equivalent width emission lines 
(with relative width $\lambda/\Delta \lambda \sim 30$), they should be easily detected 
even with the low spectral resolution of the J-PAS survey ($\lambda/\Delta\lambda\sim 40-80$). 
We illustrate how such detections can be made by taking
the spectra of two SDSS AGNs, and simulating the low-resolution spectra as they would be 
observed through the J-PAS narrow-band filter system. In Fig. \ref{fig:qso} we show the results for 
a type-I QSO at $z\sim 3$, and a Seyfert 2 at $z\sim 2.6$. Although type-II AGNs have much 
narrower spectral features than type-I, it will still be possible to detect those with high 
equivalent width \citep[see also][]{bongiovanni10}. Assuming $\Delta \lambda\sim 100 {\rm \AA}$, the EW$_{min}$ 
that could be measured in the rest frame for narrow-line sources is $\sim 140${\rm \AA}.

We estimated the number of QSOs that will be detected in the J-PAS survey using both
the LFs and the surface densities of small-area but fairly complete surveys. 
A first crude estimate can be made using QSO surface density determinations made by 
\citet{beck-winchatz07} on high-latitude HST fields. At a limiting magnitude of 
23 we obtain, for an area of 8500 square degrees, $\sim 2.1\times 10^6$ QSOs. 
For type-I quasars, the luminosity function is reasonably well-known for luminous objects 
($M_I < -22$) and redshifts $\lesssim 2.5$ \citep[e.g.][]{hopkins07a,croom09}. 
Based on these luminosity functions, the number of QSOs that we expect to identify with 
J-PAS is around $2.5\times 10^6$ objects up to $z=2.5$. For higher redshifts we used the 
luminosity function by \citet{willott10} and estimate, in the range $4 < z < 7$, 
around $2.1\times 10^4$ QSOs. 

Given the broad spectral features of type-I quasars, the vast majority of these objects will 
have very accurate photometric redshifts ($\sigma_z \sim 0.0015 (1+z)$) and a low rate of 
outliers \citep{2012MNRAS.423.3251A}. The purity of the J-PAS dataset and the astrometric accuracy 
imply that this will be, by far, the largest and most complete sample of quasars and AGNs 
at the time of completion. With such a large sample of AGNs many astrophysical problems can be addressed with 
unprecedented detail and accuracy. 

The luminosity function of QSOs and AGNs as a function of redshift constrains the evolution 
of AGN populations and their effects on their environment. The shape of the quasar LF at high 
redshifts, particularly its faint end, represents a critical observational constraint on the 
early formation history of massive black holes, on their contribution to the reionization, and on
feedback processes which affect the formation of their hosts galaxies \citep{glikman10}. The 
true shape of the quasar LF is not well known at  $z\gtrsim4$ due to flux limits of 
large-area surveys. The J-PAS survey will allow us to determine the LF at high redshift.

Depending on the cadence of the J-PAS survey, it can also provide variability information on a 
large sample of AGNs, which, through reverberation mapping, may allow us to determine the size of the broad line region (BLR)  
for thousands of objects. Variability is also an important tool to cross-check contamination of
quasars by blue stars (a potential problem mainly at $z \lesssim 1$).

QSOs can be competitive tracers of large-scale structure \citep{2012MNRAS.423.3251A,sawangwit11}. Large quasar groups (LQG) and QSO pairs are also indicators of the 
underlying galaxy overdensity \citep{haberzettl09,sodre09} up to 
$z\sim1-2$. These observations can probe the interplay between the environment at halo 
and super-halo scales, the galaxy stellar populations, and the QSO/AGN phase of galaxy 
evolution. 

During the past two decades, a sample of $>$100 strongly lensed QSOs (SLQSOs) and 
binary QSOs (e.g. CASTLeS \footnote{http://www.cfa.harvard.edu/castles/}) has been harvested 
from different surveys, 
either in the optical (e.g. SDSS, SLACS) or radio (e.g. CLASS). It is 
expected that this catalog will increase by about 2 orders of magnitude with 
ongoing and planned wide-field surveys. A large enough sample of SLQSOs can be used 
to constrain the CDM power spectrum on small scales, as well as to test different dark matter candidates 
\citep{maccio08}. On the other hand, follow-up measurements of time delays (e.g. 
COSMOGRAIL) of a large sample of multiple-image SLQSOs can allow a direct measurement of $H_0$ 
\citep[see][for a review]{jackson07}, yielding strong and independent constraints on cosmic expansion,
the dark energy equation of state parameter $w$, and the flatness of the universe $\Omega_k$
despite the limitations of this method \citep[see][]{falco85,wucknitz02}. Following the 
predictions of \citet{oguri10}, J-PAS should be able to detect about 1000 SLQSO up 
to $z\sim5$. By time-monitoring this sample, a $<1$\% level 
precision in the Hubble constant could be reached \citep{coe09}.

\paragraph{Radio-galaxies and radio-loud quasars}
\label{sec:radio}

Radio continuum surveys such as the Faint Images of the Radio Sky at Twenty centimeters (FIRST) survey \citep{becker95}, the National Radio Astronomy Observatory (NRAO) Very Large Array (VLA) Sky Survey \citep[NVSS][]{condon98}, and the Westerbork Northern Sky Survey \citep[WENSS][]{rengelink97} cover large fractions of the sky while probing the extra-galactic radio source population out to significant cosmological distances \citep[$z\sim6$;][]{debreuck00}. The clustering of radio sources, such as radio galaxies and radio-loud quasars, is an important probe for both cosmology and the evolution of galaxies and the large-scale structure. Radio continuum surveys, however, do not provide redshift information on the detected radio sources. Redshifts must be obtained by spectroscopic follow-up or by cross-correlating the radio-selected sample with surveys performed at other wavelengths and with known redshifts. Historically, radio source redshift distributions have been obtained statistically based either on an extrapolation of the redshift distribution determined for much smaller subsamples selected at relatively high flux densities \citep[e.g.][]{dunlop90}, or by observing a complete sample down to low flux densities in a small area (typically a few square degrees) of the sky \citep{waddington01,best04,brookes08}.

More recently, large-area sky surveys in the optical such as the SDSS and 2dF surveys have enabled the identification of large samples of radio sources from NVSS and FIRST through the cross-correlation with objects with known spectroscopic or photometric redshifts \citep[e.g.][]{magliocchetti04,best05,donoso09,passmoor13}. Such cross-correlations have allowed us to study the properties and clustering of primarily the low redshift end of the radio source redshift distribution. For example, \citet{best05,best12} used the SDSS spectroscopic sample to study the properties of radio sources at a median redshift of $z\sim0.1-0.2$. \citet{donoso09,donoso10} was able to study the clustering properties of 14,000 radio sources at $0.4<z<0.8$ identified with luminous red galaxies (LRGs) from the SDSS, finding that the radio sources are more strongly clustered on scales of less than 1 Mpc compared to the parent LRG sample. They also found that the excess clustering scales with radio power, and that radio galaxies and radio-loud quasars are similar only at the highest radio luminosities. This suggests that the AGN triggering mechanism may be linked to environment. \citet{chen13} compared optical spectra of massive radio loud AGN at $z\sim0.2$ and $z\sim0.6$, also from SDSS, finding evidence that the history of the gas accretion, star formation and black hole growth is regulated by feedback from the radio-jets. \citet{fine11} compare the clustering of NVSS and FIRST radio sources in three samples at $z\sim0.35$, $z\sim0.55$ and $z\sim0.68$, finding that the halo mass of radio AGN hosts must be constant over this redshift range, similar to that found for QSOs albeit at a much higher clustering amplitude (indicating that radio AGN have much more massive hosts than optical QSOs). \citet{overzier03} measured strong clustering of NVSS/FIRST radio sources at an (inferred) median redshift of $z\sim1$, finding that radio galaxies trace the most massive structures in the universe, possibly the progenitors of rich Abell-type clusters. 

It has been observed that the optical spectra of radio galaxies either show strong high-excitation emission lines, or low-excitation emission lines \citep{laing94,jackson97,tadhunter98}. These two classes of radio galaxies, high-excitation galaxies (HEGs) and low-excitation galaxies (LEGs), are, according to the current paradigm, interpreted as being powered respectively by cold gas on a high rate and radiatively efficient accretion process, and by radiatively inefficient accretion flows of hot gas at low accretion rates \citep[e.g.][]{buttiglione10,best12,janssen12}. Alternatively, or in addition to, the energy released through black hole spin is also a strong candidate to explain this class separation \citep[e.g.][]{mcnamara11,martinez11}. Given that the jet launching mechanism in radio galaxies is conjectured to be strongly associated with the accretion disk emission \citep[e.g.][]{rawlings91}, the HEGs and LEGs radio and disk-related emission can   pose important constraints and insights on the link between jets and accretion, which remains an important issue in AGN models.

Moreover, `standard' radiatively efficient accretion, and advection-dominated radiatively inefficient accretion are associated with different AGN activity modes, and thus with different AGN feedback mechanisms. Whereas the standard accretion mode is associated with quasar activity, the radiatively inefficient mode leads to little radiated energy, but it can produce highly energetic jets, also known as the `radio mode' \citep[e.g.][]{hine79,hardcastle07}. `Radio mode' activity is taken as the main mechanism to switch off star formation in the most massive galaxies, thereby explaining the shape of the observed luminosity function. This mode of AGN feedback is an important ingredient in  semi-analytic models of galaxy formation, although its physical implementation is not exactly clear \citep[e.g.][]{bower06}. The current procedure is therefore to adjust the feedback parameters in order to match observed luminosity functions. LEGs, being powered by this mode of AGN activity, are the perfect laboratory to study the `radio accretion mode' and to quantify its physical parameters. A statistically significant study of LEGs is thus utterly compelling. Analytic models also predict that whereas the 'radio mode' of AGN activity should dominate over a wide range of luminosities at $z=0$, the `standard mode' is expected to have a much more significant weight at $z\sim1$ \citep{hopkins06}. Being a strongly evolutionary population, the study of the space densities of HEG and LEG radio galaxies provides important clues as to how and when the two forms of accretion are triggered and their evolution with redshift. \citet{best12} presented a sample of $\sim$7300 SDSS classified sources, with a median redshift of $z=0.16$. \citet{fernandes13} have performed the highest redshift study of HEGs and LEGs at $z\sim1$ with a sample of 27 galaxies. J-PAS should 
be able to robustly identify large numbers of these sources up to a 
redshift of $z\lesssim 0.8$.

By probing to much fainter magnitudes with very accurate photometric redshift information for individual sources, J-PAS will be used to identify much larger and less biased samples of radio sources, and out to much higher redshifts, than allowed by the existing data sets.  The more accurate redshift distribution resulting from J-PAS will also allow a better assessment of the integrated Sachs-Wolfe effect through cross-correlation of the clustering signal detected on large angular scales in the NVSS \citep{overzier03} with the CMB \citep{HernandezMonteagudo2010}, and may even be employed toward constraining dark energy \citep{camera12}. At high redshifts, radio sources can be used as probes of cluster environments \citep{best03,venemans07}, and evolution observed in their typical environments as a function of redshift and radio power offers insight into the typical feeding and feedback mechanisms of radio-loud AGN \citep{hill91,hart11}. In the relatively nearby universe, we will be able to focus on the properties and environment of the large number of faint radio sources in which the radio continuum emission may arise also from starbursts rather than (or besides) an AGN.

The NVSS survey has a source density of 15 and 1.5 $/\sq\degr$ at
limiting flux densities of 10 and 100 mJy, respectively. The deeper
FIRST survey reaches a source density of 
$35/\sq\degr$ at 3 mJy. The median redshifts probed by these surveys is $z\sim1$. 
J-PAS will thus cover up to $\sim300,000$ radio sources, allowing a great many studies related to the population of radio-loud AGN over the redshift range $z=0-1.5$. J-PAS will be able to constrain fluxes and equivalent widths of the brightest lines \citep[e.g. \oii, \hb, \oiii, \sii, \nii, \oi, \ha;][]{buttiglione10}, which can be used to determine the main emission line diagnostic ratios for thousands of galaxies up to a redshift of $z\sim1.5$, while at the same time providing insights into the stellar populations, morphologies, and environment and the evolution with redshift. 

\subsection{Synergy with other surveys}
\label{sec:multi}

Naturally, J-PAS will be able to offer a wealth of information to studies performed at other wavelengths, and vice-versa. Here we will present a brief overview of a selection of other extra-galactic surveys from X-ray to radio wavelengths that will be cross-matched with J-PAS. 

Synergy between J-PAS and X-ray surveys will allow us to probe the galaxy-AGN connection, star formation, and calibrate group and cluster masses. Apart from continuing targeted observations by the {\it Chandra} and {\it XMM-Newton} X-ray observatories, eROSITA will perform the first imaging all-sky survey in the medium energy X-ray range up to 10 keV with an unprecedented spectral and angular resolution, as a part of the Russian {\it Spectrum-R\"ontgen-Gamma satellite} \citep[SRG,][]{kolodzig13a,kolodzig13b}. 

The {\it Galaxy Evolution Explorer} \citep[GALEX,][]{martin05} has
performed imaging surveys in the Far UV (FUV) $1350-1780$\AA\ and Near
UV (NUV) $1770-2730$\AA\ at a resolution of $\sim5$\arcsec\ (FWHM),
allowing an unprecedented view on the history of star formation from
$z\sim1.5$ to the present. The GALEX All-Sky Imaging Survey (AIS)
covers an area of $26,300/\sq\degr$  
down to a NUV depth of 21 (AB) mag (40 million sources). The Medium Imaging Survey (MIS) covers an area of $5,000/\sq\degr$  to a NUV depth of 23 mag (22 million sources). The Deep Imaging Survey (DIS) covers about $100/\sq\degr$  down to $\sim25$ mag. The GALEX surveys have been crucial for extending the wavelength range of the SDSS into the UV, where the youngest stellar populations dominate. Likewise, GALEX will be an invaluable resource for the J-PAS galaxy evolution survey. 

In the near-infrared domain, the J-PAS galaxy survey will benefit from the on-going, UKIDSS Large Area Survey \citep[LAS,][]{lawrence07}, that covers an area of 4000 square degrees at high Galactic latitudes, in the four bands $Y$, $J$, $H$ and $K$ to a depth of $K_{AB}=18.4$. Going significantly deeper, a patch of $\sim$750 square degrees will be observed to $5\sigma$ limiting depths of $z\approx22.4$, $Y\approx21.4$, $J\approx20.9$, $H\approx19.9$ and $K_s\approx19.3$ (all in the Vega system) by the VISTA Kilo-degree Infrared Galaxy Survey \citep[VIKING,][]{sutherland12}. An additional contribution, especially in the photometric $Y$ band (up to $\approx21.9$ AB mag), 
could come from the Pan-STARRS PS-1\footnote{From http://www.ps1sc.org/Data\_Release.shtml: All of the data, images, and catalogs taken by PS1 for the PS1 Science Mission, which is funded by the member institutions of the PS1SC, will become public at the end of the PS1 Mission (2013).} 
$3\pi$ Steradian Survey \citep[see][]{tonry12}. At longer wavelengths, the {\it Wide-field Infrared Survey Explorer} (WISE) has mapped the sky at 3.4, 4.6, 12, and 22 $\mu$m with an angular resolution of 6.1, 6.4, 6.5, and 12.0\arcsec, probing dust-enshrouded star-forming and starburst galaxies and AGN \citep{wright10}. 

In the sub-mm domain, the SCUBA-2 ``All-Sky'' Survey (SaSSy: $\sim$4800 square degrees, rms sensitivity of $\sim$30 mJy beam$^{-1}$ at 850 $\mu$m) offers a new-generation survey approach. SCUBA-2 \citep{holland13} is a 100--150 times faster instrument than the previous one, and will offer interesting possibilities for the analysis of different sub-types of luminous infrared galaxies \citep[see][for a pilot study]{mackenzie11}.

Fianlly, by combining J-PAS with existing radio surveys in the Northern sky at 1.4 GHz, such as NVSS, FIRST, WENSS, and LOFAR, will allow us to constrain the properties of the nearby star-forming population, as well as those of the population of radio-loud AGN out to cosmological distances (see Sect. \ref{sec:radio}). The Square Kilometer Array (SKA) is an internationally-funded radio telescope, which will consist of more than one thousand of $\sim$15-meter size dishes in a central area of diameter $\sim$5 km, surrounded by a similar number of dishes in an area stretching up to several thousands of km.  The SKA will be much more sensitive than any existing radio telescope, aimed at answering some of the most fundamental questions of the Universe we live in \citep{carilli04,schilizzi04}.  Since its construction is expected to be completed in 2022, several pathfinders have been, or are currently being, built. LOFAR \citep[the Low Frequency Array,][]{rottgering06} and Apertif \citep[the updated Westerbork phased-array,][]{oosterloo10} in the Northern Hemisphere, and ASKAP \citep[the Australian SKA Pathfinder,][]{johnston08} and MeerKat \citep[the South African SKA Pathfinder,][]{booth09} in the Southern Hemisphere can be considered as the SKA pathfinders. The pathfinder projects are expected to devote a significant amount of observing time to a number of ``legacy'' projects with the main aim of studying galaxy formation and evolution through cosmic time, e.g.:

WODAN, an Apertif legacy project, will observe at the relevant observing wavelength of 21 cm (continuum) all of the Northern sky at $\delta>30\deg$, and down to a 1-$\sigma$ rms noise figure of about 10--20$\mu$Jy beam$^{-1}$ (a factor of 25-50 more sensitive than the NVSS). The angular resolution will be about 15 arcsec, a factor of three better than the angular resolution of the NVSS. ASKAP has a similar legacy project, called EMU (Evolutionary Mapping of the Universe). EMU is complementary to WODAN, as it will observe also in the 21 cm continuum band the whole Southern Sky and up to $\delta=+30\deg$ \citep{norris11}.  After about 1.5 yr of observation, EMU will reach an homogeneous 1-$\sigma$ noise rms figure of about 10$\mu$Jy beam$^{-1}$, with an angular resolution of about 10 arcsec. 

Not surprisingly, the overall goals of these radio continuum surveys will match those pursued by J-PAS. In particular, EMU and WODAN will (i) trace the evolution of star-forming galaxies from z$\approx$2 to the present day, using a wavelength mostly unbiased by dust or molecular emission, and (ii) trace the evolution of massive black holes throughout the history of the universe, and understand their relationship to star formation. Since J-PAS and WODAN/EMU will observe the same areas of the sky with very deep sensitivity at their corresponding bands, those surveys will detect and unambiguously identify many star-forming galaxies and accreting black holes at the low-mass end  from the local universe up to $z \approx 2$. This region of the parameter space (low-masses and/or  high-redshift) is so far largely uncharted territory. J-PAS will be in a unique position to provide the much needed high quality (photometric) redshifts for the $\sim$100 million sources expected to be detected by WODAN/EMU, as spectroscopic redshifts for such a huge amount of galaxies is simply unaffordable. At the same time, J-PAS will provide information on the properties of the radio source host galaxies as well as their small- and large-scale environments. 

In the \hi\ line, the extension of ALFALFA \linebreak 
\citep[Arecibo Legacy Fast ALFA Extragalactic HI Survey,][]{haynes08} and its spin-offs (HIghMass, SHIELD, GASS) will cover $\sim$13,000 square degrees of sky, while AGES\footnote{http://www.naic.edu/$\sim$ages/} (Arecibo Galaxy Environment Survey) will provide a larger sample of clusters and groups of galaxies ($cz < 10000$ km s$^{-1}$) by covering 3,000 square degrees at 300 s integration time per beam \citep{giovanelli08}. Also, the combined WNSHS (APERTIF Westerbork Northern Sky \hi\  Survey\footnote{http://www.astron.nl/$\sim$jozsa/wnshs/}) and WALLABY \citep[Australian SKA Pathfinder \hi\ All-sky Survey,][]{koribalski12} programs will offer an unprecedented \hi\ image of the whole sky (sensitivity of 0.65 and 0.55 mJy per 100 kHz beam at 30\arcsec\ and 13\arcsec\ angular resolutions, respectively). 

\FloatBarrier 

\section{Scientific Goals III}
\FloatBarrier 

\subsection{Stars and the Galaxy}

J-PAS is designed to obtain low-resolution spectroscopic of $\sim$ 8000 squared degrees of the  sky up to a limiting magnitude R(AB) $\sim$23.5 with S/N ratio higher than 5. It can be considered the largest IFU survey ever carried out. 
Two different filter sets (12 for J-PLUS and 54+5 for J-PAS) define two distinct experiments with clearly differentiated telescopes and detectors and hence scientific goals, observational strategy, and timing. Between 200 and 500 million stars are expected to be listed in the final J-PAS catalogue which will provide one of the most detailed views of the Milky Way halo in the northern hemisphere sampled on 54 points of the optical spectrum. 

\subsubsection{Stellar Populations}
Stellar populations, consisting of individual stars that share coherent spatial, kinematic, chemical, or age distributions, are powerful probes of a wide range of astrophysical phenomena. They describe different evolutionary stages, interior physics and phenomenology throughout the whole life of the stars, as well as the structure and evolution of different stellar systems. 

The success of these studies will lie in the capability of J-PAS photometry to physically characterize the stellar populations and estimate their mean physical variables such as reddening, temperature, gravity and metallicity. Low resolution spectroscopy provided by J-PAS is what makes the difference compared to other photometric surveys such as SDSS, Pan-STARRS and LSST, and will allow a better taxonomy of the different stellar populations and a more accurate determination of physical properties. Experience acquired with the ALHAMBRA photometric system (Aparicio Villegas et al. 2010, 2011) will be very useful in defining the best strategy for characterizing the different stellar populations and to estimate their physical properties. Low-mass stars in the solar neighborhood; RR Lyrae variables that populate the thick disk and the halo, fantastic tracers of the structure of their respective Galactic subsystems; line-emission objects including planetary nebulae and cataclysm variables; and white dwarfs in a wide range of metallicities tracing the evolution time  of different galactic subsystems -- all these are, among others, good examples of the large variety of stellar objects that  will be found in this survey.

\paragraph{RR Lyrae stars}

Assuming the main halo metallicity $[Fe/H]=-1.5\,\rm{dex}$
\citep{2008ApJ...684..287I}, \mbox{RR Lyrae} stars mean absolute magnitudes show
very low dispersion \mbox{$\langle M_V\rangle=0.59\pm0.03$}
\citep{2003LNP...635..105C}, so they can be used to measure distances and they are bright enough to be
detected. For instance, SDSS has probed distances up to
$\sim110\,\rm{kpc}$ \citep{2010ApJ...708..717S} with limiting magnitudes more
restrictive than the present survey (see Sec 2.1), so the detection of
this type of variable stars observed in different colors is
conservatively expected (just considering the limiting magnitude
$1\,\rm{mag}$ dimmer than that of SDSS) to occur at distances so far as
$\sim190\,\rm{kpc}$. Such detections have important implications on the
formation history of our galaxy. At the same time, the survey will
provide a large amount of SEDs that will significantly improve our
understanding on the physics and the processes that take place at the
interior of this kind of variable stars.

RR Lyrae stars are easy to identify by their characteristic
light-curve, with periods from $P=0.2\,\rm{days}$ to
$P=1.2\,\rm{days}$, whenever it is sufficiently well-sampled.
Nevertheless, they are expected to be identified in the survey by means
of their color. RR Lyrae stars have been shown to be efficiently
and robustly found in the past even with two-epoch data, using accurate
multiband photometry obtained by the Sloan Digital Sky Survey (SDSS).
The identification can even be feasible with single-epoch colors, due 
to the fact that RR Lyrae stars span a very narrow range of colors; 
e.g. for the SDSS \citep{2005AJ....129.1096I}:
\begin{eqnarray}
 0.99 & < u-g < & 1.28 \\
-0.11 & < g-r < & 0.31 \nonumber \\
-0.13 & < r-i < & 0.20 \nonumber \\
-0.19 & < i-z < & 0.23 \nonumber
\end{eqnarray} 
Other ranges have been suggested more recently by \cite{2010ApJ...708..717S}. In
that regard, a prediction on expected colors with the SDSS photometric
system was conducted by \cite{2006MNRAS.371.1503M}. The approach of combining
these color criteria and the, somehow limited, time-resolved
observations will be adopted to look for this type of pulsating stars.

As it is well known, RR Lyrae have been demonstrated to be reliable
tracers of the halo; as already mentioned, they are relatively bright
and they obey a period-metallicity-luminosity relation (see
\citealt{2013MNRAS.435.3206D,2004ApJS..154..633C}), so they can therefore serve as
distance indicators, as well as kinematic and metallicity tracers. Some
surveys for RR Lyrae stars \citep{2010ApJ...708..717S, 2004AJ....127.1158V,
2005AJ....129.1096I} have already detected several substructures in the halo, and
considering the extension that will be observed by J-PAS, $8500$ square
degrees, these observations will introduce strong constraints on the theoretical
models and numerical simulations of the galactic halos formation 
(see e.g. \citealt{2010MNRAS.406..744C}).

The possibility of building a metallicity map of the galactic halo
employing the J-PAS low resolution spectra will be explored.
This is not straightforward due to the resolution of J-PAS
spectra, and the changes undergone by the stellar spectrum with the
phase of RR Lyrae's oscillation coupled with the observation strategy of
the survey.

RR Lyrae stars are not only bright stellar candles useful for drawing
the structure of the outer regions of the galactic halo, but they
represent unique probes for the study of stellar interiors and their
response to internal perturbations. The observational strategy of the
survey will provide 4 points, at least, per filter for all the stars in
the sample, which will allow the discovery of new variables and a better
coverage of the light curve for previously catalogued stars.

RR Lyrae stars populate the halo globular clusters. The mean pulsational
period of cluster RRab stars (pulsating in the fundamental mode) is
closely related to the metallicity of the cluster. If we represent the
globular clusters into the P(ab) vs. metallicity space they show a
bimodal distribution forming two separated concentrations called
Oosterhoff groups \citep{1939Obs....62..104O}. Poor metal clusters
present a mean fundamental period near to 0.65 days, while those with
metallicity above -1.4 show a value of around 0.55, separated by a gap
centered at 0.60 days. Although this phenomenon has been known for
seventy years and there are numerous studies about the possible nature
of this bimodality, its physical origin is still a matter of debate. The
theoretical understanding of RR LyraeLyrae needs observations that not only
cover the time domain but also the parameter space where this kind of
pulsation is present. It is in this context that J-PAS will be unique in
providing well sampled SEDs of these stellar pulsators.

\pagebreak 

\paragraph{White Dwarfs}

White dwarfs are the end state of all main sequence stars less massive than $8M\sun$, 
which means that $98\%$ of all stars will end up as white dwarfs. First and foremost, 
J-PAS will allow us to discover many new white dwarfs. It will go deeper than SDSS; 
most of SDSS spectroscopically confirmed white dwarfs have a magnitude below 20.5, 
while J-PAS will be complete (5$\sigma$ detections) down to 22.5 in each filter. 
So we should see white dwarfs 2.5 times farther than SDSS and therefore the total 
volume will be ($2.5^3$ - 1 = 14.6 times larger. By definition every object in J-PAS 
will be spectroscopically observed, while in SDSS only chosen objects had their 
spectra taken, so our white dwarf sample will also be much more complete than SDSS.
 
White dwarfs in the J-PAS survey can be of great value for white dwarf studies as well as for other areas of stellar astrophysics:

\begin{enumerate}
\item Individual white dwarfs can be used as a distance
  indicator. Through analysis of their J-PAS multicolor and model
  stellar atmospheres, we can determine their $T_{eff}$ and log
  g. Using the well defined mass-radius relationship we can obtain a
  luminosity for these objects and therefore a distance. These
  distances can be very useful when the white dwarfs belong to a given
  star group. 
 
\item In stellar groups we can use the cooling times of the coolest white dwarfs as an age indicator for the whole group. As the coolest white dwarfs come from the most massive main sequence stars, the main sequence lifetime for them is very short compared to the cooling time. White dwarfs can be used as chronometers in those cases when the coolest white dwarf in the group is detectable by J-PAS, which means the group distance modulus added to the absolute magnitude of the coolest white dwarfs is less than $\approx 22.5$, J-PAS  magnitude limit. In particular we will scrutinize every open cluster within J-PAS coverage area in search of white dwarfs to determine their ages as well as to study the initial-final mass relationship.

\item Metal-line white dwarfs (DZ). These objects are associated with debris disk white dwarfs. Every white dwarf presenting a dust disk also presents metal lines. Currently 20 WDs are known to have infrared excess, and therefore a debris disk. About 100 WDs present metal lines in their spectra. J-PAS should allow at least an order of magnitude increase in the number of these objects.

\item White dwarf luminosity function (WDLF) in the halo and disk. The WDLF is of extreme importance to the studies of white dwarf physics, also as stated above it is very useful to date groups of stars, from open clusters ([Koester and Reimers, 1996]) to the Milky Way ([Winget et al, 1987], [DeGennaro et al, 2008]).
The increase in the total numbers of white dwarfs obtained by J-PAS will allow a much improved WDLF of halo and disk white dwarfs as well as the luminosity function in clusters.
 
\item White dwarf mass function.  The mass distribution of white dwarfs is closely related to mass loss for both single and binary stellar evolution.  The large number of white dwarfs identified by J-PAS will allow us to improve the precision of the current white dwarf mass function.

\item Pulsating white dwarfs.  J-PAS four visits to each field, times 56 exposures in all filters will allow us to search for variability among all the white dwarfs identified by the survey.  

\end{enumerate}

\paragraph{Cataclysmic Variables}

Cataclysmic variables (CV) are an important component of the galactic population since they provide the closest and most abundant instance of accretion disks but are also a theoretical identified channel for the production of type I Supernovae. Estimations of the space density estimates of CVs vary. Cieslinski et al. (2003) suggested $\rho~\le$ 5 $\times$ 10$^{-7}$ pc$^{-3}$, for dwarf novae, the most common type of CV. The ROSAT North Ecliptic Pole Survey (Pretorius et al. 2007) gives $\rho$ = 1.1 $\times$ 10$^{-5}$ pc$^{-3}$. Rogel et al. (2008) adopt $\rho$ = 0.9 $\times$ 10$^{-5}$ pc$^{-3}$ in their model to predict the number of CVs in the galactic plane.

 We adapted the Ortiz and Lapine (1993) Galactic star counts model to
 predict the high latitude counts of CVs. The luminosity function of
 Patterson (1998) purged from the super-soft sources was used in our
 estimate. We used g = V - 0.14 for a typical CV, and the very
 conservative assumption that the total density of CVs is $\rho$ = 1.0
 $\times$ 10$^{-6}$ pc$^{-3}$. 
The corresponding cumulative counts (scaled to 8000 squared degrees) for g $<$ 22 are $\sim$ 600 and 200. Notice that these numbers would be ten times larger if the total density of CVs was $\rho$ = 1.0 $\times$ 10$^{-5}$ pc$^{-3}$, as in Pretorius et al. (2007) and Rogel et al. (2008).

{\bf Peculiar Stars} Ap stars are chemically peculiar stars in, or just off, the main sequence.  Their peculiarity is in the fact that they show a very high abundance of rare earth metals. We understand this phenomenon as being caused by radiative levitation from the inner core of an entirely radiative star.  The existence of magnetic fields makes the radiative levitation deposit the rare earths close to the magnetic poles (Michaud, 1970, ApJ, 160, 641).  The basic mechanism for the existence of these objects is understood; however, the reason why some A stars turn into Aps and others do not is still not known.  Ap stars represent approximately 2\% of all A stars, therefore we expect a large increase in the number of known Ap and other chemically peculiar stars.  Statistic studies of such a large number of chemically peculiar objects will shed new light into their origin.  In particular, studies relating the occurrence of Ap stars with the age of stellar groups, specially open clusters, are particularly important, as there is a possibility that the peculiarities increase in strength as the stars evolve.

\paragraph{J-PAS and metal-poor halo stars}
 
In the J-PAS survey a great number of stars and minor systems will be also observed, 
which will provide a unique sample of galactic halo objects and will enable 
a deep study of the galactic structure and the halo stellar population. The stellar halo 
population is basically formed by stars of population II, which are old metal-poor 
stars, and the inferior limit of the distribution of metallicities of these stars 
is not defined yet. In fact, so far the poorest-metal old stars found are :
\begin{itemize}
\item The ultra-metal-deficient red giant CD-38 245 with $[Fe/H]=-4\ldotp5$     (Bessel \& Norris 1984)
\item The ultra-metal-poor and Carbon-rich HE 00557-4840 with $[Fe/H]=-4\ldotp75$ (Norris et al. 2007)
\item The chemically ancient star HE0107-5240 with $[Fe/H]=-5\ldotp3$        (Christlieb et al. 2004)
\item The most iron-poor star HE1327-2326 with $[Fe/H]=-5\ldotp6$ (Aoki et al. 2006)
\end{itemize}

These stars are in general giants and red subgiants with V=13-15 magnitudes, representing 
the extreme halo population II.

Some surveys with different characteristics from J-PAS did already study the galactic 
halo looking for metal-poor stars, such as the HK Survey (Beers et al. 1992), a 
low resolution spectroscopic survey developed with the 2.5m telescope, Du Pont, 
and centered on the HK line of Ca II; the Hamburg/ESO Survey (Wisotzki et al. 1996), a spectroscopic 
survey of QSOs that revealed white dwarfs, horizontal branch stars and metal-poor 
stars; or the Sloan Digital Sky Survey (SDSS, York et al. 2000) and its extension for the exploration of the galaxy 
SEGUE (Yanny et al. 2009) in operation from 2005 in the Apache Point Observatory. 
The success of these surveys incite the appearance of other new projects such as the SkyMapper (Keller et 
al. 2007), the counterpart of the SDSS in the South hemisphere, which began to operate 
in March of 2010 using a 1.35m telescope in the Siding Observatory and with a photometric 
system of 6 filters, 4 of them similar to the \textit{g, r, i, z} SDSS filters, 
or the LAMOST project (Newberg et al. 2009) which was initiated in 2011 using a 4m telescope, 
and will provide low-resolution spectra for objects with $-10\text{degree} < \delta < +90\text{degree}$. 

The 54 narrow-band filters of the J-PAS photometric system could be considered as 
a very low-resolution spectroscopy. With this idea in mind, the first approach for 
the study of these metal-poor stars will be done through the comparison of these observed 
``spectra'' with theoretical J-PAS spectra from synthetic photometry, such as it is done 
in the methodology developed for the ALHAMBRA photometric system (Aparicio Villegas 
et al. 2010), the $Q$-Fit-Algorithm, a methodology for the estimation of the main physical 
stellar parameter ($T_{eff}$, $log(g)$, $[Fe/H]$ and $E_{B-V}$) of stars of a great 
variety of spectral types and luminosity classes using reddening-free $Q$-parameters:

\begin{equation}
Q_{ijkl}=(m_i - m_j) - \frac{E_{ij}}{E_{kl}}(m_k - m_l), i=1:n-3, j=i+1, k=j+1, l=k+1
\end{equation}

\noindent where $m_i$ is the magnitude in the band $i$ of the photometric system 
of $n$ bands, and $E_{ij}/E_{kl}$ is the color excess ratio, which only depends on the interstellar extinction law.

\begin{center}
\begin{figure}
\centering
\includegraphics[width=\columnwidth]{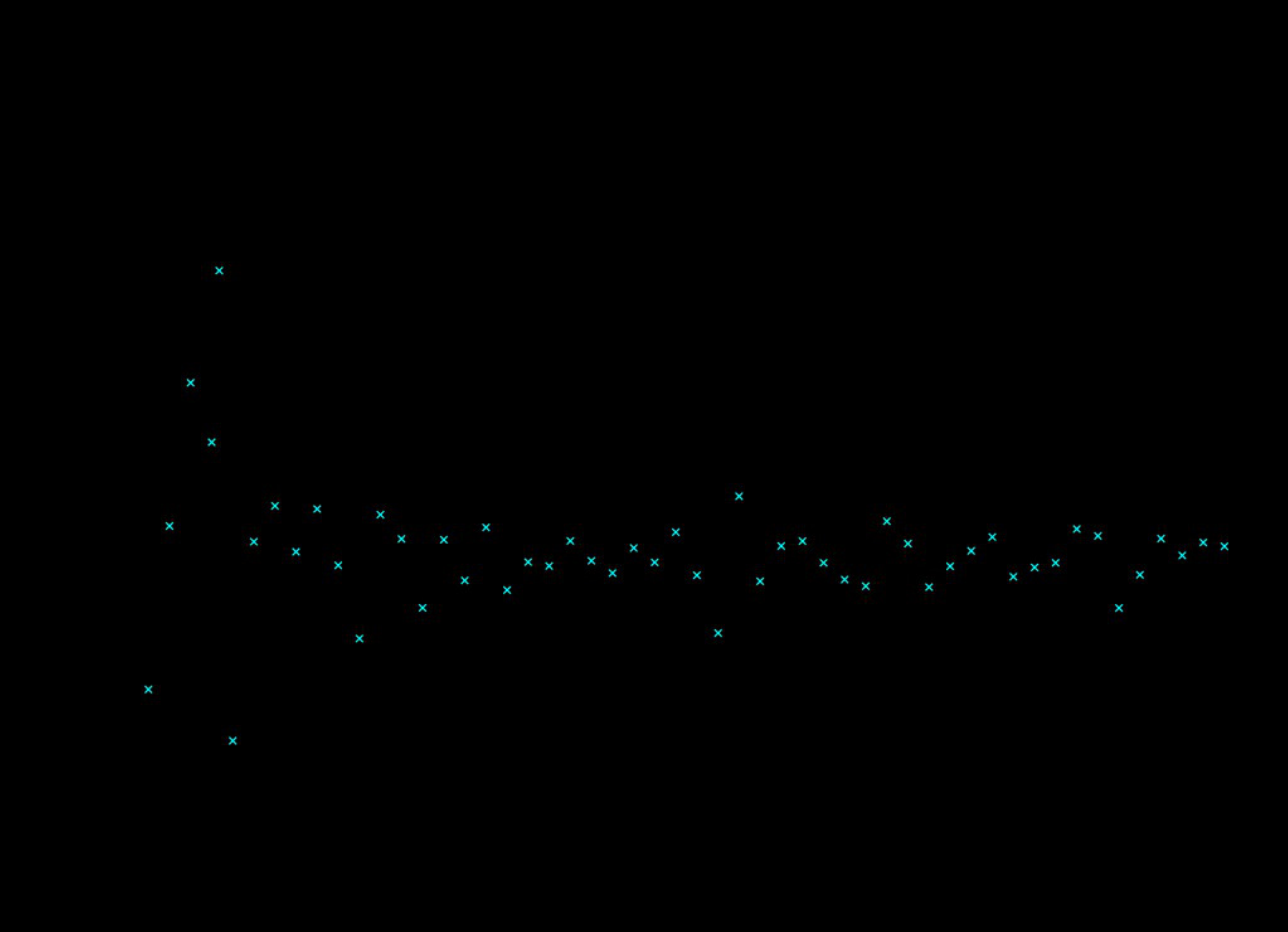}
\caption{Example of fit using 52 $Q$-parameters generated with J-PAS synthetic photometry. 
Black circles corresponds to the $Q$-parameters of a model of the AMBRE library while the red crosses are 
the $Q$-parameters of the star BD292091 of the Next Generation Spectral Library. The main physical parameters 
of the star and of the model are described at the top of the graphic together 
with the $\chi^2$ value of the fit.}
\label{bd292091}
\end{figure}
\end{center}

In Aparicio Villegas et al. (2011) it can be found a brief resume of this method 
while a more complete paper is in preparation. Figure \ref{bd292091} presents an example of one of these 
possible fits using the 52 $Q$-parameters generated with the 53 correlative colors from the 54 J-PAS 
narrow-bands. The values of these $Q$-parameters are shown in the y-axis, while the 
x-axis is just the number of the $Q$-parameter in increasing order; red crosses are the 
$Q$-parameters of the star BD292091 of the Next Generation Spectral Library (NGSL, 
Gregg et al. 2004) and in black circles are represented the $Q$-parameters of one 
model from the theoretical library AMBRE (de Laverny et al. 2012); The main physical 
parameters of both, the star and the model, are described at the top of graph, together 
with the $\chi^2$ value of the fit.

  This methodology will be applied to the J-PAS narrow-band filter system in order to analyze 
the capability of this photometric system in the estimation of the main stellar 
parameters, such as was done for ALHAMBRA, but it can be also adjusted to be focused 
in classifying these metal-poor stars. For that task, it will be necessary the selection 
a priori of the filters which provide a greater information about the chemical composition 
of these objects; this could be handled with a qualitative criterion selecting the 
filters that contain absorption lines, like the lines of ionized Calcium 
at 3933\AA{} and 3963\AA{}, which are contained in filters JSCH3900 and JSCH4000, 
the molecule CH in 4300\AA{}, situated in filter JSCH4300, the infrared triple of 
Calcium at 8498\AA{}, 8542\AA{} and 8662\AA{} contained in filters JSCH8500 and 
JSCH8600, or the Magnesium line at 5183\AA{} in filter JSCH5200. Or with a more statistical 
task, using PCA or generating linear regressions between color combinations and the 
values of stellar metallicity. The subset of filters chosen in each case would have 
a more weight in the fits in order to obtain more accurate metallicity determination.

Although this photometric system probably will not enable to determine the metallicity 
values of these stars with a spectroscopic accuracy, it will certainly allow the 
detection of these objects among all the other point-like objects that will be found 
in the J-PAS survey, and then a posterior high-resolution spectroscopic observation 
of these objects will be necessary to complete the study of their chemical composition 
and refine the results in metallicity with this methodology.

\paragraph{Halo post-AGB population: PNe and   proto-PNe}                                                        
The outer layers of the low- and intermediate-mass stars (0.8-8 M$_{\odot}$), which were enriched throughout their evolution, are ejected during the asymptotic giant branch (AGB) and post-AGB phases, and subsequently ionized by the remnant hot nucleus, forming the so-called planetary nebulae (PNe). Analysis of the ionization lines formed in the ejected envelopes of these evolved stars is one of the most important ways in which the chemical and physical characteristics of the gas are studied (for reviews, see Stasi/'nska 2002; and Magrini, Stanghellini and Gon\c calves 2012).

Such stars can be important sources of enrichment of He, N and C in the interstellar medium (ISM; e.g., Yin et al. 2010). Also, van den Hoek \& Groenewegen (1997) and Marigo (2001) predicted theoretically the possibility that (PNe) progenitor stars produce O and Ne and dredge these elements up to the stellar surface - through the third dredge-up, TDU - resulting in enrichment of the ISM. The latter is a metallicity dependent effect, which happens mainly in low-metallicity environments (Pequignot et al. 2000; Magrini \& Gon\c calves 2009). In any case, PNe and their abundances are clearly important tools to study stellar evolution. By considering elements other than those just discussed, PNe can also reveal the imprint of ISM abundances when their progenitor stars were formed, in the Galactic as well as in the extragalactic context.

About 3,000 PNe are known in the Galaxy (Parker et al. 2012), and only a few, about 20, objects have been identified as halo PNe, from their location, kinematics and chemistry (see, for a recent ref., Otsuka et al. 2010). Halo PNe are able to reveal precious information for the study of low- and intermediate-mass star evolution and the early chemical conditions of the Galaxy.

The progenitors of halo PNe are believed to be $\sim$0.8 M$_{\odot}$ stars, as most of the stars of the halo. However, some halo PNe seems to have evolved from massive progenitors (Otsuka et al. 2010). According to the current stellar evolution models (e.g., Fujimoto et al. 2000) the TDU must take place in the late AGB phase. The TDU efficiency increases with increasing mass and/or decreasing metallicity. At halo metallicities, it is predicted that the TDU is efficient in stars with initial masses greater than $\sim$1~M$_{sun}$ (Karakas 2010; Stancliffe 2010). Moreover, the stellar evolutionary models predict that the post-AGB evolution of a star with an initial mass $\sim$0.8~M$_{\odot}$ proceeds too slowly for a visible PN to be formed. Therefore, clearly, the origin and evolution of halo PNe are still one of the unresolved big problems of stellar evolution.

So, interestingly enough from the point of view of the halo J-PAS survey, the study of the halo PNe - by determining their elemental abundances and ejected masses - will certainly enlighten not only the Galactic chemical evolution at early phases, but also our knowledge of the stellar evolution at environments other than the Galactic disk.

The characteristic low continuum and intense line emissions of PNe make them detectable at the halo distances. For example, the halo PNe BoBn~1, DdDm~1 and PS~1, located somewhere between 11 and 24~kpc from the Sun, have B magnitudes of 16, 14 and 13.4, respectively (Otsuka et al. 2010; Otsuka et al. 2009; Rauch et al. 2002). Such values are easily encompassed by J-PAS, since the typical limit magnitude of the survey will be about 22-23. Given the low number of halo PNe known so far, we plan a follow-up study for any possible candidate identified by the survey. 

\begin{center}
\begin{figure}
\centering
\includegraphics[width=0.8\textwidth]{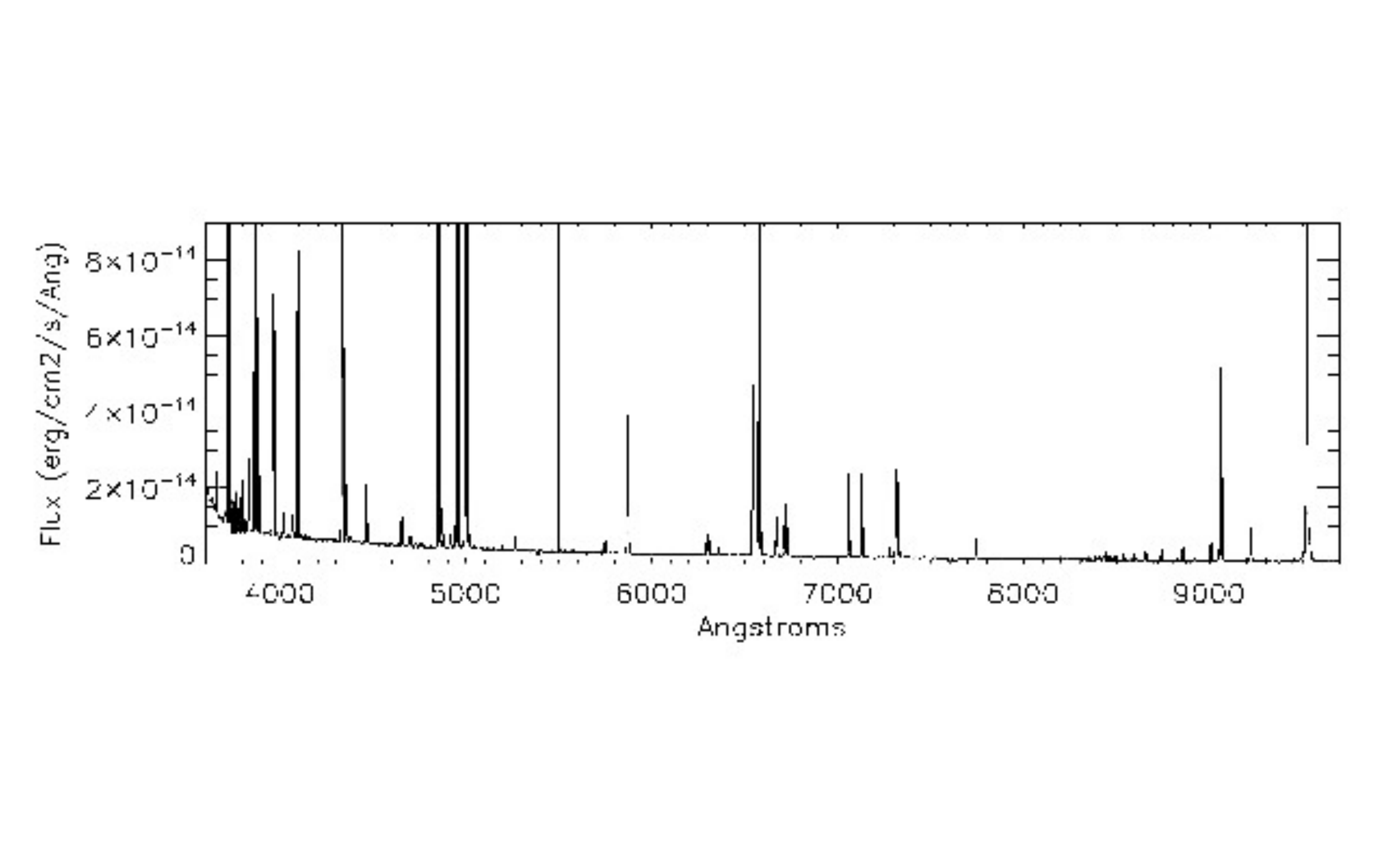}
\caption{Observed medium-resolution optical spectrum of DdDm~1.}\label{DdDm1}
\end{figure}
\end{center}

To explore the possibility of using the J-PAS survey to detect halo PNe, we have convolved typical halo PNe spectra to simulate the corresponding J-PAS ``spectrum'' of these nebulae. In Figure~4.5 we show such a convolution, based on the medium-resolution optical spectrum of DdDm~1 (kindly provided by C. Pereira, and observed with the a configuration similar to that of Pereira \& Miranda 2006), shown in Figure \ref{DdDm1}

\begin{center}
\begin{figure}

\centering
\includegraphics[width=0.8\textwidth]{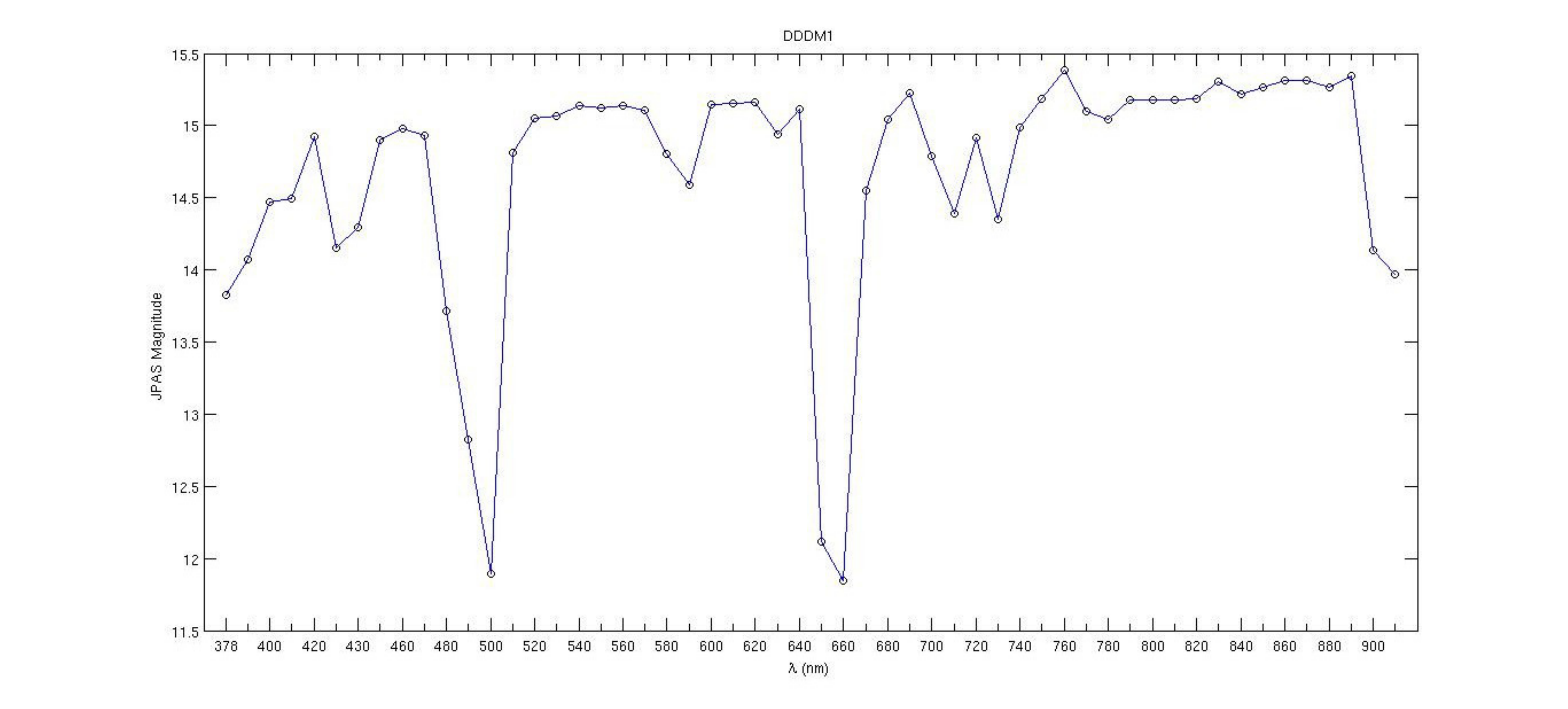}
\caption{J-PAS convolved spectrum of DdDm~1 (J-PAS magnitudes versus wavelength).}\label{DdDm2}
\end{figure}
\end{center}

From Figure \ref{DdDm2} it is clear that the 54+5 J-PAS filters can easily detect the 
low-continuum, strong emission-line objects such as PNe. The intensity of the lines vary substantially with excitation, but using the halo PNe spectra available for our study so far, all of them are detectable by the J-PAS filter set.

There is another interesting class of post-AGB stars that is also likely detectable by survey, the proto-PNe. Theoretical calculations predict that post-AGB stars evolve with constant bolometric luminosity, while their radius shrinks and their effective temperature rises. When the temperature is high enough ($T_{eff}$ $\sim 20,000$ K) to ionize the circumstellar nebula, the emission lines become detectable by spectral sky surveys, owing, first of all, to the presence of the H$_{alpha}$ emission line. Hot post-AGB stars, or proto-PNe stars, are the immediate progenitors of the central stars of PNe. Hot candidate proto-PNe possess dust shells with temperatures ranging from 100 to 250 K, display spectra of early B-type stars with signs of a supergiant plus emission lines, and are usually located outside the Galactic plane (e.g. IRAS 19336-0400; Arkhipova et al. 2012).

On the other hand, in globular clusters, high luminosity F-G type post-AGB stars (yellow post-AGB) were observed. The most famous of those is RAO24 in $\Omega$ Cen (e.g., Gonz\'alez \& Wallerstein 1996). In the GC NGC5986, two A-F supergiants were discovered (Alves et al. 2001). Since the large Balmer jump of these stars is well traceable with a good choice of filters, more extensive specific surveys should discover post-AGB stars in the halo and external galaxies. Bond \& Alves (2001) report upon the detection of yellow post-AGB stars in M31 and its dwarf elliptical NGC 205, yielding a good reproduction of the Cepheid distance to M31. Therefore, these stars are important as tracers of halo structure and dynamics, for studying advanced evolutionary stages of low-mass stars, and to provide new standard candle for measuring extragalactic distances.

We are now working on the construction of a grid of PNe and post-AGB nebular spectra, to widely prove their detectability using J-PAS photometry. Moreover, the latter can be better explored by applying appropriated color-color diagrams (Viironen et al. 2009), as well as performing the Q-fit algorithm, described in Apar\'\i cio-Villegas et al. (2010). The latter will provide the parameters (line ratios, for instance) that can distinguish these nebulae from other strong emission-line objects of the halo. Another important application of the synthetic grid of spectra we are creating is to identify the J-PAS filter ratios that can at least give limits to the abundance of the nebulae detected from the J-PAS spectra, even before the spectroscopic follow-up.

\paragraph{Hunting  Tidal Stream Stars in the Milky Way with J-PAS}[a]
 
Within the hierarchical framework for galaxy formation (e.g., White \& Rees 1978), the stellar bodies of massive galaxies are expected to form and evolve not only through the inflow of cold gas, but also via the infall and successive mergers of low-mass, initially bound systems (commonly referred to as satellites) that span a wide mass range.  As a consequence, the stellar halos of these galaxies should contain a wide variety of diffuse structural features, such as stellar streams or shells, which result from interactions and mergers with dwarf satellites. The most spectacular cases of tidal debris  are long, dynamically cold stellar streams, that wrap around the host galaxy's disk and  roughly trace the orbit of the disrupted progenitor satellite.

$\Lambda$-CDM simulations predict that stellar streams may be detected nowadays, with sufficiently deep observations, in the outskirts of almost all nearby galaxies (Bullock \& Johnston 2005; Cooper et al. 2010). Although the most luminous examples of diffuse stellar streams and shells   around massive elliptical galaxies have been known for many decades (e.g.  Arp 1966), recent studies have shown that fainter analogues of  these structures are common around spiral galaxies in the local Universe (Martinez-Delgado et al. 2010), including the Milky Way (MW) and
Andromeda (Belokurov et al. 2006; McConnachie et al. 2009).

Though it is now clear that these minor mergers likely played a prominent role in creating the halo of our Galaxy, the chemical abundance patterns of current MW "surviving" satellites are typically very different that those of halo stars (Venn et al. 2004)), and the reason for these differences remains controversial (Majewski et al. 2002; Font et al. 2006).The J-PAS survey
plans to test the bridge from dwarf galaxy to halo stars {\it directly} by exploring (for first time photometrically) the chemical trends of an extensive sample of M-star  {\it bona-fide} members of known MW tidal streams (e.g. Sagittarius, Monoceros, Virgo). The membership of these stream
stars was previously vetted by medium  resolution radial velocity surveys. By selecting suitable J-PAS color indices sensitive to abundance patterns of the stream stars (e.g [$\alpha/Fe$], we plan to design a new method to identify new stream stars in the halo field and in the solar neighborhood and to search for abundance variations among known streams. This new census of stream stars will provide unprecedented information on the contribution of these merger events in the formation and chemical  evolution of the Galactic halo.

\subsubsection{Additional topics}

\paragraph{Blue objects in the Galactic halo} 

Just to mention further details on the blue objects in the galactic halo: it would be nice to search for rare stars with B-type spectra (blue objects) in the halo and then select candidates for a spectroscopic follow-up in order to derive elemental abundances and then classify those objects as 1) B stars formed in situ; 2) runway stars ejected from the disk; 3) evolved stars (BHB, hot post-AGBs, ..).

\subsection{Stellar Variability}

The timescale in which the brightness of astrophysical objects vary encompasses a considerably broad time-domain frequency range. These variations are observed with the most diverse shapes and amplitudes, sometimes affecting the entire electromagnetic spectrum, a specific band, a group of lines from chemical species or even a single line. In some cases the variability is periodic (such as on RR~Lyr, pulsating stars and eclipsing binaries) but there are a number of cases where they are not (e.g. eruptive stars).

Despite the large amount of characteristics, the variability on astrophysical objects is classified either as intrinsic or extrinsic. Intrinsic variations are those related to changes in the physical properties of the object itself. These may be related to changes in the shape, temperature, instabilities and even magnetic properties which results in flux variability. Examples of intrinsically variable objects are solar-type and low mass active stars, pulsating stars, dwarf-novae, novae and supernovae.  As the name suggests, extrinsic variations are those related to changes in the medium, surroundings or line of sight to the object. Eclipses and occultations in general are the classical cases of extrinsic variability.

Most variable objects display both types of variability and identifying and characterizing them is a fundamental step in understanding the underlying physics. These tasks are quite complementary and rely strongly on the observational strategy as well as on the properties of the variability itself (amplitude, period/recurrence and shape). Identifying a variable source is a much easier task than characterizing its variability. Usually, identification of a variable source requires only that the system is observed a sufficient number of times such that the scatter in flux, caused by variability, is larger than the scatter caused by uncertainty in the measurement. On the other hand, characterizing the variability - namely, obtaining periods/recurrence time, amplitude and shape - requires a much more detailed analysis. For instance, in case of periodic signals it is required that at least two cycles are observed with moderate time sampling (at least a $1/10$ of cadence) in order to obtain reliable results.

As already mentioned, the J-PAS survey strategy will provide up to
four epochs of observations in each of the 56 filters for each field,
possibly separated by $20-30$ days (see Section~\ref{sec:SN}). In order to estimate the probability of finding variable sources with this setup we performed some Monte-Carlo simulations of variability detection and characterization. We separated the different types of variabilities into three groups;
\begin{description}
\item[Eclipsing binaries:] In this case the variability is caused by occultation of one of the sources of the system by its companion. Since less light arrives at the observer when light is blocked, we observe a decrease in the flux of the system. A typical light curve of an eclipsing binary can be characterized by the duration and depth of the eclipses and the orbital period of the system. 
\item[Pulsating stars:] Most commonly observed in RR~Lyr and specific stages of white-dwarfs cooling evolution, pulsation is modeled by sinusoidal variations, single or multi-periodic, which are characterized by amplitude and period. 
\item[Transients:] This type of variability may represent a vast class of objects, that display eruptive behavior. Here we can include solar and low-mass active stars during flares, dwarf novae and novae during eruption and much more. Also, supernovae may be considered as an eruptive event but this will be discussed in Section~\ref{sec:SN}.  Although the details in the shape of the variability can be quite different, it is possible to roughly model them by considering a simple two-parameter curve, represented by peak luminosity and a decay time ($\rm \Delta T$). The decay is modeled as a Gaussian, and therefore, the decay time is its FWHM.  Since eruptive events are not periodic and may occur in the most different timescales we decided to considered only one such event for each run.
\end{description}

\pagebreak

\begin{center}
 \begin{figure}
 \centering
 \includegraphics[width=\columnwidth]{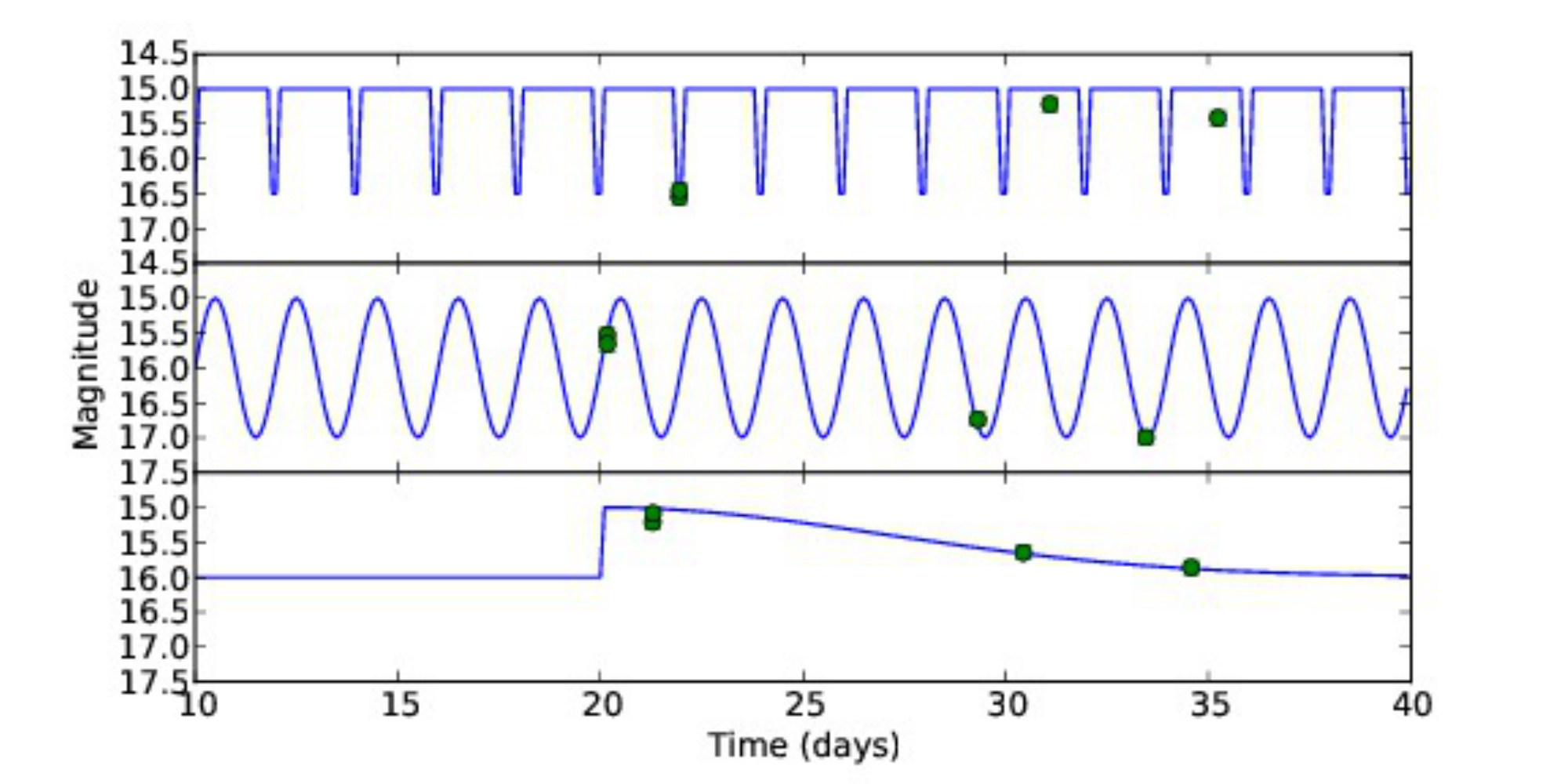}
 \caption{Example of model light curves (solid lines) representing the three types of variability used during the characterization of the survey strategy detectability efficiency. The points with error bars are the modeled observations, taking into consideration the survey strategy for each filter independently.}\label{fig:var01}
 \end{figure}
\end{center}

In Figure \ref{fig:var01} we show an example light curve for each of these groups, with solid-lines representing the models and dots representing the observations. In order to estimate the detectability we generated a grid of light curves representing the three groups. For each light curve from a specific group we selected a different set of parameters, shown on Table~\ref{tab:varpar}. We then introduced uncertainty in the data to represent observations of stars with S/N from $5$ to $220$ (the saturation limit, \url{http://jpaswiki.astro.ufsc.br/mediawiki/images/6/65/JPLUS_Exposure_Times.pdf}).

\begin{table}[ht]
\centering
\begin{tabular}{cccccc}
\hline
\textbf{Var. Type} & \textbf{Parameter} & \textbf{start}& \textbf{end}   & \textbf{step}  & \textbf{\#}\\
E                                & Orbital period                        &        0.04 days                &        1.0 days                &         0.1 days                &        10                \\
E                                & Orbital period                        &        1.0 days                &        10.0 days                &         0.25 days                &        36                \\
E                                & Eclipse Duration                &0.01 1/$\rm P_{orb}$&0.1 1/$\rm P_{orb}$&         0.01                        &        10                \\
E                                & Eclipse Depth                &        0.1        mag                &         1.0        mag                &        0.1        mag                &        10                \\
\hline
P                                & Pulsation period                &        0.04 days                &        1.0 days                &         0.1 days                &        10                \\
P                                & Pulsation period                &        1.0 days                &        10.0 days                &         0.25 days                &        36                \\
P                                & Amplitude                        &        0.01        mag                &         1.0        mag                &        0.02        mag                &        50                \\
\hline
T                                & Decay time                        &        0.04 days                &        1.0 days                &         0.1 days                &        10                \\
T                                & Decay time                        &        1.0 days                &        10.0 days                &         0.25 days                &        36                \\
T                                & Amplitude                        &        0.01        mag                &         1.0        mag                &        0.02        mag                &        50                \\
\hline
\end{tabular}
\caption{Set of parameters used to generate the model light curves for the detectability simulation. Note that the time grids, orbital/pulsation period and decay time, are separated in two sub-grids with different time resolution. This is important so we can better sample the lower end of the grid.}
\label{tab:varpar}
\end{table}

\pagebreak

For each set of parameters and S/N we produced 1000 light curves with different  observing epochs (following the observational strategy of four epochs, two back-to-back, another  $\sim1-$ and  $\sim2$ months sub-sequentially), which resulted in a total of 9\,200\,000 simulated light curves. Finally, if the standard deviation of the modeled observations is larger than $3-\sigma$ of the expected standard deviation, for each S/N (e.g. the magnitude of the object), it is possible to identify the variability of the target. By measuring the proportion between the number of targets for which variability is detected with those that are not, for each set of parameters, we estimate the probability of detecting variable sources. 

\begin{center}
 \begin{figure}
 \centering
 \includegraphics[width=0.3\columnwidth]{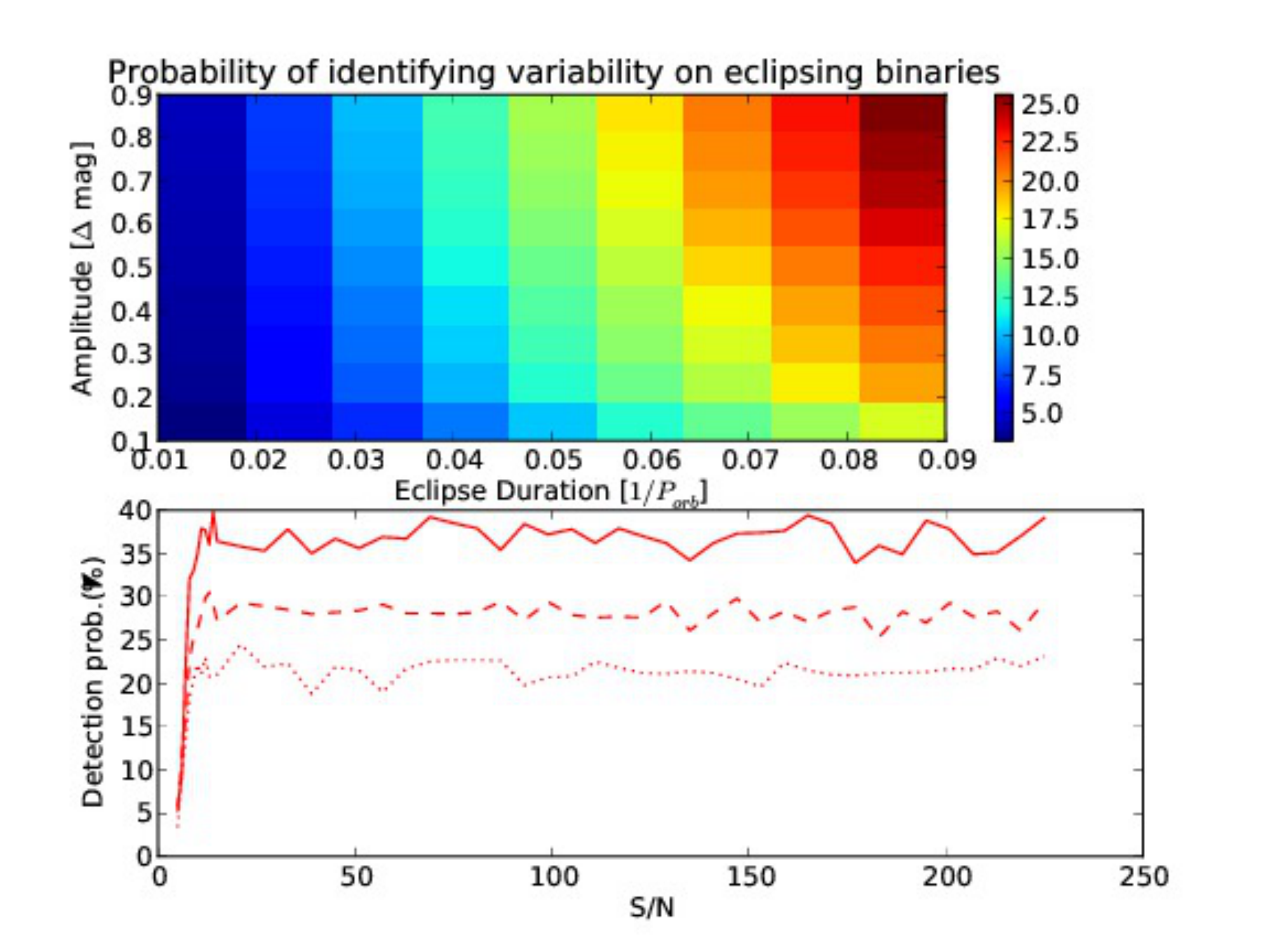}
  \includegraphics[width=0.3\columnwidth]{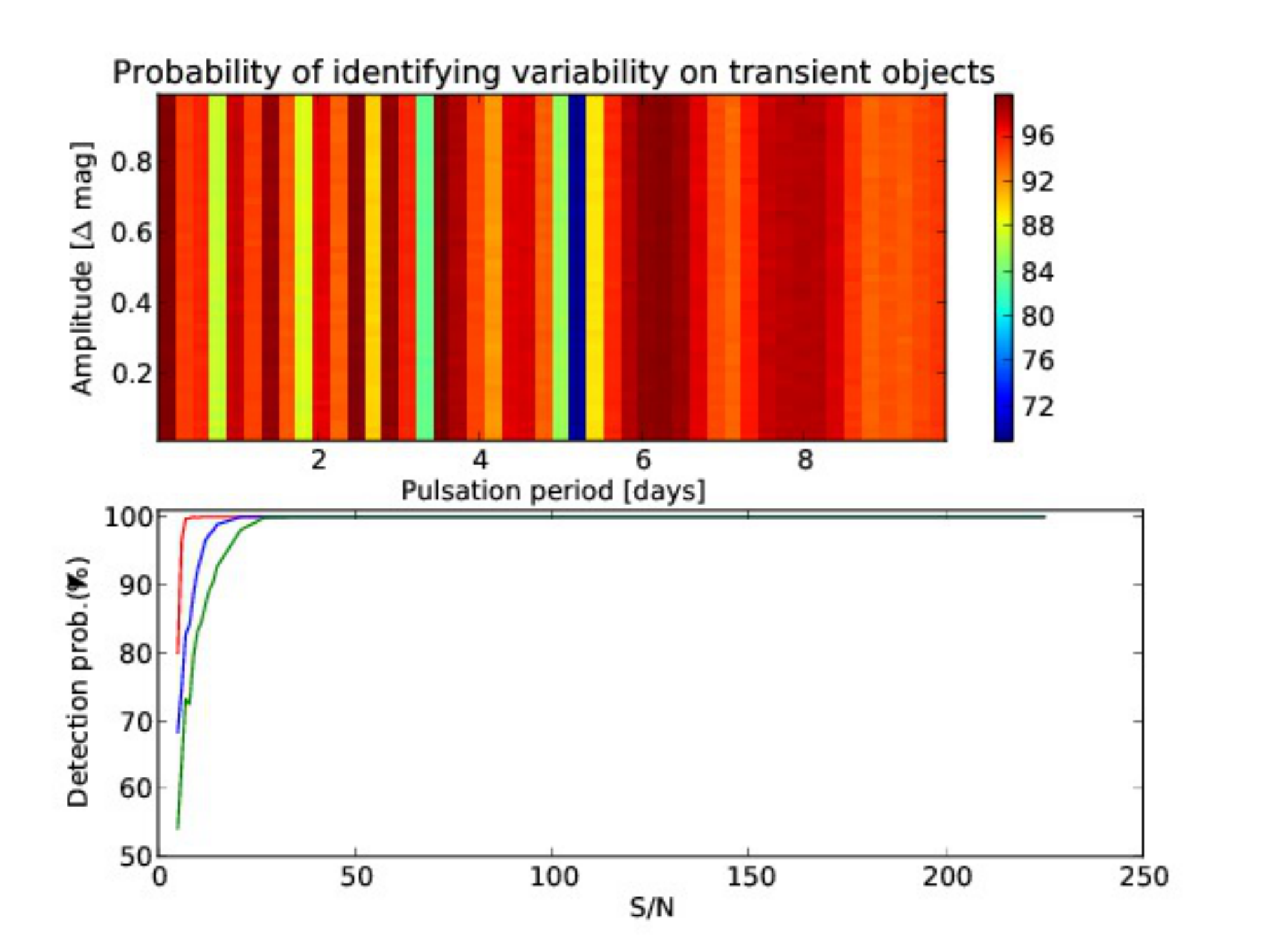}
   \includegraphics[width=0.3\columnwidth]{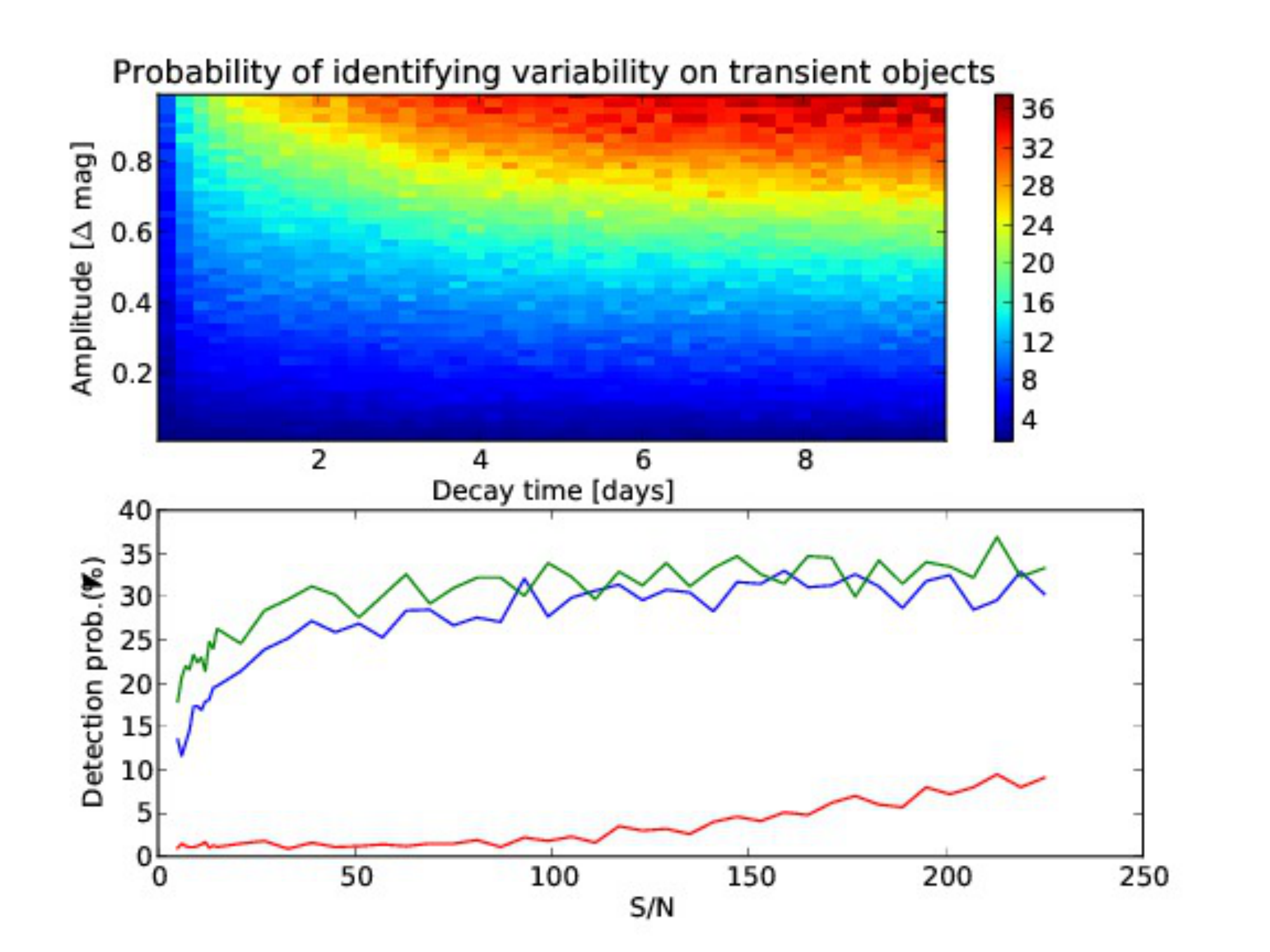}
 \caption{ {\bf Top panels:} Map of the averaged probability of detecting variability, obtained by the Monte-Carlo simulation. From left to right we show the results for eclipsing binaries, pulsations and transients, respectively (see text for details). These maps are averaged over S/N and, in the case of eclipsing binaries, also in orbital period. {\bf Bottom panels:} S/N dependence of the identification probability for a set os parameters.  For eclipsing binaries the solid, dashed and dotted lines are for orbital periods of $0.04 d$,  $1d$ and $9.5d$.}\label{fig:var02}
 \end{figure}
\end{center}

The top panels of Figure~\ref{fig:var02}  show a two-dimensional probability map of identifying variability for the three different types of light curves, for some selected set of parameters, and averaged over S/N. For eclipsing binaries, top left panel on Figure~\ref{fig:var02}, we show the probability averaged over orbital period as well as on S/N. Here average probability ranges from $\sim 4\%$ in the worst case scenarios (shallow and short eclipses) to up to $\sim 25\%$ in the best cases (deep and long eclipses). The case is much better for pulsating binaries where (top middle panel on Figure~\ref{fig:var02}) where we obtain virtually 100\% detection probability for a large number of cases. Surprisingly, there is very little dependency of detecting pulsating objects 
 with respect to the amplitude of the pulsation and a very strange pattern behavior with respect to pulsating period. This is likely to be caused by some sort of resonance between the observations and the pulsating period. For transients, the case is much similar to that obtained by eclipsing binaries where we have better average probability of finding large-amplitude and slow-decay events than finding small-amplitudes fast-decay, namely $\sim 36\%$ and $\sim 3\%$ respectively.

As expected there is a strong dependency with respect to the S/N of the data, as can be seen on the bottom panels of Figure~\ref{fig:var02} for some selected set of parameters. For the eclipsing binary case, bottom left panel,  the different curves are for different orbital periods and same eclipse depth and relative eclipse duration. Here the solid lines are for a $0.04d$ orbital period, and has the larger identification probability. As can be seen, shorter orbital period (which means more frequent eclipses) results in easier variability identification. Typically, the detectability is practically zero for $\rm S/N \lesssim 10$, increasing sharply between $10  \lesssim S/N  \lesssim 20$ and then asymptotically for higher S/N.

Furthermore, in our simulations we considered only the observations of a single filter. It is not straightforward to consider the impact in the identification probability for observations of each target in all the 56 filters. The easiest approach would be to consider the observations individually. In this case the result  would be more or less to increase the detection probability with the square root of the number of filters, e.g. a factor of $\sim \times 7$. A more refined analysis could be to evaluate the impact of variability on each filter for each kind of object. This task is very object-specific and which is beyond the scope of this work.

Overall, and even though the J-PAS survey was not designed for discovering variability, it will definitively provide means to identify a huge amount of sources. Most importantly, even though the variability in those cases won't be characterized (amplitudes, periods, etc.) the sources themselves will have a low resolution J-PAS spectra. This means that, in most cases, it will be possible to characterize automatically the source of the variability making the follow-up work much easier. 

\subsection{Minor planet science with J-PAS} The observation of a large
number (tens of thousands) of minor planets is an added science bonus
to large sky surveys like J-PAS. There are basically two science
cases for minor solar system bodies observed in such surveys: the
detection of previously unknown bodies, and  characterization of the
physical properties of those  bodies. 

\subsubsection{Detection}
The efficiency of the first science case is highly dependent on the
cadence of the survey.  However, even
with the cadence constraints imposed by the main goal of the survey, 
J-PAS images would still be useful for detection of minor bodies. The
2+1+1  observation strategy  that is to be adopted for most of the
survey has implications for its ability to detect and track minor bodies.

During a visit, a pointing position will be imaged repeatedly for 16
(single exposures) or 30 minutes (double exposures), with individual
exposures of 60 seconds.  Therefore it will be possible to detect
main belt and near earth objects within each  visit. Slow-moving
objects, like TNOs, on the other hand will be detected by comparing
positions of the objects on the subsequent, one-month apart,
visits. Considering all filter trays, each pointing position will be
visited 13 times during the 7 years of the survey.

The individual frames, taken at different filters,  have all limiting
magnitudes around $m_{AB}=22.1$ for single visits and $m_{AB}=22.5$
for the visits with two exposures per filter. This last value is close
to the detection limit of the PS1 Pan-STARRS telescope 
\citep{2007DPS....39.0802J}. However, the cadence of Pan-STARRS and
other synodic surveys makes them more efficient than J-PAS for the
discovery of minor bodies. That said, J-PAS can play an important role
in the discovery of minor bodies in general and potentially hazardous
objects in particular, since at this time the efforts to discover
asteroids are based on observations made mostly in North America and
Hawaii, with no major detection facility operating on Europe.  Taken
together with the detections of other surveys, the observations of
minor bodies provided by J-PAS can significantly improve the chance of
detection of potentially hazardous objects.

\subsubsection{Physical properties: phase parameters}

The variation of the observed magnitudes of minor bodies - phase
curves - is a function of both the texture and of the composition of
their surfaces. It is an important parameter, which however is still
not known accurately for most objects.

The phase curve can be measured using  the reduced magnitude
($M_{\lambda}(1,1,\alpha)$, i.e., calibrated
magnitude of the moving object, corrected from distance to the Sun
($r$) and Earth ($\Delta$) against the phase angle ($\alpha$).

The reduced magnitude is easily obtained as
$M(1,1,\alpha)_{\lambda}=M_{\lambda}-5\log{(r\Delta)}$, while
$\alpha$ is the angle, on the
moving object, that subtends the distance Earth-Sun, readily obtained
from the object's orbital data. The technique is fairly simple:
Observe
calibrated magnitudes for different angles $\alpha$ and fit the widely
used $H-G$ relationship \citep{1989aste.conf..524B}
$$H_{\lambda} =
M_{\lambda}(1,1,\alpha)+2.5\log{((1-G)\Phi_1(\alpha)+G\Phi_2(\alpha))},$$
where $H_{\lambda}$ is called the absolute magnitude,
$\Phi_i$ are known function of the phase angle, and $G$ a free
parameter. Note that $H_{\lambda}$ is $M(1,1,0)_{\lambda}$, an
impossible geometrical configuration.

The absolute magnitude is related to both size, $D$, and geometric
albedo, $p_{\lambda}$, of the moving target through
$$D~[km]=C_{\lambda}\times10^{(3-H_{\lambda}/5)}p_{\lambda}^{-0.5}$$,
where $=C_{\lambda}$ is a known constant.
Thus obtaining a large number of $H_{\lambda}$ will allow us to impose
constrains on the size distribution of the populations of minor
bodies.

Typical ranges of phase angles for minor bodies are many dozen degrees
for Near Earth Objects, a few dozen degrees for Main belt asteroids,
and
no more than ten degrees for Centaurs and trans-Neptunian objects. In
this last case the $H-G$ relation could be
simplified to
$$H_{\lambda} = M_{\lambda}(1,1,\alpha)+\beta\alpha$$
\citep[see]{2008AJ....136.1502R}
 because it reaches a nearly-linear region.

The cadence of J-PAS (2+1+1) spread in time could allow to obtain phase
curves for TNOs and Centaurs provided they do not
escape the field. In principle we will only have a maximum of three
points in three different position which can then be fitted using the
simplified version of the $H-G$ relationship. For other populations,
with larger non-sidereal motions, we will rely on serendipitous
observations
of the same object in (likely) different fields in the same filter
and, in this case, fit the full $H-G$ relation due to the, probable,
larger span in phase angle coverage.

\subsection{A target of opportunity program for the T80: Responding to
GRB alerts provided by {\it Swift} and {\it FERMI}}

The T80 mount reaches a maximum angular speed of 8$^{\circ}/s$ with an
acceleration  of 1$^{\circ}/s^2$,  so  it can  point  towards any  sky
direction in less than $\sim50$  seconds. This makes the T80 telescope
very suitable  to respond to rapidly fading  objects, mostly Gamma-Ray
Burst (GRBs).   To this end, it  will be necessary to  connect the T80
control    system    to     the    Gamma-ray    Coordinates    Network
(GCN\footnote{http://gcn.gsfc.nasa.gov/}) through  a dedicated socket.
The  time   necessary  for  the  {\it   Swift}  high-energy  satellite
\citep{2004ApJ...611.1005G}  to  detect  a  GRB  and  disseminate  its
coordinates  through   the  GCN   network  is  $\sim15$   seconds  for
observatories  sited  in  the Iberian  Peninsula  \citep{Jelinek2012}.
Therefore  the  T80  telescope  could  start acquiring  data  at  most
$\sim65$ seconds after the GRB is detected.

The error  boxes provided  by {\it Swift}  seconds after the  GRB have
error boxes  radii of a few arc  minutes, so they are  well covered by
the T80 field of view. In order  to reduce the chip read out time (the
$10k \times 10 k$ CCD takes  20 second to read out) the detector could
be windowed  around the {\it  Swift} GRB error  box.  One of  the most
interesting aspects  of the  T80 poses on  its capabilities  to respond
also to alerts coming from  {\it FERMI(LAT)} and {\it IPN} spacecraft.
The  large field  of view  of the  T80 would  cover most  of  the {\it
  FERMI(LAT)}  and  {\it IPN}  error  boxes  with  a single  pointing,
avoiding inefficient mosaics.

The  GRB observations  would be  carried out  through the  $griz$ SDSS
filters that would  be always mounted at the T80.   Just after the GCN
notice has arrived, a series of $griz$-band loops will be activated at
the T80.   The typical  exposure time  of the first  loop would  be of
$\sim$15s,  $\sim$20s,  $\sim$30s   and  $\sim$60s  in  $griz$  bands,
respectively.  These exposure times would yield AB limiting magnitudes
of  $\sim22$ in  $gri$ and  21.3 in  $z$-band,  respectively (assuming
$S/N$  ratios of  5 in  all bands,  a seeing  of $0.7^{\prime\prime}$,
airmass=1.1 and dark time).  Given that the optical source is expected
to fade, the exposure times  of the following loops would be enlarged.
Thus, in  each band, the exposure  time of the  subsequent $n$-th loop
will be  obtained scaling  the exposure  time of the  first loop  by a
$(t_n/t_1)^2$ factor  (being $t_{n}$ and  $t_1$ the time  elapsed from
the  $n$-th and  first loops  to the  gamma-ray  event, respectively).
This strategy is designed to  keep roughly constant the $S/N$ ratio of
afterglows.

The  combination of  these deep  $griz$-band limits  with the  T80 low
reaction time  are consistent with  detecting a large fraction  of the
afterglows'  population  \cite[see   the  compilation  of  light-curves
  by][]{2010ApJ...720.1513K,2011ApJ...734...96K}.    The   afterglows
detected  in the $griz$-bands  with the  T80 would  allow to  test the
spectral  indexes  and  light-curves  predicted  by  both  the  reverse
\citep{1997ApJ...476..232M}     and     forward    shock     scenarios
\citep{1998ApJ...497L..17S}.

For  long lasting GRBs  (durations $>$  1 min)  the rapid  T80 slewing
capabilities could enable  us to detect also the  prompt optical flash
contemporaneous to the gamma-ray emission \citep{1999PhR...314..575P}.
The non detections would also be useful to constrain the population of
dark   GRBs  \citep{2004ApJ...617L..21J}   and  their   host  galaxies
\citep{2012A&A...545A..77R}.

Considering the  current GRB detection  rate by {\it  Swift} ($\sim90$
GRBs$/yr$)  and assuming the  real-time GRB  visibility of  Calar Alto
\citep[$\sim20\%$      based     on      a      close     geographical
  location,][]{2010AdAst2010E..55G,2012ASInC...7..303G},   we   expect
$\sim 18$ GRBs/yr occurring at Javalambre's night.  If we add the GRBs
detected by {\it FERMI(LAT)} we would expect $\sim 19$ GRBs/yr. Taking
into  account  the  number  of  nights  with  clouds  less  than  50\%
\citep{Moles2010}  we  foresee to  acquire  prompt data  for
$\sim12$ GRBs/yr.

It is interesting to note that  for GRB redshifts in the $3 \lesssim z
\lesssim 6.5$  range, the Lyman-$\alpha$  dropout would be  covered by
the $griz$  bands, so a rough  redshift estimate could  be possible in
the   first   hours   after   the   GRB,   as   done   in   the   past
\citep{2006Natur.440..181H,2005A&A...443L...1T,2009Natur.461.1258S,2009Natur.461.1254T}.
That  would  allow rapid  and  efficient  triggers  at large  aperture
telescopes  (GTC,  VLT,...)   in order  to  determine  spectroscopic
redshifts    based   on    afterglow    metallic   absorption    lines
\citep{2009ApJS..185..526F}.

We  do not  discard that  the  T80 could  also be  triggered by  other
high-energy events that  might mimic GRBs, as already  happened in the
past                           with                          Magnetars
\citep[GRB070610/ SwiftJ195509.6+261406,][]{2008Natur.455..506C}, tidal
disruption                                                       events
\citep[GRB110328A/ SwiftJ164449.3+573451,][]{2011Sci...333..199L},
thermal high-energy mergers \citep[GRB101225A,][]{2011Natur.480...72T}
or                bright                X-ray               transients \linebreak
\citep[GRB120916A/ SwiftJ174510.8-262411,][]{2013MNRAS.432.1133M}.    In
the   future   we   might    also   implement   T80   activations   to
high-energy/gravitational-wave  alerts provided  by  {\it FERMI(GBM)},
{\it LIGO} or {\it VIRGO}.

\FloatBarrier 

\newpage
\section{Photometric calibration}
\FloatBarrier 

  Large scale structure analysis is very demanding on the homogeneity of
the photometry on the whole area of the survey. This imposes strong
requirements on the relative and absolute photometric calibration of
the J-PAS photometry.

 Large sky surveys carried out with large field of view cameras like
J-PAS have to face a series of difficulties in their quest for
obtaining accurate and homogeneous photometry. Some of the problems
are connected with the instrumental setup and others are related with
the variability of the observing conditions along the duration of the
whole survey, which is of the order of several years in the case of
J-PAS. 

 Among the problems connected with the instrumental setup that have to
be controlled and corrected as much as possible we can list the
following ones:
\begin{itemize}
\item Difficulties in obtaining suitable flat-fields. Given the large
  field of view, it is difficult to obtain homogeneously illuminated
  screens for dome flat-fields whereas for sky flat-fields taken at
  twilight the problem is the gradient of the sky illumination.
\item Plate scale variations. Variations in the plate scale (or solid
  angle subtended by single pixels) from centre to edge 
  introduce photometric distortions.
\item Variations of the point spread function (PSF) along the whole
  camera. Although JST and JPCam have been designed from scratch to
  meet the best image quality in the whole focal plane, it is expected
  that some residual variations of the PSF will remain. They 
  have to be taken into account when performing object photometry. 

\item Variations in the transmission curves of the filters. A
  particular problem in J-PAS connected with the large format of the
  narrow band filters is the presence of variations of the central
  wavelength across the filter. This means that objects at different
  positions on the filters are seen through slightly difference
  bandpasses.
\item Pupil ghost. J-PAS filters are very narrow and, therefore, 
  reflect most of the incident light, this light reaches the field 
  corrector and is reflected back to the camera but, generally speaking, 
  on to a different filter with a different bandpass, hence allowing 
   the corresponding wavelength range to pass to the CCD, creating a 
   pupil ghost. After having detected this problem,
  several changes have been introduced in the manufacturing of the
  filters, the coating of the field corrector and the distribution of
  the filters in the filter trays which have almost, but not totally, 
   eliminated this effect. 
\item Temporal variations of the performance of the different
  components (CCDs, filters, mirror coating,...). The stability of the
  system is another important factor in reaching an homogeneous final
  photometric catalog. Any change in the properties of any of the
  components either in the optical path or the detectors will be
  translated in a change in the effective transmission of the overall
  system and, hence, it will impact the photometry of the
  objects. Periodical tests of the performance of the different
  elements will help to control and mitigate these variations.
\end{itemize}

Many of these problems are dealt with during the reduction process
(flat-fielding, photometric flat), nevertheless, some residuals should
be taken into account in the calibration process.

Other problems related with the observational conditions are well known: 

\begin{itemize}
\item Variations of the transparency of the atmosphere. They can appear, 
 with different intensity and wavelength distribution, as a
  result of changes in the concentration of particular molecules
  ($O_2$,$O_3$, $H_2O$) in the atmosphere, dust concentration, or even 
 barometric variations. To illustrate
  the importance of these effects, \citet{Padmanabhan2008} blame ``the
  unmodeled atmospheric variations at Apache Point Observatory'' as
  the main culprit of the limiting errors that they obtained in their
  ubercalibration of the SDSS data.
\item Variations in the sky brightness. Observations at different moon
  phases will have different depth with equal exposure times just
  because of the reduction of the signal to noise due to the increase
  in the background flux. 
\item Seeing variations. The quality of the sky above the OAJ
  in the Pico del Buitre is superb with a median seeing of $\approx 0.5 \arcsec$, but 
our observations will span a wide range of different seeing conditions and they have
  to be combined in an optimal way to impact as little as possible 
  the quality of the photometry.
\end{itemize}

Finally, the ability of the photo-z codes to obtain accurate
photometric redshifts or the success of any spectral fitting procedure
depends on the absolute calibration of the photometry (although in
many cases what really matters is the relative calibration between
filters, i.e.. the accuracy of the colors). This means how accurate is
the transformation from the observed instrumental magnitudes to
calibrated magnitudes, which in the case of J-PAS are AB
magnitudes. AB magnitudes are defined as~\citep{Oke1974, Bessell2012}:

\begin{equation}
  \label{eq:abmag}
  m_{AB} = -2.5\log\frac{\int{f_\nu(\nu) S(\nu) d\nu/\nu}}{\int{S(\nu)
      d\nu/\nu}} - 48.60 =-2.5\log\frac{\int{f_\lambda(\lambda) S(\lambda)
      \lambda d\lambda}}{\int{S(\lambda) \lambda d\lambda}} - 48.60
\end{equation}

where $f_\nu(\nu)$ is the observed absolute flux in erg cm$^{-2}$
s$^{-1}$ Hz$^{-1}$, $f_\lambda(\lambda)$ is the observed absolute flux
in erg cm$^{-2}$ s$^{-1}$ \AA$^{-1}$; and $S(\nu)$ and $S(\lambda)$ is
the transmission curve from energy flux to photon flux. The advantage
of the AB system is that magnitudes are directly related with the flux
in physical units while other systems, as the Vega magnitudes, are
based on the arbitrary fixed value set for the reference source (in
that case, the magnitudes of the star Vega or $\alpha$ Lyr).

In practice, the calibration of the photometry, i.e.. the computation of
the transformation from the instrumental magnitudes to the calibrated
ones or, equivalently, the computation of the zero points, is based on
the observation of standard stars (i.e.. stars with known calibrated
magnitudes) allowing to compare their instrumental magnitudes with the
calibrated ones and, afterward, applying the found zero points to the
rest of objects with unknown calibrated magnitudes. Traditionally,
this has been done interleaving observations of the target fields with
observations of standard stars. However, several reasons make this
procedure unfeasible for a project like J-PAS:
\begin{itemize}
\item Filters in the J-PAS filter system are non standard
  filters. This means that there aren't catalogs of standard stars
  available like there are, for example, for the Johnson
  system~\citep{Landolt1992,Landolt2009} or the SDSS
  system~\citep{Smith2002}. The only alternative is to used
  spectrophotometric standard stars (SPSS) to perform synthetic
  photometry~\citep{Howell1986}. However, SPSS are scarce and even
  scarcer those that have spectrum available in the whole J-PAS
  wavelength range. This will improve in the near future thanks to the
  sample of SPSS that it is being compiled for the \textit{Gaia}
  survey~\citep{Pancino2012}. However, still the density of SPSS will
  be very low and this would mean that the telescope would have to
  move large angular distances from target fields to the closest SSPS
  and this is highly not recommended if one wants to keep the optical
  system as stable as possible.
\item Many of the SPSS are quite bright for a 2.5m telescope.
\item Large CCDs and large FoV cameras are prone to spatial variations
  of their sensitivity, and therefore, one would need to map that
  variation with the SPSS in order to calibrate different areas of the
  focal plane.
\end{itemize}

Already with the SDSS, it was realized that the photometric
calibration of large sky surveys should rely on auxiliary systems
~\citep[like the Photometric Telescope of the SDSS,][]{Hogg2001}. Also
the experience of the SDSS has shown that the problem of calibrating a
large survey with a lot of overlapping exposures can be split in two
steps: first, a relative calibration of the overlapping exposures or
ubercalibration~\citep{Padmanabhan2008}; and then an absolute
calibration which, in an ideal case, would be just a common zero point
for all the objects in the survey. This procedure has been already
applied in current large survey like Pan-STARRS1~\citep{Schlafly2012}.

The calibration procedure that J-PAS will incorporate try to overcome
most of the difficulties described above using a mixture of the
techniques already applied in SDSS and other large surveys. First, an
auxiliary smaller telescope (an 83cm aperture telescope named
Javalambre Auxiliary Survey Telescope or JAST) with a large field of
view camera will image in advance the same area covered by J-PAS using
a special set of filters. The goal of this preliminary survey will be
to identify and classify millions of stars that will serve as
secondary standard stars (SSS) for J-PAS. This preliminary survey will
allow to perform other kind of scientific studies, especially of
galaxies in the local universe, and has been named \textbf{Javalambre
  Photometric Local Universe Survey} or \textbf{J-PLUS} (see
Section~\ref{sec:jplus}). With these SSS we will be able to calibrate
each single exposure of J-PAS as well as tiles made of a combination
of exposures. When a enough large area of J-PAS has been observed in
any filter with at least 4 exposures, we will apply the
ubercalibration to homogenize the relative calibration in that
filter. Given the particular disposition of the filters in the JPCam,
this situation will happen with at least 14 filters close in
wavelength. This fact in combination with the use of SPSS falling in
the observed area, will be used to tied the relative calibration
between filters. And additional technique that will be used is that of
the stellar locus in the version developed
by~\citet{Kelly2012}. Finally, during the computation of the
photometric redshift, another tool that will help to improve the
photometric calibration (for photo-z estimation) will be the procedure
developed by~\citet{molino13}, who making used of the galaxies
identified as emission line galaxies by the BPZ code~\citet{benitez00}
compute offsets in the zero points which improve the resulting
photo-z's providing values close to those obtained calibrating with a
sample of spectroscopic redshifts.

In the following sections we will describe in more detail the key
points in the calibration procedure for J-PAS.

\subsection{Calibrating J-PAS with J-PLUS\label{sec:jplus}}

\subsubsection{Description of J-PLUS}

 The \textbf{Javalambre Photometric Local Universe Survey}
(\textbf{J-PLUS}) will be a preliminary survey that will be carried
out from the OAJ with the main goal of producing a catalog of millions
of stars in the same area of J-PAS with accurate spectral type and,
hence, accurate synthetic J-PAS magnitudes. As a side effect, the data
resulting from J-PLUS will be used for many other astrophysical
research with particular attention in the study of galaxies in the
local Universe.

J-PLUS will be carried out with a 83cm-aperture telescope (the
\textit{Javalambre Auxiliary Survey Telescope} or \textit{JAST}) with
a field of view of 1.7$^\circ$ (diameter) with full performance and
2.0$^\circ$ if some vignetting is allowed. The camera mounted in the
telescope (T80Cam) will have a large format CCD providing a plate
scale of 0.55''/pixel.

The key feature of J-PLUS will be the set of filters. The J-PLUS
filter system (J-PLUS FS, hereafter) will consist of 4 Sloan filters
(g, r, i ,z) and 8 especially designed filters with different
purposes:
\begin{itemize}
\item an $u_J$ filter which is a modification of the Sloan u for
  better performance at these wavelengths;
\item 5 filters located in particular absorption features: Ca HK lines
  ($\lambda_c=3950\AA$), H$\delta$ ($\lambda_c=4100\AA$), G band
  ($\lambda_c=4300\AA$), Mgb-Fe band ($\lambda_c=5150\AA$) and Ca
  Triplet ($\lambda_c=8610\AA$);
\item 2 filters in two regions directly related with the star
  formation in local galaxies: [OII] ($\lambda_c=3780\AA$) and
  H$\alpha$ ($\lambda_c=6600\AA$).
\end{itemize}

The election of filters located in the absorption features was based
on the preparatory work for the \textit{Gaia}
survey~\citep{Jordi2006a} in which it is explained how these bands and
combinations between them can be used to determine different stellar
parameters.

Meanwhile, the [OII] and H$\alpha$ filters were selected for studies
of the star formation in galaxies in the local Universe. Be aware that
the filter at $\lambda_c=5150\AA$ can be used also for this purpose
for galaxies with $0.001<z<0.0485$ using the [OIII] line at
$\lambda=5007\AA$.

 Monte Carlo simulations with a library of theoretical stellar spectra
have shown that with S/N per filter larger than 50 it is possible to
recover the spectral type with enough accuracy. 

The exposure times on the J-PLUS filters have been chosen to reach at
least $m_{AB}=18$ with S/N=50 in all the filters. For some filters
with additional scientific value, apart from the calibration purposes,
we have increased the exposure times:
\begin{itemize}
\item For the $u_J$ we aim to reach $m_{AB}=23$ with S/R=3.
\item For g and r we will reach $m_{AB}>23.2$ (S/N=3). For these 2 filters we
  will obtain 3 sets of 3 additional exposures with a time  gap
  between each set of one week. This will help to study transient
  objects and to reach higher depth.
\item For $H\alpha$ filter we will aim to $m_{AB}\sim22.6$ (S/N=3).
\item For $i$ band the goal is $m_{AB}\sim22.3$ (S/N=3).
\item For $z$ band the goal is $m_{AB}\sim21.5$ (S/N=3).
\end{itemize}

\subsubsection{Calibration of J-PLUS}

The J-PLUS survey strategy will be focused on minimizing the variation
of observing conditions of all the exposures on the different filters
in each single pointing. For this reason, the imaging of each patch of
the sky will consist of series of 3 exposures with small dithering in
each filter (except additional exposures in the g and r filters that
will be done at three additional epochs to detect transient objects).

The calibration of the J-PLUS images will rely on several exposures
each night of one or more spectrophotometric standard stars that will
be used to set the photometric zero points. In principle, the
determination of the atmospheric extinction coefficients could be done
with the same SSS given that they are observed at different
airmasses. However, we decide to use an specific extinction monitor
system. The next section describes the procedure used to determine the
atmospheric extinction coefficients.

\subsubsection{Determination of the atmospheric extinction}

The monitoring of the atmospheric extinction is one of the key issues
in the current era of large sky surveys. \citet{Padmanabhan2008} point
out the unmodeled variations of the atmospheric conditions as the
main source of the remaining uncertainties of the ubercalibration that
they applied to the SDSS DR8 data.

The main components affecting the transmission of the atmosphere in
the optical wavelength are three~\citep{Hayes1975}:
\begin{itemize}
\item Rayleigh scattering.
\item Absorption by molecules, in particular ozone and water.
\item Mie scattering by small aerosol particles.
\end{itemize}

They have different dependence with the wavelength and with different
atmospheric parameters. For a given location we can simplify their
functional dependence with the wavelength like this:
\begin{itemize}
\item Rayleigh scattering: $\kappa_R \sim \lambda^{-\alpha_R}$
\item Ozone absorption: $\kappa_O \sim  k_O(\lambda)$.
\item Aerosols scattering : $\kappa_{ae} \sim \lambda^{-\alpha_{ae}}$
\end{itemize}

where $\kappa$ are the extinction coefficients of each component,
$\alpha_R$ and $\alpha_{ae}$ condense the wavelength dependence of the
Rayleigh and aerosols scattering and $k_O(\lambda)$ is the shape of
the ozone absorption band. Once $\alpha_R,\, \alpha_{ae}$ and
$k_O(\lambda)$ are determined for a given location they are assumed to
be constant or with a small temporal variation. The determination of
the wavelength dependence of each component can be done
observationally taken spectra of stars with known spectra outside the
atmosphere or can be modeled using a code of atmospheric radiative
transfer with the suitable parameters\footnote{An example of open source radiative transfer code
  is libRadtran (http://www.libradtran.org/).}.

\begin{figure}[htb]
  \centering
  \includegraphics[scale=0.5]{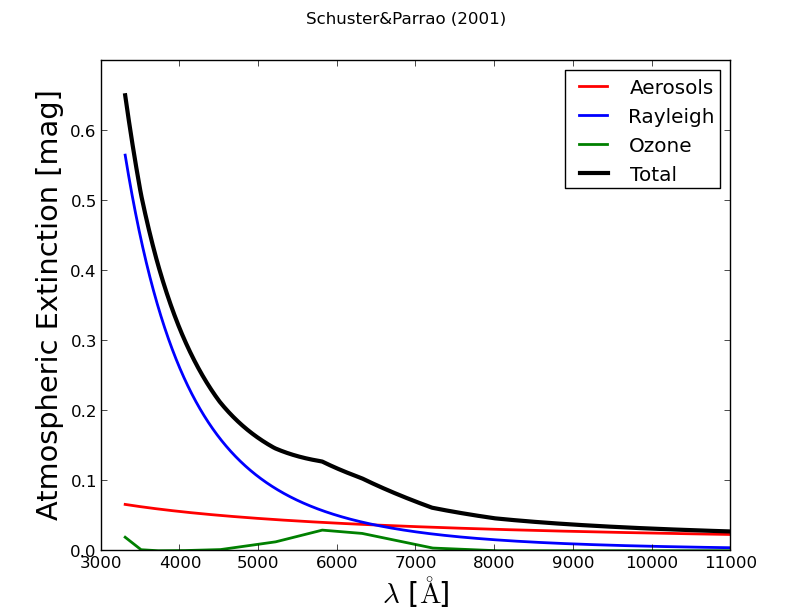}
  \caption{Illustration of the 3-components atmospheric extinction
    curve derived from \citet{Schuster2001}. The blue, green and red
    lines corresponds to the three components: Rayleigh scattering,
    ozone absorption and aerosols' scattering. The black line is the
    total combined absorption. }
  \label{fig:schuster2001}
\end{figure}

The total atmospheric extinction curve will be the combination of
these 3 components (see Fig.~\ref{fig:schuster2001}):
\begin{equation}
  \label{eq:atm_model}
  \kappa_{Total}(\lambda) = A_R \lambda^{-\alpha_R} +  A_O k_O(\lambda)
  + A_{ae} \lambda^{-\alpha_{ae}}
\end{equation}

The coefficients $A_R$, $A_O$ and $A_{ae}$ represent the relative
importance of each component and their variation will be the
responsible of the changes in the overall atmospheric extinction. In
the case of $A_R$ the main factor affecting it would be the local
atmospheric pressure while for $A_O$ and $A_{ea}$ will be the
concentration of ozone and aerosols in the atmosphere. 

The task of the extinction monitor will be to allow the determination
of the three parameters $A_R$, $A_O$ and $A_{ae}$.

Currently, we have installed in the OAJ an extinction monitor
developed by Dr. J. Aceituno called EXCALIBUR (\textit{EXtinction
  CAmera and LumInance BackgroUnd Register}). This consists on a
commercial 11'' aperture telescope with a SBig ST10 2184$\times$1472
pixels camera and a set of 10 filters. The filter set includes Sloan
u,g,r,i and 6 additional medium band filters at centered at
wavelengths 4250\AA, 4800\AA, 5200\AA, 5900\AA, 7000\AA~and
8700\AA. The goal of this set of filters is to suitably sample the
atmospheric extinction curve.

The nightly procedure will consist in the observation of standard
stars at different airmasses with the full filter set. This will allow
to compute the extinction coefficients in each of the bands
corresponding to the filter set. Then, Equation~\ref{eq:atm_model}
will be fitted to the observed extinction coefficients and the fit
will provide the nightly values of $A_R$, $A_O$ and $A_{ae}$ and,
therefore, characterizing the nightly atmospheric extinction
curve. 

Once $\kappa_{Total}(\lambda)$ is known, it is possible to obtain the
extinction coefficients in any other band like those of J-PLUS or even
J-PAS (although some filters of the J-PAS will be highly affected by
narrow absorption telluric lines, especially from water, that cannot
be determined with this procedure).
 
\subsubsection{Stellar spectral fitting}

At the end of J-PLUS we will have absolute calibrated photometry in
the J-PLUS filter system (FS) for millions of stars in the same area
as J-PAS. However, except for the H$\alpha$ and the [OII] filters,
J-PLUS FS and J-PAS FS are different and we need a way to transport
the calibration in the J-PLUS FS to the J-PAS FS. To do this we will
rely on the spectral fitting of the J-PLUS photometry of the SSS with
a library of stellar spectra. Nowadays, there are several empirical
and theoretical stellar libraries that can be used for this
task. Empirical libraries like \citet{Pickles1998} and
MIUSCAT~\citep{Vazdekis2012,Ricciardelli2012} have the advantage of
being based on real data, however, they don't cover the parameter
space as much as the theoretical libraries like those of
\citet{Coelho2005}. Simulations will be carried out to check which one
of the two possibilities or even an hybrid solution provide the most
reliable reconstruction of the real spectra of the stars from the
J-PLUS photometry.

The output of the spectral fitting will then be used to compute the
synthetic magnitudes in the J-PAS filter system.

\subsubsection{Calibration of the J-PAS exposures}

The final calibration of the J-PAS exposures will be done identifying
the SSS from J-PLUS in each J-PAS exposure. The zero point for each
exposure will be a robust average  of the difference between the
synthetic magnitudes of the SSS and their instrumental magnitudes:
\begin{equation}
  \label{eq:jpas_cal}
  ZP = \langle m^{SSS}_{synth} - m^{SSS}_{instr}  \rangle
\end{equation}

In the construction of the final tiles in which several different
exposures are combined, instead of calibrating each single image
independently, the procedure will consist in, first, matching the flux
of all the exposures using stars in common in different exposures
(given the survey strategy this is more complex than just combine 4
exposures), and after the combination it will proceed with the
calibration of the final tile using the same procedure as with the
single exposures.

\subsection{Improvements to the J-PAS calibration}

With the advent of large sky surveys, and especially the SDSS, new 
procedures to improve the overall calibration of the survey have been
developed. Two of these have been already applied with success and
will be included in the calibration procedure of J-PAS: the
ubercalibration~\citep{Padmanabhan2008} and the stellar
locus~\citep{High2009}. And additional procedure developed by members
of the J-PAS team within the ALHAMBRA project will also help to
improve the final quality of the photometric redshifts provided by
J-PAS~\citep{molino13}.

We describe briefly each of these procedures below. 

\subsubsection{Ubercalibration}

The ubercalibration~\citep{Padmanabhan2008} is an \textit{a
  posteriori} calibration procedure which takes advantage of the
overlapping of many exposures in the survey. The result is an
improvement of the relative calibration of the exposures included in
the analysis and, for example, these authors reach a $\sim 1\%$
precision in the relative calibration in all the bands of the SDSS
DR8, except for the $u$ band where the uncertainties in the
atmospheric extinction where larger and the precision was $\sim 2\%$.

Summarizing the description done for these authors, the photometric
model that relates the \textit{relative} calibrated magnitudes ($m$)
with the instrumental ones ($m_{ADU}$) can be expressed mathematically
as:
\begin{equation}
  \label{eq:photmodel}
  m = m_{ADU} + a(t) -k(t)x + f(i,j;t) + ...,
\end{equation}

where $a(t)$ is the photometric zero point and describes the temporal
behavior of the optical response of the system, while $f(i,j;t)$
describes the spatial variation where $i,j$ are the coordinates in
the CCD. On the other hand, $k(t)$ is the atmospheric coefficient
extinction per unit airmass ($x$). The actual functional form of any
of these terms and the number of parameters to be determined in the
model will depend on the amount of effects that the model should
describe and their complexity. In the most simple model, $a(t)$ and
$f(i,j;t)$ would be constant (for a perfect system) and $k(t)$ will
change from night to night. In more real systems, $a(t)$ will be
constant for rather long periods and $f(i,j;t)$ will have a spatial
dependency but not a temporal one. In this case, $a(t)={a_\alpha}$
where $a_\alpha$ is the photometric zero point in each of the period
of constant value, $k(t)={k_\beta}$ where $k_\beta$ is the extinction
coefficient for each night and $f(i,j)={f_\gamma}$ where $f_\gamma$ is
the value in particular regions of the CCD (one would compute $f(i,j)$
in a grid instead of in each single pixel).

With this information, we can solve for the parameters
${a_\alpha,k_\beta,f_\gamma}$ with a $\chi^2$ minimization procedure
given that:
\begin{equation}
  \label{eq:chi2}
  \chi^2[a_\alpha,k_\beta,f_\gamma] = \sum^{n_{star}}_i \chi^2_i,
\end{equation}
and $\chi_i^2$ is:
\begin{equation}
  \label{eq:chi2_i}
  \chi^2_i = \sum_{j\in O(i)}\left[ \frac{m_i - m_{j,ADU}-
      a_{\alpha(j)}+k_{\beta(j)}x - f_{\gamma(j)}}{\sigma_j}\right],
\end{equation}

where $j$ runs over the multiple observations, $O(i)$, of the $i$th
star and $\sigma_j$ is the error in $m_{j,ADU}$.

The success of the procedure will depend on:
\begin{itemize}
\item the goodness of the model to represent the real behavior of the
  system;
\item the complexity of the system (increasing the number of
  parameters);
\item the stability of the system (less stable implies the need of
  more parameters in the model);
\item the amount of observations and, in particular, the amount of
  observations of the same star in different conditions like different
  CCD location, different night, flat field, etc.
\end{itemize}

To provide an idea of the degree of complexity of a photometric model
similar to the one that would be needed for J-PAS we can take a look
to the work done for the first 1.5 years of Pan-STARRS1
data~\citep{Schlafly2012}. For a model that includes the
characterization of the photometric zero point (one parameter for each
night), the atmospheric extinction (one parameter for each night), the
illumination correction (8 parameters for each of the 60 CCDs) and an
additional correction for the seeing (2 parameters for a quadratic
model), the authors construct a model with $\sim900$ parameters for
each independent filter. The authors show that it is possible to reach
accuracies of 10mmag with this procedure.

\subsubsection{Stellar locus and spectro-ubercalibration}

The shape of the spectral energy distribution (SED) of most of the
stars is basically that of a blackbody with a temperature given by the
effective temperature of the star (modified by the content in
metals). This makes that the colors of most of the stars follow tight
relations when plotted in color-color diagrams. The location of the
majority of the stars in these color-color diagrams is called the
stellar locus. \citet{Covey2007} computed the location of this locus
in the color space defined by the ugriz Sloan bands and the JHK$_S$
2MASS bands for a sample of more than 300,000 stars in common in both
surveys. \citet{High2009} developed a full calibration procedure
making used of the stellar locus. \citet{Kelly2012} extended the
stellar locus method to the case in which the observed photometric
bands differ from those used to compute the standard stellar locus
(mainly the SDSS bands) with the help of a library of stellar spectra.
This will allow to apply the stellar locus procedure to the particular
filter set of J-PAS.

However, J-PAS will go beyond the color-color diagrams and it will
obtain accurate stellar spectral classifications. The comparison
between the observed J-spectra of millions of stars with libraries of
stellar spectra will help to obtain a relative calibration between
different bands including the effects of the galactic extinction. The
result will be a spectro-ubercalibration.

In the spectro-ubercalibration, white dwarves (WDs) will play an 
important role because of their relative simple spectra and the degree
of accuracy that have been reached by synthetic models~\citep{Oke1974,
  Bohlin1996, Koester2010}. \citet{Kleinman2013} have identified and
classified more than 10,000 WDs. J-PAS will be able to
identified many more WDs (this means several WDs per square degree)
providing a grid of reliable spectroscopic anchors for the J-PAS
calibration.

\subsubsection{Photometric redshifts}

  One of the main goals of the photometric calibration is to reduce as
much as possible the uncertainties in the computation of the
photometric redshifts. The calibration of photometric redshifts can be
improved by adjusting the photometric zero-points with the help of 
a subsample of galaxies with known spectroscopic redshift
\citep{coe06}. However, in a survey as large as
J-PAS it is not trivial to have  enough galaxies with a wide enough
spectroscopic redshift range in all the pointings~\citep[regarding this issue
for future large photometric surveys, see][]{Newman2013}.
Fortunately, as ~\citet{molino13} have shown, it is possible to accurately 
calibrate the photometric zero points using the galaxies that are 
identified as emission line galaxies by the BPZ code. The offsets computed with these 
emission line galaxies not only improve their photo-z (by construction) 
but also improve the determination of the photo-z of galaxies with very different SEDs like 
early type galaxies. 


\FloatBarrier 

\section{Survey Operation}
\FloatBarrier 

\subsection{Introduction}

The daily operation related to the J-PAS survey is strictly related with 
the daily activities of the OAJ and the UPAD (see appropriate sections in this
paper).
In general terms, one can think about each night as a normal observing run,
with its main components: target selection, acquisition of the proper calibration
frames, scientific observations and data reduction.
Nevertheless, the large amount of pointings required in the survey and the 
enormous data rate make it impossible to manage the whole survey in ``classical
mode'' and the workflow has been streamlined in such a way that it can be 
automated as much as possible.
This section describes the operation flow. Since the flow has a period of
24\,hours, the description arbitrarily starts with the day-time operations.

\subsection{Daytime Operation}

\subsubsection{Data Transfer and Analysis}

Immediately after an observation has been taken, the science frame is 
stored in the OAJ/CPD. The same image is also sent via radio-link to the
UPAD, where the data is going to be reduced and analyzed (see data reduction
section of this paper).

\subsubsection{Data Validation}

As in any observation, weather or technical issues (e.g. focus
change) can affect an observation. Therefore, the validation of
the data is crucial for the completion of the survey.
It is important to note that, since observations are obtained every
(about) two minutes, it is almost impossible to check all of them 
for people on the mountain (more in the ``night time operation'' 
section). It is foreseeable that only a sample of images will be
visually inspected even in Teruel. 
To automate the data validation, 
when an image has been fully reduced, the system checks
basic parameters of the image: if it fulfills the quality criteria set for the
survey (both in terms of seeing and depth), the observation is 
considered as ``valid'' otherwise, it is set as ``to be repeated''.
For operational reasons,
we estimate that all the data obtained during a night have been validated
by noon of the following day.

\subsubsection{Night Scheduling}

After the data validation, the scheduling of the following night
can be started. This deals with checking the ``survey progress''
table, which not only includes the ID and coordinates of the 
fields to be observed during the project but also information on
when the field was observed and a quality flag (basically a field
can be either ``observed'', ``to be observed'' or ``to be repeated'').

This task is performed by a software (the ``scheduler'') 
which runs through all the targets and computes a series of
figures of merits taking into account the minimum airmass of
a field during the night, moon distance, and need for repeated 
observation (see 
Ederoclite et al. 2012, SPIE, 8448, 1).
The total figure of merit is obtained by combining the partial figures of
merit.
The fields are then ordered on the basis of their figures of merit (the 
fields which are not observable are not included) and added to the
``night target list'', which is then sent to the observatory (more in the
``night-time operation'' section).

\subsubsection{OAJ Telescopes' Afternoon Calibrations}

Every day, bias frames are obtained before the opening of the telescope.
At sunset (or sunrise), twilight flat fields are obtained on a daily basis (depending
on cloud coverage). Dome-flats are not foreseen as the field of view of the OAJ
telescope is too large to guarantee a uniform illumination. More in the ``calibration
plan'' section.

\subsection{Night-time Operation}

\subsubsection{The Weather and the Astronomical Conditions Monitoring}

For normal astronomical observations, it is of the highest importance 
to monitor the weather and the ``astronomical conditions'' (seeing, extinction,\ldots).
For this purpose, at any given time, the telescope operators have continuous
access to a webpage with the weather conditions (temperature, humidity,
wind speed and direction). Humidity and wind speed are the two most 
relevant values for observations, as the telescopes must close with 90\%
humidity and with 18m/s of wind.

The astronomical conditions are constantly monitored with a RoboDIMM,
an extinction monitor (``Excalibur''), and an all-sky camera (``AstMon''),
which will provide the seeing, the atmospheric extinction and the cloud
coverage.

\subsubsection{The observing queue}

It is of the highest importance to be able to take real-time decisions
during an observing run. 
During a large survey, it is important that the decisions take into account
a series of factors which have to do with contingencies (like the weather)
but also strategical (like the possibility to re-observe a field, depending on
the observation strategy which is foreseen).
In a project which deals with thousands of pointings it is not possible to 
carry out such an effort manually (not even after the ``scheduler''
has reduced the amount of observable targets for a night).
A software (the ``sequencer'') is therefore in charge of preparing the
observing queue taking into account: the current pointing of the 
telescope, the time, the position of the moon (if present) and the weather/astronomical
conditions as given by the monitors (see ``Weather and the Astronomical Conditions Monitoring''
section).
The sequencer takes a few seconds to run and is executed every 
hour (TBD) and prepares the observing queue for the following hour.
Obviously, the more often the software is executed, the better will
be the choice of the targets. This is done in order to take into account
the possibility of weather change during the night which, obviously, make
pointless the definition of an observing queue only once at the beginning
of a night.

\subsubsection{Observation Execution}

The observing queue is ordered in such way that the first target of the
list is the most suitable target for the observation. 
The observatory control system gets the information from the observing queue
and moves to the field. 

In general terms, the telescope is going to point to another target only few times 
during a night. Most of the observations will be comparatively small offsets.

When the T250/JST moves to a pointing, the telescope gets to position, it opens
the shutter, sets up the guiding and the wavefront sensing.  In fact, when the
shutter opens, it allows some light into the science CCD array, the wavefront sensors
and the guide camera.

\subsubsection{Data Quick Look}

When a scientific observation is obtained, it is moved from the computer control
workstation to the OAJ/CPD. Here among other things (see UPAD section), the 
image is reduced for quick look. The main difference between this reduction
and the ``final'' data reduction is the availability, in the UPAD, of the ``main
calibration frames''.

This quick look is used for the telescope operators to judge the performance
of the telescope/camera system.
The quick look will also continuously report a series of basic measurements
like FWHM (to be compared with the seeing measured through the DIMM, see 
``Weather Station'' section), the sky background, and the ellipticity.

The analysis of the image $i$ of a night happens during the image $i+1$
and, therefore, no reaction can happen before image $i+2$.

\subsubsection{The calibration plan}

As in any observatory, the telescopes at OAJ have a calibration plan,
which is meant to deal with both the scientific data reduction and the
health-check of the telescope/camera systems.
Each of the values being monitored will be accessible through a 
dedicated webpage.

Bias frames are obtained daily, before opening the telescopes. Sky flat fields
are obtained on a daily basis as well (depending on the cloud coverage).

The readout noise is measured from the difference of bias images. Since bias
frames are obtained daily, we plan to perform this measurement on a daily basis
as well.

The gain, readout noise and linearity of the CCD are fundamental parameters in 
data reduction. We do not expect these value to vary significantly, and therefore
the monitoring will happen on a weekly basis. It is important to notice that, for these images,
dome-flats are normally used. Nevertheless, a ``proper'' dome-flat 
(i.e. useful to flatten science images) is not viable for our telescopes and, therefore,
the ``dome-flats'' will be used by projecting light to the dome, which is a good
enough approach to perform an illumination test, where the stability of the light
source is more important than its uniformity.

Dark images are not expected to be required in a nitrogen-cooled CCD. For 
monitoring purposes, we will take darks once per month.

It is assumed that the shutter will not deteriorate quickly and, therefore, 
a test of the shutter opening is only foreseen to happen every three months.
Roughly with the same cadence, a fringing frame is going to be created.

 Bright time (less than 2\,days from full moon), which is not suitable for the 
survey, is going to be used for calibration purposes. 

The image quality (i.e. the fwhm, the ellipticity of the images and the sky-background) is going to be measured for each science image (both at the 
stage of quick look and of ``final reduction''). Each science image will also be ``astrometrized'' with respect to the USNO catalogue.

The photometric calibration is treated in a dedicated section.

\subsection{Data Publication}

A project like J-PAS, with many researchers distributed all across the world, needs
a way to distribute the data. Moreover, the success of such a project is directly related
with the amount of researchers, not directly involved, who use the data for their
research.

For this purpose, the collaboration is highly committed with the use of tools
of the Virtual Observatory (VO).
All the reduced images and the final catalogue will be accessible through 
VO-protocols. 

Internal data releases will allow the researchers of the collaborations to 
take advantage of the observations. 

\FloatBarrier 

\section{Data Management}
\FloatBarrier 

An important part of the OAJ project is the deployment of a
data center UPAD (Unit for Data Processing and Archiving) for handling, analyzing and storing the
significant amount of data produced by the OAJ telescopes during the survey
development. J-PAS and J-PLUS surveys will produce about
$\sim2.5$ PB of information accounting for the raw and processed data. 
The two telescopes may produce up to $\sim$ 1.5 TB per
night. The processing and archiving of these data will be done in the UPAD datacenter, which is located in
 Teruel about 30 km away from the OAJ. For the transmission of the
data from the OAJ to the UPAD, there is a radio-link with
bandwidth 700 Mbps which allows to download the data as soon they are
produced.
More details concerning the data management
pipelines, and the hardware solutions to
 store and process the survey data are given in the next sections.
(see also \cite{2012SPIE.8451E..16C}).

\subsection{Image Format, Data Rates and Data Volumes}

JPCam is a wide field camera of (3 deg diameter) that will be installed at the
JST/T250. The focal plane includes a mosaic of 14 large format CCDs $\sim 9.2
\times 9.2$ kpix. The pixel size is 10$\mu$m producing a pixel scale of 0.23
arcsec/pix. The total image size produced by one of the CCDs considering
overscan and pre-scan areas is $\sim$ 180 MB. The J-PAS survey will be carried
out using a set of 54 narrow band filters, 1 medium band and 1 broad band one that are arranged in four filter trays. Observations through an additional tray containing a broad band filter will be made in advance to serve, other than for specific scientific purposes, as reference for astrometric matching and source detection. In terms of data processing the images acquired by each CCD are processed independently since the light reaching each CCD pass through a different filter. 

The JAST/T80 has a FoV of 2 deg diameter and the plate scale is 55.5
6arcsec/mm. The JAST/T80 camera is equipped with a detector of $\sim 9.2  
\times 9.2$ kpix of 10$\mu$m that yields a pixel scale of 0.55 arcsec/pix . The
raw images will be stored in 16 bits which gives an image size (considering the over and pre-scan) of 236 MB. 

\begin{figure}[H]
\begin{center}
\includegraphics[width=14cm]{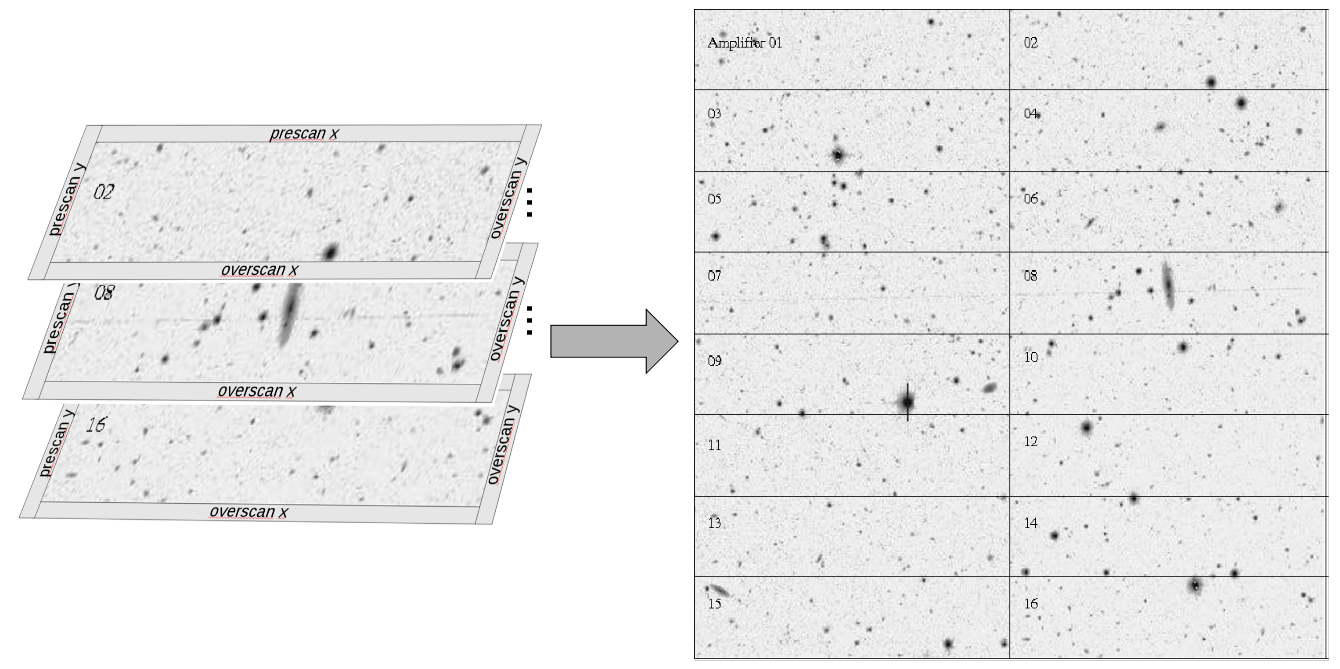}
\end{center}
\caption{Schematics of the image format with the 16 amplifiers in different
HDUs and how the images are reassembled using a single HDU after pre-scan or/and
overscan correction and trimming.
\label{Fig:overscan}}
\end{figure}

The data collected by each CCD in the cameras will be stored different FITS
files. Each CCD has 16 amplifiers and the raw frame will contain overscan and pre-scan sections in both directions. To store this information the raw FITS files will have 17 Header and Data Units (HDUs), the primary HDU will only contain a header with the common metadata. The other HDUs will store the data corresponding to each amplifier and the headers that describe the electronics of the amplifier and the data organization inside the HDU. After the overscan or pre-scan correction and the trimming of the overscan and pre-scan areas, the image is reformatted to a single HDU containing the whole image and joining properly the information coming from the different amplifiers as is shown in Fig.~\ref{Fig:overscan}.

\subsection{Data Rates}

The two telescopes at the OAJ will be mostly dedicated to the J-PLUS (JAST/T80) and J-PAS (JST/T250) surveys. 
For JPLUS observations, with individual exposures of 35 secs, the
data rate after considering the overheads due to filter changes and telescope
movements is $\sim$ 13.8 GB/h. 

The JST/T250 will capture simultaneous exposures in 14 CCDs producing 2.53 GB
of data per reading. The JPCam camera at the JST/T250 telescope has four filter
trays where the 56 J-PAS filters are distributed plus an additional tray
containing the broad band reference filter. For J-PAS, the 3 bluer filter trays
will be exposed $\sim$240 secs, whereas the reddest tray will have a longer
exposure time ($\sim$ 480 secs). The expected data rate without
binning is 120 GB/h. Every day a set of 30 calibration frames (bias, and flat fields) has to be
collected. Added with the science data each night the telescope produce 1.3 TB of
data. When binning of 2$\times$2 pixels is used, which
is foreseen for the narrow band filters, the rate is 34GB/h and the total data $\sim340$TB.

The amount of data per year depends on the final observational strategy for
J-PAS defining how the time along the year is allocated for the different
trays. The exposure time in each tray is divided in four sub-exposures per sky
position. The actual survey strategy plan is to collect the first two
individual exposures of a pointing contiguously and revisit it twice, after one
and two months. Considering an useful time of 1800h/year we expect $\sim 230$
TB/year on raw images during the first year of observations with the reference broad band filter.
After that, the data of the J-PAS survey will be collected using a binning of
2$\times$2 pixels and produce about $\sim 62$ TB/year.

\begin{figure}[h!]
\begin{center}
\includegraphics[width=.8\textwidth]{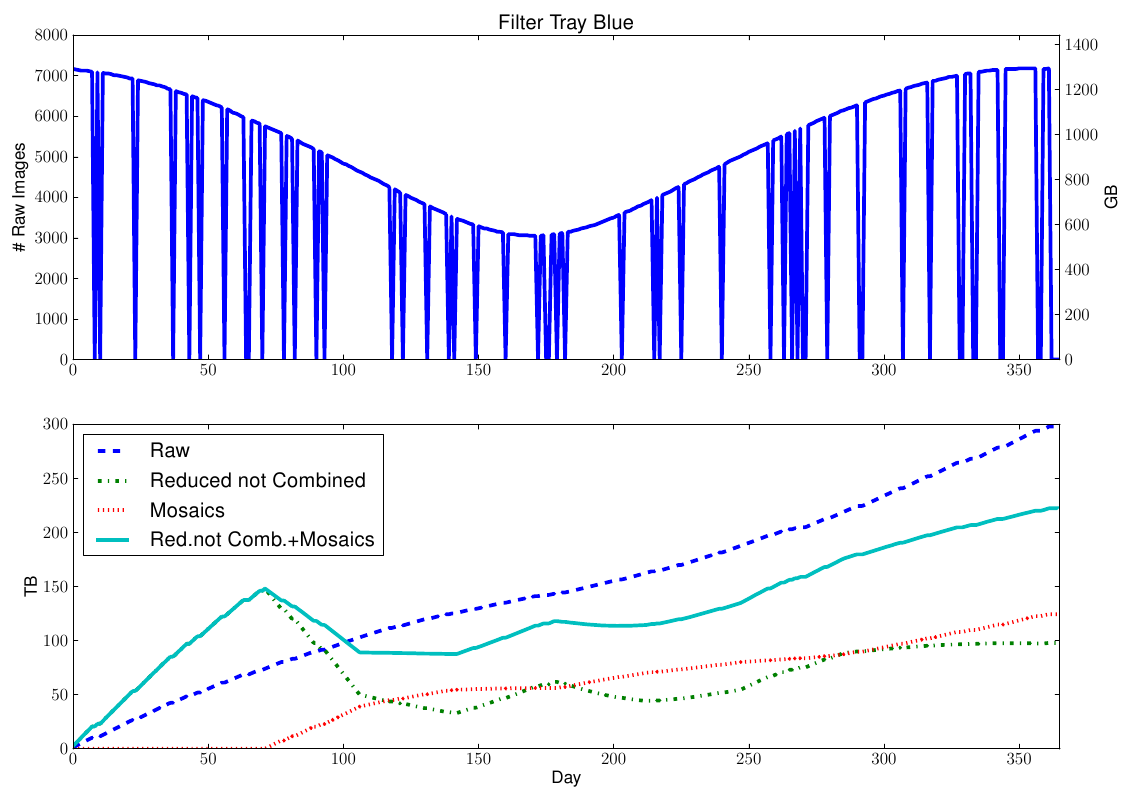}
\end{center}
\caption{ Estimated data rates obtained with the NB filter trays used with JPCam. The panels correspond to the number of images
obtained by night at different epochs during the year ({\it top panel}), 
 and the
 cumulative data volume during the year assuming that individual processed
 images are combined in deep mosaics and deleted from the storage system 
 as soon as the 4 images are collected (\it bottom panel). \label{Fig:DataCollectionRates}}
\end{figure}

Figure \ref{Fig:DataCollectionRates} shows the data collection rates. In the figure no binning it is considered. Periods of bad weather (15\% of the time) are inserted randomly. The top panel shows the number of images (note that 14 images are produced in a single JPCam exposure) collected per night. The bottom panel shows the cumulated data volumes in raw, individual processed images and mosaics. It is assumed that as soon as the final deep mosaics are combined, the individual reduced images are deleted from the disk system.

\subsubsection{Total data volumes}

The JPCam pixel size is $0.23\arcsec/$pixel, which implies that a survey of
$\sim 8500\sq\degr$  will contain a total of $\approx 2$ Tpix which is $\approx 4.0$ TB of
information for a sky pass. The total survey is observed through 59 filters
with a total of 308 individual exposures per sky position. 
Considering the pre and over-scan areas and an increase of $10\%$ of
inefficiency due to the camera geometry, and that binning is being used in 56
filters the whole survey raw data (including calibration frames) amount to
$\sim520$ TB. 

The coadded images (4 bytes image + 2 bytes weight map) amount to 760 TB of
information considering a $10\%$ increase due to geometrical issues. The
initial approach is to produce the averaged calibration frames per month, this
will lead to 2.4 TB in master calibration images during the six years of J-PAS data collection. 

Taking all this data into account J-PAS needs $\sim$2.5 PB to store the
raw data plus 2 versions of processed data and auxiliary products. 
A ratio 2:1 of lossless compression is possible for the science raw data. 
Using this ratio of raw data the final number will be $\sim1.9$ PB. Table
\ref{Tab:storagenumbers}
gives the data volumes expected by the J-PLUS and J-PAS surveys.

\begin{table}[!h]
\begin{threeparttable} 
\caption{ Storage needs for the J-PLUS and J-PAS surveys. Only images \label{Tab:storagenumbers}} 
\begin{tabular}{| l | l | l | l | l |}
\hline
  & J-PLUS & J-PLUS compressed & J-PAS & J-PAS compressed\\
  \hline
  &&&\multicolumn{2}{|c|} {not binned | binned} \\

\hline
Night 10h & 158 GB & 79 GB & 1290 | 340 GB & 644 | 170 GB\\   
Year raw data \tnote{1} & 31.6 TB & 15.8 TB & 232 | 62 TB & 116 | 31 TB\\ 
\hline\hline
Total raw data \tnote{1} & 42.8 TB & 21.4 TB & 520 TB & 260 TB\\ 
Coadded data (1DR) & 32.2 TB & - & 760 TB \tnote{2}& - \\ 
Coadded calibration & 204 GB & - &  2.4 TB & - \\
\hline
\end{tabular}
\begin{tablenotes}
\footnotesize \item[1] Include calibration frames 
\footnotesize \item[2] The weight map  stored in 2 bytes.
\end{tablenotes}
\end{threeparttable}
\end{table}

\subsection{OAJ-UPAD Data Flow}
At the OAJ after data acquisition the images will be converted to the FITS
format defined as input for the pipelines. The image headers will be filled
with information from the telescope, the monitors and the meteorological
station. This information will be provided by the Observatory Control System
through a historic database. The image headers apart from the standard FITS
keywords, will be complemented with HIERARCH special keywords containing the information needed for the image processing and quality check. The data processing pipeline will use this HIERARCH keywords to add some information during data processing. 

The images are sent to the OAJ/CPD computing nodes to perform a quick data
validation. The raw data will be archived archived in the storage system at OAJ
and transferred to UPAD/CEFCA using the existing radio-link. The OAJ storage
will contain a buffer of the 2 or 3 last months of operations. During day time
two copies of the raw data in magnetic tapes will be done as backup.

\begin{figure}[h!]
\begin{center}
\includegraphics[width=14cm]{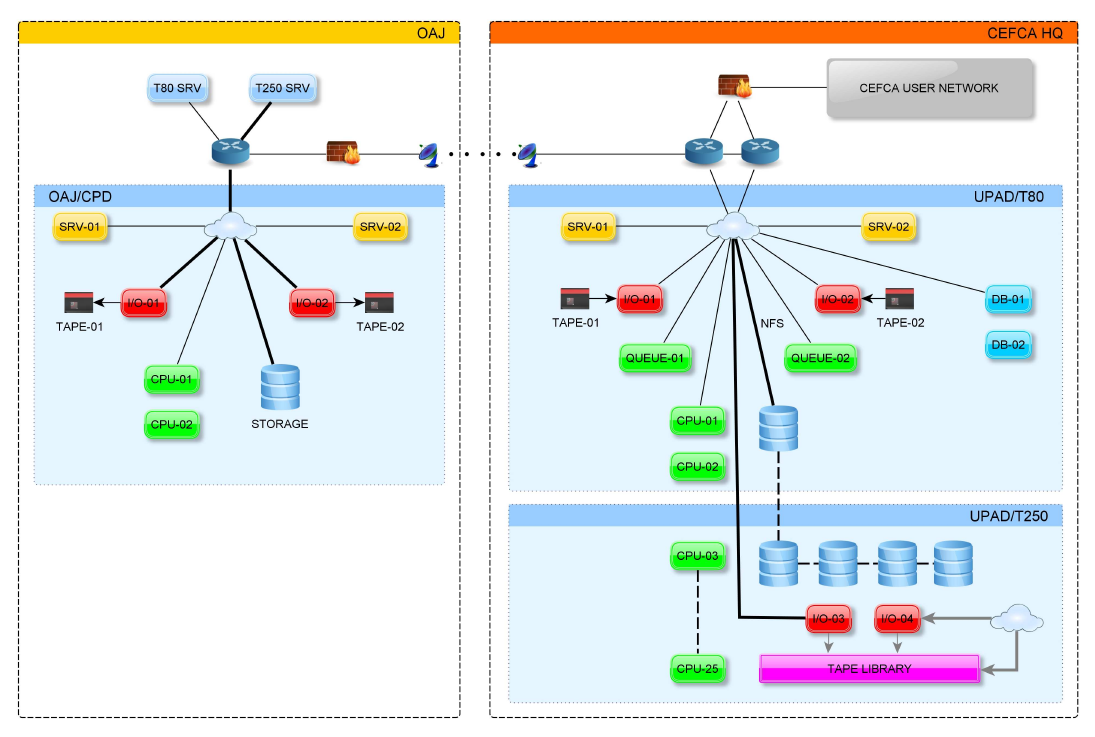}
\end{center}
\caption{Schematics of the hardware involved in the data movement inside the
Observatory and data transmission to the OAJ. On the right side there are two
different phases. Above, the deployment hardware for data
transmission, service machines and data processing for the JAST/T80 data it is shown.
In the bottom part, it is shown the hardware dedicated to data reduction and
catalog generation for the JST/T250 data. \label{Fig:hardw}}
\end{figure}

In Fig.~\ref{Fig:hardw} it is shown the foreseen hardware to manage the data flow inside the OAJ, the data transmission to the UPAD, and the processing nodes. 

\subsection{Image processing pipeline}

The processing of the images is divided in two main stages. The first one is
related with the instrumental correction, the creation of the pixel mask for
the individual frames, and also includes a first astrometric and photometric
calibration. The second stage deals with the creation of the final tiles by
combining the individual corrected frames and weight maps, the source
extraction, and the insertion of the catalogs in the science database. Both
parts of the pipeline are implemented in Python and use software from the astronomical community that have been integrated through Python wrappers.

\subsubsection{Daily pipeline} 

\begin{figure}
\begin{center}
\includegraphics[width=12cm]{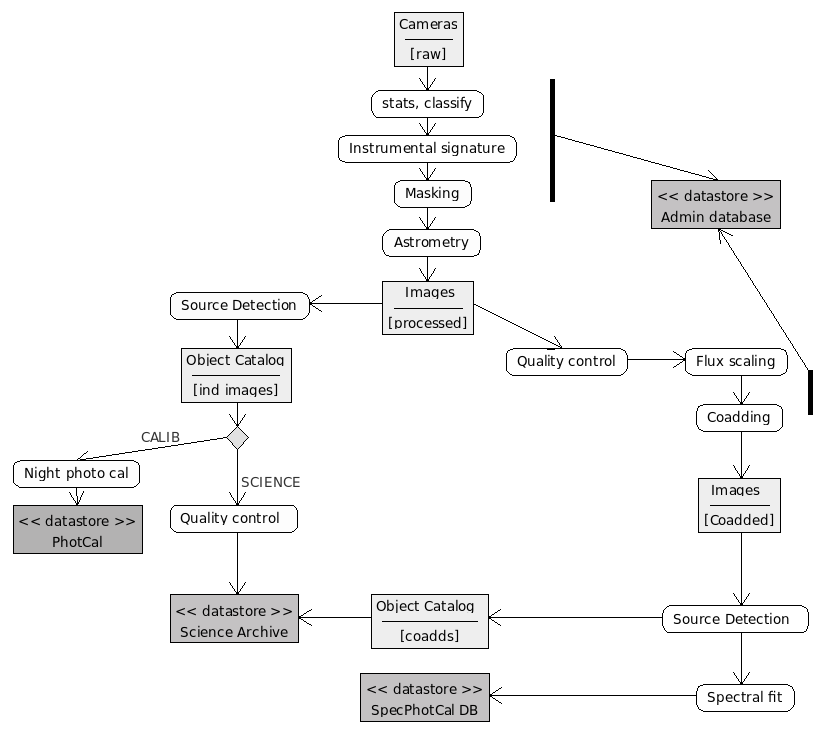}
\end{center}
\caption{J-PLUS pipeline data flow. The calibration will be
different for J-PAS which will make use of the previously calibrated
stars, with their spectral type determined from J-PLUS, to calibrate each individual image (Gruel et al. (2012)). \label{Fig:dataflux}}
\end{figure}

The main steps performed by the pipeline are summarized in
Fig.~\ref{Fig:dataflux}. The processing of each image is controlled through an
administrative database. The Operative Archive is a table of this database which store the information about the raw frames and serve as the input to search for new images to process. 

The daily pipeline use previously generated master calibration frames
(bias, flat field, fringing pattern, etc). These frames are generated for
predefined observing blocks. The master calibration frames metadata are stored in a table inside the
administrative DB. As the OAJ telescopes are used mainly for a dedicated
program the definition of each observing block will be done under the basis of
instrumental stability. 

The calibration images are created by an automatic
algorithm that selects the individual frames acquired during the target
observing block from the Operative Archive table. Then create the target master
calibration frame, and finally apply it to each one of the input individual
frames. The analysis of the residuals will indicate if the master calibration
frame is appropriate to be used to correct the science images collected during
the period or if the observing block shall be split in different dates. The time interval when each calibration frame is valid is uploaded to the database. 

The operations done by the daily pipeline are independent for each image and
can be managed by a batch-queue control system. In the first processing of a
science image, just after acquisition, the master calibration frames from a
previous observing block will be used for quality assessment. Once the current
observing block is concluded the image processing will be repeated with the
proper master calibration frames. All the operations performed on the images
are stored in the Reduction Control table of the administrative DB. 

After correcting the instrument signature the daily processing includes the
creation of the mask associated to each individual frame. To produce the final
image mask the pipeline uses the following strategies. The bad pixels are
located through the hot and cold pixel mask associated with the bias and
flat-field frames, respectively. To compute the cosmic ray mask SExtractor is
used with a retina filter. 
The retina filter is a neural network which is trained using the cosmic ray
mask created with LACOSim (\cite{2001PASP..113.1420V}).

The linear patterns produced by satellite trajectories are identified and
located using the Hough transformation to locate aligned detections among the
centers of the sources computed in an initial catalog (computed using
SExtractor \cite{1996A&AS..117..393B}). 

The pixel mask is formatted as a compressed image in the second HDU of the
individual processed images. A binary code is used to identify any problem 
affecting each pixel (bad pixel, cosmic ray, satellite trajectory,
saturation) or if it is associated with an object. The compressed pixel masks
and individual images catalogs occupy a factor $\sim$200 and $\sim$1200,
respectively, less than the associated raw data image. Considering that
recomputing the pixel masks and catalogs requires a high CPU cost, those will
be kept in the storage system for each data release. Eventually, keeping these
products in the archive, together with the fact that the arithmetic operations
and the identification of the master calibration frames applied over the raw
data are stored in the Reduction Control Table, allows for reprocessing of the
raw images or the recovery of the individual processed images whenever required. 

The last step done by the daily pipeline is to compute the astrometric solution
for each individual image, to this extent we are using SCAMP
(www.astromatic.net \cite{2002ASPC..281..228B}). The astrometric solution computed independently for each exposure can be improved when the final mosaics are combined by using the broadband image as a reference to produce the internal solution.

\subsubsection{Tile combination}

The goal of this part of the pipeline is to create a data-cube for each survey
tile. The data-cube refers to the image mosaics in all the filters aligned so
each pixel in the different filter images samples the same sky position. For
the J-PLUS, as T80Cam has an unique CCD, the tile will be defined as the high S/N area after creating the combined images. For J-PLUS the dithering pattern will be a small fraction of the CCD to allow the correction of the bad zones of the CCD, and the removal of stellar haloes when computing background or fringing master frames. The telescope displacement between adjacent pointings will produce an overlap among the final tiles to allow for perform a zero point transportation. When creating the combined image we will use one of broad band images as a reference to re-calibrate the internal astrometric solution using SCAMP. 

SWarp (www.astromatic.net \cite{2002ASPC..281..228B}) is used to generate the mosaics from the individual images. Before
producing the average mosaic, each image is scaled to match the photometry of a
reference image in the tile. To compute the scaling factors the pipeline will
make use of the calibration information generated for each observing night/run. 
It is foreseen to perform a relative calibration among the individual images using the stars inside the field. The results of both methods will be compared. 

In the case of the J-PAS, due to the camera geometry and the fact that the
light reaching each CCD pass through a different filter, it is not possible to
define a pattern which is covered in the same way by the 14 filters arranged in
a tray. The approach for the J-PAS tiles is to define a grid
in the sky defined by the centers, pixel scale and image size. Each tile in the
grid will be covered first by a broad band filter that will serve as an internal
reference for the astrometric solution. The tile mosaic image in each filter
will be created as soon as the four individual exposures are obtained. At this stage the
individual image flux scaling is computed from the zero point assigned to them
and using relative calibration. The procedure for the individual image calibration for the J-PAS data is
described in \cite{2012SPIE.8448E..1VG}. 
\subsection {Final products for J-PAS}\label{Sec:finalproducts}

The final products of J-PAS in terms of images will be provided as data-cubes.
Data-cubes will be provided for each final sky tile an consist on the registered images for each filter.
Along with the tile image those
directories will contain the associated weight map, individual image mask, catalogs and validation plots (e.g astrometric calibration).

The size of the tiles will be close to the size of each CCD (0.6 x 0.6 sq degrees). A certain overlap area among adjacent tiles
will be reserved for calibration (as SDSS ubercalibration) purposes. A summary
of the data to be computed and preserved for each tile or data-cube is given in
table \ref{Tab:storageproducts}. The J-PAS survey have $\sim
23700$ tiles (each in 59 filters). Considering the images, weight maps, catalogs, masks and plots
produced for the individual images the total volume of a data release is $\sim
817$ TB.

\begin{table}[h!]
\begin{threeparttable} 
\caption{ Size of the final coadded tiles and products for J-PAS. \label{Tab:storageproducts}} 
\begin{tabular}{| l | l |}
\hline
\hline
\multicolumn{2}{|l|}{\bf Per tile per filter} \\
\hline
  tile image + weight map & 545 MB \\
  tile catalog & 10 MB \\
  individual images plots (4 exposures) & 8 MB \\
  rendered images  ( " ) &  2.5MB \\
  individual image catalogs ( " ) & 20 MB \\
  individual image masks( " ) & 0.08 MB \\ 
{\bf Total products per tile (1 filter)} &  585.3 MB\tnote{1}\\
\hline
\hline
\multicolumn{2}{|l|}{\bf  Data-cube per tile} \\
\hline
Images + weight maps + catalogs (59 filters) & 32.73 GB\\  
Individual image (catalogs + rendered frames + masks + plots) & 1.80 GB\\
{\bf Total products per data-cube} &  34.53 GB\tnote{1}\\
\hline
\hline
\multicolumn{2}{|l|}{\bf Whole survey ($\sim23700$ tiles)} \\
\hline
  Images + weight maps + catalogs & 775 TB \\
{\bf Total products whole survey (1 DR)}  &  817 TB\tnote{1} \\ 
\hline
\end{tabular}
\begin{tablenotes}
\footnotesize \item[1] Decimal units are used in the text (1KB = 1000B)
\end{tablenotes}
\end{threeparttable}
\end{table}

\subsection{Photometric pipeline}
  To reduce the J-PLUS and J-PAS data we will use the ALHAMBRA pipeline,
which has been developed by J-PAS members and well tested with the
ALHAMBRA dataset, which comprises $4\sq\degr$ of imaging with $23$
medium band filters and which, apart from its scientific value, it is
a perfect testbed to develop the methods required for J-PAS. See
\citet{cristobal2009}, \citet{molino2013} and Crist\'obal et al., 
in preparation, for a detailed description.  

\subsection{Storage and Processing Facilities (UPAD)}

The UPAD is the data center where the image archiving and data processing of the data coming from the OAJ telescopes is going to be performed.

The hardware at the UPAD will be deployed in two main phases. The first one
provide the infrastructures to do the data transmission between the OAJ and the
UPAD through the dedicated radio-link, disk storage system for the OAJ, OAJ
computing nodes and the OAJ backup system. During this phase, the batch-queue
manager, the processing nodes and the storage devices to manage the J-PLUS data
will be implemented. Those systems are shown in Fig.~\ref{Fig:hardw} under the
names OAJ-CPD and UPAD/T80. For J-PLUS the disk archive at UPAD have a capacity
of  $\sim$100 TB. This storage system will allow to archive the raw and processed data for the J-PLUS survey until the whole infrastructure for the UPAD is deployed. 

The second phase of the UPAD implementation will increase the storage
capabilities and processing power to archive and process the data coming from
the JPCam at JST/T250. Due to the large volumes of information that has to be
stored, the storage system is divided in at least two different tiers. A
storage disk system, that provides fast access to the hot data, and a tape
library to do the data archive and an additional backup which will allow to
increase the storage space to several PB.

\paragraph{Data compression}

The raw images from both telescopes may be lossless compressed using Fpack
(\cite{2009PASP..121..414P}).
A study with ALHAMBRA raw data and J-PAS simulated data yields that a 2:1
compression rate on science images with low background (noise) can be achieved. The numbers obtained using this compression rates on the raw data are given in table \ref{Tab:storagenumbers}

{\bf Compression of processed science frames}

The rate of lossless compression on the raw data can not be acquired on the processed frames
containing floating point values. It is possible to acquire a higher rate of compression $\sim4:1$ on the final
processed images using a lossy algorithm. It is has to be evaluated whether it
is admissible to add some additional noise ($< 1$ADU) on the final processed
images in order to save a significant volume of storage.

{\bf Storage needs: Disk Storage System and Robotic Tape Library combination}

As it it explained above the amount of data that has to be stored by the J-PAS
survey is about 2.5 PB. It is pointed out that the volume of raw data to
be stored for J-PAS is $\sim 520$TB and the amount of information in processed
combined images, catalogs and other products is $\sim 820$ TB. 

To define the storage architecture it should be considered that:

\begin{itemize}
\item The amount of data to store, this is the raw data and several data revisions. The final data volume may grow up to several PB. 
\item That data will not be accessed at the same rate during their lifetime. For example raw data will be required by the processing nodes at a higher rate during the two months after data collection, which is the timescale to combine them into final mosaics due to observing strategy, and after that only eventually in a potential data reprocessing. Also old data releases will not be accessed frequently once the catalog are available.
\item Due to the size of the images, the tier 1 storage shall provide a quick access to the data.
\end{itemize}

\begin{figure}
\begin{center}
\includegraphics[width=14cm]{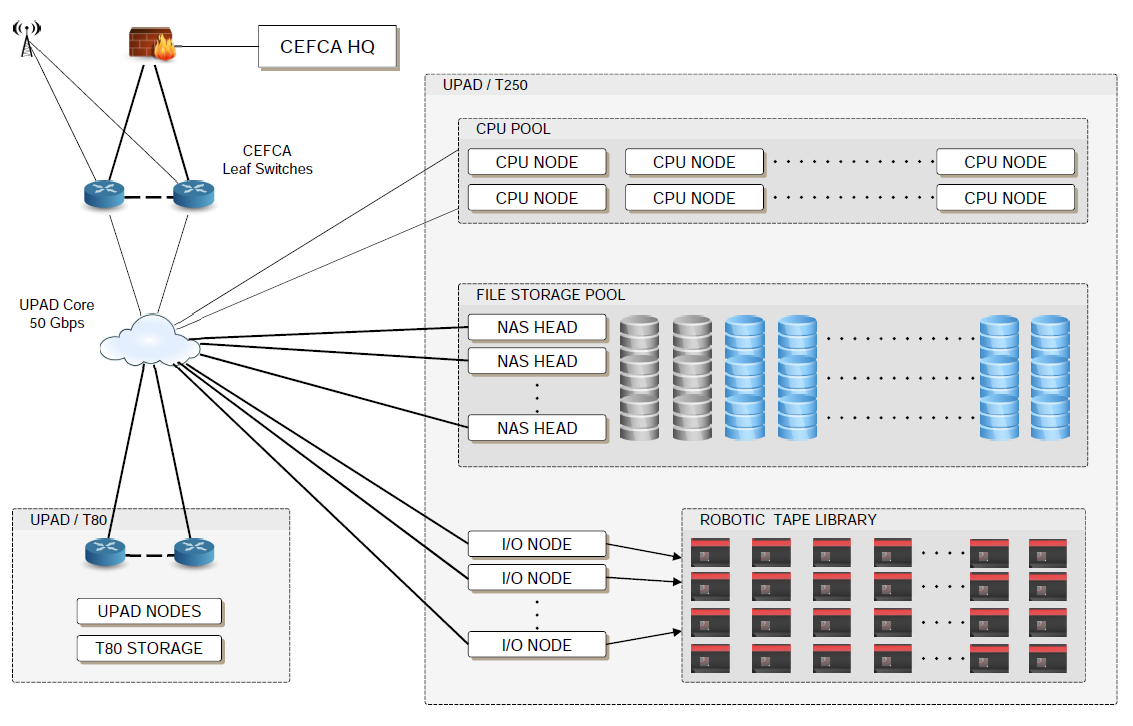}
\end{center}
\caption{Schema of the storage and processing systems.\label{Fig:schema}}
\end{figure}

In this context, to store the raw data and products of the J-PAS and J-PLUS
surveys a combination of a disk storage system and a robotic tape library will
be used. Data will be stored in the adequate media depending on the access frequency.
A schema of the architecture is shown in Fig~\ref{Fig:schema}

The disk storage system will be used as and big cache to do the data
processing. The processing of J-PAS data is limited by I/O operations
against storage. To hold the daily processing rate (and
potential reprocessing) the storage has to provide an aggregated bandwidth of
$\sim$ 50Gbps. With this bandwidth between the storage and the processing
nodes, the time to read an image of 200MB is less than 20 secs in the case of a high
network contention. Which produce an overhead of $< 10\%$ in the individual image processing.
 The network hardware shall be defined to provide
this rates and that filesystem will be distributed. 

The robotic library will be used as a near-line storage. It provides a more dense media to store
the data at a lower initial and running cost.
The tape library is a work space allowing
the movement of big blocks of data (several days/months of raw data or sky areas of processed data
to the disk storage). The robotic library will store a
copy of the total raw data, the final combined images, associated
information, and products (catalogs) for several data releases. 

In the next section it is described which data is planned to be set in disks and in the tape library.

{\bf Disk storage system}
The disk storage will be dimensioned to store the data that are accessed frequently:

\begin{itemize}
\item {\it Raw and processed data collected during the last two months:} This data will be coadded into tiles and
deleted from the disk space. Once they are removed from the disk system the
copy in the library will allow automatic
access to those data if needed. In case that we need to re-process an important
part of the survey, some policies shall be defined to move data massively to
the disks storage.    

\item {\it The last revision of the data-cubes and auxiliary data:} Include the
tiles and weight maps per filter, and other data as the individual image masks,  
catalogs, and diagnostic plots.

\item {\it A copy of the combined master frames:}. Bias, Flats and Fringing patterns. The master frames for older data releases can be also stored here to allow some data re-processing.

\item {\it A copy of the auxiliary data and products (catalogs) from previous releases}. 

\end{itemize}

Table \ref{Tab:storageneeds} show the volumes needed in the disk storage system to archive the different products. 
Binning ($2 \times 2$) in acquisition is considered in the 56 narrow band filters. 
Two kind of compression is used to estimate the
volumes. 1) Lossless compression (2:1) only applied to the data stored in
integers. 2) Lossless compression on integer images plus lossy compression (4:1) on the processed frames. 

\begin{center}
\begin{table}[h!]
\begin{threeparttable} 
\caption{ Size of storage in DISK SYSTEM needed for J-PAS. Numbers in TB. \label{Tab:storageneeds}} 
\begin{tabular}{|l | l | l | l|}
\hline
&\multicolumn{3}{|l|}{binning in 56 filters} \\
\hline
& No Compression & Compression lossless  & Compression lossy  \\
\hline
Raw data (2 months)         & 20.4 & 10.2 &10.2\\
\hline
Processed data (2 months)  & 36.5 & 36.5  &9.2\\
\hline
Final tiles               & 760.7 \tnote{3}& 633.9 \tnote{4} & 253.6\\
\hline
Comb master CF (6DR)    & 4.2 & 4.2 & 1.1\\
\hline
Aux Products (3DR)    & 114.1 & 114.1 & 114.1 \\
\hline
Sum                        & 935.7 & 798.7 & 388.0  \\
\hline
\end{tabular}
\begin{tablenotes}
\footnotesize \item[1] Decimal units are used in the text (1KB = 1000B)
\footnotesize \item[2] Numbers for 59 filters
\footnotesize \item[3] No compression used in Weight maps, stored in 16 bits.
\footnotesize \item[4] Compression used in Weight maps, stored in 16 bits.
\end{tablenotes}
\end{threeparttable}
\end{table}
\end{center}

\paragraph{Robotic tape library}

As mentioned before, the robotic library will be used to archive all the raw
data and products. The raw data and older data releases not
frequently requested will be stored here and moved to the disk storage if needed.

It is considered to store in the robotic library the following data.
\begin{itemize}
\item A near-line copy of all the raw data.
\item A copy of the last released data-cubes. Also a copy of the images of some previous releases.
\item A copy the master frames for all the releases.
\item A copy of the products for all releases: individual image catalogs,
validation plots, individual image masks, log files, etc.
\end{itemize}

Table \ref{Tab:storageneeds2} shown the volumes needed to store the enumerated data. In different columns is reflected the decrease in data volumes in case of conversion of the processed data to integers in 16 bits.  Note that $2\times2$ binning during acquisition in the narrow band filters is assumed to compute the data volumes.

\begin{center}
\begin{table}[h!]
\begin{threeparttable} 
\caption{ Size of storage in TAPE LIBRARY needed for J-PAS. Numbers in TB. \label{Tab:storageneeds2}} 
\begin{tabular}{| l | l | l | l | }
\hline
 & \multicolumn{3}{|l|}{binning in 56 filters} \\
\hline
& No Compression & Compression (lossless)  &  Compression (lossy) \\
\hline
Raw data (J-PAS)       &   518.5 & 259.3 &  259.3\\
\hline
Final tiles               &  760.7 \tnote{3}& 633.9 \tnote{4}& 253.6\\
\hline
Comb Master CF (6DR)    &   4.2 & 4.2 & 1.1\\
\hline
Products (6DR)    &    339.9 & 339.9 & 339.9\\
\hline
Sum (2 DR images)   &    2383.8 & 1871.0 & 1107.3  \\
\hline
\end{tabular}
\begin{tablenotes}
\footnotesize \item[1] Decimal units are used in the text (1KB = 1000B)
\footnotesize \item[2] Numbers for 59 filters
\footnotesize \item[3] No compression used in Weight maps, stored in 16 bits
integers.
\footnotesize \item[4] Compression used in Weight maps, stored in 16 bits
integers.
\footnotesize \item[5] Considered the sum of raw data plus 2 data releases of
tiles and 6 releases of Master CF and products (catalogs, pixel mask, ...). 

\end{tablenotes}
\end{threeparttable}
\end{table}
\end{center}

\paragraph{UPAD computing needs}

The UPAD has to be equipped with computing power that allows the following
actions:
\begin{enumerate}
\item Create the averaged Master Calibration Frames. During the instrument
commissioning the UPAD team will verify the frequency for Calibration Frame
updates. 
\item \label{2} Reduce and calibrate daily the individual frames obtained by the two telescopes.
\item \label{3} Generate the averaged tiles from the individual frames once a sky area have
been completed in a set of filters.
\item Extract the final catalogs from the final tiles.
\item Be able to reprocess the survey data with a sustained rate.
\item Execute other software developed by the collaborators.
\end{enumerate}

The execution times to perform the first three task above have been evaluated
using the alpha version of the J-PAS data processing pipeline and simulated images 
with similar characteristics that the ones collected for J-PAS.
The test have been done in a development system with 3 nodes with similar 
characteristics than the target ones 3.2GHz processor, 12 cores, 48 GB RAM. 
Sun Grid Engine is used as batch-queue software. 
The average data reduction for one image is  $\sim200$ secs, 
and a tile of 4 images is combined in  $\sim400$ secs. Compute the final
catalogs may take $\sim120$ secs per frame. 
Taking into account that $\sim8000$ science images will be collected per
night, and that $\sim2000$ tiles shall be combined, all this amount to 733 hours of CPU in daily processing.

The two surveys amount to  $7.5\times10^6$  frames and  $1.5\times10^6$  tiles so
$\sim 6.3\times10^5$ CPU hours. A system with ~300 cores shall be able to process daily the data 
in 4-5 hours and hold an adequate rate of data reprocessing.

There are some parts of the code that have not been taking into account like:
\begin{itemize}
\item Compute catalogs to run the photometric redshift code (BPZ) and photometric redshifts code.
\item Other software developed by the J-PAS collaboration.
\end {itemize}

\paragraph{Science Database and External Access system}
Once the final catalogs are computed the information frequently used by
the scientific community will be ingested in a database. When J-PAS survey
is finished the information stored in database will be $\sim20$TB. After studying different alternatives, a clustered SQL database engine is the current approach.
The SQL data model is appropriate for the kind of data to store, the SQL
queries are amply used by the the astronomical community, and this approach
allow an easier integration with the Virtual Observatory protocols.
To share the processed data (images and catalogs) with the scientific community
a system dedicated to the data distribution “on demand” is being considered
(EDAM, External Data Access Machine). This system will contains a replica of
the internal databases with the validated data that according
with the release policies shall be public available.
The EDAM will also provide access to final images or raw data under some restrictions on the demanded data volumes.

\FloatBarrier 

\section{The Observatorio Astrof\'{\i}sico de Javalambre}

\FloatBarrier 

\subsection{Site and infrastructures}

The {\it Observatorio Astrof\'{\i}sico de Javalambre} (hereafter OAJ)
is the new astronomical facility at the {\it Pico del Buitre}, in the
{\it Sierra de Javalambre} (Teruel, Spain), where J-PAS will be
conducted from. The site has an altitude of 1957\,m above the sea
level, with excellent astronomical characteristics in terms of median
seeing of $0.71$\,arcsec in $V$ band, a fraction of totally clear nights
of $\sim 53\%$ ($\sim 75\%$ with
at least $30\%$ of the night clear) and a remarkable darkness, a
feature quite exceptional in continental Europe
\cite[see more details in][]{Moles2010}.

The OAJ is a facility specifically designed to carry out large sky
surveys with two unique telescopes of unusually large fields of view
(FoV). The main one is the Javalambre Survey Telescope (JST/T250)
an innovative Ritchey-Chretien, alt-azimuthal,
large-etendue telescope with a primary mirror diameter of 2.55\,m and
3\,deg (diameter) FoV. The JST/T250 is the telescope devoted to
conduct J-PAS, making use of a unprecedented panoramic camera, JPCam,
that will be described in Section~\ref{jpcam}.  The second largest telescope at the OAJ 
is the Javalambre Auxiliary Survey Telescope (JAST/T80), a
Ritchey-Chretien, german-equatorial telescope of 83\,cm primary mirror
and 2\,deg FoV. The primary goal of JAST/T80 is to perform the
Javalambre Photometric Local Universe Survey (J-PLUS), that will be
described in a different paper. In short, J-PLUS will cover the same
sky area of J-PAS using 12 filters in the optical range, which are
specifically defined to allow the photometric calibration of
J-PAS. These filters are: 4 SDSS filters (g,r,i,z) which allow to
anchor the photometry to that of the SDSS, 6 filters of $200-400\,\AA$
width, centered on key absorption features (e.g. H$\delta$,
the G-band, Mg$b$/Fe lines, and the Ca triplet) for stellar
classification and stellar population studies, and 2 narrow band
filters in common with the J-PAS filter set which cover the
[OII]/$\lambda 3727$ and H$\alpha$/$\lambda 6563$ lines, for anchoring
the J-PAS calibration and also mapping the SFR in nearby galaxies
(z$<0.017$).

Overall, both JST/T250 and JAST/T80 have been particularly conceived
by CEFCA to optimize the effective etendue. As part of the OAJ
contract, the detailed design and manufacturing of the two telescopes
is led by the belgian company AMOS, under CEFCA collaboration, review,
and supervision. A detailed description of the OAJ and their telescopes can be found in \cite[]{2013hsa7.conf..862C}, \cite{2012SPIE.8448E..1AC}, \citep{2011hsa6.conf..771C} and \citep{2010SPIE.7738E..26C}.

\begin{figure}[t] 
   \centering
   \includegraphics[width=1.5in]{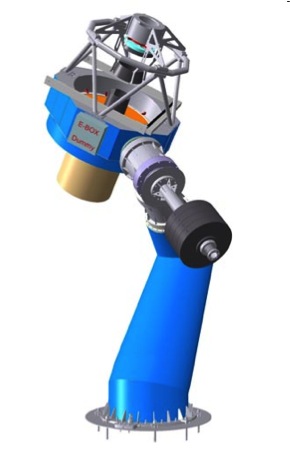} 
   \includegraphics[width=3.5in]{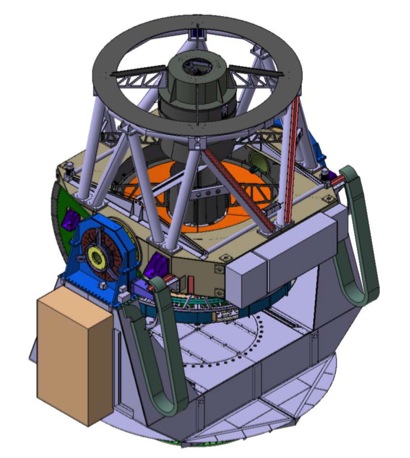} 
  \caption{An illustration of the final designs of the 2.55\,m Javalambre Survey Telescope (JST/T250; right) and the 83\,cm Javalambre Auxiliary Survey Telescope (JAST/T80; left) in true relative scales.}
   \label{fig:telescope_designs}
\end{figure}

\begin{figure}[t] 
   \centering
   \includegraphics[width=5in]{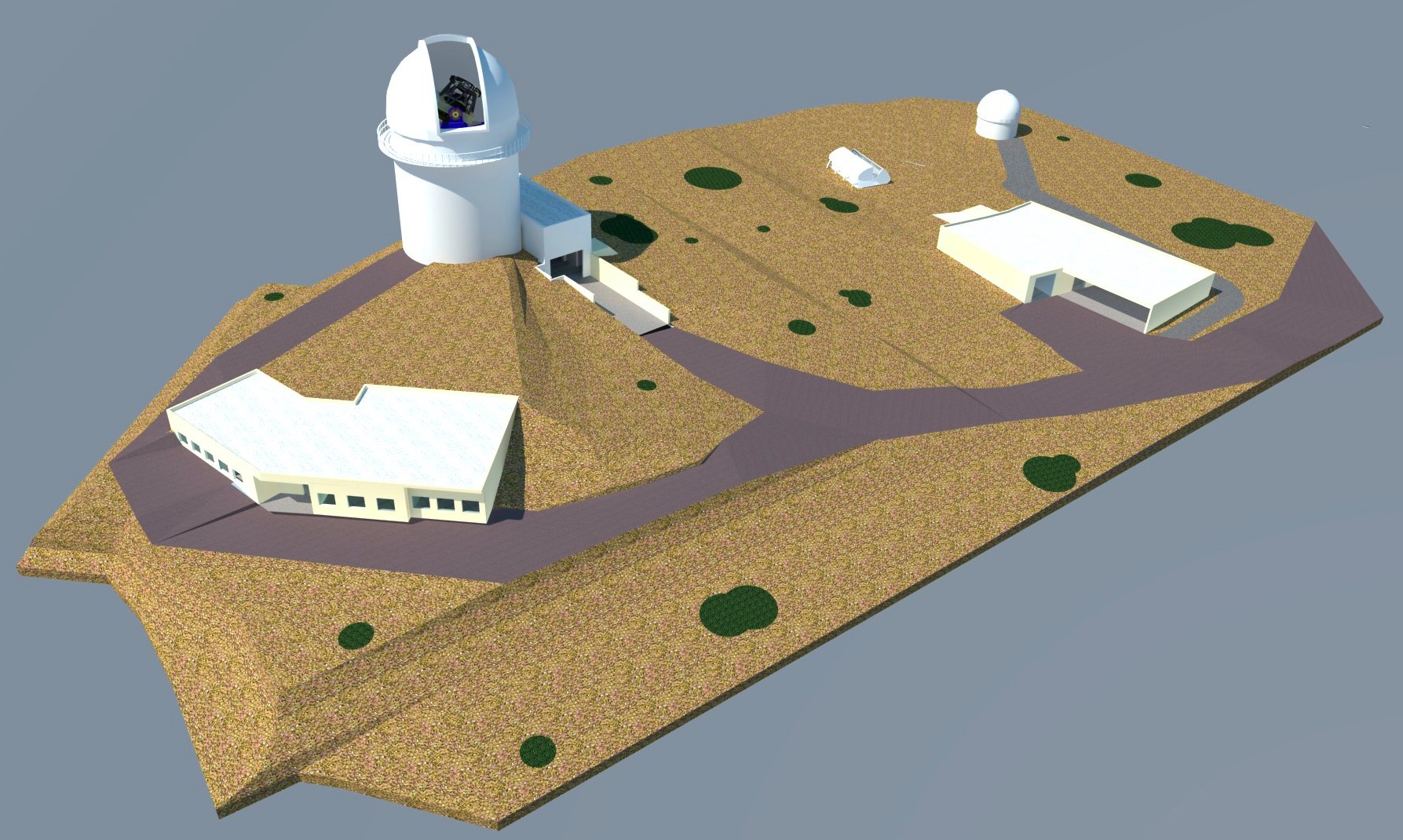} 
  \caption{Final design of the OAJ. See text for details.}
   \label{fig:OAJcw}
\end{figure}

\begin{figure}[t] 
   \centering
   \includegraphics[height=1.7in]{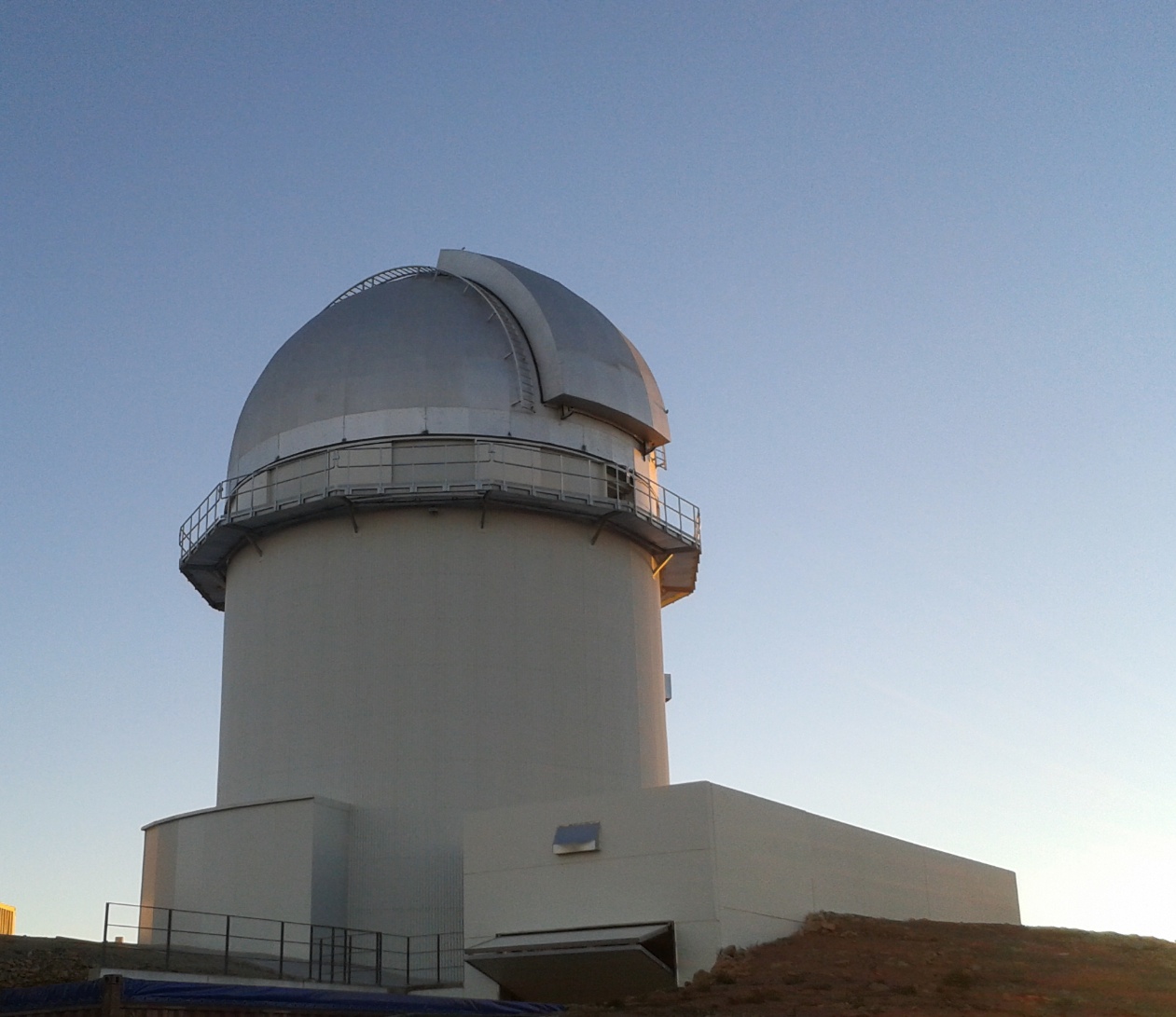} 
   \includegraphics[height=1.7in]{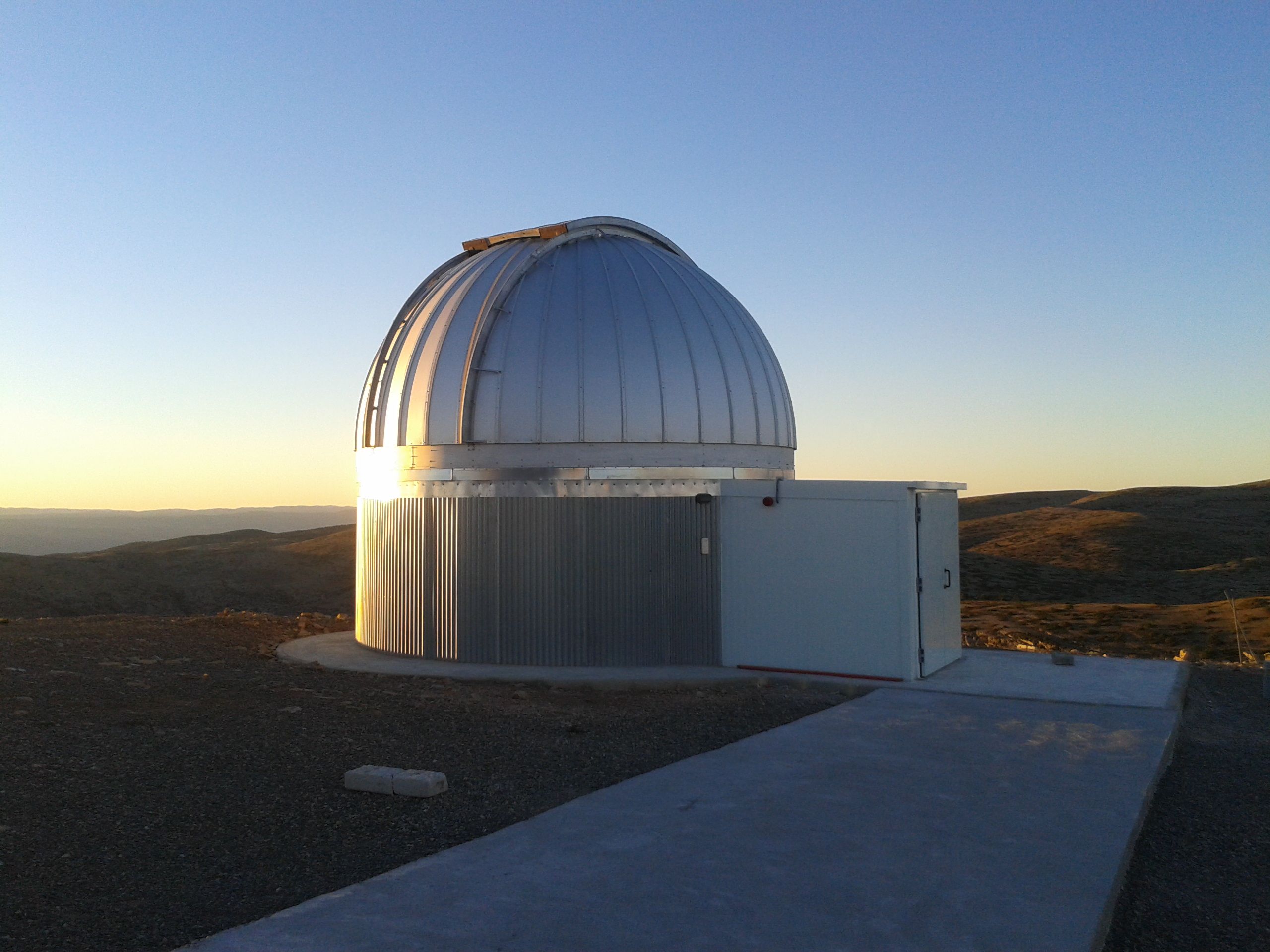} 
   \includegraphics[height=1.7in]{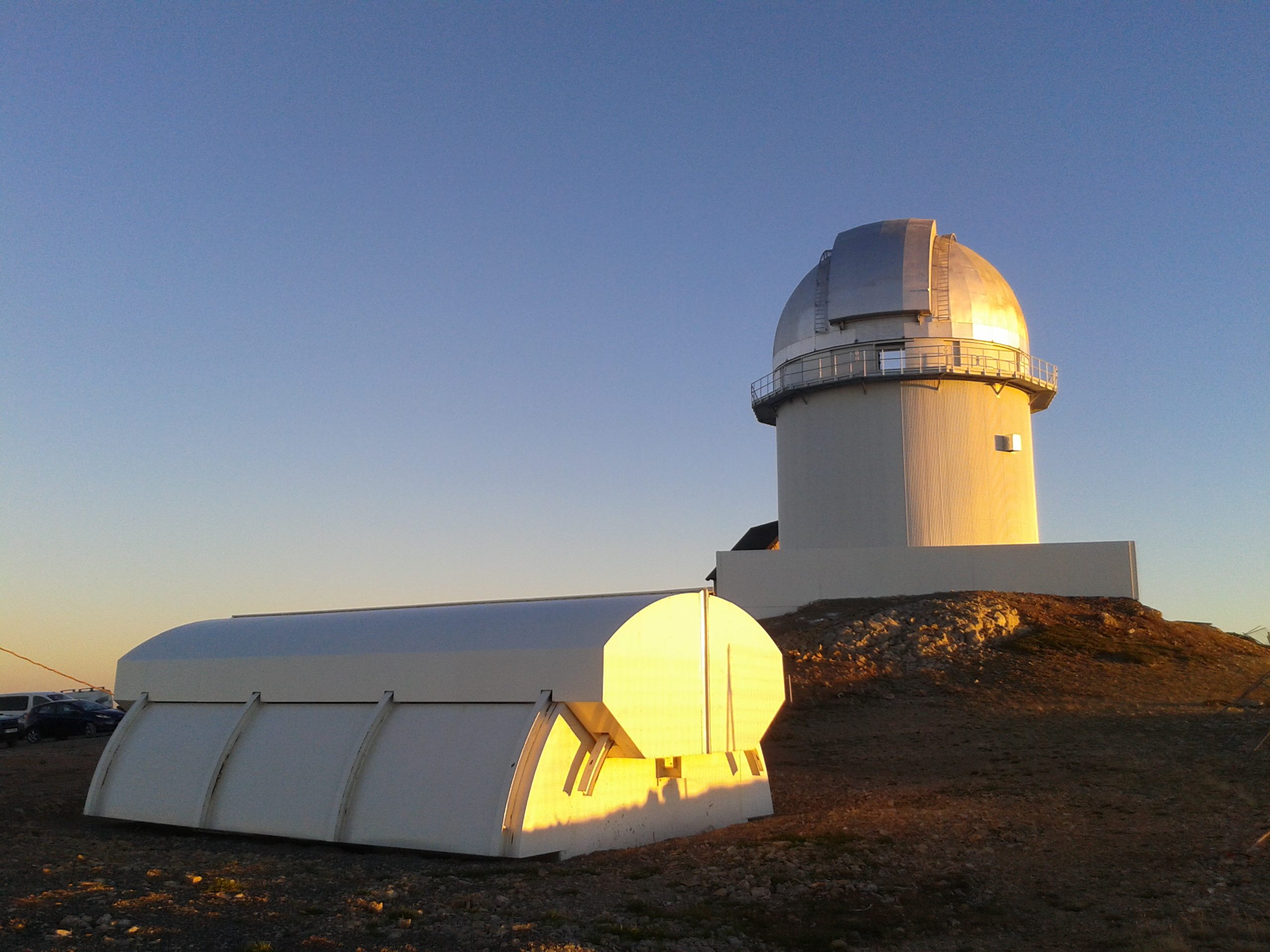} 
  \caption{Views of the main telescope buildings at the OAJ: the JST/T250 building and the maintenance area (left), the JAST/T80 building (center), and the DIMM seeing and extinction monitor building (right). See text for details}
   \label{fig:OAJbuildings}
\end{figure}

A general layout of the OAJ is presented in
Figure~\ref{fig:OAJcw}. The JST/T250 building (top-left in Fig.~\ref{fig:OAJcw}; see also Fig.~\ref{fig:OAJbuildings} left)
consists of a main cylindrical insulated building of 21\,m
total height from the ground level, including the dome, which is air-conditioned and has a
half-sphere geometry of $\sim 13$\,m diameter, a double main shutter and a wind shield. Robotic openings
around the static concrete structure of the telescope floor allows to
control the air flow inside the dome preventing air stagnation and
temperature gradients inside the dome. The building includes
three working and storage levels, including a engineering control
room. Aside the main JST/T250 building there is an underground room of $\sim6$\,m height with an overhead crane for
storage, maintenance and mirror aluminizing procedures. A platform of 9\,m$^2$
with capacity for up to 15\,tons is used to transport the mirror in its cell from
the telescope camera to the underground aluminizing area.

The JAST/T80 building  (top-right in Fig.~\ref{fig:OAJcw}; see also Fig.~\ref{fig:OAJbuildings} center)
consists of the telescope floor, with a dome of 6.2\,m diameter, an underground floor dedicated to
storage and maintenance and an underground engineering control room. The
monitor building (top-center in Fig.~\ref{fig:OAJcw}, between the two main telescope
buildings; see also Fig.~\ref{fig:OAJbuildings} right)
contains the DIMM seeing monitor, the extinction monitor, and other devices for monitoring the
sky night quality. The enclosure of this building consists of two
openings in semicylindrical shape that deploy in opposite directions
allowing observations for altitude values larger than $\sim$20 deg.

The rest of buildings in Fig.~\ref{fig:OAJcw} are devoted to host the general installations of the OAJ, the astronomical control room,
laboratories, the data center and the residence. All the buildings communicate each other
through underground tunnels, hence guaranteeing the safe use of the observatory in case of bad weather conditions as well as an efficient maintenance of the general OAJ installations.

\subsection{The OAJ Telescopes}

\begin{figure}[t] 
   \centering
   \includegraphics[height=3.0in,angle=-90,origin=c]{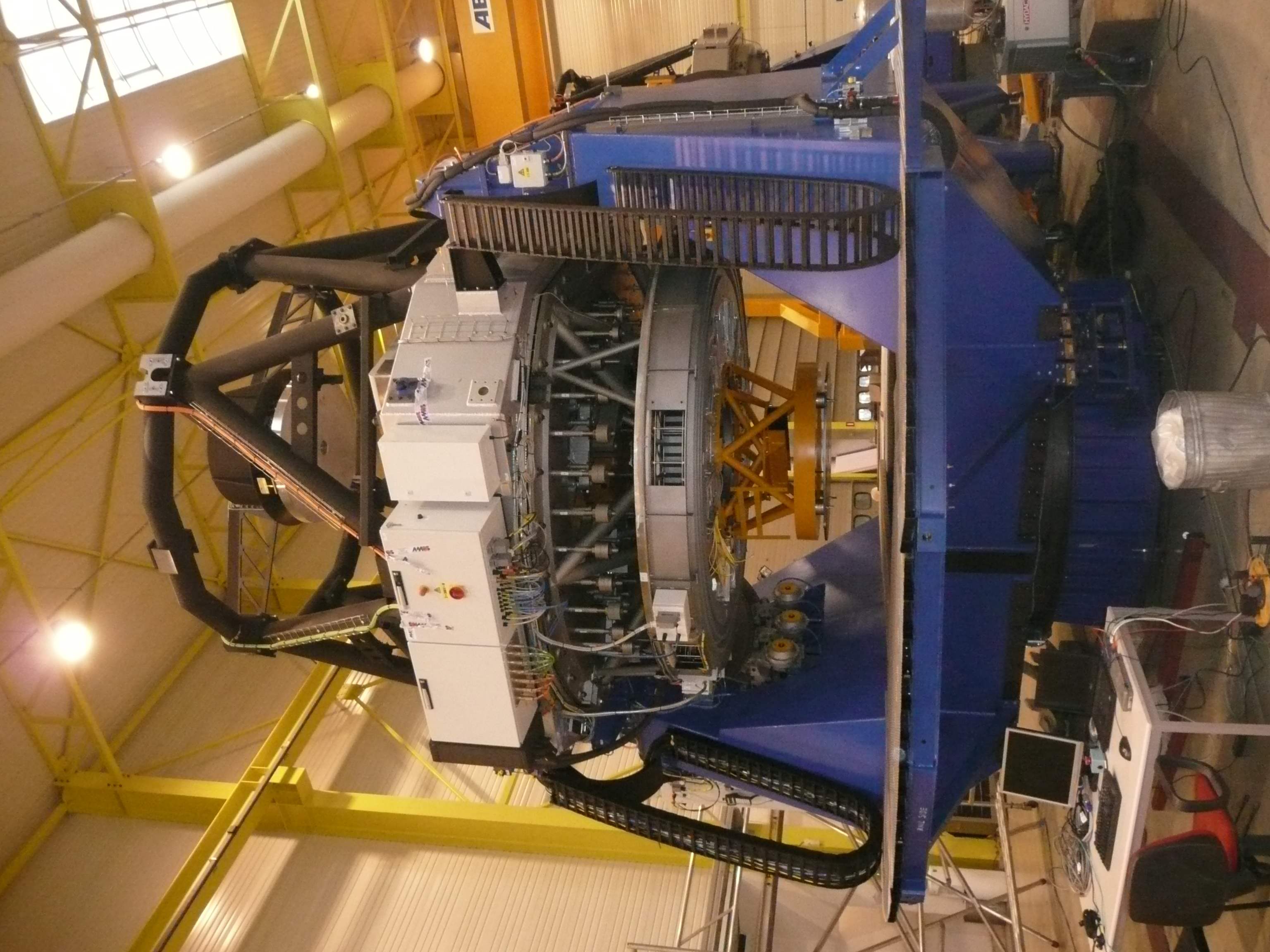} 
   \includegraphics[height=3.0in,angle=-90,origin=c]{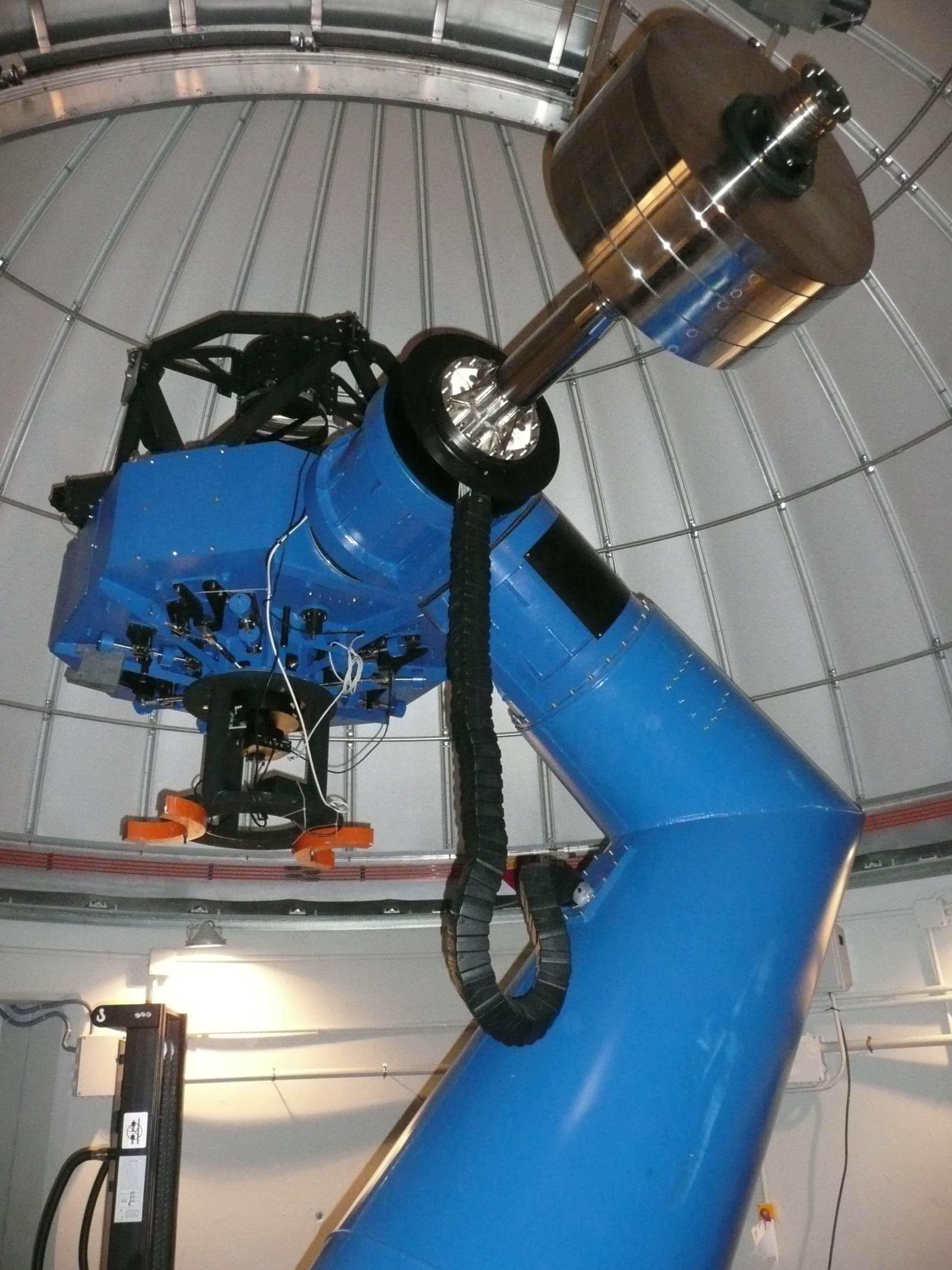} 
  \caption{The OAJ main telescopes: JST/T250 at the integration hall in AMOS headquarters (left) and JAST/T80 at the OAJ (right).}
   \label{fig:OAJtelescopes}
\end{figure}

\subsubsection{JST/T250}

The centerpiece of the OAJ is the JST/T250 (Figure~\ref{fig:OAJtelescopes}; left).
JST/T250 and its panoramic camera JPCam are driven by the scientific
requirements to conduct J-PAS. Motivated by the need of optimizing the etendue,
JST/T250 is a very fast optics telescope (F\#3.5) with a plate scale
of 22.67\,arcsec/mm. This leads to a very compact mechanical design,
with a distance between the primary (M1) and secondary (M2) mirrors of just $\sim$2.2 m. The focal plane of JST/T250 is flat and
corresponds to a Cassegrain layout. The effective
collecting area of JST/T250 is 3.75m$^2$, yielding an etendue of
26.5m$^2$deg$^2$. It is designed to support
instruments at the Cassegrain focus of up to 1300\,kg.

JST/T250 is optimized to provide a good image quality (EE50 diameter $< 10$\,$\mu$m) in the
optical spectral range ($330-1100$\,nm) all over the 48\,cm diameter focal
plane ($\sim7\sq\degr$). To achieve this goal JST/T250 includes a unique field corrector of 3
lenses of fused silica, with 4 aspherical surfaces and diameters in the range $50-60$\,cm. 
The geometry and optical performance of the
J-PAS filters and the entrance window of JPCam were taken into account during the final optical design
to simplify the aspheric departure of the field corrector of
lenses. The lenses in the field corrector are supported inside a barrel
of low carbon steel specifically designed to keep their relative
positions during operation using a passive hexapod structure made of
INVAR. The barrel is rigidly connected to the fixed flange of the
instrument rotator and sealed to the entrance window of JPCam, keeping
a dry, slightly hyperbaric atmosphere around the lenses.

The JST/T250 guiding system consists of a set of 4 auxiliary CCDs at
the edges of the JPCam focal plane. Also, in order to keep the system
in focus and preserving the image quality during the survey execution,
8 additional CCDs located at the edges of the focal plane in extra and
intra focal positions allow to perform wave-front curvature sensing
corrections in real time. The control system of JST/T250 allows to
work in continuous closed loop, analyzing the defocused images at the
auxiliary CCDs and providing the Zernike coefficients to the telescope
control system that converts it in M2 corrections in piston, tip and
tilt to the M2 hexapod.

For the alignment, testing and verification of JST/T250, the OAJ will make use
of two independent verification cameras that can be independently
displaced to any position of the focal plane by means of a set of $x$
and $z$ translation stages. Having two independent cameras allows for
simultaneous testing of the image quality and telescope performance in
different regions of the focal plane, which is essential given the
large focal plane of JST/T250.

\subsubsection{JAST/T80}

The JAST/T80 has an 83 cm diameter M1 with a FoV of $2\sq\degr$. It is also a fast optics telescope (F\#4.5) driving a plate scale at the Cassegrain focal plane of 55.56\,arcsec/mm. Mechanically, JAST/T80 has a German-equatorial mount (Fig.~\ref{fig:OAJtelescopes}; right). The optical tube assembly has also a very compact layout, with just $\sim830$\,mm between M1 and M2. With a weight of around 2500\,kg, JAST/T80 supports instruments at the Cassegrain focus of up to 80\,kg. Together with its panoramic camera, T80Cam, JAST/T80 will be primarily devoted to perform J-PLUS and the photometric calibrations of J-PAS. Nevertheless, its large FoV and high sensitivity makes it ideal for many other scientific goals. 

Like JST/T250, the optical design is based on a Ritchey-Chretien
configuration plus a field corrector of three lenses of fused silica,
in these case with just spherical surfaces and diameters are in the
range $15-17$\,cm. The whole system is optimized to work in the
optical range, yielding a polychromatic image quality better than
9.0\,$\mu$m (EE50; diameter) inside the 13\,cm diameter focal plane
($\sim 3.1\sq\degr$), after having accounted for all error sources in
the error budget. The design is also optimized to account for the J-PLUS filters and the T80Cam entrance window in the optical path. 

Guiding at JAST/T80 is carried by means of a 20\,cm piggy-back telescope with an additional CCD camera. Because of the large FoV, keeping the system in focus and free of aberrations all over the FoV is expected to require small M2 corrections through the hexapod every few hours. For this reason, a specific procedure for wave-front curvature sensing has been designed at CEFCA, making use of the scientific CCD of T80Cam. Since JAST/T80 is not so demanding of this type of corrections as JST/T250 is, no auxiliary CCDs are necessary in this case. 

\FloatBarrier 

\section{The J-PAS Cameras}
\label{jpcam}
\FloatBarrier 

  To carry out the J-PAS and J-PLUS surveys, the OAJ telescopes will be equipped with 
JPCam and T80Cam, two panoramic cameras designed to exploit survey capabilities 
of the JST/T250 and the JAST/T80, respectively. As the overall effective etendue 
of a telescope plus instrument system is ultimately determined by the CCD filling 
factor, JPCam and T80Cam have been designed to maximize the telescopes' focal plane 
coverage while maintaining the high image quality requirements.

JPCam and T80Cam are direct imaging instruments designed to work in a fast convergent 
beam at the Cassegrain foci that are based on state-of-the-art, large format CCDs. 
T80Cam will include a low-noise 9.2k $\times$ 9.2k, $10\mu$m pixel,
high efficiency CCD 
manufactured by e2V, providing a useful FoV of $2.1\sq\degr$  (65\% focal plane coverage) 
with a plate scale of 0.55 arcsec/pix. JPCam, on the other side, will include a 
mosaic of 14 9.2k $\times$ 9.2k, $10\mu$m pixel CCDs specially developed by e2V for J-PAS, 
providing a useful FoV of $4.7\sq\degr$  (67\% focal plane coverage) with a plate scale 
of $0.2267 \arcsec$/pix. Moreover, JPCam will include 12 auxiliary detectors at the 
focal plane for guiding and wave front sensing. 

The cameras are equipped with a filter unit designed to mount the complete J-PLUS 
and J-PAS filter sets. T80Cam includes two filter wheels with 7 positions each, 
this configuration allows the 12 J-PLUS filters to be permanently installed on 
the camera so no night-to-night filter exchange is required. Following the same 
low maintenance strategy, JPCam has been equipped with a filter tray magazine that 
includes up to five filter trays, each one mounting 14 filters. This filter unit 
design permits the 70 J-PAS physical filters to be permanently installed on JPCam. 

The definition and procurement of JPCam and T80Cam is lead by a consortium of several 
funding institutions from Spain (CEFCA and IAA-CSIC) and Brazil (ON, IAG/USP, and 
CBPF). The funding of JPCam and T80Cam is guaranteed by that consortium, including 
the filter sets for both instruments. The commissioning of T80Cam and JPCam is planed 
for Q3 2013 and Q3 2015, respectively. 

\subsection{T80Cam, the wide field camera for the JAST/T80 telescope}

  JAST/T80 will be equipped with an instrument designed to exploit the telescope 
survey capabilities, the T80Cam. The JAST/T80 and T80Cam primary goal is to perform 
the photometric calibration of the JST/T250 surveys by means of the Javalambre-Photometric 
Local Universe Survey (J-PLUS). J-PLUS will image $\sim 8500\sq\degr$ of Northern Sky using 
12 filters in the optical range. These are a combination of narrow- and broad-band 
filters carefully optimized to retrieve stellar parameters ($T, log(g), [Fe/H]$) 
through flux calibrated stellar models fitting. 

 T80Cam is a wide field, direct imager that will be installed at the Cassegrain 
focus of the JAST/T80. It is equipped with an STA 1600 backside illuminated detector. 
This is a 9.2k$\times$9.2k, $10\mu$m pixel, high efficiency CCD that is read from 16 
ports simultaneously, allowing read times of 12s with a typical read
noise of 3.5 electrons (RMS). This full wafer CCD covers a large fraction of the JAST/T80's 
FoV with a pixel scale of 0.55\texttt{"}/pixel. Table \ref{TCam1} summarizes T80Cam performances. 

\newpage

\begin{table}
\centering
\label{TCam1}
\begin{tabular}{|l|l|}
\hline
FoV  & 	$\diameter =1.7\degr$ (full performance) \\
      &  $\diameter = 2.0\degr$ (reduced performance)\\
\hline 
EE50 &  $\diameter =9 \mu$m \\
EE80 & $\diameter =18 \mu$m\\	 
\hline
CCD format &	$9216\times9232$ pix \\
            &        $10 \mu$m/pix\\
\hline 
Pixel scale &	$0.55\arcsec$/pix \\
\hline 
FoV coverage &	$2.0\sq\degr$(fill factor $65\%$) \\
\hline 
Read out time &	$12s$ \\
\hline 
Read out noise & $3.5e^{-}/$pixel \\
\hline 
Full well &	$123ke^{-}$ \\
\hline 
CTE&	0.99995 \\
\hline 
Dark current &	$0.0008e^{-}/$pixel $s^{-1}$ \\
\hline 
Number of filters  &	12 \\
\hline 
\end{tabular}
\caption{T80Cam parameters}
\end{table}

\begin{figure}
\centering
\includegraphics[width=0.8\textwidth,keepaspectratio]{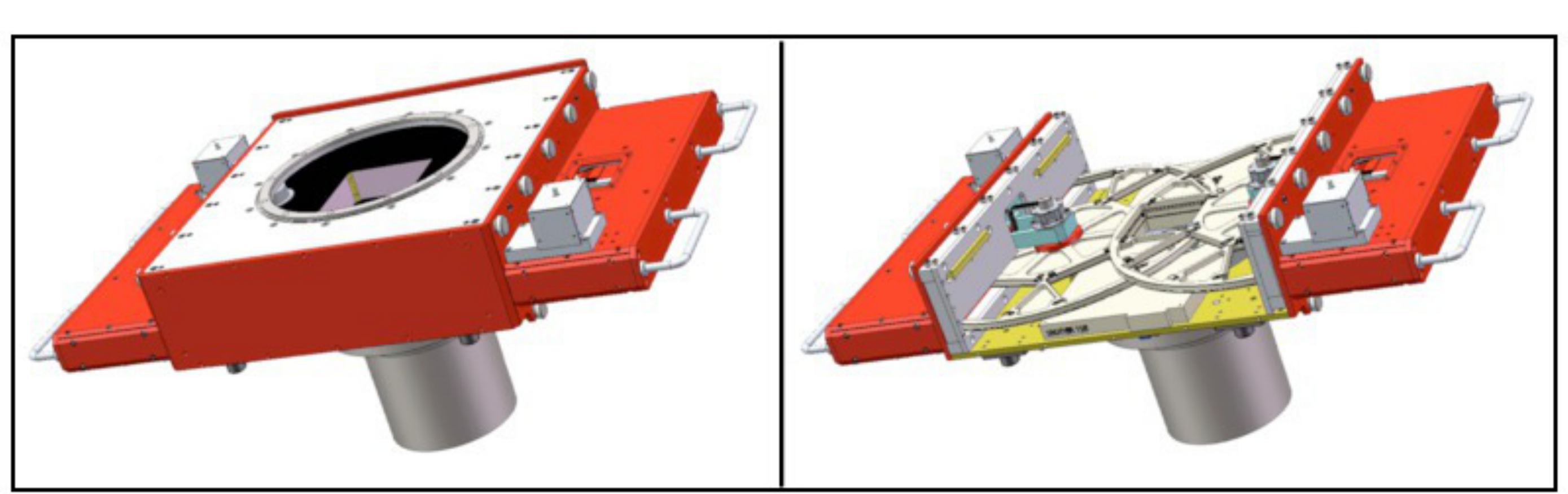}
\caption{T80Cam design. Upper panel shows the complete assembled instrument. 
The two main subsystems can be clearly identified. The top part of the instruments 
represents the FSU containing the shutter and the two filter wheels. The cylindrical-shaped 
object underneath the FSU represents the camera subsystem. Lower panel shows the 
same view of T80Cam after cover removal to show the two filter wheels and the shutter.}
\label{fcam1}
\end{figure}

The instrument consists of two main subsystems: the filter and shutter unit (FSU) 
and the camera subsystem (see Figure \ref{fcam1}). The FSU holds two removable filter 
wheels and the shutter. The camera subsystem, below, comprises the cryostat, the 
cooling and vacuum systems, the CCD, an optically powered entrance window and the 
detector electronics.

\subsubsection{The Camera System}

\begin{figure}
\centering
\includegraphics[width=0.8\textwidth,keepaspectratio]{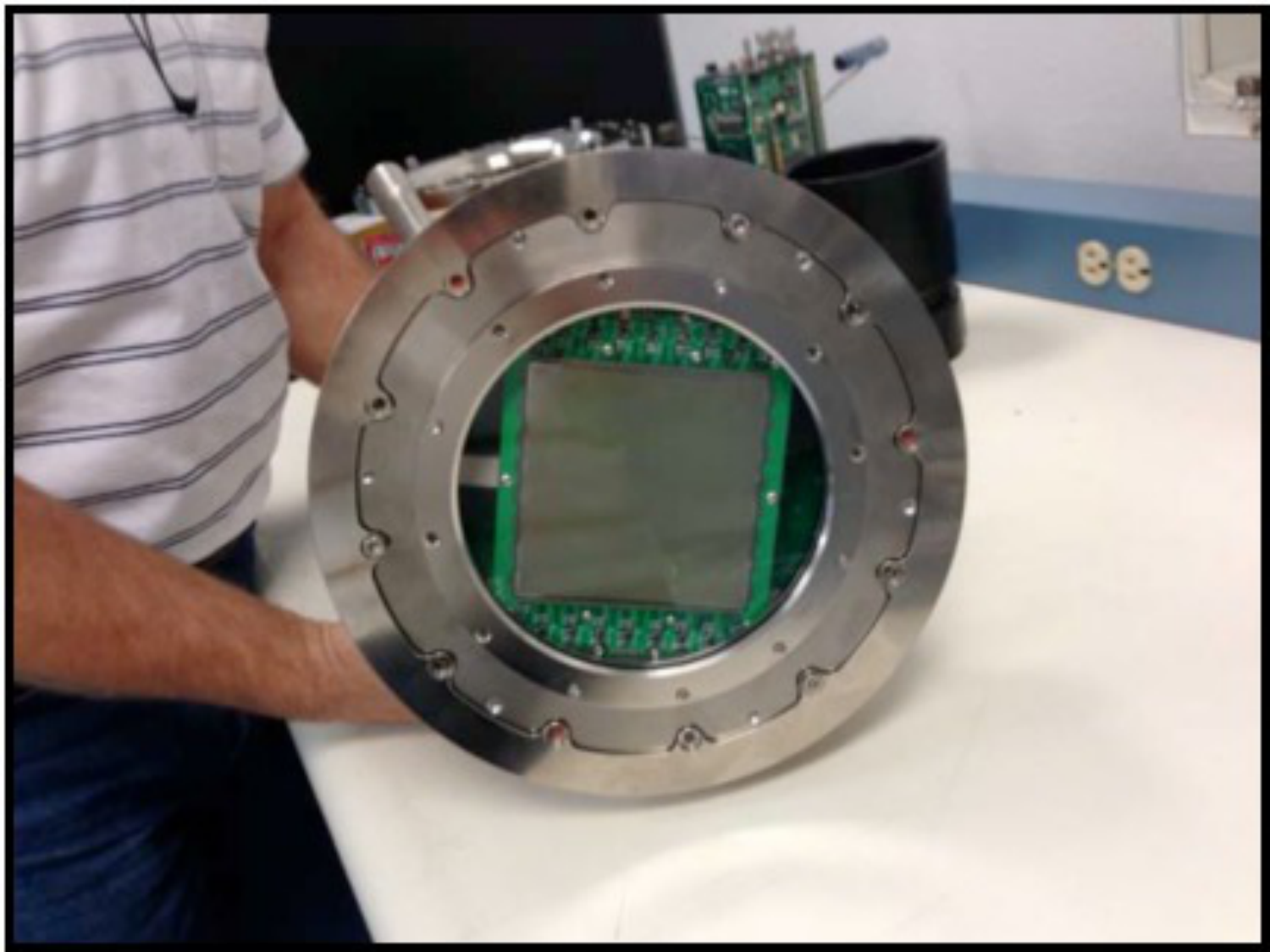}
\caption{1110S camera equipped with an engineering CCD during its 
manufacture at the Spectral Instruments premises}
\label{fcam2}
\end{figure}

 The camera system is an 1110S camera manufactured by Spectral Instruments (Tucson, 
AZ, USA). It is equipped with a grade-1, backside illuminated E2V CCD, a 9.2k$\times$9.2k, 
$10\mu$m pixel high efficiency CCD. This CCD has an image area of 92.16mm x 93.32mm 
and has a broadband AR coating for optimized performance from 380 to 850nm. Figure \ref{fcam2} 
 shows the 1110S camera with an engineering CCD mounted during its manufacture 
at Spectral Instruments premises.

 The sensor is cryo-cooled to an operating temperature between $-100\degr$C and $-110\degr$C 
with a cryo-tiger refrigeration system, a closed-cycle Joule-Thomson effect cryogenic 
refrigerator system. The chamber will be evacuated to a level of
$10^{-4}$ Torr using a turbo dry vacuum pump. 

The camera entrance window is in fact the fourth element of the JAST/T80 field corrector, 
and together with the filters, it is part of the telescope optical design optimization. 
The window is a 10mm thick, weakly powered field-flattener with an 8mm distance 
between its inner surface and the focal plane. The entrance window has been manufactured 
by Harold Johnson Optical Laboratories (Gardena, CA, USA).

\subsubsection{The FSU System}

The FSU has being designed and manufactured by the Instituto Nacional de Pesquisas 
Espaciais (INPE) and CEFCA. It includes the filter unit, the shutter and the cryostat 
support flange. The later allows for fine alignment of the camera system with respect 
to the telescope optical axis. The shutter is a 125mm clear aperture shutter that 
has been acquired through the Argelander Institut fur Astronomie. The FSU holds 
two removable filter wheels, each one capable of holding 6 filters plus an empty 
position. As it can be seen, filters are operating in a converging beam and close 
to the detector. Figure \ref{fcam3} shows the FSU assembly during its finals stages of 
integration and commissioning at CEFCA. 

\begin{figure}
\label{fcam3}
\centering
\includegraphics[width=0.8\textwidth,keepaspectratio]{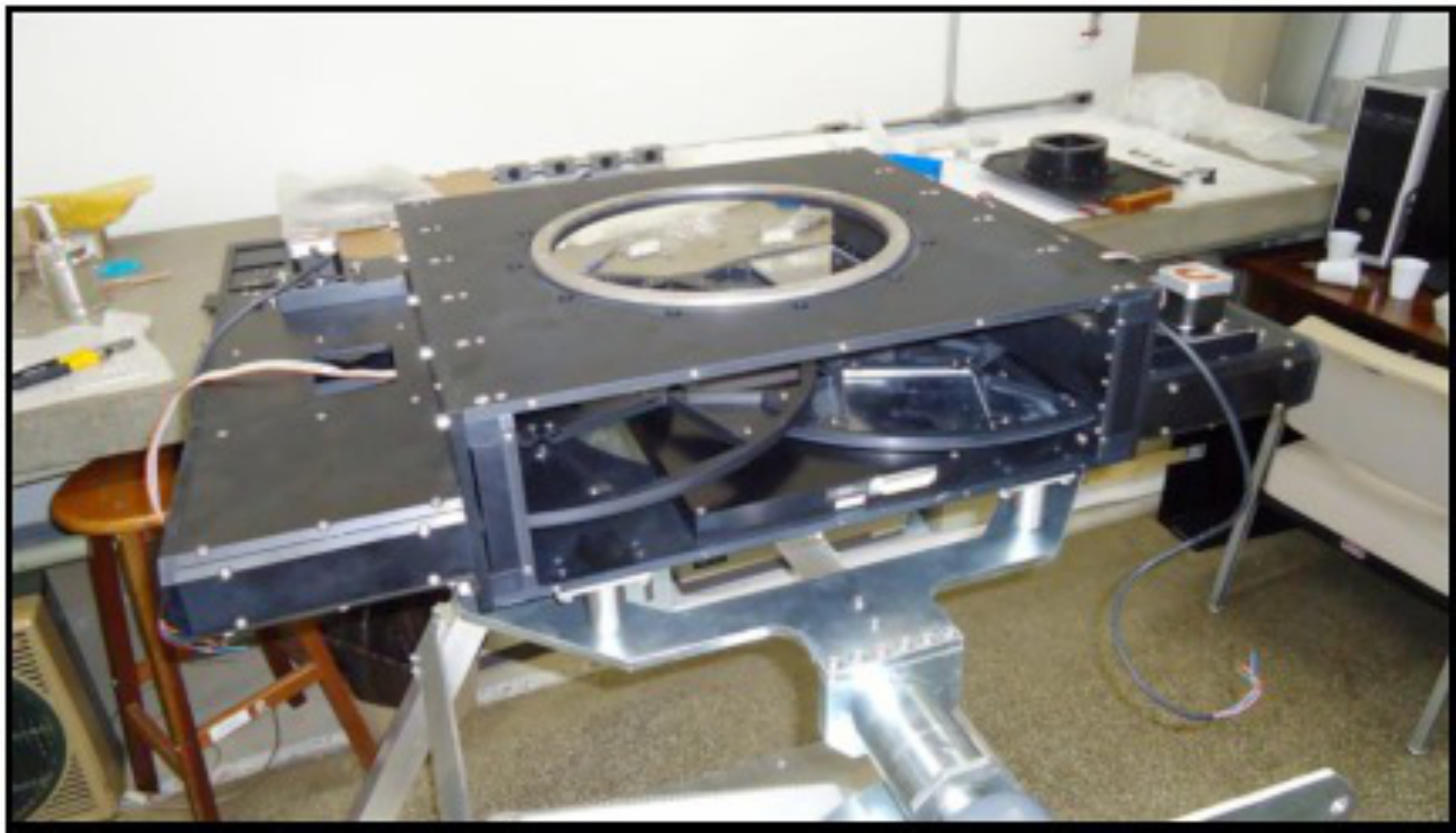}
\caption{FSU during its AIV at CEFCA laboratory}
\label{fcam3}
\end{figure}

 Summarizing, T80Cam design has been driven by the main science goals defined by 
the Science Working group inside the J-PAS collaboration. The instrument has therefore 
been optimized to develop the J-PLUS survey. The commissioning and acceptance of 
T80Cam is planed for the end of 2013. In this time scale, the J-PLUS survey to 
be performed with this camera will start in early 2014.

\subsection{JPCam, the 1.2Gpixel camera for the J-PAS survey}

 The main scientific instrument for JST/T250 is JPCam, a 1.2 Gpixel camera that 
will be installed at the Cassegrain focus. JPCam has been designed to perform the 
J-PAS survey, so maximizing the efficiency of FoV and wavelength coverage while 
guaranteeing a high image quality over the whole focal plane have been the main 
instrument design drivers. 

JPCam is a wide field, direct imager equipped with a mosaic of 14 9.2k$\times$9.2k, 
10$\mu$m pixel backside illuminated, low noise detectors from e2V. Each science CCD 
is read from 16 ports simultaneously, allowing read times of 12s with a typical 
read noise of $4e^{-}$ (RMS, goal). Its filter unit has been designed to admit 
5 filter trays, each mounting 14 square filters corresponding to the 14 CCDs of 
the mosaic. Each CCD will view only its corresponding filter avoiding optical cross-talk 
from their neighbors. The filters will operate close to, but up-stream from, the 
dewar window in a fast converging beam. With this configuration, JPCam will cover 
$4.7\sq\degr$ (67\% focal plane coverage) with a plate scale of $0.2267\arcsec$/pix 
and will allow all the 70 required filters (56 main J-PAS filters plus several copies 
of the broad-bands) to be permanently installed on the camera, so no night-to-night 
filter exchange will be required.

Because of the JST/T250 telescope's very wide FoV combined with the confirmed excellence 
of the OAJ's intrinsic site seeing, JPCam is required to fully optimize and maintain 
the image quality across the full focal plane of the mosaic. Optical analysis reveals 
that it is necessary, not only to guide the telescope and keep it optically aligned 
by adjusting the position of its secondary mirror, but also of the focal plane 
itself. To perform this task JPCam includes an hexapod actuator system that is 
controlled thanks to a set of wavefront sensors in the periphery of the instrument's 
FoV. So the JPCam 14 science CCD mosaic is complemented with 12 auxiliary detectors, 
4 for auto-guiding (AG) and 8 for  wavefront sensing (WFS) tasks. The auxiliary 
detectors are E2V frame-transfer devices fed by light from broad-band filters mounted 
in the edges and corners of each filter tray. Table \ref{TCam2}
summarizes JPCam performances. 

\begin{table}
\label{TCam2}
\centering
\begin{tabular}{|l|l|}
\hline
 FoV  & 	$\diameter =1.7\degr$ \\
\hline  
EE50 &  $\diameter =11 \mu$m \\
EE80 & $\diameter =22 \mu$m\\	 
\hline  
CCD format  &  Science (14X) $9.216k\times9.232k$pix$^2$,  $10 \mu$m/pix\\
            &  Guiding (4X) $1.024k\times1.024k$ pix$^2$,  $13 \mu$m
            (frame transfer)\\
            &  Wavefront sensing (8X) $2.048k\times2.048k$pix$^2$, $15
            \mu$m (frame transfer)\\
\hline 
Pixel scale &	$0.2267\arcsec$/pix \\
\hline 
FoV coverage &	$4.7\sq\degr$(fill factor $65\%$) \\
\hline 
Read out time &	$12s$ \\
\hline 
Read out noise & $4e^{-}/$pixel (goal) \\
\hline 
Full well &	$130ke^{-}$ \\
\hline 
CTE&	0.99995 \\
\hline 
Dark current &	$0.0006e^{-}/$pixel $s^-1$ \\
\hline 
Number of filters  &	70 \\
\hline 
\end{tabular}
\caption{JPCam parameters}
\end{table}

\begin{figure}
\centering
\includegraphics[width=\textwidth,keepaspectratio]{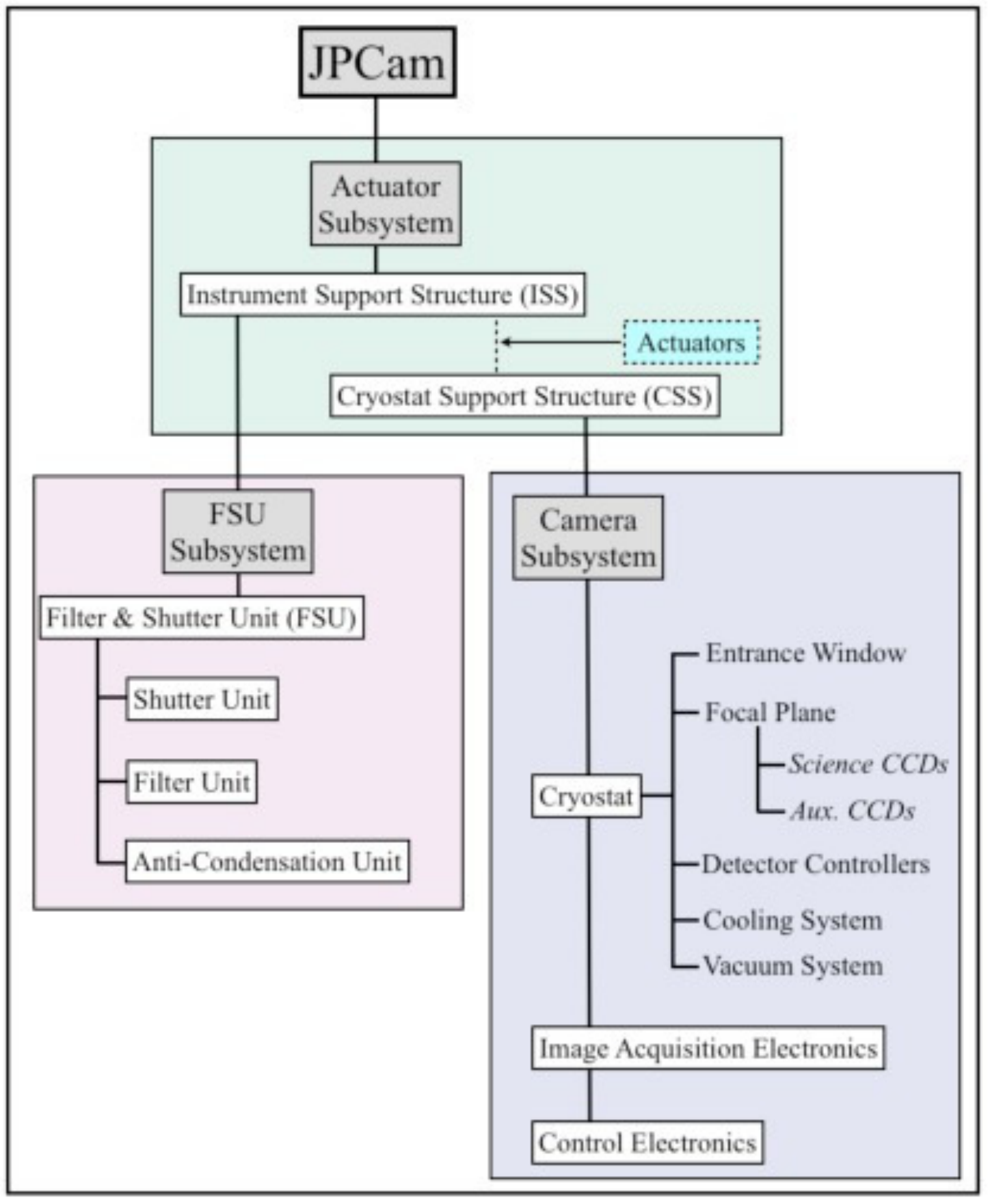}
\caption{JPCam product tree}
\label{fcam4}
\end{figure}

Therefore, JPCam's final design includes the following three main subsystems (Figure \ref{fcam4}):

- Actuator Subsystem: The hexapod actuator system (HAS) attach the cryostat to 
the Instrument Support Structure (ISS) through the Cryostat Support Structure (CSS) 
and provides the required focus and tilt adjustments to the focal plane. The HAS 
is being designed and manufactured by the company NTE-Sener (Barcelona, Spain).

- Filter and Shutter Unit Subsystem (FSU): The FSU is mounted directly to the ISS 
and comprises the filter tray exchange mechanism and the shutter. FSU is being 
designed and constructed by a Brazilian consortium led by INPE (Instituto Nacional 
de Pesquisas Espaciais). The massive 515mm aperture shutter is supplied by the 
Argelander-Institut für Astronomie, Bonn.

- Camera Subsystem (CryoCam): The CryoCam comprises the dewar entrance window, 
the CCD mosaic and their associated controllers, the cooling and vacuum systems 
and the image acquisition electronics and control software. The CryoCam is being 
supplied by e2v under contract to J-PAS.

\begin{figure}
\centering
\includegraphics[width=0.8\textwidth,keepaspectratio]{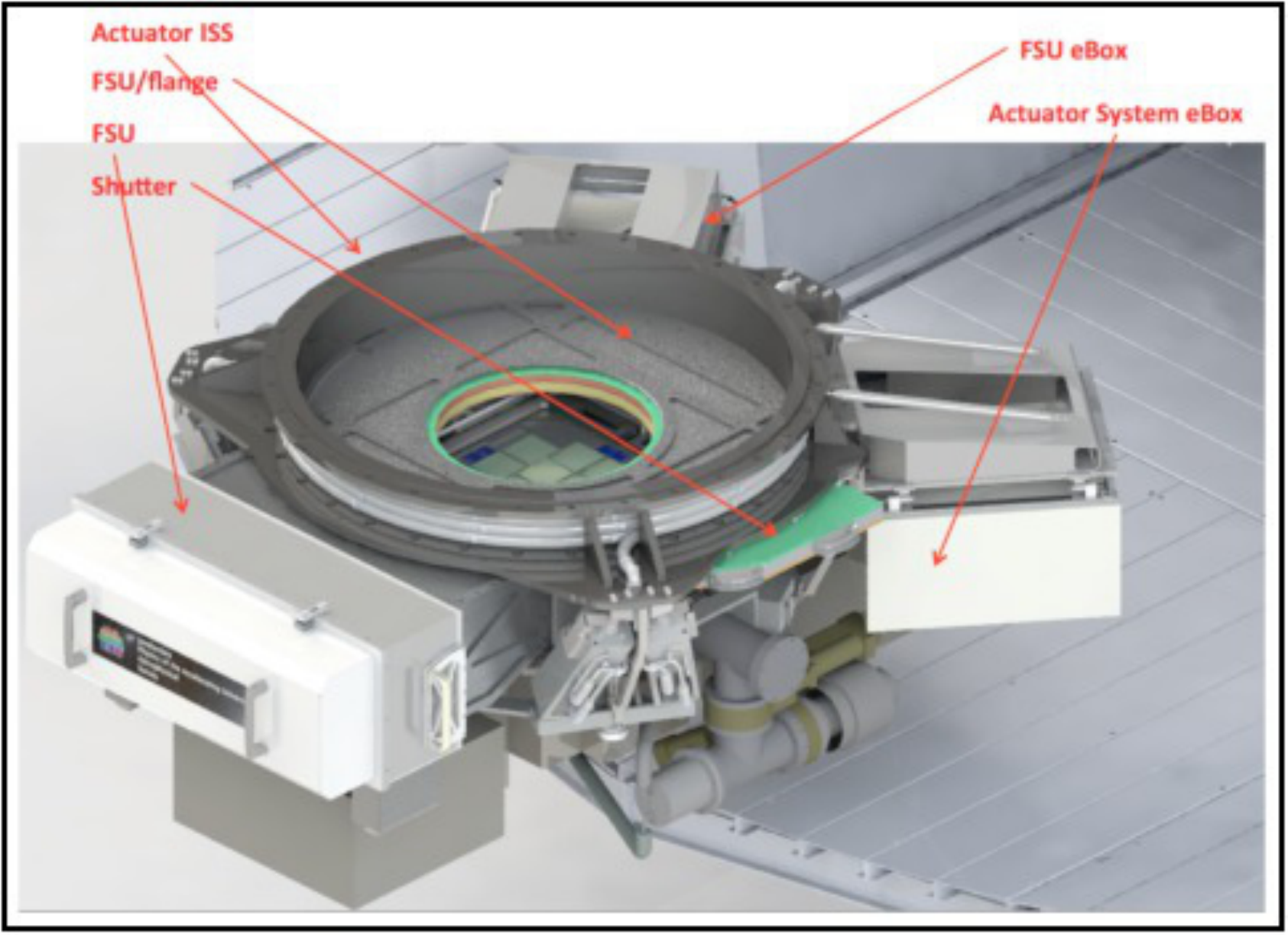}
\caption{A view of JPCam fully assembled where the different subsystems 
are identified.}
\label{fcam5}
\end{figure}

The three main subsystems are identified in the fully assembled JPCam design, as 
shown in Figure \ref{fcam5}.

\subsubsection{The Hexapod Actuator System (HAS)}

\begin{figure}
\plottwo{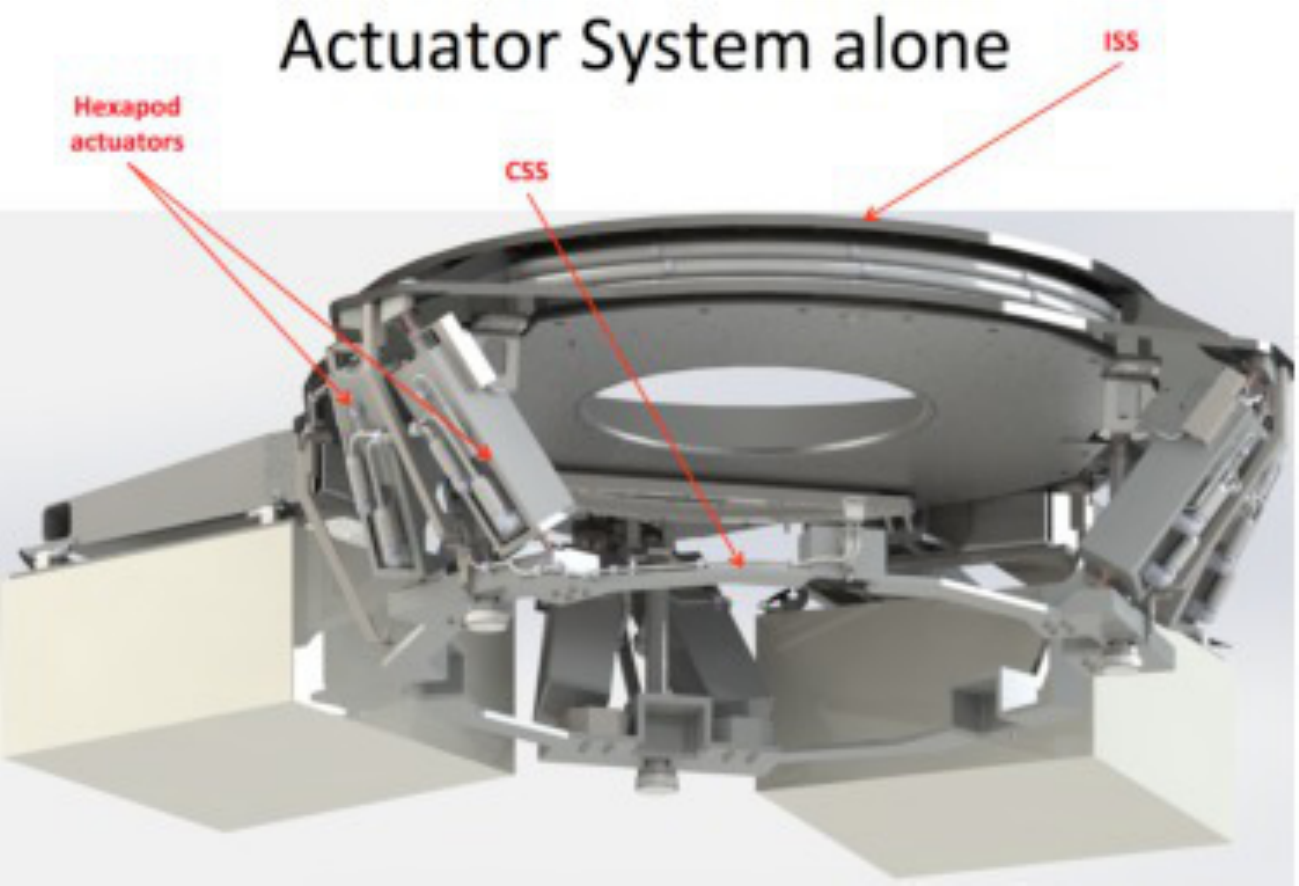}{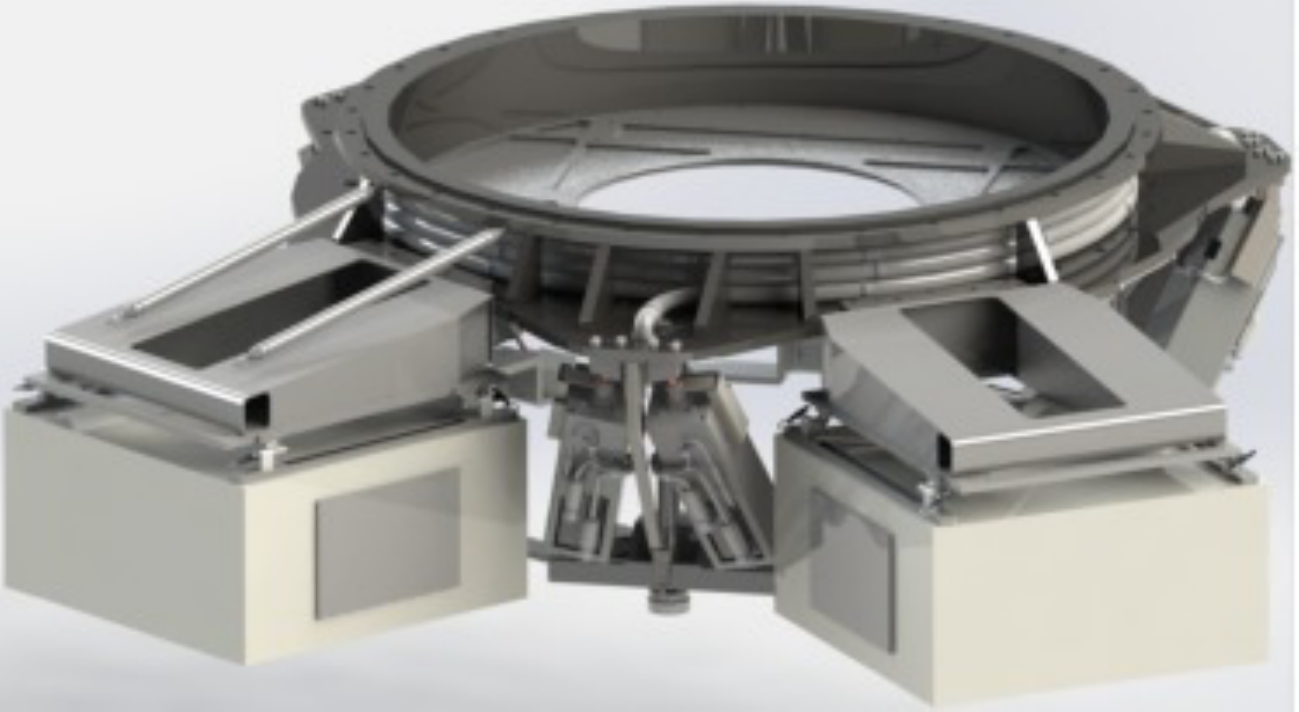}
\caption{JPCam HEX}
\label{fcam6}
\end{figure}

The HAS is providing the CryoCam with focus and tip-tilt movement aimed to compensate 
the telescope deformation produced by the gravity and/or temperature changes. It 
will  be able to move the Cryocam, whose weight is about 600Kg, with an accuracy 
of 4µm (focus) and 20 arcsec (tip-tilt). The main elements of the HAS are (Figure \ref{fcam6}):

- ISS: it interfaces with the Telescope flange, holds the FSU, and is the attachment 
to a set of actuators. 

- CSS: supported by the set of actuators, it is in charge of keeping in place the 
CryoCam. It is the moving part of the HAS. 

- Hexapod System and hexapod control electronics: Six actuators assemblies attached 
to both the ISS and CSS configure the hexapod system.

\subsubsection{The Filter and Shutter Unit (FSU)}

\begin{figure}
\plottwo{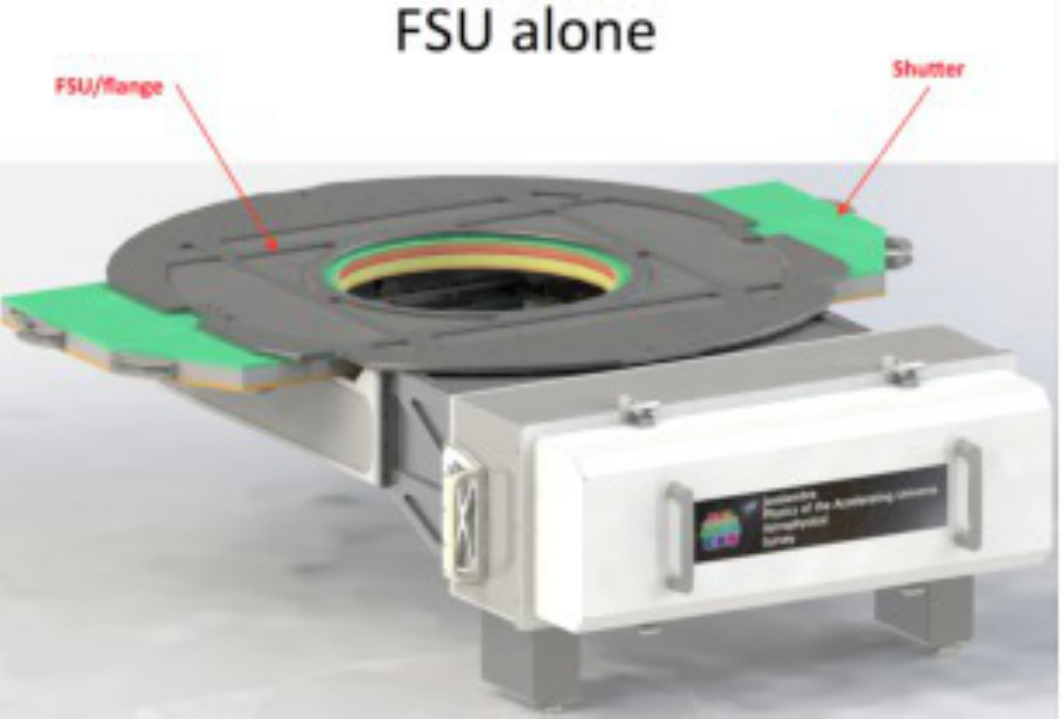}{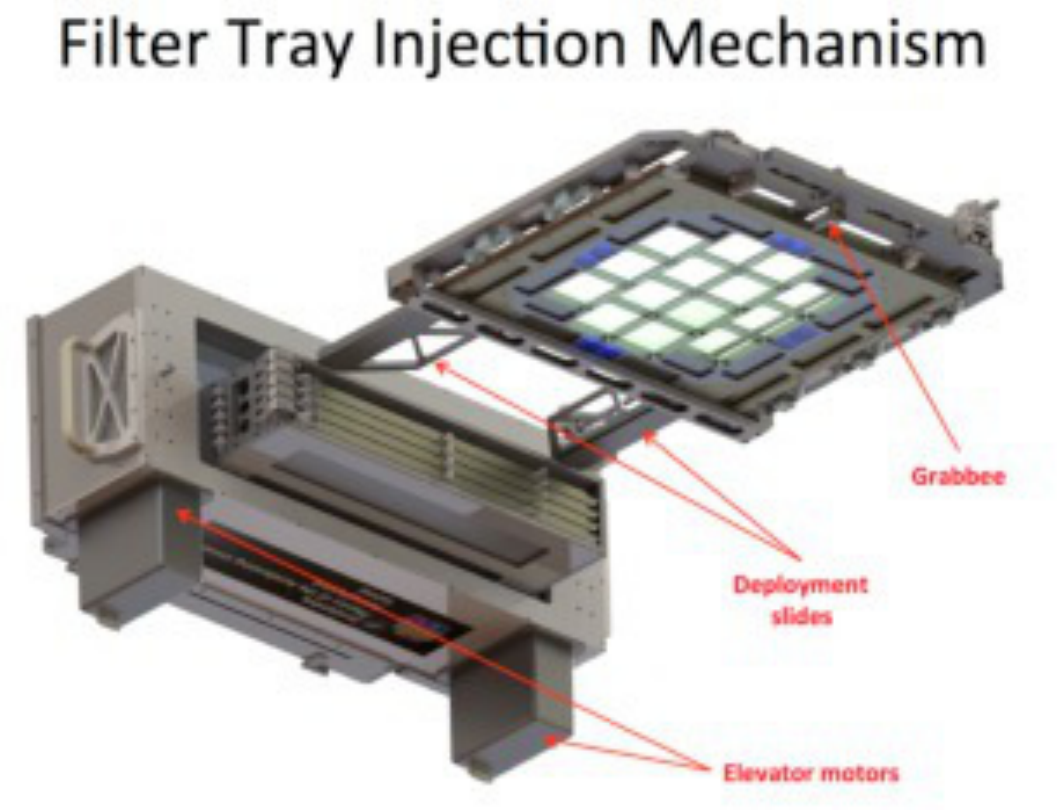}
\caption{JPCam FSU (upper panel) and a closer view to the filter tray 
injection mechanism (lower panel)}
\label{fcam7}
\end{figure}

The FSU is designed to admit 5 filter trays. All five of which contain 14 square 
filters each corresponding to the 14 CCDs of the detector mosaic. Additionally, 
the filter trays also have filter holders for broad-band filtering of the 12 auxiliary 
WFS and AG chips. Details of the FSU are shown in Figure \ref{fcam7}.  

The 5 filter trays are selectable remotely so the FSU will include the motors and 
encoders and the control system needed for their operation. Each filter tray is 
designed to be easily and manually removable and exchangeable from the closed frame. 
Individual filters can be manually removable from their tray once the tray has 
been removed from the module. 

  The focal-plane of the T250 telescope is non-telecentric and hence, in order to 
retain the steepness of each intermediate-band filter bandpass profile and the 
uniformity of its wavelength centering, the filter must be held in each tray so 
as to induce a differential tilt in each of the 14 filters of the mosaic, so that 
each filter is perpendicular to the chief ray at its centre. This amounts to a 
maximum tilt of 3.5deg for the outer filters of the mosaic equivalent to a 6mm 
departure from a flat surface. Furthermore, in order to minimize the peripheral 
vignetting of the CCD by its corresponding filter, the distance between filters 
and CCDs is required to be as close as practical. A nominal gap of 4mm between 
the filters and the dewar window has been chosen to allow for filter tray deployment 
and the necessity of positioning the mosaic focal plane with the HAS. 

 The JPCam has a 515mm diameter aperture and is supplied by the Argelander-Institut 
für Astronomie, Bonn. It is a ``two-curtain'' shutter that guarantees an homogeneous 
illumination of the focal plane. It allow for exposures as short as 10ms with an 
exposure uniformity better than 1ms over the full FoV.

Finally, in order to prevent frost and/or condensation from forming on the large 
(about 550mm diameter) dewar window, the FSU will be sealed and over-pressured 
with N2.

\subsubsection{The Camera System(Cryocam)}

\begin{figure}
\includegraphics{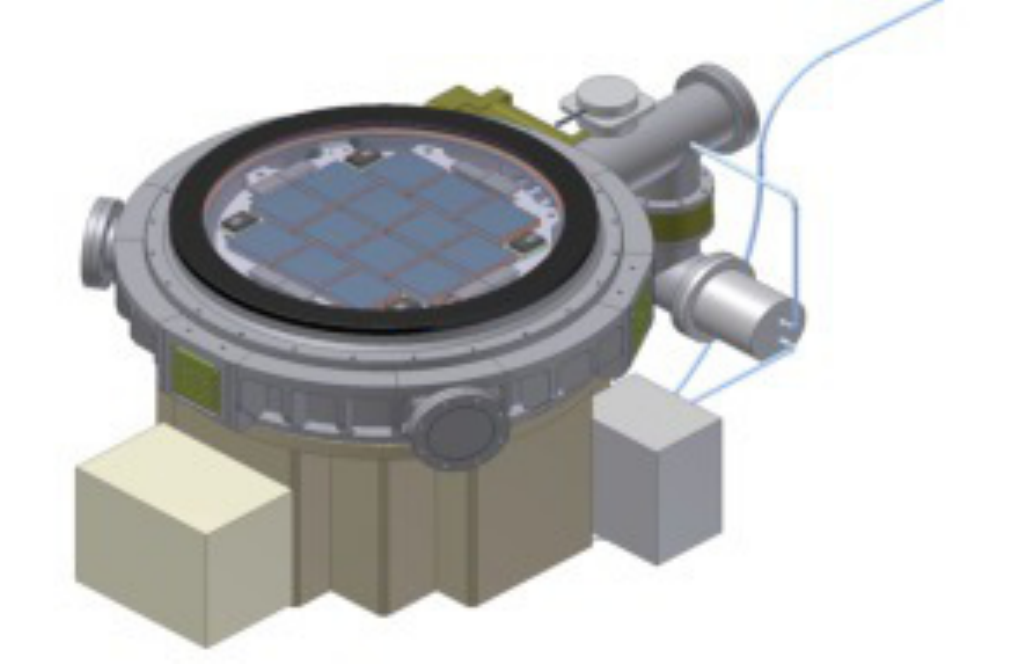}
\caption{JPCam CryoCam design}
\label{fcam8}
\end{figure}

\begin{figure}
\includegraphics{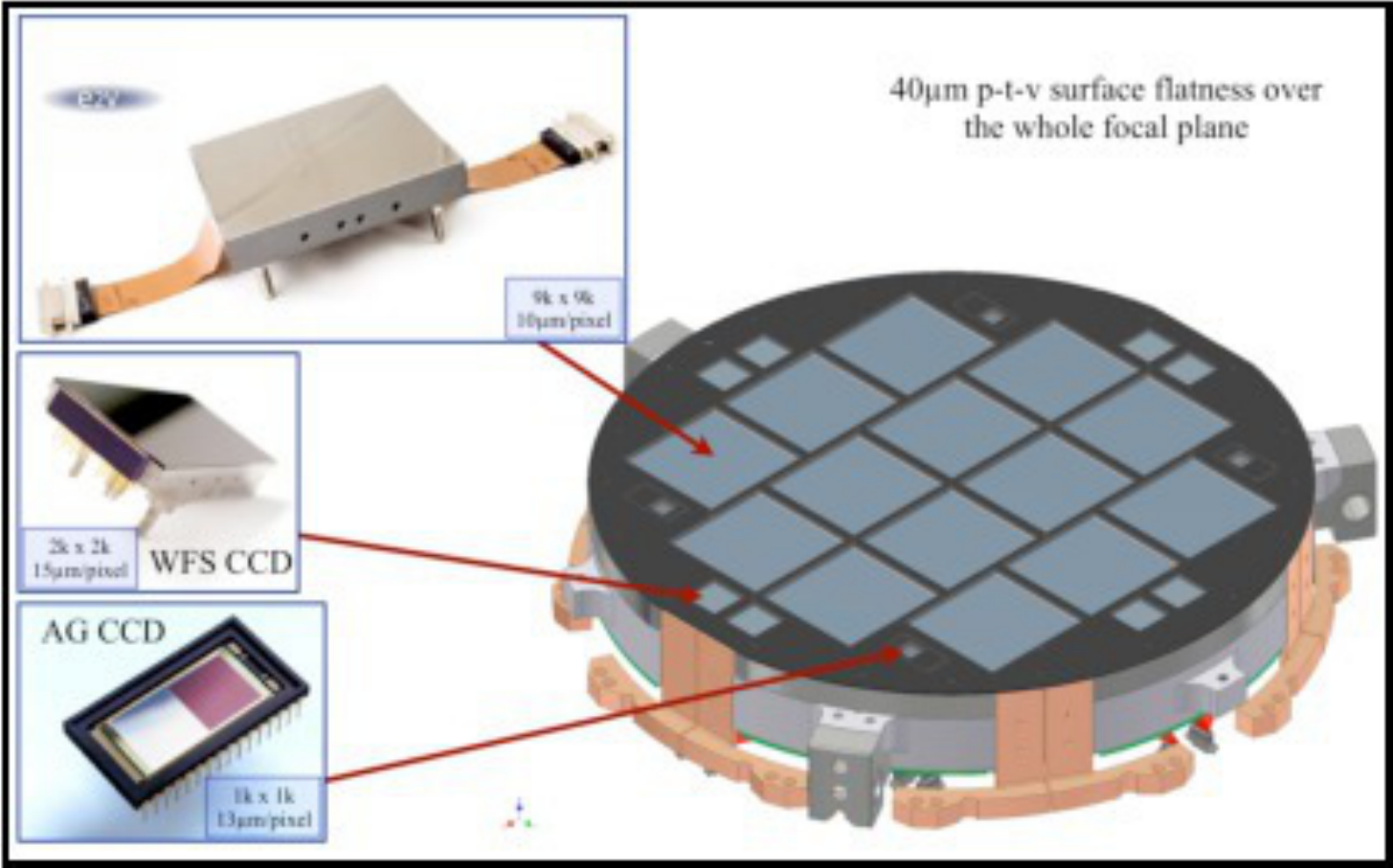}
\caption{JPCam's focal plane layout as supplied by e2v. The 14 loosely 
packed, full-wafer, e2v science sensors are shown mounted on the FPCP. In the periphery 
are mounted 4, 1k2 frame-transfer (FT) guide CCDs and 4 pairs of 2k2 FT WFSs. The 
black surface on top of the CCD mosaic if the light baffle intended to remove any 
undesired reflections inside the CryoCam. Thermal links are also shown.}
\label{fcam9}
\end{figure}

The CryoCam is being supplied by e2v. The CryoCam design is shown in Figure \ref{fcam8}, 
while the layout of the focal plane cold plate is given in Figure \ref{fcam9}, where 
the 14 science sensors and 12 auxiliary guide and WFS CCDs are shown. 

The data from the sensors is gathered in the Detector Electronics Box and transmitted 
via four Camera Link ports to three PCs, the science data, the AG and the WFS PCs. 
Two fibre optic channels are used to transmit the Science CCD data to reduce the 
data transmission time. The time to transfer a full image from JPCam to the science 
data PC is 8.4s, lower than the read out time.

\begin{figure}
\includegraphics{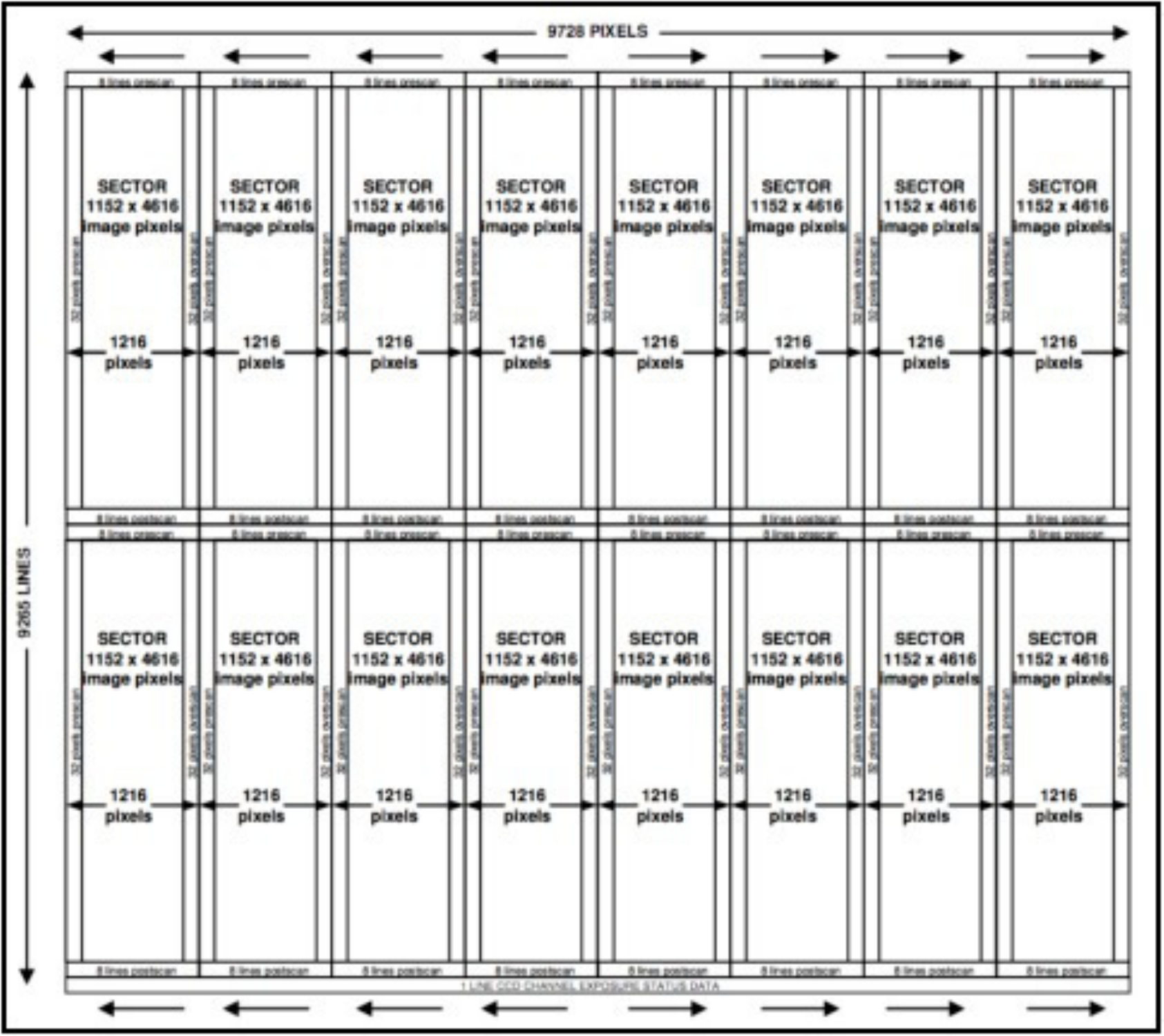}
\caption{Science CCD array Camera Link frame format (no binning).}
\label{fcam10}
\end{figure}

The format of one science CCD image data is shown in Figures \ref{fcam10}. The entire 
CCD is currently read out as if it were a single large image (9728 pixels wide 
x 9265 pixels high) and an additional line of status data is appended on the end.

 The cryostat will cryo-cooled to an operating temperature between -100°C and -110°C 
with  cold nitrogen gas. A large LN2 tank, mounted on the telescope fork, will 
feed the cryostat through routing of the flexible cooling lines via the telescope 
cable wraps, as required to accommodate both cassegrain and altitude rotation.

The chamber will be evacuated to a level of 10\textsuperscript{-6} Torr using a 
cryostat mounted turbo-pump  that will run continuously into a Mini-Roots-dry-backing-pump 
mounted at some distance at the telescope fork where the liquid nitrogen tank is 
mounted. The two pumps will be connected through small bore flexible Stainless 
Steel Tubing routed through the telescope cable wrap.

The camera entrance window is in fact the forth element of the JST/T250 field corrector, 
and together with the filters, it is part of the telescope optical design optimization. 
The window is a 545mm diameter and 27mm thick, weakly powered field-flattener with 
an 8mm distance between its inner surface and the focal plane mosaic. 

Summarizing, JPCam design has been driven by the main science goal defined by the 
J-PAS collaboration, that is, the J-PAS survey. The commissioning and acceptance 
of JPCam is planed for mid 2015, so the J-PAS survey should will start in late 
2015.

\FloatBarrier 

\section{The J-PAS Collaboration}
\FloatBarrier 

\subsection{A Brief History of J-PAS}

  J-PAS stands for Javalambre-PAU Astrophysical Survey, and its
  starting point was in 2006, when the Spanish Government opened a
  call for proposals focused on large projects, 
the so-called ``Consolider'' grants. A collaboration of Spanish
groups, composed of High Energy physicists and astrophysicists, 
which called itself PAU (Physics of the Accelerated Universe), led by
Enrique Fern\'andez, a high energy physicist from IFAE, applied for
funding for several Dark Energy projects in which they were
involved. The proposal was received positively, but the referee
appointed by the Ministry required the groups to focus on a new, 
single large Dark Energy project instead of scattering the grant money
into different efforts.

  At that time the most obvious niches in the field of Dark Energy
  Surveys were already filled, with PanSTARRS and DES covering the
  broad imaging field and BOSS, the spectroscopic options, so it was
  not trivial to find a new observational alternative which was truly
  competitive. Narciso (Txitxo) Ben\'\i tez,  working at the Instituto de
  Matem\'aticas and F\'\i sica Fundamental (CSIC) in Madrid at the
  time, realized that it was feasible to use narrow band photometric
  redshifts to reach the precisions required to measure radial
  Baryonic Acoustic Oscillations $0.033(1+z)$, one of the most
  promising Dark Energy probes, and in early 2007 proposed the PAU
  collaboration carrying out a $8000\sq\degr$ survey using a set of $40-50$ narrow
  band filters. This was accepted as the basis of the new proposal,
  which was submitted to the Ministry, and received 5M\euro of
  funding. The paper \citet{B2009} describes the rationale behind 
  that proposal and still gives a pretty accurate description of much 
  of what J-PAS intends to do.  

   Almost simultaneously there was a proposal, led by Mariano Moles, of
  a new observatory at  Javalambre, Teruel, Spain, to take advantage of the superb
  astronomical characteristics of the site, which he had identified in
  the early 90's. Both proposals supported each other, PAU providing a
  competitive scientific case  and the OAJ the required 
  astronomical and technical infrastructure.

  In 2009, the PAU collaboration split in two. A part of the
  collaboration decided to use the bulk of the PAU funds to build a
  $<1\sq\degr$ camera for the William Herschel Telescope to carry out
  a $100-200\sq\degr$ survey  with a similar observational set-up as the one
  described in \citet{B2009}, but with an area $80$ times smaller and  
  different scientific goals \citep{gazta2012}. They retained the name
  of  PAU-Survey. The remainder of the collaboration, centered in Granada, Teruel and
 Valencia, decided to continue with the original survey idea and
 develop the project  from Javalambre, where a dedicated 2.5m
 telescope is being specifically build for this project. The project
 was open to the Brazilian Astronomical community as equal partners in
 2009 and founded again as J-PAS. Most of the funding for the
 construction of the JPcam has been provided by Brazilian grants 
led by Renato Dupke, Claudia Mendes de Oliveira y Laerte Sodr\'e. 
The first of the  J-PAS biannual meetings was held in Granada, in 
October of 2010. 

\newpage

\subsection{Organizational structure}

 The Figure shows the management structure that is being used by the Collaboration 
to carry out the J-PAS project and by the Collaboration Board to oversee the J-PAS Project, 
including their interfaces with OAJ facilities. Finally, it is being used to organize 
and coordinate the scientific work of the Collaboration.

\subsubsection{The Collaboration Board}  

 The Collaboration Board is composed of one member from each Institution providing 
financial resources involved in the Collaboration. It conducts periodic reviews of the J-PAS Projects, 
costs, goals and scientific. The current members of the Collaboration
Board are Narciso Ben\'\i tez (IAA-CSIC), Renato Dupke (ON), Mariano Moles (CEFCA) and Laerte Sodr\'e (IAG-USP). 

\subsection{The Survey Management Committee}

 The Survey Management Committee (SMC) is the body to organize,
 articulate and coordinate all the necessary activities to achieve the goals of the Collaboration. The SMC 
brings the skills and efforts of the Members and Participants into the Projects and 
assists the Project Managers in coordinating the contributions of the Collaboration and the Collaborating Institutions.

  The OAJ Project Manager and the Cameras Project Manager are responsible for the preparation of the 
documents on the Installation and Commissioning Plans for the T80Cam and JPCam and for the Operations and Maintenance Plan. 

  Apart from the members of the Collaboration Board, which are natural
 part of the SMC, the other members are  Javier Cenarro (CEFCA), Jordi
 Cepa (IAC), Alberto Fern\'andez-Soto (UV), Antonio Mar\'\i n (CEFCA),
 Claudia Mendes de Oliveira (IAG-USP) and Keith Taylor (ON). 
 
\subsubsection{Scientific Directors}

  The Scientific Directors coordinate the activities at the systems interfaces of the 
J-PAS Projects and the contributions of the Collaboration for the installation, commissioning and operation 
phases of the Survey; serve as the principal point of contact between 
the Survey Management Committee and the Collaboration Board and 
represent the Collaboration in interactions with the Collaboration 
Board and the Collaborating Institutions. The also are responsible for 
coordinating the scientific activities of the Science Working Groups
(SWGs) and appointing the SWG and Science Groups (SG) heads. 
 
 The Scientific Directors of J-PAS are Narciso Ben\'\i tez (IAA-CSIC) and Renato Dupke (ON). 

\subsubsection{Science Working Groups and Science Groups}

  The scientific activities of J-PAS are divided into broad Science
 Working Groups (SWG), namely Observational Cosmology, Theory, Galaxy
 Evolution, Resolved Stellar Populations, Transients and Solar
 System. The heads of the SWGs are responsible for assessing, 
assisting the SGs and evaluating the timetable to achieve 
the collaboration’s scientific immediate practical goals.  
The heads of these SWGs propose how to organize the research 
within each of these areas into smaller Science Groups, and nominate
 the SG heads. 

\begin{figure}
\includegraphics[width=0.8\textwidth,keepaspectratio]{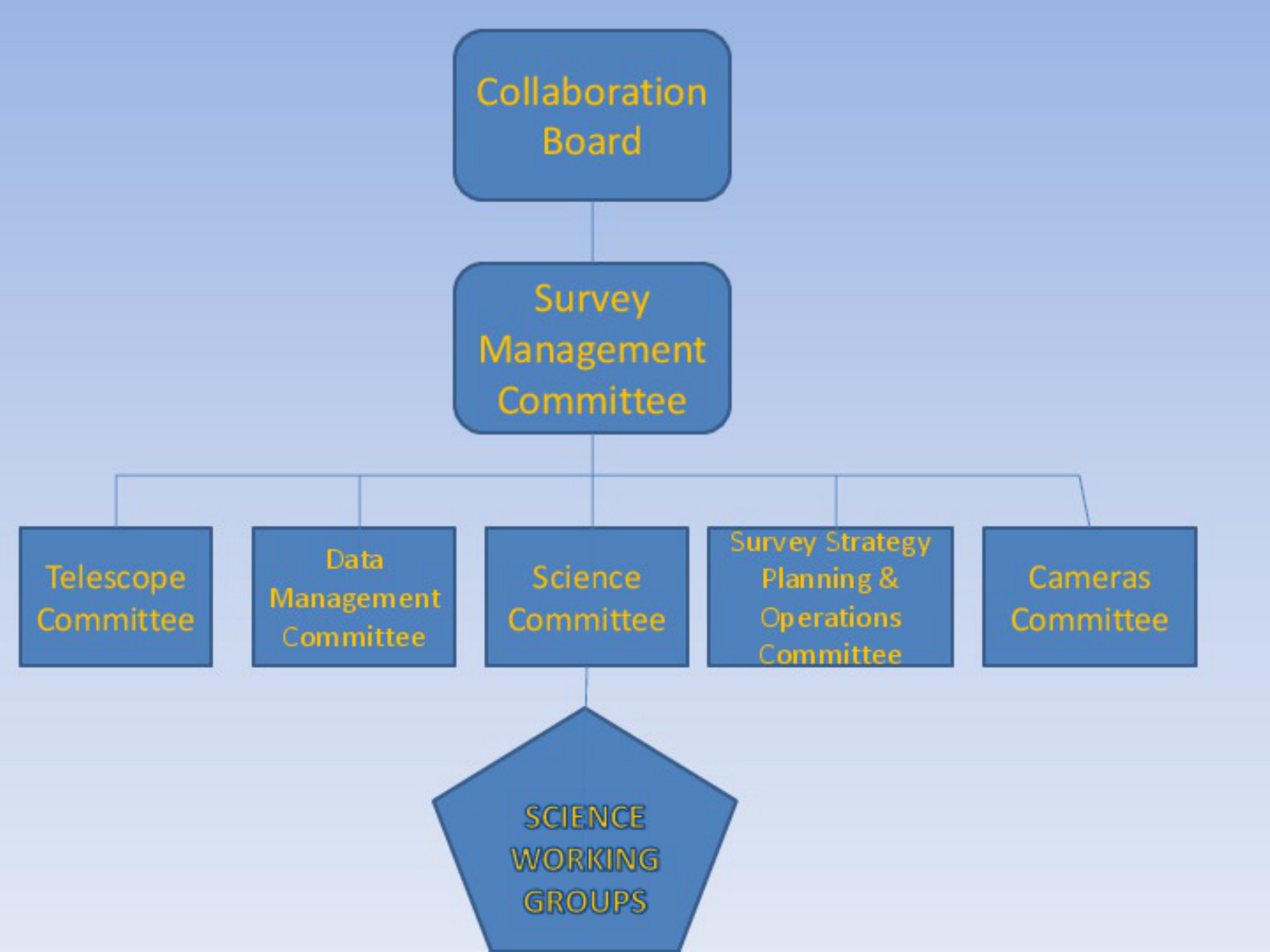}
\caption{}
\label{fig:MG}
\end{figure}

\begin{figure}
\includegraphics[width=0.8\textwidth,keepaspectratio]{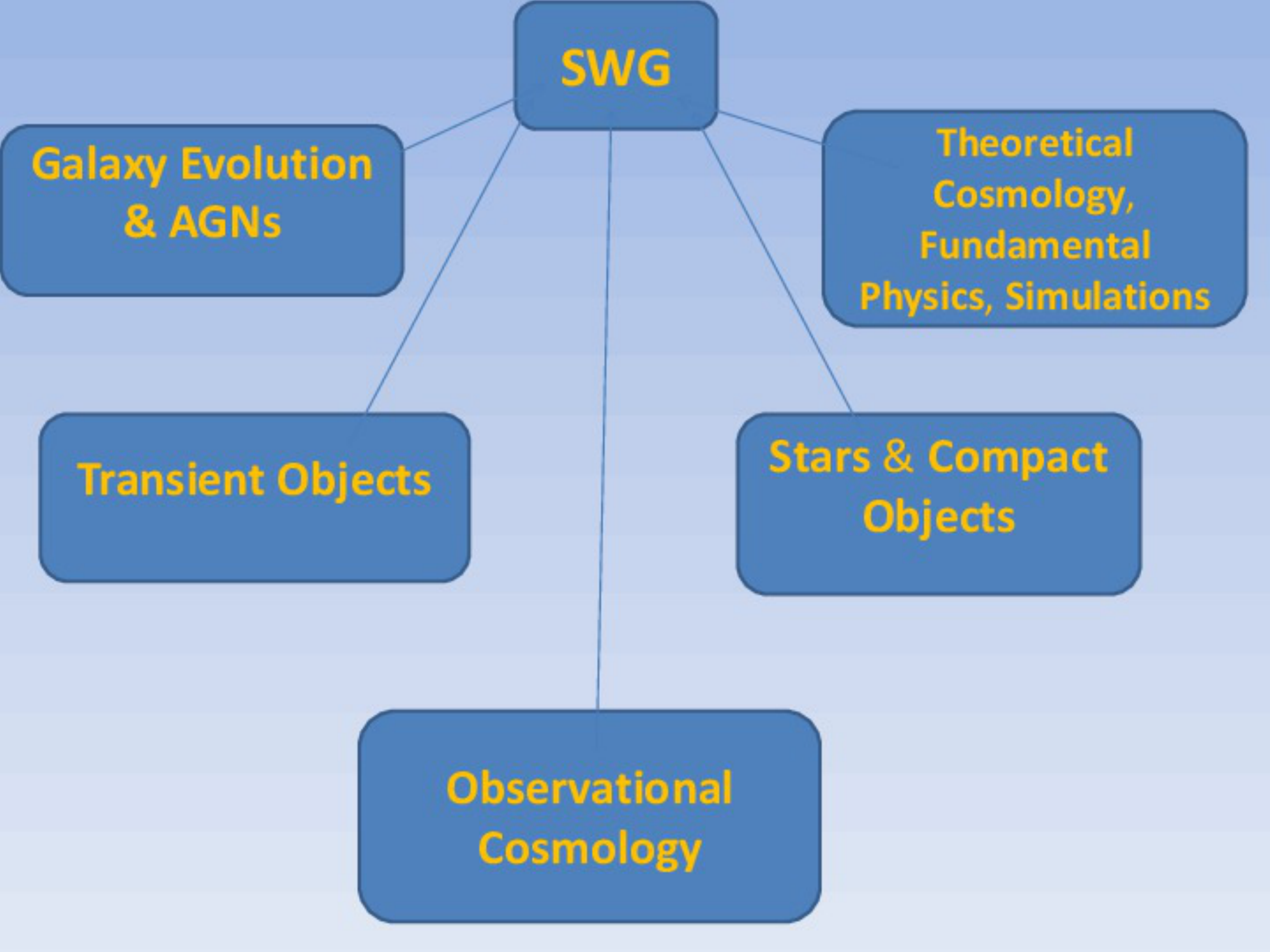}
\caption{}
\label{fig:SM}
\end{figure}

\subsection{Authorship Policy}

The J-PAS collaboration recognizes three types of papers:
\begin{itemize}
\item core papers -- Presentation paper and data releases. These are major which present a fundamental J-PAS aspect or dataset. The CB decides which papers are considered core papers. Any member has the right to sign a core paper. External Collaborators can sign the core papers if the coordinator of the corresponding SG can confirm their contribution
\item regular papers -- These papers are defined by the SWGs and SGs. Any member of the collaboration can propose a paper on a given subject and any other member of the collaboration can ask to be part of that publication.

SG heads in agreement with SWG heads decide who should lead the papers
and the co-authors list,always encouraging the collaboration between
the members; in case of doubt, or if a dispute arises regarding
authorship, the Scientific Directors and ultimately the CB has the
last word on issues of authorship. In cases of doubt, preferences will be give to junior collaborators.

Generally speaking, these papers can be signed only by the group that actually did the work (including technical contributions). In the case of papers led by graduate students, this rule should be applied even more stringently.

Builders have also the right to sign the 1st paper of each SG.

The SGs are responsible for determining the time at which a regular paper should be announced to the SG/WG and the rest of the collaboration. In principle, once first results are obtained, the paper should be announced at the SG/WG level. At first draft the collaboration should be aware and given the chance to comment and participate (at the minimum the collaboration shall have 2 weeks to review/comment)

\item  conference papers. Conference papers are defined as such by
  the SWGs and aim to publish J-PAS results in scientific
  meetings. Conference papers as well as seminars must be approved by
  the SG/SWGs; in case of doubt or dispute, the CB has the last word
  about who can speak on behalf of the collaboration. Conference papers can be signed by a few people + on behalf of J-PAS Collaboration.
\end{itemize}

 Papers should be circulated to the SGs/SWGs and CB at least two weeks
prior to submission or to making the paper public in any form. The
authorship order in a paper should give preference to who did the
work, including junior collaborators. 

  Members are not allowed to divulge or discuss either preliminary or final results, as well as any sensitive information relating to the core or parallel science of the project, without express and written agreement from the SGs/SWGs or the CB, if relevant.

\subsection{Data Policy}

  Rights to scientific images and data, including, but not limited to, images, databases, catalogs, and scientific works will be determined in accordance with internal applicable data rights policies.

\section{Acknowledgments}

The main agencies supporting J-PAS are the Gobierno de Arag\'on, the Spanish MINECO, the Ministry of Science Technology and Innovation (MCTI) of Brazil , The National Observatory/MCTI, FAPERJ, FAPESP and CNPq. 

\newpage
\bibliographystyle{elsarticle-harv}

\end{document}